# Population of Merging Compact Binaries Inferred Using Gravitational Waves through GWTC-3

R. Abbott *et al.*[*]
(LIGO Scientific Collaboration, Virgo Collaboration, and KAGRA Collaboration)

(Received 4 February 2022; revised 28 October 2022; accepted 19 December 2022; published 29 March 2023)

We report on the population properties of compact binary mergers inferred from gravitational-wave observations of these systems during the first three LIGO-Virgo observing runs. The Gravitational-Wave Transient Catalog 3 (GWTC-3) contains signals consistent with three classes of binary mergers: binary black hole, binary neutron star, and neutron star–black hole mergers. We infer the binary neutron star merger rate to be between 10 and 1700 $\mathrm{Gpc^{-3}\,yr^{-1}}$ and the neutron star–black hole merger rate to be between 7.8 and 140 $\mathrm{Gpc^{-3}\,yr^{-1}}$, assuming a constant rate density in the comoving frame and taking the union of 90% credible intervals for methods used in this work. We infer the binary black hole merger rate, allowing for evolution with redshift, to be between 17.9 and 44 $\mathrm{Gpc^{-3}\,yr^{-1}}$ at a fiducial redshift ($z = 0.2$). The rate of binary black hole mergers is observed to increase with redshift at a rate proportional to $(1+z)^\kappa$ with $\kappa = 2.9^{+1.7}_{-1.8}$ for $z \lesssim 1$. Using both binary neutron star and neutron star–black hole binaries, we obtain a broad, relatively flat neutron star mass distribution extending from $1.2^{+0.1}_{-0.2}$ to $2.0^{+0.3}_{-0.3} M_\odot$. We confidently determine that the merger rate as a function of mass sharply declines after the expected maximum neutron star mass, but cannot yet confirm or rule out the existence of a lower mass gap between neutron stars and black holes. We also find the binary black hole mass distribution has localized over- and underdensities relative to a power-law distribution, with peaks emerging at chirp masses of $8.3^{+0.3}_{-0.5}$ and $27.9^{+1.9}_{-1.8} M_\odot$. While we continue to find that the mass distribution of a binary's more massive component strongly decreases as a function of primary mass, we observe no evidence of a strongly suppressed merger rate above approximately $60 M_\odot$, which would indicate the presence of a upper mass gap. Observed black hole spins are small, with half of spin magnitudes below $\chi_i \approx 0.25$. While the majority of spins are preferentially aligned with the orbital angular momentum, we infer evidence of antialigned spins among the binary population. We observe an increase in spin magnitude for systems with more unequal-mass ratio. We also observe evidence of misalignment of spins relative to the orbital angular momentum.

## I. INTRODUCTION

The first three observing runs of the Advanced LIGO [1] and Advanced Virgo [2] gravitational-wave observatories were undertaken between September 2015 and March 2020. During that time, gravitational-wave (GW) signals from 90 mergers of binaries comprised of black holes (BHs) and neutron stars (NSs) were observed. The Gravitational-Wave Transient Catalog 3 (GWTC-3) [3] combines observations from the first three observing runs (O1, O2 [4], and O3 [3,5,6]). In this paper, we use those observations to infer the populations of NS and BH binaries in the Universe. To reduce contamination from events of nonastrophysical origin, we restrict our attention to 76 events which have a false alarm rate (FAR) of less than one per year, presented in Table I. Sixty-nine events are identified as binary black holes (BBHs), four events are neutron star–black holes (NSBHs), two events are binary neutron stars (BNSs), and one event, GW190814, is either a NSBH or BBH [7]. With this expanded catalog, we can start to probe the detailed characteristics of the populations, such as the distributions of component masses and spins, as well as investigate possible correlations between source properties.

The results presented here expand our understanding of the Universe relative to our previous merger census [20], which was based upon events observed in O1, O2, and the first half of O3 (O3a) as summarized in the Gravitational-Wave Transient Catalog 2 (GWTC-2) [5]. In addition to an increased total number of events, our census now contains an entirely new class of events. The first GW observation was a BBH merger [8], and the first BNS observation,

[*]Full author list given at the end of the article.

TABLE I. A table of GW events that meet the criteria for inclusion in this work. Events are separated by a horizontal line into sections of $\mathrm{FAR}_{\mathrm{min}} < 0.25\ \mathrm{yr}^{-1}$ and $\mathrm{FAR}_{\mathrm{min}} \geq 0.25\ \mathrm{yr}^{-1}$ (lower), where $\mathrm{FAR}_{\mathrm{min}}$ is the smallest FAR reported over all pipelines. Within these sections, events are listed by the date they were detected. Columns provide the FAR, $p_{\mathrm{astro}}$ (from the pipeline with the smallest FAR), and previously reported estimates of selected parameters; see Appendix E for discussion of how these estimates could be reassessed when adopting the results of this work. Events with at least one low-mass component are classified as BNSs or NSBHs following the analysis described in Sec. V. The low-significance event GW190531 is not included, lacking parameter inferences.

| Name | $\mathrm{FAR}_{\mathrm{min}}$ $(\mathrm{yr}^{-1})$ | $p_{\mathrm{astro}}$ | $m_1/M_\odot$ | $m_2/M_\odot$ | $\mathcal{M}/M_\odot$ | $\chi_1$ | $\chi_2$ | $\chi_{\mathrm{eff}}$ | $\cos(\theta_1)$ | $\cos(\theta_2)$ | Type | Ref. |
|---|---|---|---|---|---|---|---|---|---|---|---|---|
| GW150914 | $<1\times10^{-5}$ | >0.99 | $35.6^{+4.7}_{-3.1}$ | $30.6^{+3.0}_{-4.4}$ | $28.6^{+1.7}_{-1.5}$ | $0.3^{+0.6}_{-0.2}$ | $0.3^{+0.5}_{-0.3}$ | $-0.01^{+0.12}_{-0.13}$ | $0.03^{+0.74}_{-0.76}$ | $-0.13^{+0.81}_{-0.68}$ | BBH | [8] |
| GW151012 | $7.92\times10^{-3}$ | >0.99 | $23.2^{+14.9}_{-5.5}$ | $13.6^{+4.1}_{-4.8}$ | $15.2^{+2.1}_{-1.2}$ | $0.3^{+0.5}_{-0.3}$ | $0.4^{+0.5}_{-0.4}$ | $0.05^{+0.31}_{-0.20}$ | $0.23^{+0.62}_{-0.86}$ | $0.10^{+0.71}_{-0.82}$ | BBH | [9] |
| GW151226 | $<1\times10^{-5}$ | >0.99 | $13.7^{+8.8}_{-3.2}$ | $7.7^{+2.2}_{-2.5}$ | $8.9^{+0.3}_{-0.3}$ | $0.6^{+0.4}_{-0.4}$ | $0.5^{+0.4}_{-0.5}$ | $0.18^{+0.20}_{-0.12}$ | $0.47^{+0.40}_{-0.49}$ | $0.25^{+0.60}_{-0.92}$ | BBH | [10] |
| GW170104 | $<1\times10^{-5}$ | >0.99 | $30.8^{+7.3}_{-5.6}$ | $20.0^{+4.9}_{-4.6}$ | $21.4^{+2.2}_{-1.8}$ | $0.3^{+0.5}_{-0.3}$ | $0.4^{+0.5}_{-0.4}$ | $-0.04^{+0.17}_{-0.21}$ | $-0.12^{+0.78}_{-0.65}$ | $-0.12^{+0.82}_{-0.69}$ | BBH | [11] |
| GW170608 | $<1\times10^{-5}$ | >0.99 | $11.0^{+5.5}_{-1.7}$ | $7.6^{+1.4}_{-2.2}$ | $7.9^{+0.2}_{-0.2}$ | $0.3^{+0.5}_{-0.3}$ | $0.4^{+0.5}_{-0.4}$ | $0.03^{+0.19}_{-0.07}$ | $0.19^{+0.60}_{-0.68}$ | $0.15^{+0.65}_{-0.78}$ | BBH | [12] |
| GW170729 | $1.80\times10^{-1}$ | 0.98 | $50.2^{+16.2}_{-10.2}$ | $34.0^{+9.1}_{-10.1}$ | $35.4^{+6.5}_{-4.8}$ | $0.7^{+0.3}_{-0.5}$ | $0.5^{+0.4}_{-0.5}$ | $0.37^{+0.21}_{-0.25}$ | $0.73^{+0.23}_{-0.52}$ | $0.48^{+0.44}_{-0.95}$ | BBH | [4] |
| GW170809 | $<1\times10^{-5}$ | >0.99 | $35.0^{+8.3}_{-5.9}$ | $23.8^{+5.1}_{-5.2}$ | $24.9^{+2.1}_{-1.7}$ | $0.3^{+0.5}_{-0.3}$ | $0.4^{+0.5}_{-0.4}$ | $0.08^{+0.17}_{-0.17}$ | $0.26^{+0.57}_{-0.81}$ | $0.17^{+0.65}_{-0.86}$ | BBH | [4] |
| GW170814 | $<1\times10^{-5}$ | >0.99 | $30.6^{+5.6}_{-3.0}$ | $25.2^{+2.8}_{-4.0}$ | $24.1^{+1.4}_{-1.1}$ | $0.4^{+0.5}_{-0.4}$ | $0.4^{+0.5}_{-0.4}$ | $0.07^{+0.12}_{-0.12}$ | $0.23^{+0.56}_{-0.68}$ | $0.14^{+0.64}_{-0.78}$ | BBH | [13] |
| GW170817 | $<1\times10^{-5}$ | >0.99 | $1.46^{+0.12}_{-0.10}$ | $1.27^{+0.09}_{-0.09}$ | $1.186^{+0.001}_{-0.001}$ | $0.02^{+0.02}_{-0.02}$ | $0.02^{+0.02}_{-0.02}$ | $0.00^{+0.02}_{-0.01}$ | $0.22^{+0.62}_{-0.79}$ | $0.13^{+0.68}_{-0.80}$ | BNS | [14] |
| GW170818 | $<1\times10^{-5}$ | >0.99 | $35.4^{+7.5}_{-4.7}$ | $26.7^{+4.3}_{-5.2}$ | $26.5^{+2.1}_{-1.7}$ | $0.5^{+0.5}_{-0.4}$ | $0.5^{+0.5}_{-0.4}$ | $-0.09^{+0.18}_{-0.21}$ | $-0.20^{+0.74}_{-0.53}$ | $-0.23^{+0.79}_{-0.58}$ | BBH | [4] |
| GW170823 | $<1\times10^{-5}$ | >0.99 | $39.5^{+11.2}_{-6.7}$ | $29.0^{+6.7}_{-7.8}$ | $29.2^{+4.6}_{-3.6}$ | $0.4^{+0.5}_{-0.4}$ | $0.4^{+0.5}_{-0.4}$ | $0.09^{+0.22}_{-0.26}$ | $0.26^{+0.58}_{-0.86}$ | $0.11^{+0.71}_{-0.84}$ | BBH | [4] |
| GW190408_181802 | $<1\times10^{-5}$ | >0.99 | $24.6^{+5.1}_{-3.4}$ | $18.4^{+3.3}_{-3.6}$ | $18.3^{+1.9}_{-1.2}$ | $0.34^{+0.47}_{-0.31}$ | $0.36^{+0.53}_{-0.32}$ | $-0.03^{+0.14}_{-0.19}$ | $-0.17^{+0.94}_{-0.73}$ | $-0.02^{+0.88}_{-0.85}$ | BBH | [5] |
| GW190412_053044 | $<1\times10^{-5}$ | >0.99 | $30.1^{+4.7}_{-5.1}$ | $8.3^{+1.6}_{-0.9}$ | $13.3^{+0.4}_{-0.3}$ | $0.44^{+0.16}_{-0.22}$ | $0.49^{+0.44}_{-0.43}$ | $0.25^{+0.08}_{-0.11}$ | $0.70^{+0.20}_{-0.47}$ | $0.25^{+0.67}_{-1.01}$ | BBH | [15] |
| GW190413_134308 | $1.81\times10^{-1}$ | 0.99 | $47.5^{+13.5}_{-10.7}$ | $31.8^{+11.7}_{-10.8}$ | $33.0^{+8.2}_{-5.4}$ | $0.58^{+0.38}_{-0.52}$ | $0.50^{+0.44}_{-0.45}$ | $-0.03^{+0.25}_{-0.29}$ | $-0.06^{+0.82}_{-0.76}$ | $-0.13^{+1.00}_{-0.77}$ | BBH | [5] |
| GW190421_213856 | $2.83\times10^{-3}$ | >0.99 | $41.3^{+10.4}_{-6.9}$ | $31.9^{+8.0}_{-8.8}$ | $31.2^{+5.9}_{-4.2}$ | $0.46^{+0.47}_{-0.41}$ | $0.46^{+0.47}_{-0.42}$ | $-0.06^{+0.22}_{-0.27}$ | $-0.15^{+0.94}_{-0.75}$ | $-0.16^{+0.96}_{-0.76}$ | BBH | [5] |
| GW190425_081805 | $3.38\times10^{-2}$ | 0.78 | $2.0^{+0.6}_{-0.3}$ | $1.4^{+0.3}_{-0.3}$ | $1.44^{+0.02}_{-0.02}$ | $0.27^{+0.51}_{-0.25}$ | $0.28^{+0.51}_{-0.25}$ | $0.06^{+0.11}_{-0.05}$ | $0.26^{+0.61}_{-0.65}$ | $0.16^{+0.70}_{-0.87}$ | BNS | [16] |
| GW190503_185404 | $<1\times10^{-5}$ | >0.99 | $43.3^{+9.2}_{-8.1}$ | $28.4^{+7.7}_{-8.0}$ | $30.2^{+4.2}_{-4.2}$ | $0.34^{+0.51}_{-0.31}$ | $0.44^{+0.48}_{-0.40}$ | $-0.03^{+0.20}_{-0.26}$ | $-0.16^{+0.96}_{-0.75}$ | $-0.02^{+0.88}_{-0.87}$ | BBH | [5] |
| GW190512_180714 | $<1\times10^{-5}$ | >0.99 | $23.3^{+5.3}_{-5.8}$ | $12.6^{+3.6}_{-2.5}$ | $14.6^{+1.3}_{-1.0}$ | $0.17^{+0.44}_{-0.16}$ | $0.36^{+0.51}_{-0.32}$ | $0.03^{+0.12}_{-0.13}$ | $0.07^{+0.82}_{-0.93}$ | $0.16^{+0.75}_{-0.98}$ | BBH | [5] |
| GW190513_205428 | $<1\times10^{-5}$ | >0.99 | $35.7^{+9.5}_{-9.2}$ | $18.0^{+7.7}_{-4.1}$ | $21.6^{+3.8}_{-1.9}$ | $0.30^{+0.51}_{-0.28}$ | $0.43^{+0.48}_{-0.39}$ | $0.11^{+0.28}_{-0.17}$ | $0.41^{+0.53}_{-1.10}$ | $0.25^{+0.68}_{-1.05}$ | BBH | [5] |
| GW190517_055101 | $3.47\times10^{-4}$ | >0.99 | $37.4^{+11.7}_{-7.6}$ | $25.3^{+7.0}_{-7.3}$ | $26.6^{+4.0}_{-4.0}$ | $0.86^{+0.13}_{-0.35}$ | $0.58^{+0.38}_{-0.52}$ | $0.52^{+0.19}_{-0.19}$ | $0.83^{+0.16}_{-0.37}$ | $0.61^{+0.36}_{-1.21}$ | BBH | [5] |
| GW190519_153544 | $<1\times10^{-5}$ | >0.99 | $66.0^{+10.7}_{-12.0}$ | $40.5^{+11.0}_{-11.1}$ | $44.5^{+6.4}_{-7.1}$ | $0.60^{+0.33}_{-0.50}$ | $0.54^{+0.41}_{-0.48}$ | $0.31^{+0.20}_{-0.22}$ | $0.65^{+0.32}_{-0.74}$ | $0.50^{+0.46}_{-1.17}$ | BBH | [5] |
| GW190521_030229 | $<1\times10^{-5}$ | >0.99 | $95.3^{+28.7}_{-18.9}$ | $69.0^{+22.7}_{-23.1}$ | $69.2^{+17.6}_{-10.6}$ | $0.73^{+0.25}_{-0.63}$ | $0.58^{+0.38}_{-0.52}$ | $0.03^{+0.32}_{-0.39}$ | $0.05^{+0.78}_{-0.89}$ | $-0.02^{+0.86}_{-0.86}$ | BBH | [17] |
| GW190521_074359 | $1.00\times10^{-2}$ | >0.99 | $42.2^{+5.9}_{-4.8}$ | $32.8^{+5.4}_{-6.4}$ | $32.1^{+3.2}_{-2.5}$ | $0.31^{+0.42}_{-0.28}$ | $0.42^{+0.39}_{-0.37}$ | $0.09^{+0.10}_{-0.13}$ | $0.17^{+0.71}_{-0.97}$ | $0.29^{+0.62}_{-1.03}$ | BBH | [17] |
| GW190527_092055 | $2.28\times10^{-1}$ | 0.85 | $36.5^{+16.4}_{-9.0}$ | $22.6^{+10.5}_{-8.1}$ | $24.3^{+9.1}_{-4.2}$ | $0.47^{+0.47}_{-0.43}$ | $0.50^{+0.44}_{-0.45}$ | $0.11^{+0.28}_{-0.28}$ | $0.28^{+0.65}_{-1.02}$ | $0.16^{+0.76}_{-1.02}$ | BBH | [5] |
| GW190602_175927 | $<1\times10^{-5}$ | >0.99 | $69.1^{+15.7}_{-13.0}$ | $47.8^{+14.3}_{-17.4}$ | $49.1^{+9.1}_{-8.5}$ | $0.38^{+0.51}_{-0.34}$ | $0.50^{+0.44}_{-0.45}$ | $0.07^{+0.25}_{-0.24}$ | $0.18^{+0.72}_{-0.99}$ | $0.19^{+0.73}_{-1.01}$ | BBH | [5] |
| GW190620_030421 | $1.12\times10^{-2}$ | 0.99 | $57.1^{+16.0}_{-12.7}$ | $35.5^{+12.2}_{-12.3}$ | $38.3^{+8.3}_{-6.5}$ | $0.61^{+0.34}_{-0.50}$ | $0.56^{+0.40}_{-0.50}$ | $0.33^{+0.22}_{-0.25}$ | $0.68^{+0.29}_{-0.84}$ | $0.50^{+0.47}_{-1.22}$ | BBH | [5] |
| GW190630_185205 | $<1\times10^{-5}$ | >0.99 | $35.1^{+6.9}_{-5.6}$ | $23.7^{+5.2}_{-5.1}$ | $24.9^{+2.1}_{-2.1}$ | $0.26^{+0.37}_{-0.23}$ | $0.38^{+0.46}_{-0.34}$ | $0.10^{+0.12}_{-0.13}$ | $0.29^{+0.63}_{-1.02}$ | $0.33^{+0.60}_{-1.08}$ | BBH | [5] |
| GW190701_203306 | $5.71\times10^{-3}$ | 0.99 | $53.9^{+11.8}_{-8.0}$ | $40.8^{+8.7}_{-12.0}$ | $40.3^{+5.4}_{-4.9}$ | $0.40^{+0.50}_{-0.36}$ | $0.44^{+0.48}_{-0.40}$ | $-0.07^{+0.23}_{-0.29}$ | $-0.22^{+1.01}_{-0.70}$ | $-0.17^{+1.00}_{-0.74}$ | BBH | [5] |
| GW190706_222641 | $<1\times10^{-5}$ | >0.99 | $67.0^{+14.6}_{-16.2}$ | $38.2^{+14.6}_{-13.3}$ | $42.7^{+10.0}_{-7.0}$ | $0.55^{+0.39}_{-0.48}$ | $0.49^{+0.45}_{-0.44}$ | $0.28^{+0.26}_{-0.29}$ | $0.66^{+0.31}_{-0.39}$ | $0.37^{+0.58}_{-1.16}$ | BBH | [5] |
| GW190707_093326 | $<1\times10^{-5}$ | >0.99 | $11.6^{+3.3}_{-1.7}$ | $8.4^{+1.4}_{-1.7}$ | $8.5^{+0.6}_{-0.5}$ | $0.24^{+0.47}_{-0.21}$ | $0.31^{+0.52}_{-0.28}$ | $-0.05^{+0.10}_{-0.08}$ | $-0.19^{+0.97}_{-0.68}$ | $-0.19^{+1.02}_{-0.71}$ | BBH | [5] |
| GW190708_232457 | $3.09\times10^{-4}$ | >0.99 | $17.6^{+4.7}_{-2.3}$ | $13.2^{+2.0}_{-2.7}$ | $13.2^{+0.9}_{-0.6}$ | $0.22^{+0.52}_{-0.20}$ | $0.30^{+0.50}_{-0.28}$ | $0.02^{+0.10}_{-0.08}$ | $0.07^{+0.80}_{-0.88}$ | $0.14^{+0.77}_{-0.97}$ | BBH | [5] |
| GW190720_000836 | $<1\times10^{-5}$ | >0.99 | $13.4^{+6.4}_{-3.0}$ | $7.8^{+2.3}_{-2.2}$ | $8.9^{+0.5}_{-0.8}$ | $0.40^{+0.40}_{-0.35}$ | $0.51^{+0.43}_{-0.45}$ | $0.18^{+0.14}_{-0.12}$ | $0.54^{+0.42}_{-0.97}$ | $0.31^{+0.62}_{-1.11}$ | BBH | [5] |
| GW190727_060333 | $<1\times10^{-5}$ | >0.99 | $38.0^{+9.5}_{-6.2}$ | $29.4^{+7.1}_{-8.4}$ | $28.6^{+5.3}_{-3.7}$ | $0.46^{+0.47}_{-0.42}$ | $0.45^{+0.46}_{-0.41}$ | $0.11^{+0.26}_{-0.25}$ | $0.30^{+0.61}_{-1.02}$ | $0.20^{+0.72}_{-1.04}$ | BBH | [5] |
| GW190728_064510 | $<1\times10^{-5}$ | >0.99 | $12.3^{+7.2}_{-2.2}$ | $8.1^{+1.7}_{-2.6}$ | $8.6^{+0.5}_{-0.3}$ | $0.32^{+0.37}_{-0.28}$ | $0.39^{+0.50}_{-0.35}$ | $0.12^{+0.20}_{-0.07}$ | $0.49^{+0.47}_{-1.13}$ | $0.30^{+0.56}_{-1.16}$ | BBH | [5] |
| GW190803_022701 | $7.32\times10^{-2}$ | 0.94 | $37.3^{+10.6}_{-7.0}$ | $27.3^{+7.8}_{-8.2}$ | $27.3^{+5.7}_{-4.1}$ | $0.41^{+0.51}_{-0.37}$ | $0.45^{+0.49}_{-0.40}$ | $-0.03^{+0.24}_{-0.27}$ | $-0.07^{+0.91}_{-0.80}$ | $-0.06^{+0.93}_{-0.83}$ | BBH | [5] |
| GW190814_211039 | $<1\times10^{-5}$ | >0.99 | $23.2^{+1.1}_{-1.0}$ | $2.59^{+0.08}_{-0.09}$ | $6.09^{+0.06}_{-0.06}$ | $0.03^{+0.05}_{-0.03}$ | $0.52^{+0.41}_{-0.46}$ | $0.00^{+0.06}_{-0.06}$ | $0.01^{+0.87}_{-0.90}$ | $-0.03^{+0.87}_{-0.83}$ | BBH | [7] |
| GW190828_063405 | $<1\times10^{-5}$ | >0.99 | $32.1^{+5.8}_{-4.0}$ | $26.2^{+4.6}_{-4.8}$ | $25.0^{+3.4}_{-2.1}$ | $0.44^{+0.45}_{-0.40}$ | $0.41^{+0.46}_{-0.37}$ | $0.19^{+0.15}_{-0.16}$ | $0.51^{+0.44}_{-1.09}$ | $0.37^{+0.57}_{-1.14}$ | BBH | [5] |
| GW190828_065509 | $<1\times10^{-5}$ | >0.99 | $24.1^{+7.0}_{-7.2}$ | $10.2^{+3.6}_{-2.1}$ | $13.3^{+1.2}_{-1.0}$ | $0.28^{+0.43}_{-0.26}$ | $0.42^{+0.49}_{-0.38}$ | $0.08^{+0.16}_{-0.16}$ | $0.26^{+0.63}_{-0.99}$ | $0.24^{+0.69}_{-1.06}$ | BBH | [5] |
| GW190910_112807 | $2.87\times10^{-3}$ | >0.99 | $43.9^{+7.6}_{-6.1}$ | $35.6^{+6.3}_{-7.2}$ | $34.3^{+4.1}_{-4.1}$ | $0.34^{+0.50}_{-0.30}$ | $0.37^{+0.51}_{-0.33}$ | $0.02^{+0.18}_{-0.18}$ | $0.08^{+0.79}_{-0.93}$ | $0.04^{+0.83}_{-0.88}$ | BBH | [5] |

*(Table continued)*

TABLE I. *(Continued)*

| Name | $\mathrm{FAR}_{\min}$ $(\mathrm{yr}^{-1})$ | $p_{\mathrm{astro}}$ | $m_1/M_\odot$ | $m_2/M_\odot$ | $\mathcal{M}/M_\odot$ | $\chi_1$ | $\chi_2$ | $\chi_{\mathrm{eff}}$ | $\cos(\theta_1)$ | $\cos(\theta_2)$ | Type | Ref. |
|---|---|---|---|---|---|---|---|---|---|---|---|---|
| GW190915_235702 | $<1\times10^{-5}$ | >0.99 | $35.3^{+9.5}_{-6.4}$ | $24.4^{+5.6}_{-6.1}$ | $25.3^{+3.2}_{-2.7}$ | $0.55^{+0.39}_{-0.49}$ | $0.48^{+0.46}_{-0.43}$ | $0.02^{+0.20}_{-0.25}$ | $0.06^{+0.73}_{-0.82}$ | $0.02^{+0.85}_{-0.90}$ | BBH | [5] |
| GW190924_021846 | $<1\times10^{-5}$ | >0.99 | $8.9^{+7.0}_{-2.0}$ | $5.0^{+1.4}_{-1.9}$ | $5.8^{+0.2}_{-0.2}$ | $0.24^{+0.43}_{-0.21}$ | $0.35^{+0.51}_{-0.32}$ | $0.03^{+0.30}_{-0.09}$ | $0.19^{+0.76}_{-0.97}$ | $0.15^{+0.77}_{-1.00}$ | BBH | [5] |
| GW190925_232845 | $7.20\times10^{-3}$ | 0.99 | $21.2^{+6.9}_{-3.1}$ | $15.6^{+2.6}_{-3.6}$ | $15.8^{+1.1}_{-1.0}$ | $0.36^{+0.51}_{-0.32}$ | $0.43^{+0.49}_{-0.39}$ | $0.11^{+0.17}_{-0.14}$ | $0.33^{+0.59}_{-1.02}$ | $0.27^{+0.66}_{-1.05}$ | BBH | [6] |
| GW190929_012149 | $1.55\times10^{-1}$ | 0.87 | $80.8^{+33.0}_{-33.2}$ | $24.1^{+19.3}_{-10.6}$ | $35.8^{+14.9}_{-8.2}$ | $0.64^{+0.32}_{-0.54}$ | $0.49^{+0.45}_{-0.44}$ | $0.01^{+0.34}_{-0.33}$ | $0.02^{+0.65}_{-0.83}$ | $0.04^{+0.86}_{-0.93}$ | BBH | [5] |
| GW190930_133541 | $1.23\times10^{-2}$ | >0.99 | $12.3^{+12.4}_{-2.3}$ | $7.8^{+1.7}_{-3.3}$ | $8.5^{+0.5}_{-0.5}$ | $0.39^{+0.40}_{-0.35}$ | $0.42^{+0.49}_{-0.37}$ | $0.14^{+0.31}_{-0.15}$ | $0.47^{+0.49}_{-1.08}$ | $0.31^{+0.63}_{-1.09}$ | BBH | [5] |
| GW191105_143521 | $1.18\times10^{-2}$ | >0.99 | $10.7^{+3.7}_{-1.6}$ | $7.7^{+1.4}_{-1.9}$ | $7.82^{+0.61}_{-0.45}$ | $0.23^{+0.53}_{-0.21}$ | $0.34^{+0.54}_{-0.31}$ | $-0.02^{+0.13}_{-0.09}$ | $-0.07^{+0.90}_{-0.78}$ | $-0.03^{+0.89}_{-0.83}$ | BBH | [3] |
| GW191109_010717 | $1.80\times10^{-4}$ | >0.99 | $65^{+11}_{-11}$ | $47^{+15}_{-13}$ | $47.5^{+9.6}_{-7.5}$ | $0.83^{+0.15}_{-0.58}$ | $0.65^{+0.32}_{-0.58}$ | $-0.29^{+0.42}_{-0.31}$ | $-0.61^{+0.85}_{-0.36}$ | $-0.26^{+1.04}_{-0.66}$ | BBH | [3] |
| GW191127_050227 | $2.49\times10^{-1}$ | 0.49 | $53^{+47}_{-20}$ | $24^{+17}_{-14}$ | $29.9^{+11.7}_{-9.1}$ | $0.66^{+0.31}_{-0.58}$ | $0.54^{+0.41}_{-0.49}$ | $0.18^{+0.34}_{-0.36}$ | $0.37^{+0.56}_{-1.00}$ | $0.22^{+0.71}_{-1.06}$ | BBH | [3] |
| GW191129_134029 | $<1\times10^{-5}$ | >0.99 | $10.7^{+4.1}_{-2.1}$ | $6.7^{+1.5}_{-1.7}$ | $7.31^{+0.43}_{-0.28}$ | $0.25^{+0.37}_{-0.22}$ | $0.34^{+0.52}_{-0.31}$ | $0.06^{+0.16}_{-0.08}$ | $0.27^{+0.64}_{-0.95}$ | $0.30^{+0.63}_{-1.07}$ | BBH | [3] |
| GW191204_171526 | $<1\times10^{-5}$ | >0.99 | $27.3^{+11.0}_{-6.0}$ | $19.3^{+5.6}_{-6.0}$ | $19.8^{+3.6}_{-3.3}$ | $0.52^{+0.43}_{-0.47}$ | $0.49^{+0.46}_{-0.44}$ | $0.05^{+0.26}_{-0.27}$ | $0.17^{+0.70}_{-0.97}$ | $-0.01^{+0.87}_{-0.88}$ | BBH | [3] |
| GW191215_223052 | $<1\times10^{-5}$ | >0.99 | $24.9^{+7.1}_{-4.1}$ | $18.1^{+3.8}_{-4.1}$ | $18.4^{+2.2}_{-1.7}$ | $0.47^{+0.44}_{-0.43}$ | $0.44^{+0.49}_{-0.40}$ | $-0.04^{+0.17}_{-0.21}$ | $-0.09^{+0.87}_{-0.69}$ | $-0.11^{+0.94}_{-0.78}$ | BBH | [3] |
| GW191216_213338 | $<1\times10^{-5}$ | >0.99 | $12.1^{+4.6}_{-2.3}$ | $7.7^{+1.6}_{-1.9}$ | $8.33^{+0.22}_{-0.19}$ | $0.24^{+0.36}_{-0.21}$ | $0.36^{+0.50}_{-0.32}$ | $0.11^{+0.13}_{-0.06}$ | $0.53^{+0.43}_{-1.13}$ | $0.40^{+0.55}_{-1.16}$ | BBH | [3] |
| GW191222_033537 | $<1\times10^{-5}$ | >0.99 | $45.1^{+10.9}_{-8.0}$ | $34.7^{+9.3}_{-10.5}$ | $33.8^{+7.1}_{-5.0}$ | $0.38^{+0.50}_{-0.34}$ | $0.41^{+0.50}_{-0.38}$ | $-0.04^{+0.20}_{-0.25}$ | $-0.16^{+0.97}_{-0.73}$ | $-0.09^{+0.95}_{-0.81}$ | BBH | [3] |
| GW191230_180458 | $5.02\times10^{-2}$ | 0.95 | $49.4^{+14.0}_{-9.6}$ | $37^{+11}_{-12}$ | $36.5^{+8.2}_{-5.6}$ | $0.51^{+0.44}_{-0.46}$ | $0.49^{+0.45}_{-0.44}$ | $-0.05^{+0.26}_{-0.31}$ | $-0.10^{+0.92}_{-0.78}$ | $-0.16^{+0.99}_{-0.76}$ | BBH | [3] |
| GW200105_162426 | $2.04\times10^{-1}$ | 0.36 | $8.9^{+1.2}_{-1.5}$ | $1.9^{+0.3}_{-0.2}$ | $3.41^{+0.08}_{-0.07}$ | $0.08^{+0.22}_{-0.08}$ | $0.32^{+0.55}_{-0.29}$ | $-0.01^{+0.11}_{-0.15}$ | $-0.09^{+0.96}_{-0.83}$ | $-0.03^{+0.82}_{-0.88}$ | NSBH | [18] |
| GW200112_155838 | $<1\times10^{-5}$ | >0.99 | $35.6^{+6.7}_{-4.5}$ | $28.3^{+4.4}_{-5.9}$ | $27.4^{+2.6}_{-2.1}$ | $0.34^{+0.46}_{-0.31}$ | $0.39^{+0.50}_{-0.34}$ | $0.06^{+0.15}_{-0.15}$ | $0.14^{+0.75}_{-0.91}$ | $0.25^{+0.67}_{-1.02}$ | BBH | [3] |
| GW200115_042309 | $<1\times10^{-5}$ | >0.99 | $5.9^{+2.0}_{-2.5}$ | $1.44^{+0.85}_{-0.29}$ | $2.43^{+0.05}_{-0.07}$ | $0.32^{+0.50}_{-0.29}$ | $0.44^{+0.48}_{-0.39}$ | $-0.15^{+0.24}_{-0.42}$ | $-0.61^{+1.26}_{-0.35}$ | $-0.32^{+1.13}_{-0.63}$ | NSBH | [18] |
| GW200128_022011 | $4.29\times10^{-3}$ | >0.99 | $42.2^{+11.6}_{-8.1}$ | $32.6^{+9.5}_{-9.2}$ | $32.0^{+7.5}_{-5.5}$ | $0.58^{+0.38}_{-0.51}$ | $0.50^{+0.44}_{-0.45}$ | $0.12^{+0.24}_{-0.25}$ | $0.31^{+0.60}_{-0.95}$ | $0.12^{+0.78}_{-0.96}$ | BBH | [3] |
| GW200129_065458 | $<1\times10^{-5}$ | >0.99 | $34.5^{+9.9}_{-3.2}$ | $28.9^{+3.4}_{-9.3}$ | $27.2^{+2.1}_{-2.3}$ | $0.53^{+0.42}_{-0.47}$ | $0.49^{+0.44}_{-0.42}$ | $0.11^{+0.11}_{-0.16}$ | $0.13^{+0.74}_{-0.92}$ | $0.41^{+0.54}_{-1.21}$ | BBH | [3] |
| GW200202_154313 | $<1\times10^{-5}$ | >0.99 | $10.1^{+3.5}_{-1.4}$ | $7.3^{+1.1}_{-1.7}$ | $7.49^{+0.24}_{-0.20}$ | $0.22^{+0.45}_{-0.20}$ | $0.33^{+0.53}_{-0.30}$ | $0.04^{+0.13}_{-0.06}$ | $0.18^{+0.72}_{-0.95}$ | $0.22^{+0.68}_{-0.97}$ | BBH | [3] |
| GW200208_130117 | $3.11\times10^{-4}$ | >0.99 | $37.8^{+9.2}_{-6.2}$ | $27.4^{+6.1}_{-7.4}$ | $27.7^{+3.6}_{-3.1}$ | $0.36^{+0.51}_{-0.32}$ | $0.43^{+0.49}_{-0.39}$ | $-0.07^{+0.22}_{-0.27}$ | $-0.24^{+1.03}_{-0.68}$ | $-0.15^{+1.00}_{-0.76}$ | BBH | [3] |
| GW200209_085452 | $4.64\times10^{-2}$ | 0.95 | $35.6^{+10.5}_{-6.8}$ | $27.1^{+7.8}_{-7.8}$ | $26.7^{+6.0}_{-4.2}$ | $0.52^{+0.44}_{-0.46}$ | $0.49^{+0.45}_{-0.44}$ | $-0.12^{+0.24}_{-0.30}$ | $-0.24^{+0.97}_{-0.67}$ | $-0.30^{+1.08}_{-0.63}$ | BBH | [3] |
| GW200219_094415 | $9.94\times10^{-4}$ | >0.99 | $37.5^{+10.1}_{-6.9}$ | $27.9^{+7.4}_{-8.4}$ | $27.6^{+5.6}_{-3.8}$ | $0.47^{+0.46}_{-0.43}$ | $0.48^{+0.46}_{-0.43}$ | $-0.08^{+0.23}_{-0.29}$ | $-0.23^{+0.96}_{-0.69}$ | $-0.17^{+0.99}_{-0.75}$ | BBH | [3] |
| GW200224_222234 | $<1\times10^{-5}$ | >0.99 | $40.0^{+6.9}_{-4.5}$ | $32.5^{+5.0}_{-7.2}$ | $31.1^{+3.2}_{-2.6}$ | $0.46^{+0.45}_{-0.41}$ | $0.43^{+0.48}_{-0.39}$ | $0.10^{+0.15}_{-0.15}$ | $0.28^{+0.62}_{-0.96}$ | $0.18^{+0.71}_{-1.02}$ | BBH | [3] |
| GW200225_060421 | $<1\times10^{-5}$ | >0.99 | $19.3^{+5.0}_{-3.0}$ | $14.0^{+2.8}_{-3.5}$ | $14.2^{+1.5}_{-1.4}$ | $0.59^{+0.35}_{-0.51}$ | $0.42^{+0.49}_{-0.39}$ | $-0.12^{+0.17}_{-0.28}$ | $-0.30^{+0.86}_{-0.59}$ | $-0.22^{+1.04}_{-0.71}$ | BBH | [3] |
| GW200302_015811 | $1.12\times10^{-1}$ | 0.91 | $37.8^{+8.7}_{-8.5}$ | $20.0^{+8.1}_{-5.7}$ | $23.4^{+4.7}_{-3.0}$ | $0.37^{+0.51}_{-0.34}$ | $0.45^{+0.48}_{-0.40}$ | $0.01^{+0.25}_{-0.26}$ | $-0.01^{+0.85}_{-0.85}$ | $0.11^{+0.79}_{-0.97}$ | BBH | [3] |
| GW200311_115853 | $<1\times10^{-5}$ | >0.99 | $34.2^{+6.4}_{-3.8}$ | $27.7^{+4.1}_{-5.9}$ | $26.6^{+2.4}_{-2.0}$ | $0.39^{+0.48}_{-0.36}$ | $0.41^{+0.51}_{-0.37}$ | $-0.02^{+0.16}_{-0.20}$ | $-0.09^{+0.89}_{-0.78}$ | $-0.02^{+0.87}_{-0.84}$ | BBH | [3] |
| GW200316_215756 | $<1\times10^{-5}$ | >0.99 | $13.1^{+10.2}_{-2.9}$ | $7.8^{+1.9}_{-2.9}$ | $8.75^{+0.62}_{-0.55}$ | $0.32^{+0.37}_{-0.28}$ | $0.44^{+0.47}_{-0.39}$ | $0.13^{+0.27}_{-0.10}$ | $0.47^{+0.49}_{-1.02}$ | $0.37^{+0.57}_{-1.05}$ | BBH | [3] |
| GW190413_052954 | $8.17\times10^{-1}$ | 0.93 | $34.7^{+12.6}_{-8.1}$ | $23.7^{+7.3}_{-6.7}$ | $24.6^{+5.5}_{-4.1}$ | $0.40^{+0.51}_{-0.36}$ | $0.45^{+0.48}_{-0.41}$ | $-0.01^{+0.29}_{-0.34}$ | $0.01^{+0.87}_{-0.90}$ | $-0.08^{+0.95}_{-0.82}$ | BBH | [5] |
| GW190426_152155 | $9.12\times10^{-1}$ | 0.14 | $5.7^{+3.9}_{-2.3}$ | $1.5^{+0.8}_{-0.5}$ | $2.41^{+0.08}_{-0.08}$ | $0.14^{+0.40}_{-0.14}$ | $0.009^{+0.03}_{-0.009}$ | $-0.03^{+0.32}_{-0.30}$ | $-1.00^{+2.00}_{-0.00}$ | $-1.00^{+2.00}_{-0.00}$ | NSBH | [5] |
| GW190719_215514 | $6.31\times10^{-1}$ | 0.92 | $36.5^{+18.0}_{-10.3}$ | $20.8^{+9.0}_{-7.2}$ | $23.5^{+6.5}_{-4.0}$ | $0.62^{+0.34}_{-0.53}$ | $0.55^{+0.40}_{-0.39}$ | $0.32^{+0.29}_{-0.31}$ | $0.66^{+0.31}_{-1.02}$ | $0.44^{+0.51}_{-1.19}$ | BBH | [5] |
| GW190725_174728 | $4.58\times10^{-1}$ | 0.96 | $11.5^{+6.2}_{-2.7}$ | $6.4^{+2.0}_{-2.0}$ | $7.4^{+0.6}_{-0.5}$ | $0.36^{+0.51}_{-0.33}$ | $0.51^{+0.43}_{-0.45}$ | $-0.04^{+0.26}_{-0.14}$ | $0.04^{+0.83}_{-0.82}$ | $-0.14^{+1.0}_{-0.74}$ | BBH | [19] |
| GW190731_140936 | $3.35\times10^{-1}$ | 0.78 | $41.5^{+12.2}_{-9.0}$ | $28.8^{+9.7}_{-9.5}$ | $29.5^{+7.1}_{-5.2}$ | $0.37^{+0.54}_{-0.34}$ | $0.45^{+0.48}_{-0.40}$ | $0.06^{+0.24}_{-0.24}$ | $0.16^{+0.74}_{-0.99}$ | $0.15^{+0.77}_{-0.99}$ | BBH | [5] |
| GW190805_211137 | $6.28\times10^{-1}$ | 0.95 | $48.2^{+17.5}_{-12.5}$ | $32.0^{+13.4}_{-11.4}$ | $33.5^{+10.1}_{-7.0}$ | $0.78^{+0.2}_{-0.6}$ | $0.61^{+0.36}_{-0.54}$ | $0.35^{+0.3}_{-0.36}$ | $0.66^{+0.31}_{-0.88}$ | $0.35^{+0.59}_{-1.13}$ | BBH | [6] |
| GW190917_114630 | $6.56\times10^{-1}$ | 0.77 | $9.3^{+3.4}_{-4.4}$ | $2.1^{+1.5}_{-0.5}$ | $3.7^{+0.2}_{-0.2}$ | $0.24^{+0.52}_{-0.22}$ | $0.45^{+0.48}_{-0.4}$ | $-0.11^{+0.24}_{-0.49}$ | $-0.38^{+1.02}_{-0.56}$ | $-0.13^{+0.99}_{-0.79}$ | NSBH | [6] |
| GW191103_012549 | $4.58\times10^{-1}$ | 0.94 | $11.8^{+6.2}_{-2.2}$ | $7.9^{+1.7}_{-2.4}$ | $8.34^{+0.66}_{-0.57}$ | $0.46^{+0.40}_{-0.40}$ | $0.50^{+0.44}_{-0.44}$ | $0.21^{+0.16}_{-0.10}$ | $0.54^{+0.41}_{-0.94}$ | $0.48^{+0.48}_{-1.19}$ | BBH | [3] |
| GW200216_220804 | $3.50\times10^{-1}$ | 0.77 | $51^{+22}_{-13}$ | $30^{+14}_{-16}$ | $32.9^{+9.3}_{-8.5}$ | $0.48^{+0.46}_{-0.43}$ | $0.51^{+0.44}_{-0.46}$ | $0.10^{+0.34}_{-0.36}$ | $0.24^{+0.68}_{-1.02}$ | $0.18^{+0.75}_{-1.03}$ | BBH | [3] |

GW170817 [14], which was also the first multimessenger observation, occurred in O2. It was not until O3 that NSBH binary mergers were observed for the first time: GW200105_162426 and GW200115_042309 [18]. In these two systems, the primary mass $m_1$ is larger than the maximum mass of roughly 2.2–2.5$M_\odot$ allowed by the NS equation of state [21–25], and the secondary mass $m_2$ is consistent with known NS masses (here and throughout the paper, the primary mass $m_1$ refers to the larger of the two component masses in the binary, while the secondary mass $m_2$ refers to the smaller of the two). The inclusion of NSBH events enables the first joint analysis of the entire compact object population for component masses between 1 and 100$M_\odot$, a range covering NSs and stellar-mass BHs. This enables us to clearly identify three different populations, their relative prevalences, and the possibility of gaps in the mass distribution between these populations. The availability of NSs from both NSBH and BNS events also provides enough observations to investigate the mass distribution of NSs in merging binaries.

With a larger sample of BBHs, we are able to clearly identify structures in the BH mass distribution that are not predicted by astrophysical models [26–42]. Under the interpretation that all our mergers are comprised of BHs originating from the collapse of massive stars [43–46], this novel population census is an essential benchmark for an astrophysical understanding of how such stars end their life, and of the paths that bring the remnants together [47]. If, alternatively, primordial BHs [48–52] make a significant contribution to our census of detected mergers, the BH mass function will likewise constrain fundamental physics, potentially yielding a unique window on the hot and dense early Universe [53–57]. We also demonstrate for the first time that the rate of binary mergers evolves significantly over cosmic time. In particular, the merger rate was larger in the past, a finding with direct implications for high-mass star formation and evolution.

We find that the spins of black holes are low, but nonzero. Prior to the discovery of merging black hole binaries, the majority of astrophysical predictions were for broad spin distributions, including spins close to the maximum allowed by general relativity [58–60]. After our discoveries, first-principles calculations for spins were revisited, now suggesting a preponderance of near-zero natal spins [61–63], a minority of which can be torqued to near-maximal values by binary interactions [64–67]. Neither scenario, though, fully matches our results, which prefer small but nonzero spin magnitudes with no evidence for a secondary excess of rapidly spinning systems. Another feature of our sample is the accumulation of more individual BBHs whose posteriors preferentially exhibit negative aligned spins relative to their orbital angular momentum, which provides insights into potential formation mechanisms. For the first time, we observe a correlation between the mass, or mass ratio, of BH binaries and their spins. Such correlations are not predicted in astrophysical models.

With a narrower model space and fewer events, the GWTC-2 analysis [20] identified fewer puzzling features and could be more easily reconciled with proposed source populations. Indeed, most features in that work, e.g., an overdensity at 33$M_\odot$, were grounded in or contrasted with theoretical modeling, with limited exploration of alternatives. In this work, we perform a thorough analysis of the whole mass space, using a wider range of models to account for modeling systematics. As described in Sec. II, we find several surprising new features: structures in the black hole mass distribution at a variety of masses, a broad neutron star mass distribution, an ongoing prevalence of low BH spins, and the unexpected correlation between black hole spins and mass ratio.

In order to accurately infer the population properties, we require a sample of events with a controlled, and low, level of noise contamination. The threshold for inclusion in GWTC-3 required a minimum of 50% probability of astrophysical origin for each candidate signal; as a result, over the catalog as a whole, several candidates are expected to be caused by instrumental noise. To ensure good sample purity, i.e., a low expected fraction of events caused by noise, in our analysis of these observations, we impose a further threshold on the FAR of $< 1\ \mathrm{yr}^{-1}$. At this threshold, the expected number of noise events is less than 10% of the total. We distinguish between NSs and BHs using prior information about the maximum NS mass obtained from constraints on the dense-matter equation of state that suggest that nonrotating NSs cannot be heavier than 2.5$M_\odot$ or so [25,68,69]. This classification yields 69 BBHs, two BNSs, four NSBHs, and one event, GW190814 [7], has a secondary whose mass lies just above the maximum NS mass making its classification ambiguous.

Given the small numbers of observed BNSs and NSBHs, we use a tighter FAR threshold of $< 0.25\ \mathrm{yr}^{-1}$ when inferring properties of the NS and BH populations jointly: their mass and spin distributions, their merger rates, and (potential) cosmological rate evolution. At this threshold, we have 67 events, of which two are identified as BNSs and two as NSBHs. The rate of BBH observations is significantly higher than that of BNS or NSBH mergers. Therefore, when making inferences solely about the BBH population, specifically in Secs. VI and VII, we are able to use the lower significance threshold of FAR of $< 1\ \mathrm{yr}^{-1}$, resulting in 69 confident BBH events. In order to accurately infer the astrophysical distributions, we quantify the selection effects arising from the varying sensitivity of our observatory network to different signals for a continuous range of FAR thresholds.

Our significance thresholds omit several candidates of moderate significance identified in recent work. These include candidates identified by our own search [3,6] with a probability of astrophysical origin $p_{\rm astro} > 0.5$ [70], but whose FAR lies above our threshold. For example, our chosen FAR threshold excludes some of the most massive events identified in GWTC-3 [3,6] (e.g., GW190403_051519 and GW200220_061928). In addition, other independent groups have searched the LIGO-Virgo data and identified candidate events [71–76]. We briefly discuss these events in the context of our reconstructed populations.

The remainder of this paper is organized as follows. In Sec. II we summarize the observations we reported through O3, then highlight the key conclusions obtained from them in this study. In Sec. III we describe the hierarchical method used to fit population models to the data, and steps taken to validate their results. In Sec. IV we describe analyses for the whole compact binary population, including both BHs and NSs. In Sec. V we describe our results for binaries containing one or more NSs. In Secs. VI and VII we describe our results for BBH masses and spins, respectively. In Sec. VIII we discuss the results obtained with other searches or selection criteria, comparing to the populations identified in this work. In Sec. IX we discuss the astrophysical interpretation of our observations and population inferences. In Sec. X we comment on prospects for future searches for the stochastic background of gravitational radiation from compact binary mergers during the next observing run. We conclude in Sec. XI by summarizing the significance of our results. In our appendixes, we provide the details of how we estimate sensitivity to compact binary mergers (Appendix A), a comprehensive description of the population models used in this work (Appendix B), methods we use to validate our study against prominent sources of systematic error (Appendix C), and additional details of the BBH results (Appendix D). In Appendix E, we provide revised posterior distributions for all events used in this work, each reassessed using information obtained from an estimate for the full population. In an associated data release, we provide all the analysis results and postprocessing scripts underlying our results [77].

## II. SUMMARY OF OBSERVATIONS AND RESULTS

A total of 90 compact binary coalescences (CBCs) have been detected in the first three observing runs [3]. The threshold used in GWTC-3 requires a probability of astrophysical origin of at least 50%. For the population analysis presented here, it is preferable to work with a different threshold to ensure lower contamination from signals of nonastrophysical origin, and to reduce the model dependence in assessing probabilities of astrophysical origin. Consequently, we adopt a threshold of FAR $< 1\ {\rm yr}^{-1}$, in at least one of the search analyses in GWTC-3, for all results reported in this paper. This gives 76 events with available parameter estimates, of which approximately 4.6 are expected to be nonastrophysical. This significantly expands the number of observations subsequent to GWTC-2, which included 50 events, of which 47 had FAR of $< 1\ {\rm yr}^{-1}$ and were used in our previous population analysis [20]. For analyses of binaries containing at least one NS, we use a more stringent threshold of FAR $< 0.25\ {\rm yr}^{-1}$, due to the lower number of observations. This threshold limits the number of events to 67; at this threshold, we expect approximately one event not to be of astrophysical origin. Table I shows selected properties of all events used to infer the astrophysical population of binary mergers in the Universe. The table contains all events with FAR $< 1\ {\rm yr}^{-1}$, with less significant events having FAR between 1 and $0.25\ {\rm yr}^{-1}$ which are excluded from all but the BBH analyses clearly identified. Henceforth, we abbreviate candidate names by omitting the last six digits when unambiguous.

Figure 1 shows the properties of the new observations included in this analysis [3]. The shaded regions show two-dimensional marginal distributions for individual events. For reference, the black contours show expected two-dimensional marginal distribution for observed BBH events deduced in our previous analysis of GWTC-2 (the POWERLAW+PEAK model from Ref [20]). In these plots and henceforth, we define $q = m_2/m_1$ and chirp mass

$$\mathcal{M} = (m_1 m_2)^{3/5}/(m_1 + m_2)^{1/5}. \tag{1}$$

The dimensionless spin of each black hole is denoted $\chi_i = \mathbf{S}_i/m_i^2$, where $\mathbf{S}_i$ is the spin angular momentum of the black hole, and the effective inspiral spin parameter [78]

$$\chi_{\rm eff} = \frac{(m_1 \boldsymbol{\chi_1} + m_2 \boldsymbol{\chi_2}) \cdot \hat{\boldsymbol{L}}}{m_1 + m_2}, \tag{2}$$

where $\hat{\boldsymbol{L}}$ is the instantaneous orbital angular momentum direction. Finally, $z$ is the redshift of the event, inferred from the measured luminosity distance using $H_0 = 67.9\ {\rm km\,s^{-1}\,Mpc^{-1}}$ and $\Omega_m = 0.3065$ [79]. From these plots, we make several observations that motivate the investigations and results presented in the remainder of the paper.

(i) Neutron star–black hole binaries. The two NSBH binary observations GW200105 and GW200115 [18] are apparent in Fig. 1 as two of the lowest-mass new sources. Prior to O3, GW and Galactic observations had not identified any NSBH binaries [18]. We now know that these objects exist and merge, occupying a previously unexplored region in the mass and merger rate parameter space. NSBHs form a distinct population from the BNS and most BBHs, motivating the detailed multicomponent analyses pursued in Sec. IV. For the first time, we are able to present rates for a BNS, NSBH, and BBH inferred jointly from an analysis of all observations. The NSBH merger rate is substantially larger than the BBH merger rate. As a result, our joint analyses produce a marginal mass distribution $p(m_1)$ which differs substantially from our previous work, and from analyses in this work based solely on BBHs: The NSBH merger rate overwhelms the BBH rate at low mass.

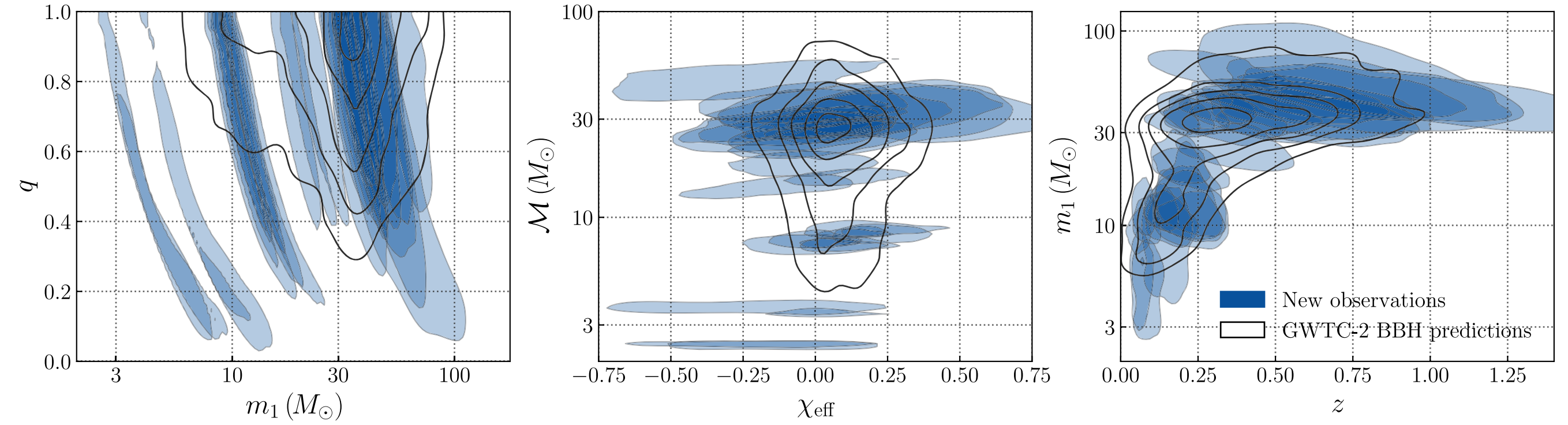

FIG. 1. New observations since GWTC-2. The measured properties of new CBC candidates announced since GWTC-2 with FAR < $\mathrm{yr}^{-1}$ and reported parameters (blue shaded regions) compared to the expected population of detected BBHs (black contours) as inferred from past analysis of GWTC-2 with the same FAR threshold [114] and the fiducial binary parameter priors described in the text. The left-hand plot shows the inferred primary mass $m_1$ and mass ratio $q$, the center plot shows the effective spin $\chi_{\rm eff}$ and chirp mass $\mathcal{M}$, and the right plot shows redshift $z$ and primary mass. The least-massive sources in this sample include NSBH events GW200105 and GW200115.

(ii) Lower mass gap. We identify a relative dearth of observations of binaries with component masses between approximately 2 and $5M_\odot$. This underabundance is visible in the spectrum of observed primary masses plotted in Fig. 1. GW and Galactic observations through O3a were consistent with a mass gap for compact objects between the heaviest NSs and the least-massive BHs [80–83]. The gap was thought to extend from roughly 3 to $5M_\odot$, potentially due to the physics of core-collapse supernova explosions [84–88]. Both Galactic and GW observations made contemporaneously with O3 challenge this assumption [7,89,90]. Most notably, the secondary in GW190814 sits just above the maximum mass that the dense-matter equation of state is expected to support [7]. The primary of GW200115 [3,18] may also lie above the maximum NS mass but below $5M_\odot$. Because of considerable uncertainty in their mass ratio, several binaries' secondaries may also hail from this gap region between 3 and $5M_\odot$. We investigate the prospect of a mass gap in Sec. IV C, treating all compact objects equivalently.

(iii) NS mass distribution. The observation of NSBH binaries enables a detailed study of the observed mass distribution of NSs, combining results from both BNSs and NSBHs. We discuss this in detail in Sec. V, comparing source classifications informed by the NS equation of state (EOS) as well as the inferred location of the lower mass gap. The inferred NS mass distribution, albeit based upon a limited sample of observations, does not exhibit a peak at $1.33M_\odot$; in contrast, radio observations of Galactic BNSs favor such a peak [91–93]. We investigate the impact of outliers in the mass distribution in Sec. V C, particularly GW190814 whose secondary mass lies above the otherwise inferred NS mass range.

(iv) Additional substructure in the BBH mass distribution. The observed masses of BBH binaries are clumped. This is most visible on the central panel in Fig. 1, where overdensities in the chirp-mass distribution from 8 to $10M_\odot$ and around $30M_\odot$ are visible. In Fig. 2, we show the one-dimensional chirp-mass distribution for BBH events. The top panel shows the observations for individual events overlaid with the observed distribution. The observations cluster in chirp mass, with about one-eighth of observed events having chirp masses within 8–$10.5M_\odot$. Compared to chirp-mass accuracy for these events ($\lesssim 1M_\odot$; see, e.g., Ref. [3]), this region is well separated from the next most massive binaries in chirp mass. There is also a significant overdensity at $\mathcal{M} \approx 30M_\odot$ and a weaker feature at $15M_\odot$. These features were previously identified using only GWTC-2 [94–97]. In the bottom panel of Fig. 2, we show the inferred astrophysical

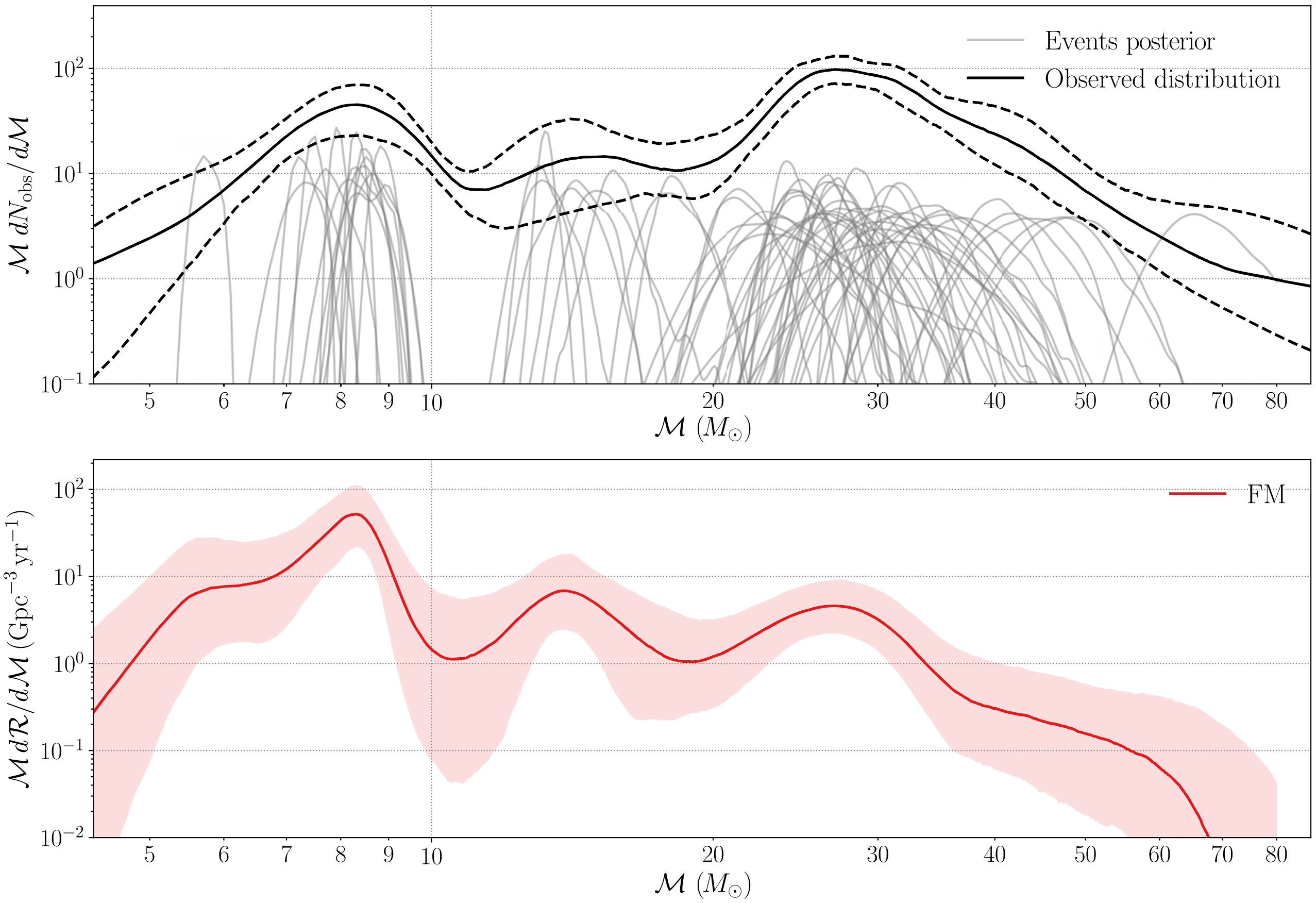

FIG. 2. Illustrating substructure in the source chirp-mass distribution for a BBH (with FAR $< 1\ {\rm yr}^{-1}$, excluding GW190814, as in Sec. VI). All event inferences shown adopt the same fiducial PE priors shown in Fig. 1 and described in the text. Top: the individual-event observations versus chirp mass (gray) and an inferred distribution of the observed chirp-mass distribution (black solid) using an adaptive kernel density estimator [115,116]. The kernel bandwidth is optimized for the local event density and a 90% confidence interval (black dashed) is obtained by bootstrapping [117]. Bottom: the solid curve is the predicted underlying source chirp-mass distribution obtained using the flexible mixture model framework (FM); see Sec. III for details. Unlike the top panel, this panel accounts for our selection effects. The distribution shows three clusters at low masses and a relative deficit of mergers in the chirp-mass range $10 - 12 M_\odot$.

distribution of chirp mass, as recovered by the same Flexible mixtures (FM) approach that first identified these modulations [94,98]. The features in the observed distribution are mirrored in the astrophysical one. In Sec. VI we show that these features are robustly identified by several independent analyses, and demonstrate that the observed structure in the mass distribution is highly significant. Since strong features correlated with chirp mass, but independent of mass ratio, are *a priori* astrophysically unlikely, these significant overdensities suggest the two-dimensional marginal distribution of the BBH population should also have significant substructure and localized overdensities. We explore this in detail in Sec. VI B

(v) BBH rate evolution with redshift. We find that the merger rate density increases with redshift. The right plot in Fig. 1 shows the distribution of events as a function of redshift. While there is a clear evolution of the observed mass distribution with redshift, this arises from the detectors' greater sensitivity to higher-mass systems. Consequently, from Fig. 1 alone, we are not able to draw inferences about the evolution of the population or merger rate with redshift. We explore these issues in detail in Sec. VI D, where we show that there is no evidence for the evolution of the mass distribution with redshift. However, the merger rate density does increase with redshift. Modeling the rate as $\propto (1+z)^{\kappa}$, we find that $\kappa = 2.9^{+1.7}_{-1.8}$. Our analysis strongly disfavors the possibility that the merger rate does not evolve with redshift.

(vi) Low BBH spins. The BBH detections exhibit effective inspiral spins distributed about a mean value $\chi_{\rm eff} \approx 0.06$, with the highest inferred spins below 0.6. The spread is consistent with expectations from GWTC-2. As identified in GWTC-2 [5], the majority of effective inspiral spins are positive, but we additionally infer the presence of events with $\chi_{\mathrm{eff}} < 0$.

## III. METHODS

### A. Data and event selection

We consider candidate events identified by our search analyses for compact binary coalescences using archival data comprising results from the GstLAL [99–101], PyCBC [102–107], and MBTA [108] analyses using template-based matched filtering techniques, and the cWB [109,110] analysis using an excess-energy search that does not assume a physically parametrized signal model. Details of these analyses and the configurations used for O3 data are given in our previous work [3,5,6]. Out of the thousands of candidates produced, only a small minority correspond to astrophysical merger signals, most being caused by instrumental noise. While methods are emerging for performing a joint population analysis including both signal and noise events [70,111–113], here we largely follow a simple procedure [20,114] of imposing a significance threshold to identify events for our population analysis and implicitly treating all events passing the threshold as true signals. The choice of threshold will then limit the expected level of noise contamination.

The analyses calculate a ranking statistic for all candidate events, which is used as the basis for estimating the events' FARs. The ranking statistic allows for sources over a broad parameter space of binary component masses and spins to be detected, without making strong assumptions on the form of the source distribution (except in the case of PyCBC-BBH, specialized for comparable-mass BBH mergers). The analyses additionally calculate an estimate of the probability of astrophysical (signal) origin ($p_{\mathrm{astro}}$) using analysis-specific assumptions on the form of the signal distribution [3,6]. Since, in this work, we explore a range of different assumptions and models for the binary merger population, we define our event set by imposing a threshold on FAR values, rather than on $p_{\mathrm{astro}}$ [20].

Our searches and event validation techniques for GW transients have so far identified 76 candidates with FAR below 1 $\mathrm{yr}^{-1}$ in LIGO and Virgo data through O3. Table I presents these events. In our analysis here, we remove candidates with probable instrumental origin (e.g., 200219_201407 [3]). Assuming our analyses produce noise triggers independently, we expect $\sum_k \mathcal{R}T_k \simeq 4.6$ false events in our sample, where $\mathcal{R}$ is the false alarm rate and $T_k$ is an estimate of the time examined by the $k$th search. For the population studies presented here, the event list can be further restricted by additional FAR thresholds to identify a high-purity list of candidates and to assess the stability of our results to changes in threshold. The choice of FAR threshold to achieve a given level of noise contamination will depend on the number of significant event candidates (and hence, likely signals) considered for an analysis. The most prominent difference concerns analyses for binaries with one or more NS components, in Secs. IV and V, as opposed to analyses which consider only BBH systems, in Secs. VI and VII. While our dataset contains many tens of confidently detected BBH mergers, there is only a handful of comparably significant BNS or NSBH events. This leads us to impose a more stringent threshold of FAR $< 0.25\ \mathrm{yr}^{-1}$ for all analyses considering NS systems.

Because population reconstruction requires careful understanding of search selection biases, we do not include candidates identified by independent analyses [71–75,118,119] of the publicly released LIGO and Virgo data [120,121]. Previous studies [112,113] suggest that our results are unlikely to change significantly with the inclusion of these events. Future analyses may be able to include events from multiple independent catalogs with a unified framework for calculating event significance independent of specific search methods [122,123]. In addition, from our intermediate mass black hole (IMBH) search focused on the detection of binaries with a total mass $100M_\odot$ or more [124,125], and a search of O2 data aimed at the detection of binaries on eccentric orbits [126], none of the additional candidates produced are significant. Thus, our assessment in the chosen mass range, or assumption on binary orbits to be only quasicircular, is expected to have a limited impact on our inference on the bulk of the binary black hole population.

Parameter estimation results for each candidate event [5] are obtained using the LALINFERENCE [127], RIFT [128,129], or Bilby [130,131] analyses. The parameter estimation analyses use Bayesian sampling methods to produce fair draws from the posterior distribution function of the source parameters, conditioned on the data and a given model for the signal and noise [132]. Unless otherwise noted, we use previously published samples for each event through GWTC-2.1 [4–6]. For GW200105 and GW200115 [18], we use the inferences reported in GWTC-3 [3]. For previously reported events through GWTC-2, we adopt the same parameter and event choices reported in our previous population study [20]. For O1 events, we use published samples which equally weight analyses with SEOBNRv3 [133,134] and IMRPhenomPv2 [135] waveforms, and for new events reported in the GWTC-2 update [5], we use published samples with higher-order modes selected by equally weighting all available higher-order mode analyses (PrecessingIMRPHM). The higher-mode analyses associated with GWTC-2 do not include calibration uncertainty. Regarding new events presented in GWTC-2.1, we use the fiducial analysis reported in that work (unless otherwise noted) comprised of merged posterior samples equally drawn from SEOBNRv4PHM [136,137] and IMRPhenomXPHM [138]. Both models implement precession and include beyond-quadrupole radiation for asymptotically quasicircular orbits. For events from the second half of O3 (O3b), newly reported in GWTC-3, we use the publicly released C01:Mixed samples [3], which equally weigh two analyses with the models SEOBNRv4PHM [137] and IMRPhenomXPHM [138]. These samples lack the impact of calibration error on the SEOBNRv4PHM analyses for GW200316, GW200129, and GW200112. A more complete description of the parameter estimation methods and waveform models used can be found in Sec. V of Ref. [5]. To avoid ambiguity where multiple versions of these samples exist, our input posterior samples adopt the $D_L^2$ prior on luminosity distance $D_L$ and have reference spins specified at 20 Hz. In the case of the BNS events GW170817 and GW190425 and the NSBH events GW200105 and GW200115, two versions of the samples are available: one that assumes component spins $\chi_{1,2} < 0.05$ for putative NSs, and a less restrictive but event-dependent bound otherwise (e.g., $\chi_{1,2} < 0.99$ for GW200105 and GW200115). We use the samples with the less restrictive spin assumption for all BNS and NSBH events considered here.

The transfer function between the observed strain and astrophysical strain is subject to a systematic calibration uncertainty. Our parameter inferences incorporate our best estimates of calibration uncertainty, as reported in previous work. Since calibration uncertainty has been incorporated independently for each event, we have implicitly assumed any consistent systematic bias applied to all events is small; we estimate less than 0.54% (1.74%) effect for LIGO (Virgo), respectively [139,140]. For O3a, the amplitude uncertainty was $\lesssim 3\%$ [141]. Because we assume the secular calibration error is much smaller than the calibration error envelope applied when analyzing individual events, we do not incorporate this calibration uncertainty into our estimates of network sensitivity. In O3, this calibration uncertainty implies $\lesssim 10\%$ systematic uncertainty in the sensitive spacetime volume and the inferred merger rate, which is subdominant to our uncertainties from Poisson counting error for most source classes and mass regions.

Each foreground event in O3 has been rigorously validated [3]. Out of the 108 triggers examined in O3 (including events not included in the final search results for this or our companion papers), only four were rejected due to the presence of instrumental noise artifacts. The number of vetoed events is comparable to or less than the expected number of false events for our fiducial analysis threshold, and far smaller than the number of events examined in this study.

### B. Population analysis framework

To infer the parameters describing population models, we adopt a hierarchical Bayesian approach, in which we marginalize over the uncertainty in our estimate of individual-event parameters; see, e.g., Refs. [142–144]. Given a set of data $\{d_i\}$ from $N_{\rm det}$ gravitational-wave detections, we model the total number of events as an inhomogeneous Poisson process, giving the likelihood of the data given population parameters $\Lambda$ as [142,143,145]

$$\mathcal{L}(\{d\}, N_{\rm det}|\Lambda, N_{\rm exp}) \propto N^{N_{\rm det}} e^{-N_{\rm exp}} \prod_{i=1}^{N_{\rm det}} \int \mathcal{L}(d_i|\theta)\pi(\theta|\Lambda) d\theta. \quad (3)$$

Here, $N_{\rm exp}$ is the expected number of detections over the full duration of an observation period for the population model $\Lambda$, $N = N_{\rm exp}/\xi(\Lambda)$ is the expected number of mergers over the observation period, with $\xi(\Lambda)$ the fraction of mergers that are detectable for a population with parameters $\Lambda$. The term $\mathcal{L}(d_i|\theta)$ is the individual-event likelihood for the $i$th event in our dataset that is described by a set of parameters $\theta$. The conditional prior $\pi(\theta|\Lambda)$ governs the population distribution on event parameters $\theta$ (e.g., the masses, spins, and redshifts) given a specific population model and set of hyperparameters $\Lambda$ to describe the model. Constraining the population hyperparameters describing the distribution of gravitational-wave signals according to different models is one of the primary goals of this paper. A notable simplification results if a log-uniform prior is imposed on $N \equiv N_{\rm exp}/\xi(\Lambda)$, the total number of events (detectable or not): One can then marginalize Eq. (3) over $N$ to obtain [143,144,146]

$$\mathcal{L}(\{d\}|\Lambda) \propto \prod_{i=1}^{N_{\rm det}} \frac{\int \mathcal{L}(d_i|\theta)\pi(\theta|\Lambda) d\theta}{\xi(\Lambda)}. \quad (4)$$

Our various population algorithms employ one or the other of these two expressions, Eqs. (3) and (4).

To evaluate the single-event likelihood $\mathcal{L}(d_i|\theta)$, we use posterior samples that are obtained using some default prior $\pi_{\varnothing}(\theta)$. In this case, we can calculate the

integrals over the likelihood with importance sampling over the discrete samples where we denote weighted averages over posterior samples as $\langle \ldots \rangle$. Equation (4), for example, becomes

$$\mathcal{L}(\{d\}|\Lambda) \propto \prod_{i=1}^{N_{\rm det}} \frac{1}{\xi(\Lambda)} \left\langle \frac{\pi(\theta|\Lambda)}{\pi_{\varnothing}(\theta)} \right\rangle, \quad (5)$$

where the factor of $\pi_{\varnothing}(\theta)$ serves to divide out the prior used for initial parameter estimation. The likelihoods are implemented in a variety of software including GWPopulation [147,148], PopModels [149], Sodapop [150], and Vamana [98]. Each code evaluates one of the likelihoods described above for population models, building a posterior with one of the emcee, dynesty, or stan packages [151–154].

In order to calculate $\xi(\Lambda)$, we simulate a reference population of compact binary mergers and reweight those that pass our detection threshold to the population described by $\pi(\theta|\Lambda)$ (see Appendix A for a technical description of this procedure). These injections cover a component mass range from 1 to $100 M_{\odot}$ and redshifts up to $z = 1.9$. The spins of the injections are sampled isotropically in orientation and uniform in magnitude. This is in contrast to the sensitivity injections we used in our previous analysis [20], which assumed an aligned-spin distribution. By fully accounting for the effects of spins on our sensitivity, we can now draw conclusions about the astrophysical distribution of BBHs which were not able to be drawn from our previous analyses. We find that this new treatment of selection effects is responsible for our new constraints on the inferred spin and redshift distributions. In particular, we find that this change gives us relatively higher sensitivity at low redshifts and relatively lower sensitivity at high redshifts with respect to the sensitivity estimates used in our previous work [20]. Additionally, Ref. [155] showed that this treatment of spin selection effects is responsible for the new spin distribution constraints rather than the inclusion of new events.

The above likelihood formulation includes uncertainty due to the finite number of samples $\theta$ per event used in the Monte Carlo integration (see, e.g., Refs. [156,157]). For details of how we alter the likelihood to mitigate this source of uncertainty, see Appendix B. In this paper, we refer to both the astrophysical distribution of a parameter—the version as it appears in nature—and the observed distribution of a parameter—what appears in our detectors due to selection effects. The posterior population distribution for a given model represents our best guess for the astrophysical distribution of some source parameter $\theta$, averaged over the posterior for population parameters $\Lambda$,

$$p_{\Lambda}(\theta) = \int \pi(\theta|\Lambda) p(\Lambda|\{d\}) d\Lambda. \quad (6)$$

The subscript $\Lambda$ indicates that we marginalize over population parameters. Meanwhile, the posterior predictive distribution refers to the population-averaged distribution of source parameters $\theta$ conditioned on detection.

### C. Population models used in this work

In this section, we briefly summarize some of the tools and ingredients we use to generate the phenomenological models $\pi(\theta|\Lambda)$ in this work. Appendix B provides a comprehensive description of the population models used in this work, including their functional form and prior assumptions.

#### 1. Parametric mass models

*Neutron star mass models.*—In the analyses that focus exclusively on the NS-containing events, we model the NS mass distribution as either a power law or a Gaussian with sharp minimum and maximum mass cutoffs. The latter shape is inspired by the Galactic double NS mass distribution [91–93]. In both models, which we call Power and Peak, respectively, we assume that the components of BNSs are drawn independently from the common NS mass distribution. For NSBHs, we assume a uniform BH mass distribution and random pairing with NSs.

*Fiducial population mass and redshift analysis.*—In the fiducial power law plus peak [Power Law+Peak (PP)] model [146,158], the mass-redshift distribution (per unit comoving volume and observer time) is assumed to be of the form $p(m_1, q, z) \propto q^{\beta} p(m_1)(1+z)^{\kappa-1}$, with $p(m_1)$ a mixture model containing two components: a power law with some slope and limits, and a Gaussian with some mean and variance. (In practice, this model as applied to GWTC-2 [5] also usually included additional smoothing parameters for the upper and lower limit of the power law.) The merger rate normalization is chosen such that the source-frame merger rate per comoving volume at redshift $z$ is given by

$$\begin{aligned} \mathcal{R}(z) &= \frac{dN}{dV_c dt}(z) \\ &= \mathcal{R}_0 (1+z)^{\kappa}, \end{aligned} \quad (7)$$

where $\mathcal{R}_0$ is the local merger rate density at $z = 0$ and $\kappa$ is a free parameter governing the evolution of $R(z)$ with higher redshift. The corresponding redshift distribution of BBHs (per unit redshift interval) is [146]

$$p(z|\kappa) \propto \frac{1}{1+z} \frac{dV_c}{dz} (1+z)^{\kappa}, \quad (8)$$

where the leading factor of $(1+z)^{-1}$ converts time increments from the source frame to the detector frame. Past analyses generally fixed the redshift distribution of binaries, assuming a source-frame merger rate that is constant and uniform-in-comoving volume; this choice corresponds to $\kappa = 0$. Our previous population studies [20,114] additionally considered an evolving merger rate with variable $\kappa$.

TABLE II. Merger rates in $\mathrm{Gpc}^{-3}\,\mathrm{yr}^{-1}$ for the various mass bins, assuming merger rates per unit comoving volume are redshift independent. BNS, NSBH, and BBH regions are based solely upon component masses, with the split between the NS and BH taken to be $2.5M_\odot$. We also provide rates for binaries with one component in the purported mass gap between 2.5 and $5M_\odot$. For all but the last row, merger rates are quoted at the 90% credible interval. For the last row, we provide the union of 90% credible intervals for the preceding three rows, as our most conservative realistic estimate of the merger rate for each class accounting for model systematics. The PDB (pair) model is distinct from the other three models due to its use of a pairing function [161] and is therefore excluded from the union of credible intervals in the final row. In Sec. VI we estimate the merger rate for BBHs alone, accounting for variation in merger rate versus redshift.

| | BNS | NSBH | BBH | NS gap | BBH gap | Full |
|---|---|---|---|---|---|---|
| | $m_1 \in [1, 2.5]M_\odot$ $m_2 \in [1, 2.5]M_\odot$ | $m_1 \in [2.5, 50]M_\odot$ $m_2 \in [1, 2.5]M_\odot$ | $m_1 \in [2.5, 100]M_\odot$ $m_2 \in [2.5, 100]M_\odot$ | $m_1 \in [2.5, 5]M_\odot$ $m_2 \in [1, 2.5]M_\odot$ | $m_1 \in [2.5, 100]M_\odot$ $m_2 \in [2.5, 5]M_\odot$ | $m_1 \in [1, 100]M_\odot$ $m_2 \in [1, 100]M_\odot$ |
| PDB (pair) | $170^{+270}_{-120}$ | $27^{+31}_{-17}$ | $25^{+10}_{-7.0}$ | $19^{+28}_{-13}$ | $9.3^{+15.7}_{-7.2}$ | $240^{+270}_{-140}$ |
| PDB (ind) | $44^{+96}_{-34}$ | $73^{+67}_{-37}$ | $22^{+8.0}_{-6.0}$ | $12^{+18}_{-9.0}$ | $9.7^{+11.3}_{-7.0}$ | $150^{+170}_{-71}$ |
| MS | $660^{+1040}_{-530}$ | $49^{+91}_{-38}$ | $37^{+24}_{-13}$ | $3.7^{+35.3}_{-3.4}$ | $0.12^{+24.88}_{-0.12}$ | $770^{+1030}_{-530}$ |
| BGP | $98.0^{+260.0}_{-85.0}$ | $32.0^{+62.0}_{-24.0}$ | $33.0^{+16.0}_{-10.0}$ | $1.7^{+30.0}_{-1.7}$ | $5.2^{+12.0}_{-4.1}$ | $180.0^{+270.0}_{-110.0}$ |
| Merged | 10–1700 | 7.8–140 | 16–61 | 0.02–39 | $9.4 \times 10^{-5} - 25$ | 72–1800 |

Power Law+Dip+Break *model (PDB).*—To fit the distribution of BH and NS masses, we use a parametrized model consisting of a broken power law with a notch filter [159,160]. The variable depth of this notch filter allows for a dearth of events between two potential subpopulations at low and high mass. It also uses a low-pass filter at high masses to allow for a potential tapering of the mass distribution at high BH masses. The component mass distribution is then

$$p(m|\lambda) = n(m|M^{\rm gap}_{\rm low}, M^{\rm gap}_{\rm high}, A) \times l(m|m_{\max}, \eta) \times \begin{cases} m^{\alpha_1} & \text{if } m < M^{\rm gap}_{\rm high}, \\ m^{\alpha_2} & \text{if } m > M^{\rm gap}_{\rm high}, \\ 0 & \text{if } m < m_{\min}. \end{cases} \tag{9}$$

Here, $l(m|m_{\max}, \eta)$ is the low-pass filter with power-law $\eta$ applied at mass $m_{\max}$, and $n(m|M^{\rm gap}_{\rm low}, M^{\rm gap}_{\rm high}, A)$ is the notch filter with depth $A$ applied between $M^{\rm gap}_{\rm low}$ and $M^{\rm gap}_{\rm high}$. In this model, the primary and secondary masses are fit by the same parameters and are related by a pairing function [161,162]. Two pairing functions are considered. The first is random pairing: Primary and secondary masses take independent values so long as $m_2 < m_1$. This model takes the form

$$p(m_1, m_2|\Lambda) \propto p(m = m_1|\Lambda) p(m = m_2|\Lambda) \Theta(m_2 < m_1), \tag{10}$$

where $\Theta$ is the Heaviside step function that enforces that primary masses are greater than secondary masses and $\Lambda$ is the full set of eight hyperparameters. The second is a power-law-in-mass-ratio pairing function, as in Ref. [161]. The full mass distribution in the power-law-in-mass-ratio model is thus described by

$$p(m_1, m_2|\Lambda) \propto p(m = m_1|\Lambda) p(m = m_2|\Lambda) q^\beta \Theta(m_2 < m_1). \tag{11}$$

Previous studies of the BBH population have shown that BBHs prefer a power-law-in-mass-ratio pairing function as in Eq. (11) [161]. We find that applying this model to the full mass spectrum causes the BBHs to dominate the inference on the value of the pairing exponent $\beta$. However, low-mass events such as BNSs and NSBHs have a pairing function that deviates significantly from that of BBHs [160]. BBHs prefer a steep positive power law as a pairing function, whereas the low-mass pairing function is best described by a shallow positive power law, though it is still consistent with random pairing. Ideally, one would introduce additional parameters in the model to allow for separate pairing functions [160]. However, this adds unwarranted complexity, as we find that the choice of pairing function only minimally impacts the rate of BBH mergers and overall shape of the mass distribution above approximately $10M_\odot$. On the other hand, the morphology of the low-mass end of the mass distribution is noticeably impacted by the pairing function, as are the rates of BNS and NSBH mergers. Because of these considerations and the fact that the low-mass pairing function has been found to be consistent with random pairing [160], we highlight the results from the random pairing model [Eq. (10)] in the remainder of this work. However, we provide results from fits to both the independent and power-law pairing models in the data release as well as in Table II.

### 2. Spin models

*Fiducial population spin analyses.*—Compact binary spins may be parametrized in several different ways. In addition to the dimensionless spin magnitudes $\chi_i$ ($i \in \{1,2\}$) and the polar tilt angles $\theta_i$ between each spin vector and a binary's orbital angular momentum [163], we often appeal to the effective spin parameters $\chi_{\rm eff}$ and $\chi_{\rm p}$. The effective inspiral spin $\chi_{\rm eff}$ characterizes a mass-averaged spin angular momentum in the direction parallel to the binaries orbital angular momentum. The effective precessing spin $\chi_{\rm p}$, meanwhile, corresponds approximately to the degree of in-plane spin, and phenomenologically parametrizes the rate of relativistic precession of the orbital plane [164]:

$$\chi_{\rm p} = \max\left[\chi_1 \sin\theta_1, \left(\frac{3+4q}{4+3q}\right) q\chi_2 \sin\theta_2\right]. \quad (12)$$

We leverage these two descriptions to explore the nature of BBH spins in two complementary ways. First, we use the Default spin model [165] to directly measure the distribution of BBH component spin magnitudes and tilts. We model component spin magnitudes as being independently and identically drawn from a Beta distribution [164], with

$$p(\chi_i|\alpha_\chi,\beta_\chi) \propto \chi_i^{\alpha-1}(1-\chi_i)^{\beta-1}. \quad (13)$$

Values of the shape parameters $\alpha_\chi$ and $\beta_\chi$ are restricted to $\alpha_\chi > 1$ and $\beta_\chi > 1$ to ensure a nonsingular component spin distribution. We describe component spin tilts, in turn, via a mixture between two subpopulations, one with isotropically oriented tilts and another with tilts preferentially concentrated about $\theta_i = 0$ [165]:

$$p(\cos\theta_i|\zeta,\sigma_t) = \frac{1}{2}(1-\zeta) + \zeta\mathcal{N}_{[-1,1]}(\cos\theta_i;1,\sigma_t). \quad (14)$$

Here, $\mathcal{N}_{[-1,1]}(\cos\theta_i;1,\sigma_t)$ is a normal distribution truncated to the interval $-1 \le \cos\theta_i \le 1$, centered at 1 with a standard deviation $\sigma_t$. The mixing parameter $\zeta$ governs the relative fraction of systems drawn from each subpopulation. The form of Eq. (14) is motivated by a desire to capture the behavior of BBHs originating from both dynamical and isolated evolution channels, which are expected to yield preferentially isotropic and aligned-spin orientations, respectively. Perfect spin-orbit alignment across the BBH population would correspond to $\zeta = 1$ or $\sigma_t = 0$, which our prior analysis on GWTC-2 ruled out at high confidence [20]. This default spin model is characterized by two parameters characterizing the spin-magnitude distribution (e.g., $\alpha$, $\beta$) and two parameters characterizing the spin misalignment mixture model (i.e., $\xi, \sigma_t$). In part because this parametrization approaches isotropy in two independent limits ($\sigma_t = \infty$ or $\zeta = 0$), it assigns high prior weight to nearly isotropic spin distributions.

*Gaussian spin model.*—Our second approach is to instead seek to measure the distribution of effective spin parameters $\chi_{\rm eff}$ and $\chi_{\rm p}$. In this case, we phenomenologically model the joint $\chi_{\rm eff} - \chi_{\rm p}$ distribution as a bivariate Gaussian [166,167]:

$$p(\chi_{\rm eff},\chi_{\rm p}|\mu_{\rm eff},\sigma_{\rm eff},\mu_p,\sigma_p,r) \propto \mathcal{N}(\chi_{\rm eff},\chi_{\rm p}|\boldsymbol{\mu},\boldsymbol{\Sigma}) \quad (15)$$

centered at $\boldsymbol{\mu} = (\mu_{\rm eff},\mu_p)$ and with a covariance matrix

$$\boldsymbol{\Sigma} = \begin{pmatrix} \sigma_{\rm eff}^2 & r\sigma_{\rm eff}\sigma_p \\ r\sigma_{\rm eff}\sigma_p & \sigma_p^2 \end{pmatrix}. \quad (16)$$

Equation (15) is truncated to the intervals $-1 \le \chi_{\rm eff} \le 1$ and $0 \le \chi_{\rm p} \le 1$ over which the effective spin parameters are defined. This second model has five parameters for spin: two mean values and three parameters describing the covariance.

### 3. Multisource mixture model

Multi source *model (MS).*—MS models all source categories in a mixture model, with one subpopulation for each class (BNS, NSBH, and BBH). The BBH subpopulation follows the MultiSpin model introduced in our previous work [20]. This model features a power-law continuum $q^\beta m_1^\alpha$ plus a peak modeled as a bivariate Gaussian in $m_1$, $m_2$. Consequently, the mass distribution is similar to the PP model. However, the spin distribution in the power-law and Gaussian subpopulations is independent, as are the primary and secondary spins, with each of the four scenarios following the Default spin model, with $\zeta \equiv 1$.

New to MS are two additional bivariate Gaussian subpopulations characterizing BNS and NSBH mergers. The BH component of NSBH follows an independent Gaussian mass distribution. As with BBHs, these BHs follow an independent Default spin model with $\zeta = 1$. All three types of NSs (two in BNS and one in NSBH) are assumed to follow the same Gaussian mass distribution. Each type of NS follows an independent Default spin model, except here the spin magnitudes are scaled down to $\chi_{\rm max} = 0.05$, and $\zeta \equiv 0$ since tilts are not well measured.

### 4. Nonparametric models

Power Law+Spline *model (PS).*—The PS model parametrizes perturbations to a simpler phenomenological primary mass model that is modeled as a cubic spline function

$$p_{\rm PS}(m_1|\Lambda,\{f_i\}) \propto p(m_1|\Lambda)\exp[f(m_1|\{f_i\})]. \quad (17)$$

Here, $f(m_1|\{f_i\})$ is the perturbation function interpolated from a set of $n$ knots fixed uniformly in $\log m_1$ space, and with heights $\{f_i\}$ [95]. In this work, we choose as a base model a truncated power law [20,168] with a low-mass taper, similar to our fiducial model but lacking a Gaussian peak in $p(m_1)$. This model has all the parameters of the truncated model in mass and spin, as well as an additional parameter that characterizes the low-mass tapering and $n$ more describing the heights of the cubic spline knots.

Flexible Mixtures *model (FM).*—Vamana, the FM model, characterizes the population as a mixture model, summing over individually separable components describing the distribution of chirp mass, mass ratio, and components of the spin angular momenta parallel to the orbital angular momentum direction [98]. Each component is composed of a Gaussian to model the chirp mass, another Gaussian to identically model the aligned spins, and a power law to model the mass ratio distribution. The weights follow a uniform prior and are proposed using a Dirichlet distribution. We choose 11 components. This choice maximizes the marginal likelihood; however, our results are robust against selecting different numbers of components.

Binned Gaussian process *model (BGP).*—We also model the two-dimensional mass distribution as a binned Gaussian process [169,170]. In this approach, while the redshift and spin distribution are fixed (here, to uniform-in-comoving volume and isotropic and uniform in magnitude, respectively), we assume the merger rate over distinct mass bins is related via a Gaussian process that correlates the merger rates of neighboring bins. We use conventional techniques provided by pymc3 [171] to explore the hyperparameters of the Gaussian process, in particular, its covariance, to optimally reproduce our data.

## IV. BINARY MERGER POPULATION ACROSS ALL MASSES

In this section, we jointly analyze the masses of all events in Table I for several reasons. First, it allows for the inclusion of all events regardless of their inferred source type. This eliminates issues of ambiguity in source classification for a number of events in O3. Second, it makes possible the detection and characterization of additional features such as a lower mass gap between the lowest-mass objects (likely though not necessarily a NS) and the more massive BH populations [159], or multiple subpopulations [170]. Third, it facilitates a self-consistent calculation of merger rates in different regions of the mass spectrum without explicitly counting the number of events in each category [70,172]. Last, it naturally produces an overall rate of compact binary coalescences that does not require combining rates produced by disjoint models which may have differing systematics. We choose a detection threshold of $\mathrm{FAR} < 0.25\ \mathrm{yr}^{-1}$ which reduces the likelihood that subpopulations and features driven by a few events are contaminated by our background.

When searching for features in the population of compact binary coalescences, we want to draw robust conclusions, stable to different choices of model and approach. We, therefore, fit three independent population models described in Sec. III. The PDB model uses a parametrized dip in the mass distribution to characterize modulations of a simple broken power law at low mass. The BGP model is a nonparametric method allowing considerable flexibility in the mass distribution, constrained only weakly by certain smoothness priors. The BGP and PDB models assume an isotropically oriented, uniform-in-magnitude spin distribution for simplicity. For most merging binaries and particularly those with component masses below $10M_\odot$, spin effects have a subdominant impact on our sensitivity and thus on our inferences about the compact binary merger rate distribution versus mass, as shown in Appendix C 1. The MS model uses a multicomponent mixture model, treating the mass, rate, and spin parameters of each component almost entirely independently. However, to be directly comparable to the BBH-only analyses presented in Sec. VI, our MS analysis omits the outlier event GW190814, whereas PDB and BGP include it. To ensure consistent estimates of spin selection effects, the MS analysis presented here employs only O3 events; however, in Appendix C, we demonstrate that our many analyses produce comparable results when including or excluding pre-O3 results.

### A. Merger rates

Models spanning all source classifications allow us to self-consistently measure the merger rates for all detected CBCs, both overall and subdivided into astrophysically interesting mass ranges, assuming they are independent of redshift. Moreover, because events can be classified into each category using mass limits with relatively high confidence, this approach also provides our fiducial BNS, NSBH, and BBH merger rates. Specifically, taking NS masses to lie between 1 and $2.5M_\odot$ and BH masses to be between 2.5 and $100M_\odot$ and taking the lowest 5% and highest 95% credible interval out of all three models, we infer merger rates between 10 and $1700\ \mathrm{Gpc}^{-3}\,\mathrm{yr}^{-1}$ for BNS, 7.8 and $140\ \mathrm{Gpc}^{-3}\,\mathrm{yr}^{-1}$ for NSBH, and 16 and $61\ \mathrm{Gpc}^{-3}\,\mathrm{yr}^{-1}$ for BBH. Our choice of $2.5M_\odot$ as a boundary between the BH and NS, albeit different than the nominal threshold of $3M_\odot$ adopted in GWTC-3, is consistent with our subsequent classification based both on EOS and mass spectrum features. Table II provides the rate estimates obtained for the three models used in this section and, in addition, shows the rates for events in the mass gap, as discussed in detail in Sec. IV C.

For most categories, our merger rate estimates are consistent with previously published estimates. For example, following GWTC-2 we infer a BBH merger rate to be $23.9^{+14.9}_{-8.6}\ \mathrm{Gpc}^{-3}\,\mathrm{yr}^{-1}$. Our knowledge of the coarse-grained mass spectrum has not significantly evolved since our previous analysis, and we find the inferred BBH rate is consistent with our previously reported rate [20], which also omitted GW190814.

We previously reported a BNS merger rate of $320^{+490}_{-240}\ \mathrm{Gpc^{-3}\,yr^{-1}}$ [16]. With data from GWTC-3, in addition to inferring the BNS merger rate by fitting various population models, we also make an estimate by fixing the mass, spin, and redshfit distributions under simple assumptions. For this rate estimate, we assume the masses of NSs in merging binaries are uniformly distributed between 1 and $2.5M_\odot$ and the merger rate is constant in comoving volume out to a redshift of $z = 0.15$. We also model component spin magnitudes distributed uniformly below 0.4, consistent with assumptions made in our previous analysis [6]. Under these assumptions, we infer a BNS merger rate of $105.5^{+190.2}_{-83.9}\ \mathrm{Gpc^{-3}\,yr^{-1}}$. This estimate tends to be lower than that which is made using the flexible population models used in this paper, as the latter use some information from the nearby BBH and NSBH regions of the mass distribution, allowing for ambiguity in source classification and uncertainty in the rate in these regions to contribute to a higher inferred rate in the BNS population. Additionally, this estimate does not include uncertainty in $\langle VT\rangle$ from the population, as the population distribution is fixed. The Poisson variance associated with the number of BNS observations (two events in GWTC-3) is therefore the main contribution to the rate uncertainty.

For BNSs, the inferred merger rate depends on the presumed mass distribution. With few observations to pin down their behavior at low mass, the three approaches adopted in this section arrive at different compact binary mass distributions between 1 and $2.5M_\odot$. Because the merger rate in this region scales $\propto \langle VT\rangle^{-1} \simeq \langle \mathcal{M}^{15/6}\rangle^{-1}$ [173] (where $VT$ denotes the sensitive 4-volume for a specific binary and the angled brackets denote averaging over objects less than $2.5M_\odot$, the upper boundary used in this section for NS masses), the three methods used in this section arrive at merger rates within each others' uncertainty but with medians differing by factors of up to approximately 15.

For NSBHs, we previously inferred a merger rate of $45^{+75}_{-33}\ \mathrm{Gpc^{-3}\,yr^{-1}}$ assuming the observed NSBHs are representative of the population or $130^{+112}_{-69}\ \mathrm{Gpc^{-3}\,yr^{-1}}$ assuming a broad NSBH population [18]. In this work, each of our joint analyses recovers and adopts different mass spectra, producing a broadly consistent rate (between 7.8 and 140 $\mathrm{Gpc^{-3}\,yr^{-1}}$, including systematics). Combined, our results for the NSBH and BNS merger rates highlight the important role of modeling systematics when drawing inferences about populations with few confident members.

To further highlight the impact of model systematics on inferred merger rates, in Table II, we present our deduced merger rates across the mass space using all three models presented in this section. For simplicity, we label mass bins with NS and BH based solely on a boundary at $2.5M_\odot$. We also provide a rate for events in the mass gap between 2.5 and $5M_\odot$ in a binary with either a NS or BH. The bin intervals here are chosen for ease of use to roughly capture features in the mass spectrum but do not reflect our methods for event classification or our inference on features such as the maximum NS mass or edges of any potential mass gaps.

The models used in this section do not model the redshift evolution of the merger rate, and instead report a constant in comoving volume merger rate density, i.e., $\kappa = 0$ in Eq. (7). For most of the mass intervals considered, our surveys to date extend to only modest redshift, so rate evolution versus redshift can be safely neglected. However, for high-mass BBHs, our network has cosmologically significant reach, over which the merger rate may evolve. Furthermore, as discussed in Sec. II, we observe structure in the mass distribution for BBHs. Therefore, in Sec. VI we provide a more detailed description of BBH merger rates, incorporating both redshift and mass dependence.

### B. Identifying subpopulations of CBCs

As discussed in Sec. II, electromagnetic observations had previously suggested a mass gap between BHs and NSs. On the one hand, astrophysical EOS inferences limit nonrotating NS masses to be below the Tolman-Oppenheimer-Volkoff (TOV) mass $M_{\mathrm{max,TOV}} \sim 2.2$–$2.5M_\odot$ [22–25,69], and studies of GW170817's remnant limit them to $\lesssim 2.3M_\odot$ [174–179]. On the other hand, until recently [7,89,90], BHs had not been observed below approximately $5M_\odot$. The sparsity of observations between approximately 2.5 and $5M_\odot$ suggested a potential lower mass gap [80–83].

Figure 3 shows the two-dimensional merger rate versus component masses for the three models used in this section, as well as the results of FM model applied to BBHs. This representation emphasizes the importance of asymmetric binaries to the overall merger rate $d\mathcal{R}/dm_1$ for masses between 1 and $10M_\odot$. The inferred merger rates further illustrate a falloff in event rate at masses above the BNS scale, with additional peaks associated with both unequal mass binaries consistent with NSBH systems as well as approximately equal-mass BBH binaries. The rate of events with at least one component between 2.5 and $5M_\odot$ (i.e., in the purported mass gap) is constrained to be lower than the rate of BNS-like events, but is consistent with the rate of BBH-like events. As further emphasis, Fig. 4 shows the merger rate versus mass for all binaries and also restricting to binaries with $q \simeq 1$ (e.g., the diagonal bins in the BGP model). The rate for approximately equal-mass binaries is significantly lower. In other words, because asymmetric mergers like NSBH occur at a much higher rate than BBH but a much lower rate than BNS, in a joint analysis they significantly impact the marginal merger rate $d\mathcal{R}/itdm_1$ at the lowest masses.

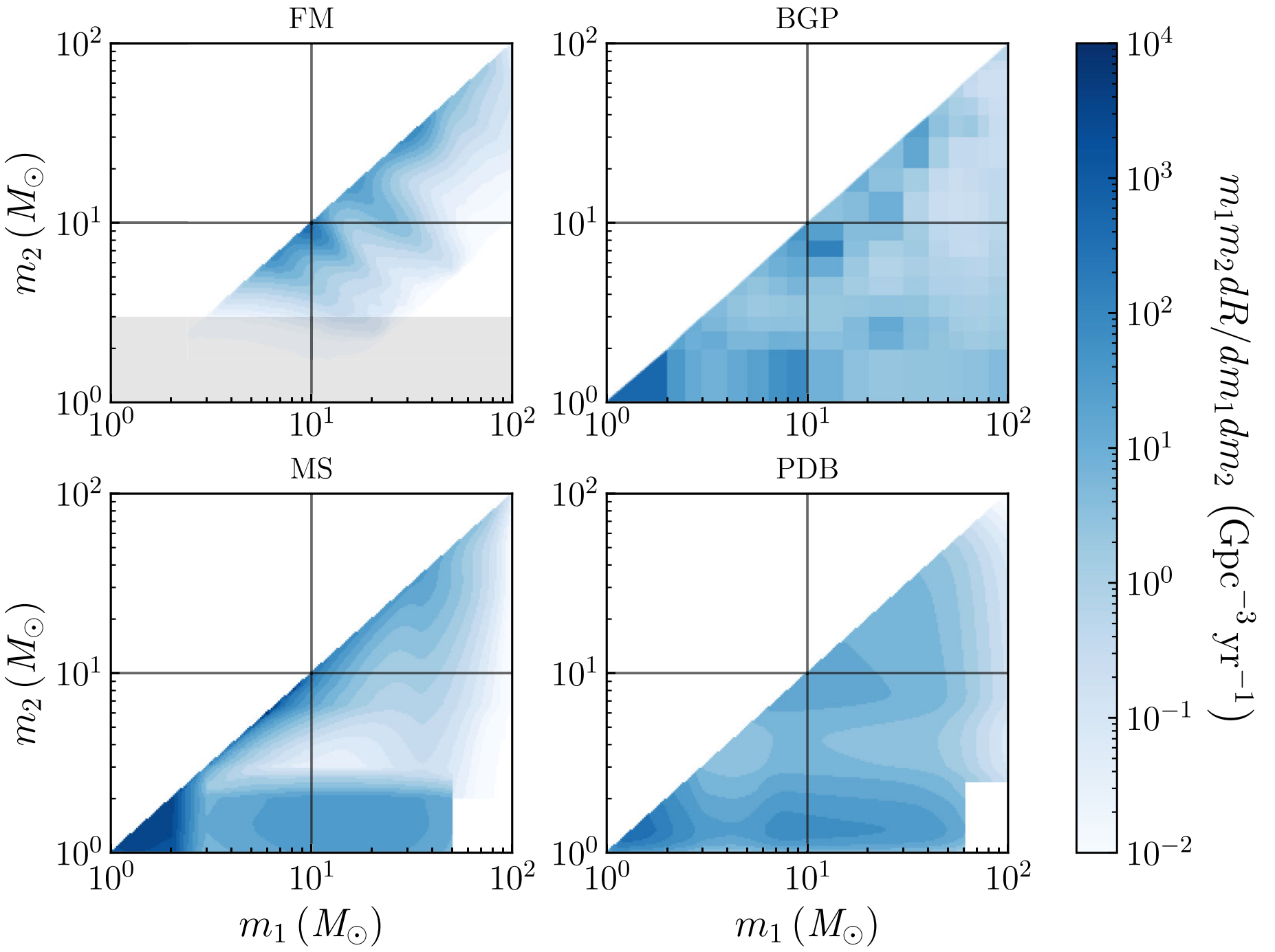

FIG. 3. Rate density versus component masses for different models inferred from events with FAR $< 0.25\ \mathrm{yr}^{-1}$, illustrating consistency on large, coarse-grained scales, but some disagreement and systematics in areas with few observed events. Top left panel: rate density computed with the FM model assuming no redshift evolution, for BBHs only. Modulations along lines of constant chirp mass are apparent. Top right panel: rate density inferred with the BGP model using all compact objects. This model can reproduce observations with localized regions of relatively enhanced rate density. In the BBH region, some regions of enhanced density are commensurate with the FM result. Bottom left panel: rate density inferred with MS. For mergers involving NSs, this model reproduces observations with broad distributions, consistent with smoothing the BGP result. For mergers involving typical BHs, this model strongly favors equal-mass mergers. Bottom right panel: rate density inferred with PDB. This model is also consistent with smoothing the FM result, producing features similar to MS, albeit with less structure in the mass ratio distribution for BBHs, and by construction lacking a peak near $30M_\odot$.

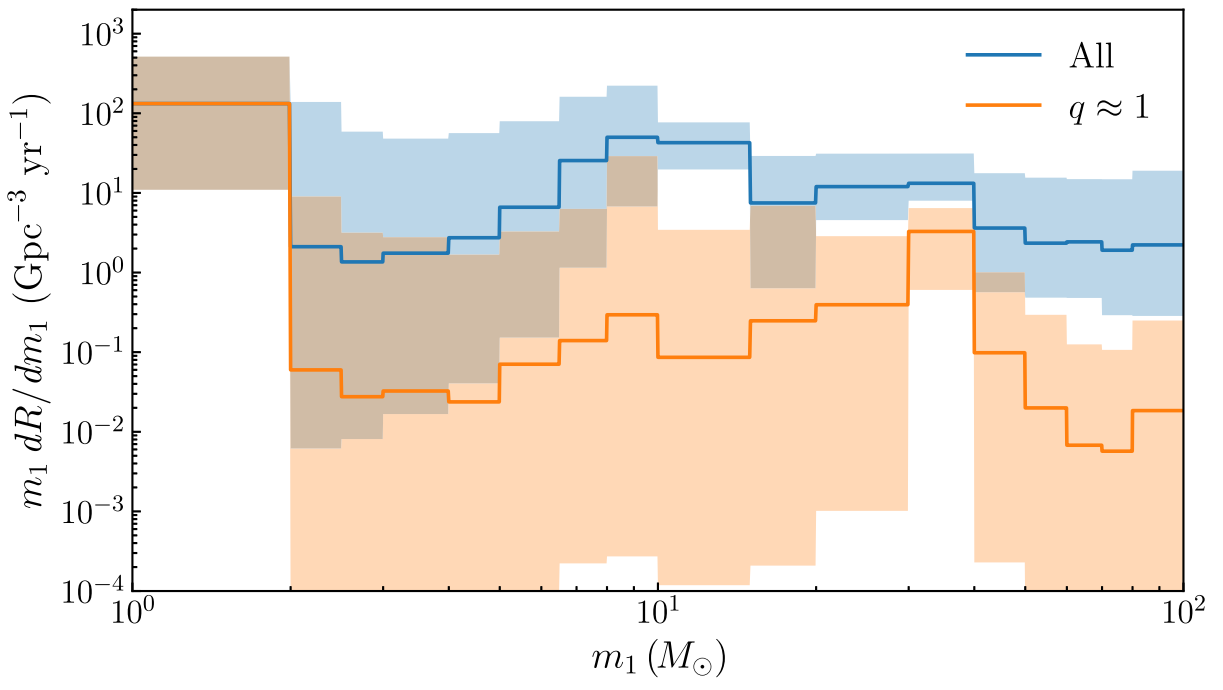

FIG. 4. Impact of asymmetric binaries on the primary mass distribution, illustrating how the depth and extent of a mass gap in any one-dimensional distribution depends on the choice of slicing or marginalization over the remaining dimension. Differential merger rate as a function of primary mass for the BGP model when considering only the diagonal $q \simeq 1$ bins in Fig. 3, i.e., $m_1 dN/dm_1 dq dV_c dt|_{q\simeq 1}$ and the population of compact binaries across all mass bins. The rate for approximately equal-mass binaries is significantly lower highlighting the contribution of asymmetric mergers like NSBHs to the marginal distribution over primary mass. The plot uses the BGP population model inferred from events passing a FAR threshold of $< 0.25\ \mathrm{yr}^{-1}$. Solid curves represent the median rate densities and shaded areas denote 90% credible regions.

This result highlights another feature. The compact binary population has (at least) three dominant populations: BNS-like systems, significantly asymmetric binaries with small $m_2$ comparable to the typical masses of NSs (i.e., including NSBHs as well as GW190814), and the main BBH population with $q$ preferentially more symmetric than $1/4$ (i.e., including GW190412 but not GW190814).

For binaries containing lower mass-gap-scale objects, our inferences about the merger rate and its dependence on mass are consistent despite considerable modeling uncertainty. For binaries containing objects between 2.5 and $5M_\odot$ and having massive BH-scale primaries ($> 5M_\odot$), the mass distribution and merger rate is informed by a few events (GW190814 in particular), thus subject to considerable uncertainty in the inferred component mass distributions. Removing GW190814 from the PDB and BGP analyses increases the statistical uncertainty in the region between 2.5 and $5M_\odot$. In particular, leaving out GW190814 decreases the lower bound on the merger rates in the BH-gap and NS-gap categories but does not affect the upper bounds, as might be expected.

Likewise, for binaries containing objects between 2.5 and $5M_\odot$ and having NS-scale companions, the merger rate is marginally informed by a few events that may not be associated with this region (i.e., GW200115), exacerbating uncertainty in the inferred NS and BH mass distributions. Providing multiple results for these two source classes explores systematic uncertainty due to potential model misspecification. The models presented in this section are subject to different sources of systematic uncertainty. For example, MS employs a Gaussian distribution to model components in BNSs, whereas PDB uses a single power law with a sharp turn-on at low masses to model all objects below the inferred lower edge of the mass gap. These differences result in considerably different BNS rates due to the limited number of detections in the NS mass range. In particular, differences in the pairing function shift the rate inference and add statistical uncertainty in the BNS region. MS and PDB (pair) both assume independent pairing of component masses [Eq. (10)]; PDB (ind) models the pairing of component masses as a power law in mass ratio [Eq. (11)], and BGP uses a piecewise-constant Gaussian process over both component masses [Eq. (B11)]. We can therefore directly compare PDB (ind) and PDB (pair) to understand the impact of assuming independent pairing. Independent pairing implies an equal number of equal-mass and asymmetric mass mergers, while there have been relatively few unequal-mass observations. Thus, a large fraction of PDBs (ind) assumed population has gone undetected, resulting in low overall rate. On the other hand, PDB (pair) finds more support for equal-mass binaries than asymmetric binaries and therefore produces a higher rate. Thus, while assuming independent pairing of component masses may seem benign, it can lead to noticeable systematic errors in the inferred merger rate if the assumption is invalid.

## C. Characterizing suppressed merger rates between NS and BH masses

Figures 3 and 4 show a reduction in the rate above NS masses. It was shown using GWTC-2 that the merger rate between 3 and $7M_\odot$ is suppressed relative to an unbroken power law extending from higher masses [20]. With additional observations, as well as models and sensitivity estimates that span the full mass range of CBCs, we can now produce a comprehensive perspective on merger rates versus mass throughout the low-mass interval 1–$10M_\odot$. In so doing, we find a dropoff in merger rates above NS-scale masses. As a result, in the detection-weighted population, objects with NS-scale mass components are well separated from objects with BH-scale masses. However, we are unable to confidently infer an absence or presence of a subsequent rise in merger rates from lower mass-gap objects. The purported lower mass gap [80–83] between the NS and BH populations would produce an extended local minimum in the merger rate versus mass, implying such a rise at the upper edge of the gap. We therefore neither find evidence for nor rule out the existence of a two-sided lower mass gap.

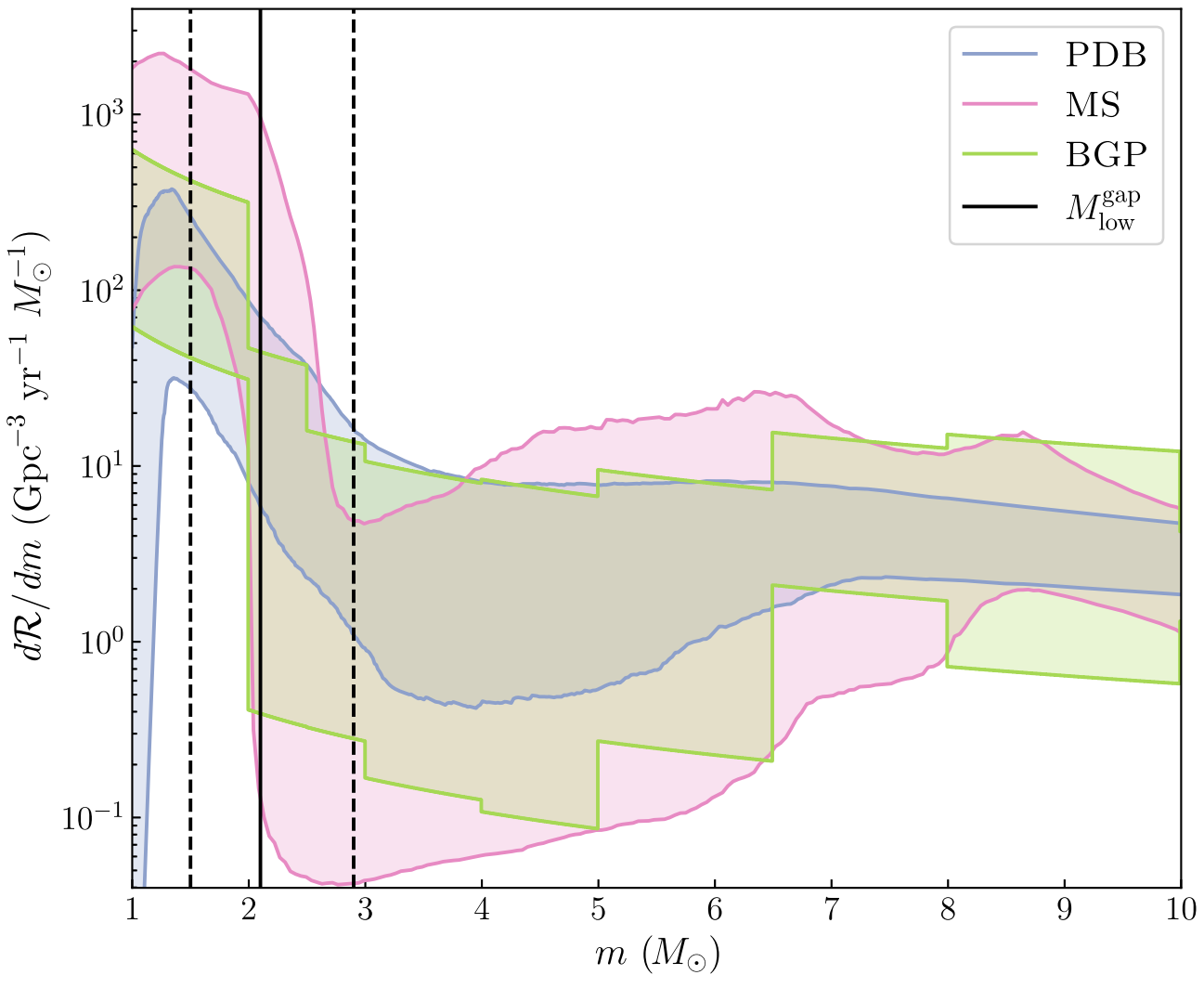

FIG. 5. Differential merger rate as a function of component mass for the PDB, MS, and BGP models. Three independent methods with different modeling assumptions agree on the merger rate versus mass, while illustrating the importance of modeling systematics on the overall rate for objects with NS-scale masses. Shaded areas denote 90% credible regions, while vertical black lines denote the median (solid) and 90% credible intervals (dashed) of the lower boundary of the mass gap $M_{\rm low}^{\rm gap}$ in the PDB model rate dropoff location.

Figure 5 shows the differential rate as a function of component mass inferred from all three models. The PDB model infers the location of this dropoff to occur at $M_{\rm low}^{\rm gap} = 2.1^{+0.8}_{-0.6} M_\odot$, as shown by black vertical lines. While the other models do not explicitly infer the location of the dropoff, they do clearly show a reduction in the rate at a similar location. The prominence of this dropoff can be characterized by comparing the rate of mergers with both masses below $2.5M_\odot$ (BNS) to that of mergers with at least one component mass between 2.5 and $5M_\odot$ (in the mass gap). For this comparison, we find that the differential merger rate of systems with at least one component in the mass gap is 1 to 2 orders of magnitude lower than the BNS rate. Thus, even in the absence of any prior knowledge of the difference between NSs and BHs, the gravitational-wave data suggest two distinct populations of compact objects. This is consistent with results initially found for GWTC-1 [159] and GWTC-2 [160].

The PDB model explicitly parametrizes the mass gap with both low- and high-mass transitions $M_{\rm low/high}^{\rm gap}$ and a gap depth $A$ (where $A = 0$ corresponds to no gap and $A = 1$ to a lower mass gap containing no events). While the posterior on $A$ peaks around 0.77, i.e., corresponding to a relatively empty mass gap, it has broad support between 0 and 1, indicating an inability to unambiguously differentiate between the presence or absence of a lower mass gap. Additionally, the Bayes factors for models with no gap ($A = 0$) or a completely empty gap ($A = 1$), relative to the parametrized model, are 0.073 and 1.4, respectively. This lack of clear preference indicates an inability to resolve the absence or existence of a clear gaplike feature in this part of the mass spectrum.

A subsequent rise in the mass distribution at $M_{\rm high}^{\rm gap}$ is also less clear to discern. The models infer mass distributions with similar support for both a mildly pronounced gap and a flat transition above $M_{\rm high}^{\rm gap}$. Both of these are consistent with the finding in our previous analyses [20] of a deviation from a single power law below primary masses of approximately $7M_\odot$.

We find that if a lower mass gap does exist, it may not be totally empty. While the merger rates show a falloff above around $2.5M_\odot$ in Fig. 5, the rate does not fall to zero. Furthermore, the component masses of six events have at least 5% posterior support between $M_{\rm low}^{\rm gap}$ and $M_{\rm high}^{\rm gap}$ when using a population-informed prior [160]. GW190814 stands out as having considerable support for its secondary being within the mass gap or below the dropoff in the rate at $M_{\rm low}^{\rm gap}$: $P(m_{2,\rm GW190814} \in [M_{\rm low}^{\rm gap}, M_{\rm high}^{\rm gap}]) = 0.76$. This event has a mass ratio $q = 0.112^{+0.008}_{-0.009}$ [7], hinting at either a potential subpopulation of low-$q$, low-$m_2$ BBHs, or a handful of NSBHs with high NS masses. The former possibility is examined in Sec. VI E, and the latter is discussed in Sec. V C. For both the NSBH systems, there is an approximately 10% probability of the secondary lying in the mass gap and, for GW200115 the primary has a 70% probability of $m_1 < M_{\rm high}^{\rm gap}$. Finally, for GW190924_021846, which is the BBH event with the lowest total mass, we find roughly equal support for the secondary being either within ($m_2 < M_{\rm high}^{\rm gap}$) or above ($m_2 > M_{\rm high}^{\rm gap}$) the mass gap.

The inferred depth of the gap does depend heavily on the assumed pairing function: A model in which objects are randomly paired with other objects regardless of mass ratio predicts a more prominent gap than one with a power-law-in-mass-ratio pairing function as in Eq. (11). Similarly, a change of the pairing function will impact the classification of various components as below, in or above the mass gap. Consequently, we do not rely on this methodology for event classification in Sec. V, and instead use EOS-informed limits on the maximum allowed NS mass, and perform leave-one-out analyses with respect to known subpopulations. The lower mass-gap-related results stated here are obtained using a random pairing model.

Though we report on our analysis with FAR $< 0.25\ \rm yr^{-1}$, to assess the stability of our results to threshold choices we repeat our analyses using all events with previously reported parameter inferences below $1\ \rm yr^{-1}$ (i.e., excluding GW190531). Even though such an analysis includes all five candidate NSBHs, our key conclusions remain largely unchanged: The derived merger rates versus mass are consistent with the error bars shown in Figs. 4 and 5, and the merger rates reported in Table II are consistent. In particular, we draw similar conclusions about the merger rate between 2 and $10M_\odot$: suppressed but likely filled, without evidence for or against a true two-sided mass gap.

## V. MASS DISTRIBUTION OF NEUTRON STARS IN BINARIES

In this section, we characterize the astrophysical population of NSs using data from the gravitational-wave events that are likely to contain at least one NS. Because of the paucity of low-mass compact binary mergers observed to date, and the difficulty in ascertaining the presence of a NS in these systems, modeling the NS population observed in gravitational waves has been challenging. In our previous population analysis through GWTC-2 [20], the rate density of BNS and NSBH mergers was estimated, but the shape of the mass distribution of the NSs in these compact binaries was not inferred. The BNS events GW170817 and GW190425 were included in a joint study [180] of the Galactic and gravitational-wave populations of BNSs which linked the two observed populations via a bimodal birth mass distribution. The confident BNS and NSBH detections made to date were analyzed in a study of the gravitational-wave population in [150], which found the observed NS masses to be consistent with a uniform distribution.

We begin by classifying the observed low-mass compact binaries as BNSs, NSBHs, or BBHs. The classifications are based on a comparison of their component masses with an EOS-informed estimate of the maximum NS mass, and are corroborated against the location of the lower mass gap between NSs and BHs as inferred in the previous section. Then, adopting these source classifications as definite and considering the BNS and NSBH detections below a FAR threshold of $0.25\ \rm yr^{-1}$, we infer the shape of the NS mass distribution in compact binaries. In contrast to Sec. IV, we do not attempt to determine the overall rate of such mergers, nor do we attempt to infer the mass distribution of BHs in coalescing NSBH systems. Our analysis makes a comparison with the observed Galactic population of NSs, and we additionally investigate the impact on the population of the event GW190814, a lower mass-gap merger whose secondary may possibly be a NS, but is more likely a low-mass BH.

### A. Events containing NSs

The gravitational-wave signal of a compact binary merger involving a NS differs from that of a BBH due to matter effects in the waveform, most notably the phasing of the gravitational waveform during the inspiral due to the tidal deformation of the NS [181]. Since none of the observations in O3b [3] yield an informative measurement of tidal deformability, the gravitational-wave data do not identify which sources contain a NS. Nonetheless, we can establish whether their components are consistent with NSs by comparing their masses to the maximum NS mass, $M_{\mathrm{max}}$, following the method described in Ref. [68].

TABLE III. Classifications for low-mass events from Table I. The probability that a component is compatible with a NS is measured by the fraction of its mass posterior lying below an estimate [25] of the maximum nonrotating NS mass, $M_{\mathrm{max,TOV}} = 2.21^{+0.40}_{-0.18} M_{\odot}$ marginalized over statistical uncertainties. We adopt a 50% threshold for classification as a NS, assuming a clean separation between NS and BH mass spectra. Probabilities are reported relative to a uniform prior on the component mass. They refer to the secondary component of all events except GW170817 and GW190425, in which case the secondary is securely below the maximum NS mass and the probability for the primary is given. The probabilities are similar and the classifications are unchanged when the component masses are compared to $M_{\mathrm{low}}^{\mathrm{gap}} = 2.1^{+0.8}_{-0.6} M_{\odot}$, the location of the lower mass gap between NSs and BHs inferred from Sec. IV's PDB analysis of the FAR $< 0.25\ \mathrm{yr}^{-1}$ events.

| Name | $\mathrm{FAR}_{\mathrm{min}}$ ($\mathrm{yr}^{-1}$) | $P(m < M_{\mathrm{max,TOV}})$ | $P(m < M_{\mathrm{low}}^{\mathrm{gap}})$ | Classification |
|---|---|---|---|---|
| GW170817 | $< 1 \times 10^{-5}$ | 0.99 | 0.97 | BNS |
| GW190425 | $3.38 \times 10^{-02}$ | 0.67 | 0.71 | BNS |
| GW190814 | $< 1 \times 10^{-5}$ | 0.06 | 0.24 | BBH |
| GW200105 | $2.04 \times 10^{-01}$ | 0.94 | 0.73 | NSBH |
| GW200115 | $< 1 \times 10^{-5}$ | 0.93 | 0.96 | NSBH |
| GW190426 | $9.12 \times 10^{-01}$ | 0.82 | ⋯ | NSBH |
| GW190917 | $6.56 \times 10^{-01}$ | 0.56 | ⋯ | NSBH |

The precise value of $M_{\mathrm{max}}$ is unknown because of uncertainty in the NS EOS. Mass measurements for the heaviest known pulsars [182,183] set a lower bound of approximately $2M_{\odot}$ on $M_{\mathrm{max}}$, while basic causality considerations imply that $M_{\mathrm{max}} \lesssim 3M_{\odot}$ [184,185]. While individual nuclear theory models for the EOS can produce maximum masses as large as approximately $3M_{\odot}$, astrophysical inferences of the EOS generally predict that the maximum mass of a nonrotating NS, the TOV mass $M_{\mathrm{max,TOV}}$, is between 2.2 and $2.5M_{\odot}$ [21–25]. Similarly, studies of GW170817's merger remnant suggest that $M_{\mathrm{max,TOV}} \lesssim 2.3M_{\odot}$ [174–179]. Rapid rotation can sustain a maximum mass up to approximately 20% larger than $M_{\mathrm{max,TOV}}$ [186]. However, the astrophysical processes that form compact binaries may prevent the EOS-supported $M_{\mathrm{max}}$ from being realized in the population; see, e.g., discussion of how supernovae engines and binary evolution may [41,187] or may not [84,188,189] populate the upper edges of the NS mass distribution.

We can therefore identify objects as NS candidates based on their mass using estimates of $M_{\mathrm{max}}$, as long as we assume a clean separation between the NS and BH mass spectra. Of the events with FAR less than $0.25\ \mathrm{yr}^{-1}$, five have at least one component mass with a 90% credible lower bound below $3M_{\odot}$, making them potentially consistent with a BNS or NSBH merger. These events are listed in Table III, and their component mass posteriors are compared to two estimates of $M_{\mathrm{max}}$ in Fig. 6.

For each of these observed low-mass events, we calculate in Table III the probability that at least one of the component masses is less than the maximum NS mass, marginalizing over statistical uncertainties and assuming a uniform component mass prior. We adopt a threshold probability of 50% for classification as a NS. Our fiducial maximum NS mass estimate is taken to be $M_{\rm max,TOV}$ from an EOS inference which is based on pulsar timing, gravitational-wave and x-ray observations of NSs [25]. That study finds $M_{\rm max,TOV} = 2.21^{+0.40}_{-0.18} M_\odot$, and the corresponding posterior distribution is shown for comparison in Fig. 6 (previous EOS inference [25] quoted the highest-posterior-density interval of 2.00–2.52$M_\odot$, while we quote the corresponding symmetric credible interval calculated from the released posterior samples [190]). Four of the FAR $< 0.25$ yr$^{-1}$ events have $P(m < M_{\rm max,TOV}) > 0.5$ for at least one component, and we deem them either BNSs if $m_1 < M_{\rm max,TOV}$ or NSBHs if only $m_2 < M_{\rm max,TOV}$. The fifth event, GW190814, has $P(m < M_{\rm max,TOV}) = 0.06$ and is therefore classified as a BBH. These source classifications do not change if, instead of $M_{\rm max,TOV}$, we compare against the rotating NS maximum mass, $M_{\rm max}(\chi)$, as calculated from an empirical relation involving the TOV mass and the component spin $\chi$ [191]. This allows for the possibility that one or more of the low-mass components is rapidly rotating.

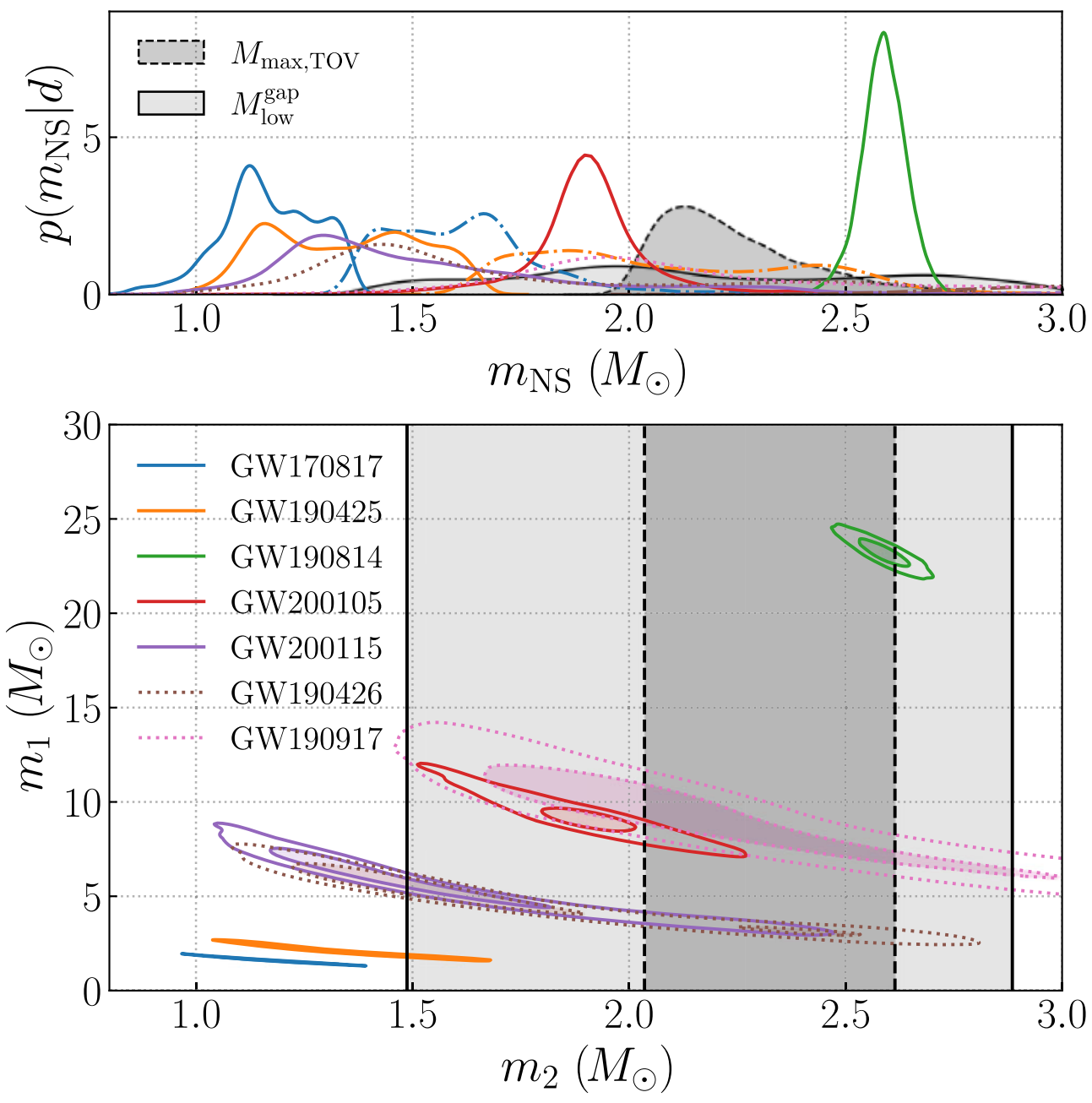

FIG. 6. Masses for events with at least one candidate neutron star. Upper panel: one-dimensional posterior distributions for the masses of the candidate NSs, as compared to estimates of the maximum NS mass based on the dense-matter EOS [25] ($M_{\mathrm{max,TOV}}$) and on the inferred location of the lower mass gap in Sec. IV's PDB analysis ($M_{\mathrm{low}}^{\mathrm{gap}}$). Primary components are shown dash-dotted. GW190814's secondary component lies above both estimates of the maximum NS mass. Lower panel: two-dimensional 50% (shaded) and 90% (unshaded) credible regions for the binary masses of each candidate NS merger. The marginal events GW190426 and GW190917 are shown dotted. The 90% credible intervals of the maximum NS mass posterior inferred from the EOS and from the lower mass-gap location are also plotted. GW190814 occupies a distinct region of the $m_1$-$m_2$ plane compared to the events deemed BNSs or NSBHs.

We draw similar conclusions about each event if we interpret the sharp decrease in merger rate near 2.5$M_\odot$ seen in the PDB analysis as the separation between NS and BH mass ranges. (This interpretation does not imply that $M_{\rm max,TOV}$ and $M^{\rm gap}_{\rm low}$ need to agree: $M^{\rm gap}_{\rm low}$ could be below $M_{\rm max,TOV}$ if the heaviest NSs the EOS can support are not realized in nature, or $M^{\rm gap}_{\rm low}$ could be above $M_{\rm max,TOV}$ if the lower mass gap occurs within the BH mass spectrum.) Following Ref. [160], we compare the component mass measurements against the inferred $M^{\rm gap}_{\rm low}$ parameter from the PDB model, as shown in Fig. 6, and list the probabilities $P(m < M^{\rm gap}_{\rm low})$ in Table III. The same four events are consistent with BNSs or NSBHs.

Figure 6 also plots the component mass posteriors for two FAR $< 1$ yr$^{-1}$ events from Table I that may contain NSs, if astrophysical in origin. In particular, GW190426 and GW190917 have masses consistent with NSBH systems [5,6]. This classification is confirmed by the $P(m < M_{\rm max,TOV})$ and $P(m < M^{\rm gap}_{\rm low})$ probabilities calculated for them in Table III.

### B. Mass distribution

Using the FAR $< 0.25$ yr$^{-1}$ events classified as BNSs or NSBHs in Table III, we infer the mass distribution of NSs in merging compact binaries. We adopt the Power and Peak parametric mass models described in Sec. III and implement a selection function based on a semianalytic approximation of the integrated network sensitivity $VT$, fixing the redshift evolution of the population and ignoring spins when estimating the detection fraction. The population hyperparameters are sampled from uniform prior distributions, subject to the condition $m_{\rm min} \leq \mu \leq m_{\rm max}$ in the Peak model, except that we assume that the maximum mass in the NS population $m_{\rm max}$ does not exceed $M_{\rm max,TOV}$. This is consistent with our use of the nonrotating maximum NS mass to classify the events, and amounts to an assumption that the NSs observed via inspiral gravitational waves are not rotationally supported. In practice, this means imposing a prior proportional to the cumulative distribution function of $M_{\rm max,TOV}$, as shown in the inset of Fig. 7 and detailed in Appendix B 1.

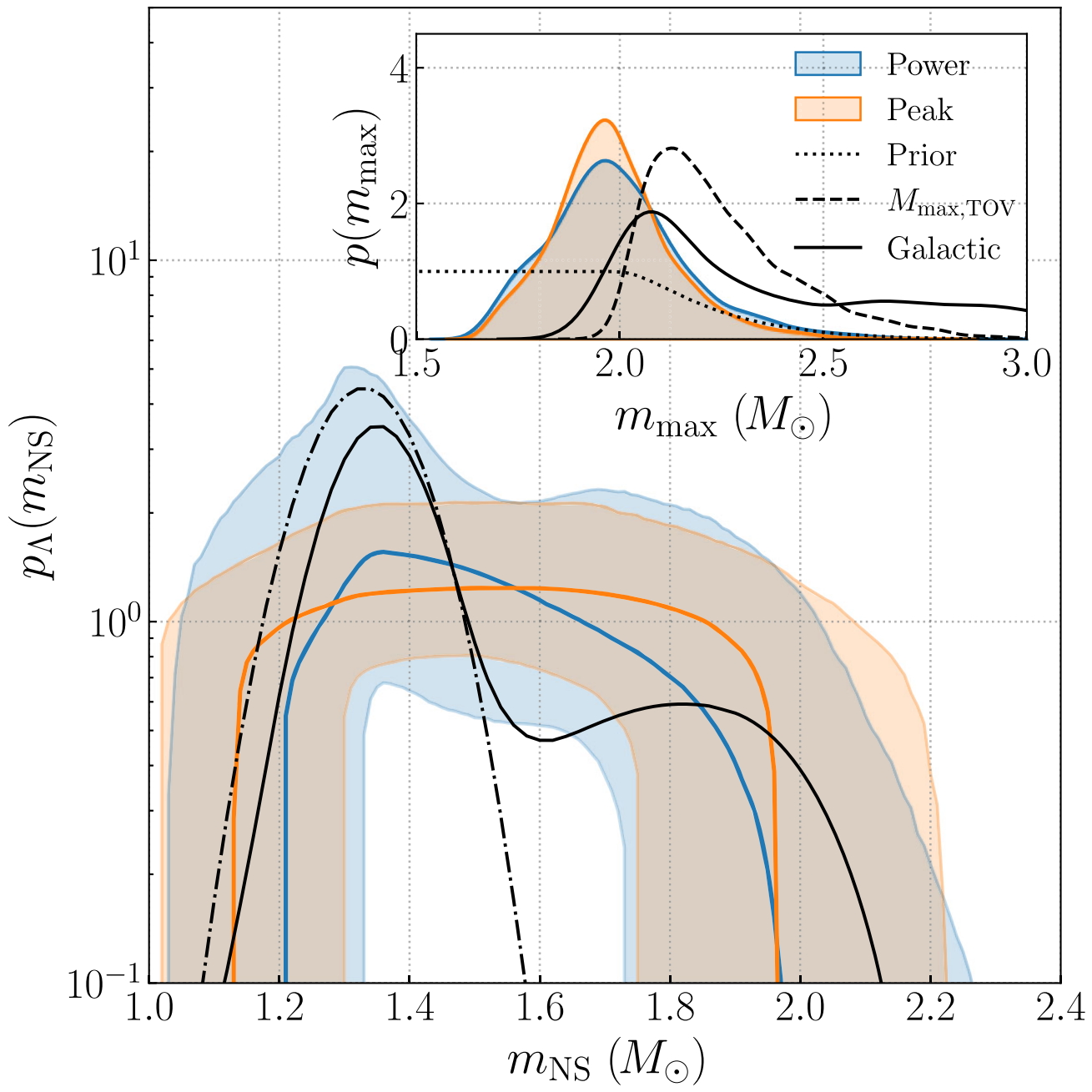

FIG. 7. Inferred neutron star mass distribution. The median mass distribution (solid) and 90% credible interval (shading) inferred for the Power (respectively, Peak) population model is shown in blue (orange), as compared to the mass distribution of NSs in Galactic BNSs [92] (dot-dashed black) and the mass distribution of all Galactic NSs [192] (solid black). The inferred gravitational-wave population has a greater prevalence of high-mass NSs, although the Galactic population of NSs extends to masses above those detected in gravitational waves thus far. The inset shows the posterior distribution for the maximum mass in the NS population for both models, as compared to the Galactic $m_{\rm max}$. The EOS-informed $m_{\rm max}$ prior, which is proportional to the cumulative distribution function of $M_{\rm max,TOV}$, is also shown in the inset (dashed). It enforces $m \leq M_{\rm max,TOV}$ using the maximum Tolman-Oppenheimer-Volkoff mass estimate from Ref. [25]. The maximum mass in the gravitational-wave population is as large as $M_{\rm max,TOV}$ within statistical uncertainties.

The inferred mass distributions for these two models are plotted in Fig. 7. The posterior population distribution for the Power model has $\alpha = -2.1^{+5.2}_{-6.9}$, consistent with a uniform mass distribution, although the median distribution is a decreasing function of mass. The power-law hyperparameter is most strongly constrained relative to the flat $\alpha \in [-12, 4]$ prior on the low end. The two bumps in the 90% credible interval visible in Fig. 7 correspond, respectively, to the minimum and maximum mass cutoffs of the population model realizations with $\alpha < 0$ and $\alpha > 0$. The median inferred PEAK distribution is relatively flat, and the peak width and location are almost entirely unconstrained relative to the prior: $\sigma = 1.1^{+0.8}_{-0.8} M_\odot$ and $\mu = 1.5^{+0.4}_{-0.3} M_\odot$ for a uniform $\sigma \in [0.01, 2.00] M_\odot$ and $\mu \in [1, 3] M_\odot$ prior subject to $m_{\min} \le \mu \le m_{\max}$. Thus, the gravitational-wave observations to date do not support a NS mass distribution with a pronounced single peak. This contrasts with the Galactic BNS subpopulation, whose mass distribution is sharply peaked around $1.33 \pm 0.09 M_\odot$ [91–93,193], as shown for comparison in Fig. 7. (We plot the best-fit Gaussian from Ref. [92]; Ref. [193] estimates statistical uncertainties of $\pm 0.03 M_\odot$ in the Gaussian parameters at the 68% credible level.) The mass distribution of NSs observed in gravitational waves is broader and has greater support for high-mass NSs; indeed, more than 93% of the mass distribution realizations from the posterior are broader than the Galactic BNS mass spectrum. The gravitational-wave population also has a greater prevalence of high-mass NSs compared to the Galactic NS population as a whole, whose mass distribution has a double-peaked shape [192,194,195].

The minimum NS mass in the gravitational-wave population is inferred to be $1.2^{+0.1}_{-0.2}$ and $1.1^{+0.2}_{-0.1} M_\odot$ in the POWER and PEAK models, respectively. The lower bound on $m_{\min}$ is a prior boundary motivated by the sensitivity model, as the gravitational-wave searches target sources above $1 M_\odot$. The maximum mass in the population is found to be $2.0^{+0.3}_{-0.3} M_\odot$ for the POWER model and $2.0^{+0.2}_{-0.2} M_\odot$ for the PEAK model, relative to the EOS-informed $m_{\max}$ prior. These values are consistent with the maximum mass inferred from the Galactic NS population $2.2^{+0.8}_{-0.2} M_\odot$ [192], as can be seen in the inset of Fig. 7. The maximum mass is the best-constrained hyperparameter in the population models. Its upper bound is more tightly constrained than the Galactic $m_{\max}$ in Fig. 7 as a result of the imposed $m_{\max} \le M_{\max,\mathrm{TOV}}$ prior, which begins tapering above $2 M_\odot$, and the strong selection bias of gravitational-wave observations toward heavier masses, which renders the nonobservation of heavier NSs informative. Nonetheless, the statistical uncertainty in $m_{\max}$ remains large, and it is expected that approximately 50 BNS detections will be needed before the maximum mass in the merging NS population can be measured to within $0.1 M_\odot$ [196].

The $m_{\max}$ value inferred from gravitational waves is also as large as $M_{\max,\mathrm{TOV}}$ within statistical uncertainties. This would not be the case if, for instance, the astrophysical processes that form coalescing compact binaries prevented $2 M_\odot$ NSs from pairing with other compact objects. Such a scenario is compatible with the EOS-informed $m_{\max}$ prior that we impose. However, we find that the difference between $m_{\max}$ and the $M_{\max,\mathrm{TOV}}$ is consistent with zero at the 90% credible level. Hence, there is no evidence that NSs as heavy as can be supported by the EOS cannot end up in merging compact binaries.

Moreover, we infer a consistent maximum mass if we adopt a uniform $m_{\max}$ prior instead of the EOS-informed one. This relaxes the assumption that the observed NS masses must be below the nonrotating maximum mass, and accounts for the possibility that rapid rotation may cause a NS's mass to exceed $M_{\mathrm{TOV}}$. Specifically, we find $m_{\max} = 2.1^{+0.8}_{-0.4} M_\odot$ in the POWER model and $2.0^{+0.8}_{-0.2} M_\odot$ in the PEAK model. The upper error bar on $m_{\max}$ extends to much higher values in this case because it is no longer subject to the tapering EOS-informed prior, which has little support above $2.5 M_\odot$. We also obtain consistent results if we expand the event list to include the two marginal NSBH detections listed in Table III, as described in Appendix C 2.

### C. Outlier events

The mass-based event classification carried out above deemed GW190814 to be a BBH merger on the basis of the maximum NS mass the EOS can support. We now further demonstrate that it is an outlier from the population of BNSs and NSBHs observed with gravitational waves. We also comment on how the interpretation of GW190814 as a NSBH merger would change the inferred NS mass distribution.

If we dispense with its $M_{\max,\mathrm{TOV}}$-based classification, and include GW190814 as a NSBH in the population analysis, the inferred maximum mass in the population is shifted up to $m_{\max} = 2.8^{+0.2}_{-0.2} M_\odot$ in the POWER model and $m_{\max} = 2.7^{+0.3}_{-0.2} M_\odot$ in the PEAK model. These values are obtained relative to a uniform $m_{\max}$ prior, since we are no longer consistently enforcing $m \le M_{\max,\mathrm{TOV}}$; all results in this subsection refer to this prior. The $m_{\max}$ posterior has support up to $3 M_\odot$, where the prior truncates and the models' fixed BH mass distribution begins. The inferred NS mass distributions with GW190814 are similar, but flatter and broader, than those depicted in Fig. 7.

To test whether GW190814 hails from the same population as GW170817, GW190425, GW200105, and GW200115, we examine the PEAK model's posterior predictive distribution for secondary masses with and without GW190814 in the event list. Figure 8 compares GW190814's measured $m_2 = 2.59^{+0.08}_{-0.09} M_\odot$ against the prediction for the largest observed secondary mass, $\max_5(m_2)$, after two BNS observations and three NSBH observations (i.e., after a sequence of observations similar to GW170817, GW190425, GW190814, GW200105, and GW200115). That is, we draw two pairs of masses from the posterior predictive distribution for BNSs and three secondary masses from the posterior predictive distribution for NSBHs, take the largest of the five secondaries, and build up the plotted distributions by performing this procedure repeatedly. The probability of observing a secondary mass at least as large as the mean of GW190814's $m_2$ in the population is only 0.2% according to the PEAK model fit that excludes GW190814; we characterize GW190814's $m_2$ by its mean, since it is measured so precisely. The equivalent probability relative to the PEAK model fit that includes GW190814 is 3.3%; we expect a rigorous, fully self-consistent calculation of a $p$-value to lie between these two numbers [197]. Hence, GW190814's secondary component is an outlier from the secondaries in BNS and NSBH systems. In the next section, we also establish GW190814 as an outlier from the BBH population observed in gravitational waves, corroborating our previous analysis [20]. These findings reinforce that it represents a distinct subpopulation of merging compact binaries.

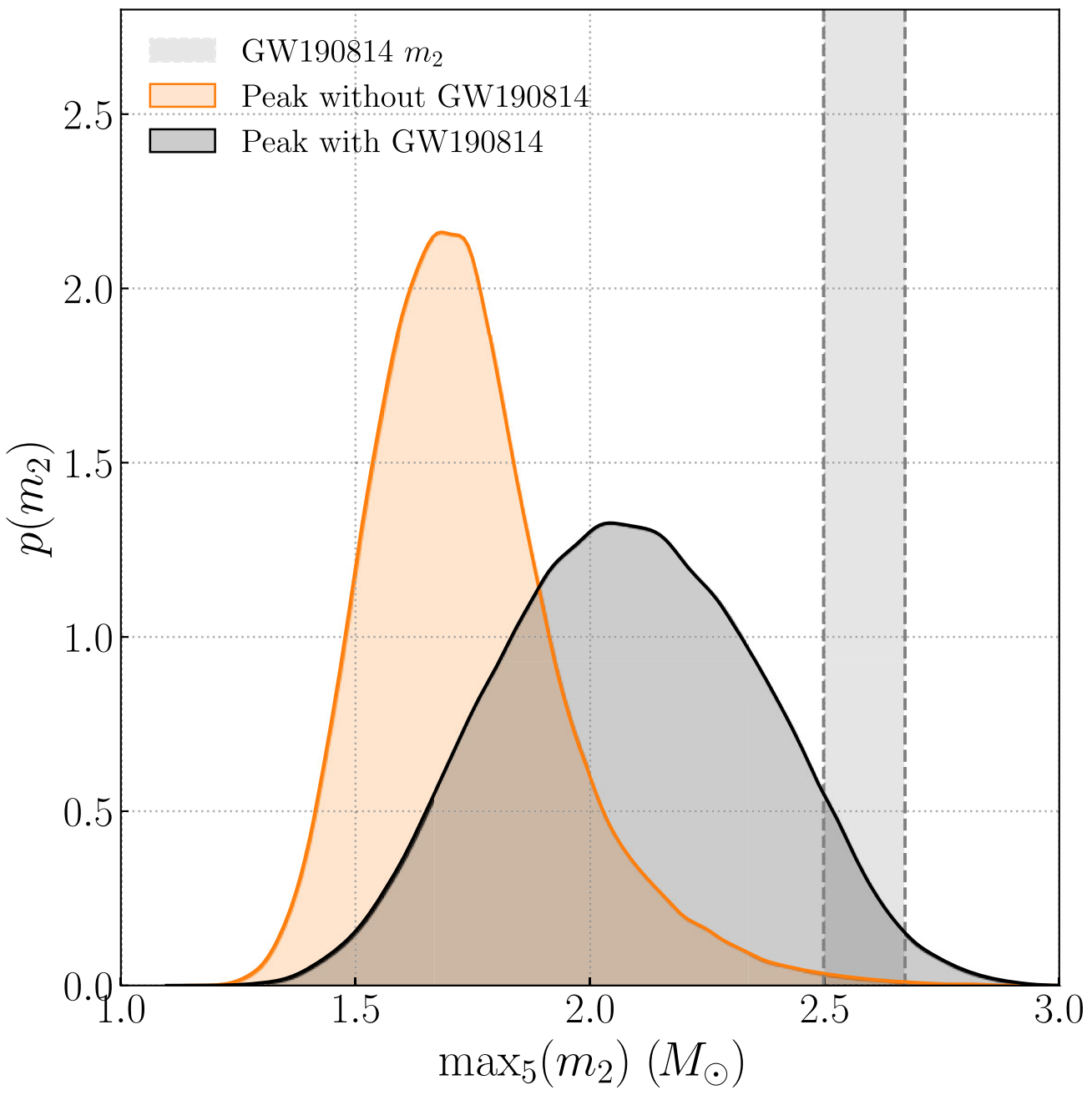

FIG. 8. Comparison between GW190814's secondary component and the largest secondary mass in the observed BNS and NSBH population. The PEAK model is fit to the population including (respectively, excluding) GW190814. The predicted distribution of the largest secondary mass $\max_5(m_2)$ observed after five detections—two BNSs and three NSBHs—is shown in orange (blue). The shaded region represents the 90% credible interval of the posterior distribution for the mass of GW190814's secondary component. GW190814's $m_2$ is a 0.2%-level outlier from the rest of the observed population of NS secondaries.

## VI. MASS DISTRIBUTION OF BLACK HOLES IN BINARIES

We find two key new conclusions about the black hole mass distribution using the GWTC-3 dataset to infer a population: that the mass distribution has a substructure reflected in clustering of detected events, and that observations are consistent with a continuous, monotonically decreasing mass distribution at masses $> 50 M_\odot$, providing inconclusive evidence for an upper mass gap. Adopting previous coarse-grained models, we find conclusions consistent with our analysis of GWTC-2 [20]. For the purposes of this section, given our large BBH sample, we adopt a FAR threshold of 1 $\mathrm{yr}^{-1}$, but we do not include the previously identified outliers GW190917 (likely a NSBH) and GW190814 (a $q \simeq 0.11$ binary) in the BBH population unless otherwise noted [6,7]. Additionally, unlike the redshift-independent results described in Sec. IV, the new analyses described in this section all account for a redshift-dependent BBH merger rate according to Eq. (7). Specifically, in this section we present results for the PP model to broadly characterize the mass spectrum and corroborate results found in GWTC-2, as well as the PS model and the BGP, as both can capture smaller-scale features in the mass distribution. All three models are described in detail in Sec. III C. We report on the same BGP analysis as performed in Sec. IV, with FAR $< 0.25$ $\mathrm{yr}^{-1}$ and without allowing for redshift dependence; by contrast, the PS, PP, and FM models allow for redshift dependence and use FAR $< 1$ $\mathrm{yr}^{-1}$. Table IV summarizes our results

TABLE IV. Merger rates in $\mathrm{Gpc}^{-3}\,\mathrm{yr}^{-1}$ for BBH binaries quoted at the 90% credible interval, for the PP model and for three nonparametric models (BINNED GAUSSIAN PROCESS, FLEXIBLE MIXTURES, POWER LAW+SPLINE). Rates are given for three ranges of primary mass $m_1$, as well as for the entire BBH population. Despite differences in methods, the results are consistent among the models. BGP assumes a nonevolving merger rate in redshift. The merger rate for PP, FM, and PS is quoted at a redshift value of 0.2, the value where the relative error in merger rate is smallest.

| | $m_1 \in [5, 20] M_\odot$ <br> $m_2 \in [5, 20] M_\odot$ | $m_1 \in [20, 50] M_\odot$ <br> $m_2 \in [5, 50] M_\odot$ | $m_1 \in [50, 100] M_\odot$ <br> $m_2 \in [5, 100] M_\odot$ | All BBHs |
|---|---|---|---|---|
| PP | $23.6^{+13.7}_{-9.0}$ | $4.5^{+1.7}_{-1.3}$ | $0.2^{+0.1}_{-0.1}$ | $28.3^{+13.9}_{-9.1}$ |
| BGP | $20.0^{+11.0}_{-8.0}$ | $6.3^{+3.0}_{-2.2}$ | $0.75^{+1.1}_{-0.46}$ | $33.0^{+16.0}_{-10.0}$ |
| FM | $21.1^{+11.6}_{-7.8}$ | $4.3^{+2.0}_{-1.4}$ | $0.2^{+0.2}_{-0.1}$ | $26.5^{+11.7}_{-8.6}$ |
| PS | $27^{+12}_{-8.8}$ | $3.5^{+1.5}_{-1.1}$ | $0.19^{+0.16}_{-0.09}$ | $31^{+13}_{-9.2}$ |
| MERGED | 13.3–39 | 2.5–6.3 | 0.099–0.4 | 17.9–44 |

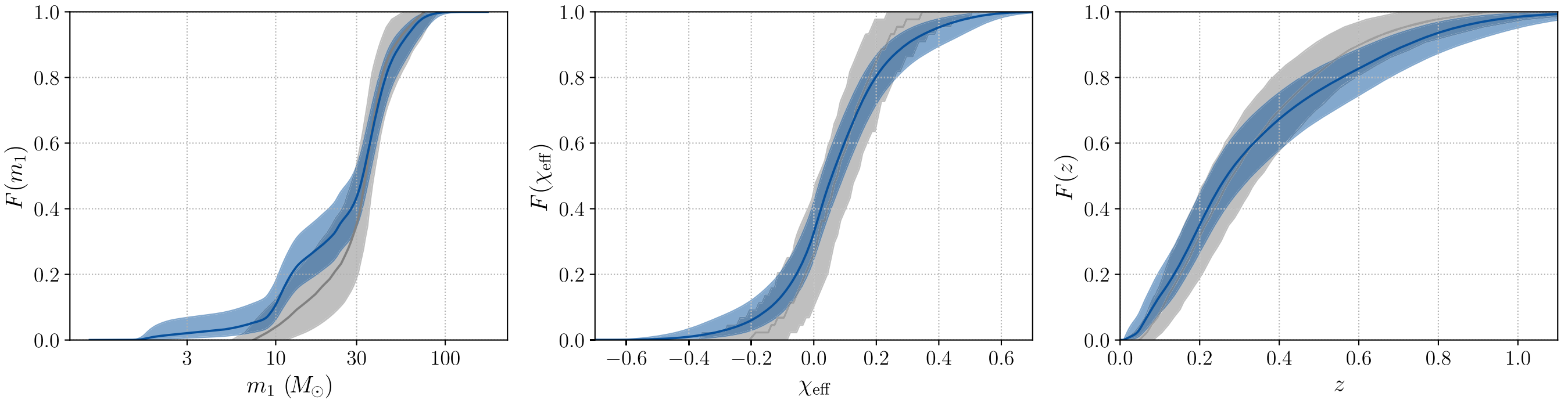

FIG. 9. The empirical cumulative density function $\hat{F} = \sum_k P_k(x)/N$ of observed binary parameter distributions [derived from the single-event cumulative distributions $P_k(x)$ for each parameter $x$] are shown in blue for primary mass (left), effective inspiral spin (center), and redshift (right). All binaries used in this study with FAR $< 0.25\ {\rm yr}^{-1}$ are included, and each is analyzed using our fiducial noninformative prior. For comparison, the gray bands show the expected observed distributions based on our previous analysis of GWTC-2 BBH [20]. Solid lines show the medians, while the shading indicates a 90% credible interval on the empirical cumulative estimate and selection-weighted reconstructed population, respectively. GW190814 is excluded from this analysis.

for the overall BBH merger rate, as well as merger rates over restricted mass intervals.

### A. Broad features of the mass spectrum

The events from GWTC-3 are broadly consistent with the previously identified population [20]. Figure 9 compares some of the expectations from our previous analysis of GWTC-2 BBHs with the comprehensive sample of GWTC-3 BBH events. The panels compare the observed and expected fractions of all events detected below a threshold in primary mass $m_1$, effective inspiral spin $\chi_{\rm eff}$, or source redshift $z$. The panels also show the Wilson score interval [198], a frequentist estimate of the uncertainty in the cumulative distribution $F$, which is approximately $\pm 1.68\sqrt{F(1-F)/N_{\rm obs}}$ when $F$ is significantly different from 0 or 1.

All the cumulative distributions in Fig. 9 are broadly consistent with our prior expectations based on coarse-grained models used in our previous work. For this reason, we begin by presenting the inferred coarse-grained mass distribution of black hole binaries, making use of the PP model [20] which best fits the population from GWTC-2.

Figure 10 shows our inference on the astrophysical primary mass (left) and mass ratio (right) distributions, using the fiducial mass model, compared to what was previously found in GWTC-2 (black). We find a power-law slope for the primary mass, $\alpha = 3.5^{+0.6}_{-0.56}$ ($2.6^{+0.79}_{-0.63}$ in GWTC-2), supplemented by a Gaussian peak at $34^{+2.6}_{-4.0} M_\odot$ ($33^{+4.0}_{-5.6}$ in GWTC-2). On the upper end, the mass of the 99th percentile $m_{99\%}$ is found to be $44^{+9.2}_{-5.1} M_\odot$. The mass ratio distribution is modeled as a power law $q^{\beta_q}$ with $\beta_q = 1.1^{+1.7}_{-1.3}$.

In contrast to our GWTC-2 population fit, the inferred mass spectrum decays more rapidly; the $m_{99\%}$ is considerably lower than $60^{+14}_{-13} M_\odot$, as was found with GWTC-2. These results are expected, given that the new observations in GWTC-3 contain a greater fraction of lower mass systems (see, e.g., Fig. 1). The fraction of BBH mergers with primary masses within the Gaussian component of the fiducial model is found to be $\lambda = 0.038^{+0.058}_{-0.026}$ ($0.1^{+0.14}_{-0.071}$ in GWTC-2), but still rules out zero. This result further highlights that the fraction of higher-mass binaries has decreased in GWTC-3. Both the mean and the standard deviation of the Gaussian component are consistent with previous inferences. Furthermore, the inferred mass ratio distribution is less peaked toward equal-mass binaries ($\beta_q = 1.1^{+1.7}_{-1.3}$) compared to GWTC-2 ($\beta_q = 1.3^{+2.4}_{-1.5}$), a result driven by the discovery of binaries with support for substantially unequal masses (see, e.g., Fig. 9).

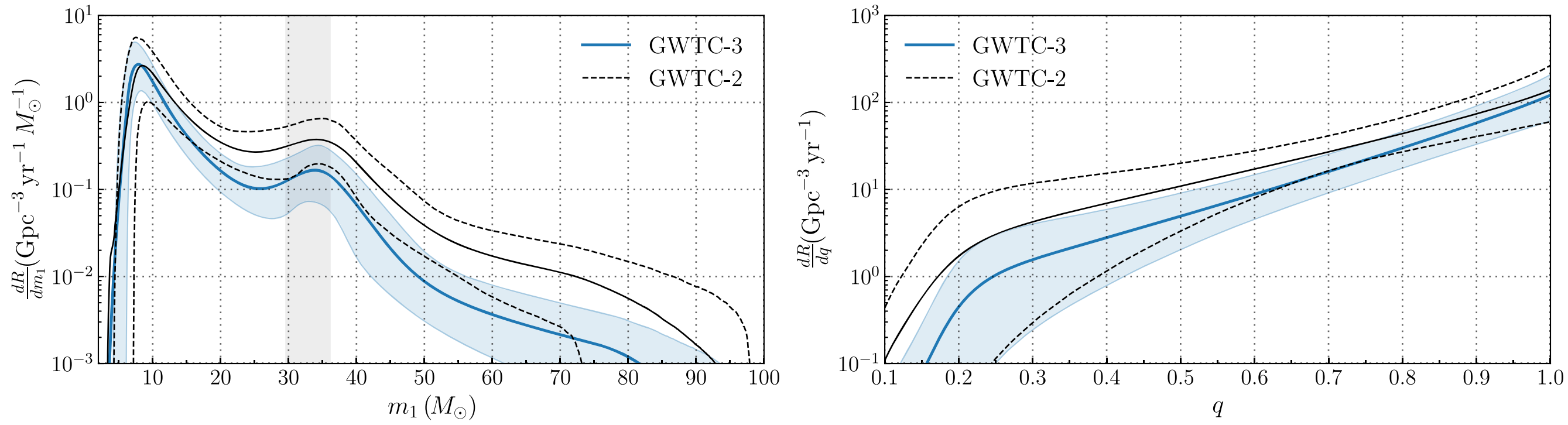

FIG. 10. The astrophysical BBH primary mass (left) and mass ratio (right) distributions for the fiducial PP model, showing the differential merger rate as a function of primary mass or mass ratio. The solid blue curve shows the posterior population distribution (PPD) with the shaded region showing the 90% credible interval. The black solid and dashed lines show the PPD and 90% credible interval from analyzing GWTC-2 [20]. The vertical gray band in the primary mass plot shows 90% credible intervals on the location of the mean of the Gaussian peak for the fiducial model.

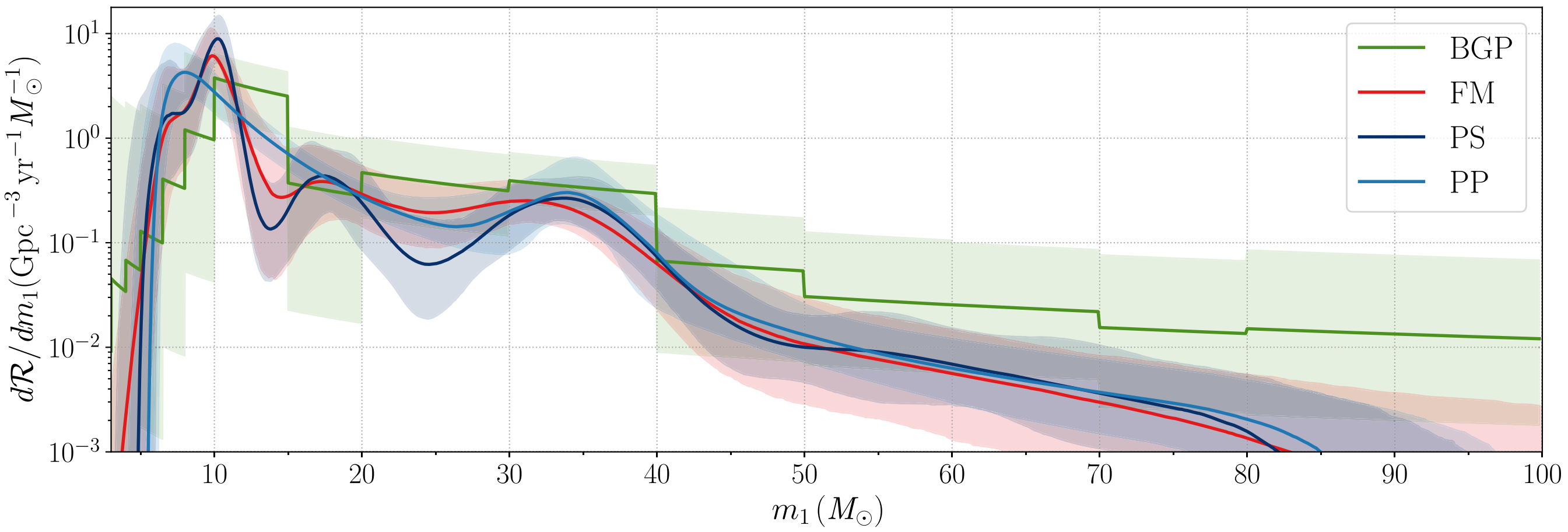

FIG. 11. The differential merger rate for the primary mass predicted using three nonparametric models compared to the fiducial PP model. Solid curves are the medians and the colored bands are the 90% credible intervals. These models offer increased flexibility compared to phenomenological models in predicting the population. The PS applies a perturbation to the primary mass in a modified version of our fiducial PP model that does not include the Gaussian peak. FM models the chirp mass, mass ratio, and aligned-spin distribution as a weighted sum of mixture components. Both of these models incorporate a single-parameter redshift evolution of the merger rate described in Eq. (7). The BGP models the two-dimensional mass distribution as a binned Gaussian process which is piecewise constant in $\log m_i$, illustrating the same analysis as presented in Sec. IV with FAR $< 0.25\ \mathrm{yr}^{-1}$. All three models infer a local maxima in the merger rate at around 10 amd $35M_\odot$. In addition, FM and PS models are consistent with a feature around $17M_\odot$, but with a limited confidence.

We previously used several other phenomenological models to interpret sources in GWTC-2. Using this broader suite of models, we draw similar conclusions to those presented above: The mass distribution is inconsistent with the single power law and has a feature at approximately 35–$40M_\odot$. The peak's location is also well separated from the largest black holes predicted by the other component: The overdensity and maximum mass are still not associated. The odd ratios discriminating between these models are modest, of order 1 in 3 to 1 in 10. Though we present an updated PP mass distribution derived from GWTC $= 3$ to illustrate consistency with the corresponding PP results obtained with GWTC-2, we cannot decisively differentiate between a peak near $35M_\odot$ versus a more generic transition toward a lower merger rate at higher mass; see Appendix D 1 for details.

In Table IV, we provide BBH merger rates for the full population, as well as split based upon the primary mass at $m_1 < 20M_\odot$, $m_1 \in [20, 50]M_\odot$, and $m_1 > 50M_\odot$ to capture the broad features of the mass spectrum: the high rate at low masses, a peak around 35–$40M_\odot$ and the falling merger rate at high masses.

### B. Mass distribution has substructure

With new discoveries in O3, we are now confident the mass distribution has substructure, with localized peaks in the component mass distribution. For example, we find overdensities in the merger rate (> 99% credibility) as a function of primary mass, when compared to a power law, at $m_1 = 10^{+0.29}_{-0.59}M_\odot$ and $m_1 = 35^{+1.7}_{-2.9}M_\odot$. At best, we have modest confidence (less than 10 : 1 odds) in additional structure. These signs of substructure were identified in O3a [94] and are corroborated by consistent observations in O3b.

We arrive at these conclusions through multiple independent analyses. Each of these model agnostic approaches attempts to reconstruct the mass distributions with minimal constraints imposed. Specifically, we employ a flexible mixture model (introduced in Sec. III C and labeled FM in tables and figures), a PS, and a BGP. Figure 11 shows the inferred rate $d\mathcal{R}/dm_1$ as a function of primary mass for each of the nonparametric models. There is a clear presence of structure beyond an unbroken single power law found when using these more flexible models, with a global maximum of the merger rate at larger masses at around $10M_\odot$ followed by a falloff to lower rates. Modulating this extended decline, the PS, FM, and even BGP show indications of additional structure. As the BGP likely cannot resolve small-scale features, we assess these features' details and significance with the remaining two models.

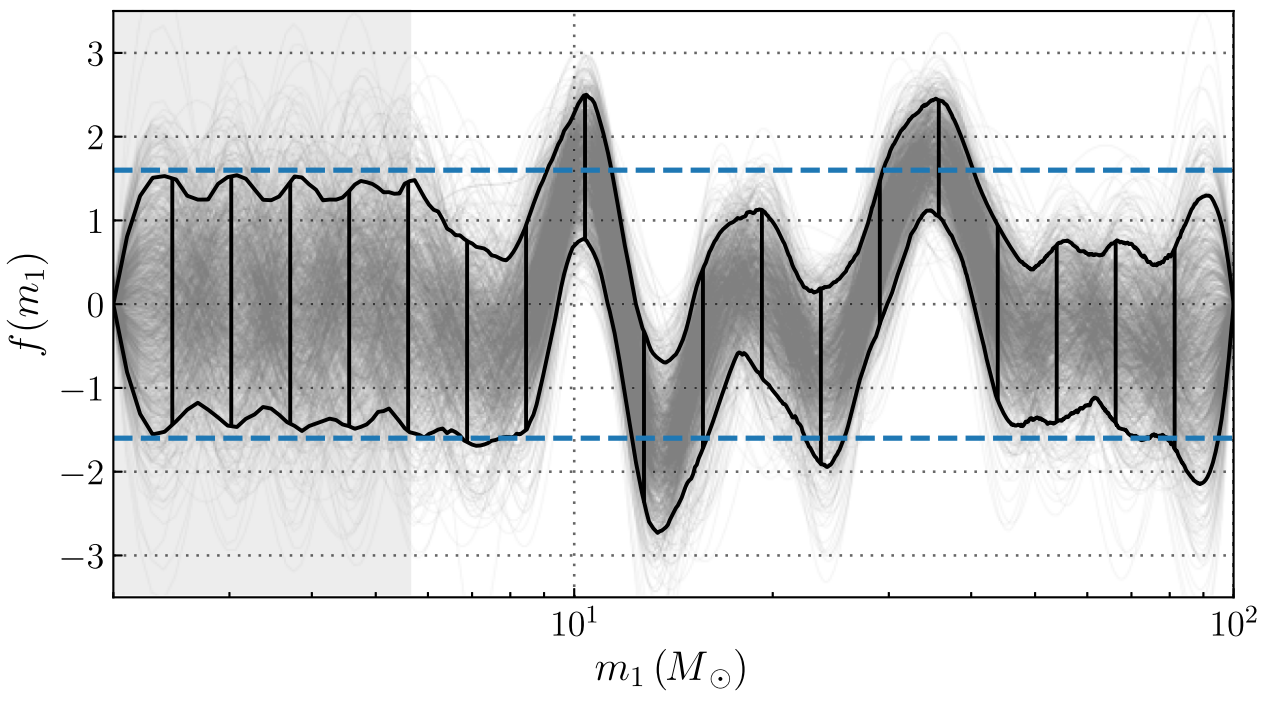

FIG. 12. The cubic spline function $f(m_1)$ describing the perturbations to an underlying power law inferred with the PS model. The thin gray lines show 1000 draws from the posterior while the black lines show the knot locations (vertical) and the 90% credible region of the posterior. The dashed blue lines mark the 90% credible bounds of the Gaussian priors (centered on zero) imposed on each knot's height. The shaded region covers any masses less than the 95th percentile of the marginal posterior distribution on $m_{\min}$. Because the low-mass region of the mass distribution is cut off and there are no observations there, the posterior in this region resembles the prior of the cubic spline function.

Figure 12 shows the results of the spline perturbation model, where 1000 posterior draws of the spline function $f(m_1)$ are illustrated, where $\exp f(m_1)$ modulates an underlying power-law distribution. The inferred perturbation $f(m_1)$ strongly disfavors zero at both the 10 and $35M_\odot$ peak, finding $f \leq 0$ at 0.216%, and $< 0.0325\%$ credibilities, respectively. Additionally, for the drop in merger rate at $14M_\odot$, the PS model finds $f \leq 0$ at 96.1% credibility.

### C. Inconclusive evidence for upper mass gap

Stellar evolution models predict a lack of black holes with masses from $50^{+20}_{-10}M_\odot$ to approximately $120M_\odot$ due to pair production instability [199–205]. The high-mass event GW190521 could have one or both components lying within this mass gap [17,206]. Other analyses of this event with independent parameter inferences have argued this event could have both components with masses outside of this range [207–209]. We define a gap as a rapid decline to zero in merger rate at some mass, followed by a rapid rise in the distribution at a higher mass. Repeating similar analyses with the full O3 dataset, we find no evidence for such a gap starting at masses $< 75M_\odot$. Following Ref. [209], we extend the PP mass model to allow for masses $> 100M_\odot$ and to include a zero-rate mass interval parametrized with the lower edge and width of the gap. Because of the uncertain origin of the previously reported peak at approximately $35M_\odot$, we do not enforce the start of the gap to be immediately following the peak in the model, which would be expected if this feature was the pileup caused by pulsational pair-instability supernovae [203,210]. With this extended model, we find minimal posterior support for the gap to start at $< 75M_\odot$ (3.1% credibility). In the region where there is posterior support for a gap starting at $< 75M_\odot$, the gap width is constrained to be $< 35M_\odot$, lower than the expected gap width of approximately $50M_\odot$ [211]. The majority of the posterior support has the gap start above $75M_\odot$, consistent with the inferred maximum mass cutoff from the PP model without including a gap. A mass gap at that range would allow both component masses of the most massive BBH in the catalog, GW190521, to fall below the cutoff; thus, we are not able to determine whether or not the mass distribution exhibits a rise again at higher masses. We find a slight preference ($\ln \mathcal{B} = 0.06$) for the PP model without a gap over one with the gap included. We report inconclusive evidence for a zero-rate upper mass gap, challenging the classical expectations for the pair-instability mass gap. Our results are consistent with either a pair-instability mass gap that starts at or higher than the upper bound of the observed catalog (i.e., $> 75M_\odot$), or the possibility that a non-negligible fraction of the high-mass binaries in the catalog were formed in a way that avoids pair instability [162,212–214].

### D. Evolution of rate with redshift

The observation of BBH mergers offers us the means of not only measuring the local merger rate per comoving volume but also the evolution of this merger rate as we look back toward larger redshifts $z$. Given the limited range of redshift to which our searches are sensitive, we parametrize the merger rate per comoving volume as a simple power law, with $\mathcal{R}(z) \propto (1+z)^\kappa$ [146].

In our previous study [20], the redshift distribution was weakly constrained, exhibiting a preference for a rate that increased with redshift but still consistent with a non-evolving merger rate. Here, in addition to new events observed in O3b, we leverage updated pipelines and our improved sensitivity models to update our inference of $\kappa$. As we discuss further in Appendix C 5, these sensitivity model refinements indicate a lower search sensitivity to high-redshift BBH mergers than previously concluded. We now confidently claim to see evolution of the BBH merger rate with redshift in our population with a FAR $< 1\ \mathrm{yr}^{-1}$, inferring that $\kappa > 0$ at 99.6% credibility. While the exact distribution of $\kappa$ does depend on the chosen mass model, we can rule out a redshift-independent merger rate at similar credible levels when adopting any of the parametrized mass distribution models used in Ref. [20].

Figure 13 shows the marginal posterior on $\kappa$ given GWTC-3 in blue, obtained while using the PP and Default mass and spin models. The dashed distribution, meanwhile, shows the previously published measurement of $\kappa$ obtained with GWTC-2. In Fig. 13 we also show our corresponding constraints on $\mathcal{R}(z)$ itself as a function of redshift. The dark blue line traces our median estimate on $\mathcal{R}(z)$ at each redshift, while the dark and light shaded regions show central 90% and 50% credible bounds. Our best measurement of the BBH merger rate occurs at $z \approx 0.2$, at which $\mathcal{R}(z = 0.2) = 19$–$42\ \mathrm{Gpc}^{-3}\,\mathrm{yr}^{-1}$. For comparison, the dashed black line in Fig. 13 is proportional to the Madau-Dickinson star formation rate model [215], whose evolution at low redshift corresponds to $\kappa_{\mathrm{SFR}} = 2.7$. While the rate evolution remains consistent with the Madau-Dickinson star formation rate model, it is not expected for these two rates to agree completely due to the time delays from star formation to merger [35,216–223].

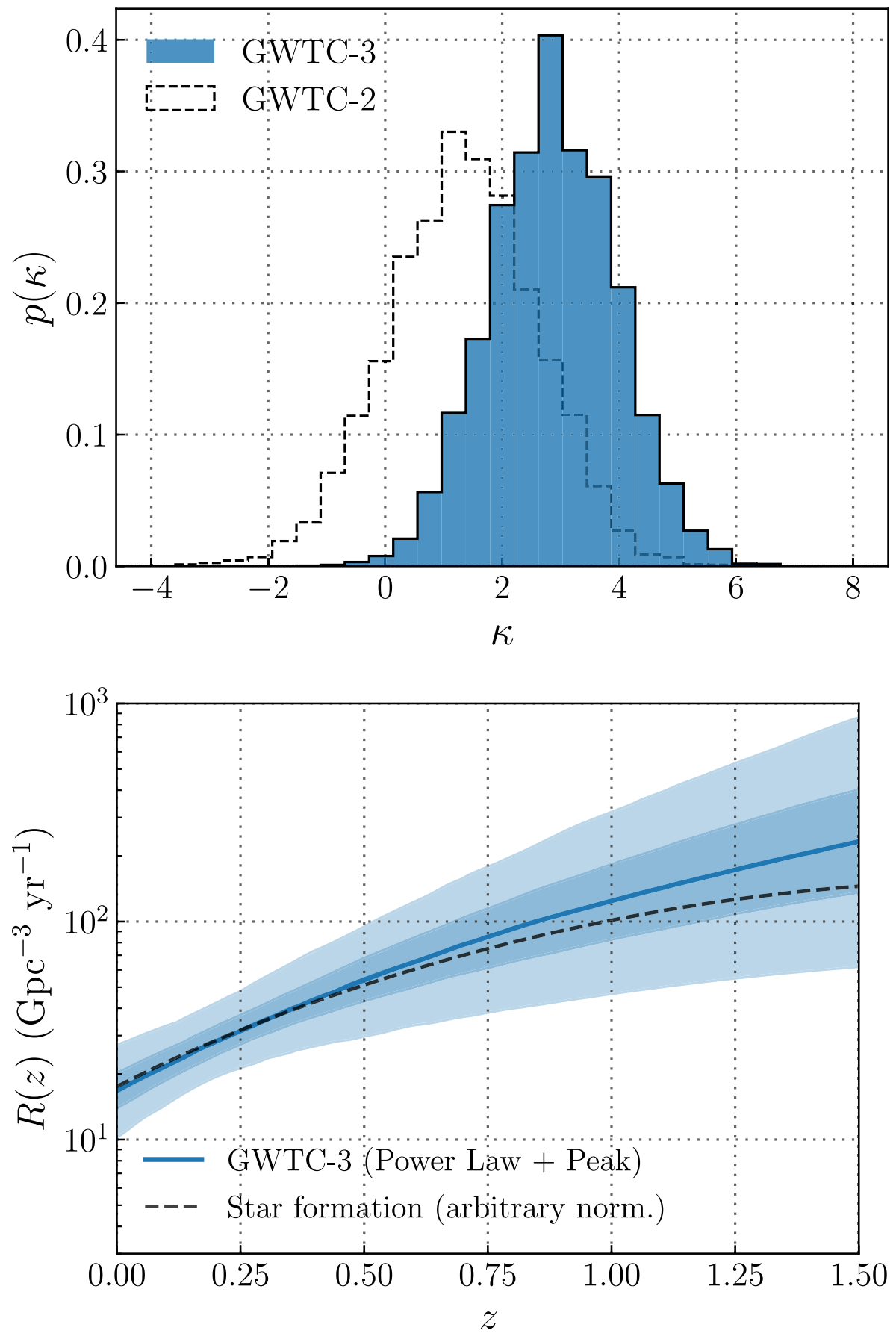

FIG. 13. Constraints on the evolution of the BBH merger rate with redshift. Top: posterior on the power-law index $\kappa$ governing the BBH rate evolution, which is presumed to take the form $\mathcal{R}(z) \propto (1+z)^\kappa$. The blue histogram shows our latest constraints using GWTC-3 ($\kappa = 2.9^{+1.7}_{-1.8}$), while the dashed distribution shows our previous constraints under GWTC-2. Bottom: central 50% (dark blue) and 90% (light blue) credible bounds on the BBH merger rate $\mathcal{R}(z)$. The dashed line, for reference, is proportional to the rate of cosmic star formation [215]; we infer that $\mathcal{R}(z)$ remains consistent with evolution tracing star formation.

In most plausible formation scenarios (e.g., if BBHs arise from stellar progenitors), we do not expect $\mathcal{R}(z)$ to continue growing with arbitrarily high $z$. Instead, we anticipate that $\mathcal{R}(z)$ will reach a maximum beyond which it turns over and falls to zero. Even in cases where the peak redshift $z_p$ at which $\mathcal{R}(z)$ is maximized lies beyond the LIGO-Virgo detection horizon, a sufficiently tight upper limit on the stochastic gravitational-wave background due to distant compact binary mergers [224–226] can be leveraged to bound $z_p$ from above, potentially yielding a joint measurement of $\kappa$ and $z_p$ [227]. As demonstrated in previous analysis [228], our current instruments are not yet sensitive enough to enable a meaningful joint constraint on $\kappa$ and $z_p$, even with the inclusion of new events in GWTC-3.

As heavy BBHs are primarily believed to arise from low-metallicity stellar progenitors [31,63,229], one might wonder if more massive BBHs are observed at systematically higher redshifts than less massive systems. Moreover, any metallicity dependence in the physics of stars, such as the maximum black hole mass imposed by pair-instability supernovae (PISN) [199,201,210], could yield redshift-dependent features in the black hole mass distribution [230,231]. Such a redshift dependence would confound efforts to leverage the PISN mass gap as a probe of cosmology. Previous investigations [232] demonstrated using GWTC-2 that redshift dependence of the maximum BBH mass would be required to fit the observations if the BBH mass distribution has a sharp upper cutoff. However, if the distribution decays smoothly at high masses, for example, as a power law, the data are consistent with no redshift dependence of the cutoff location.

We revisit this question using the latest BBH detections among GWTC-3, finding that these conclusions remain unchanged. Specifically, by modeling the high-mass tail of the distribution with a separate power-law index, we find no evidence that the distribution is redshift dependent, suggesting that the high-mass structure in the BBH mass distribution remains consistent across redshift.

### E. Outliers in the BBH population

Several systems lie at the boundary between the BH and NS categories [7,197]. So far, we have simply excluded these events from our BBH analysis. To demonstrate that this choice is internally self-consistent and well motivated, we show that these events are outliers from our recovered BBH population. Specifically, we repeat the population analysis using the PP model, highlighting the extent to which the population changes when including these events.

For a population consisting of all potential BBH events in O3, including GW190917 and GW190814, the mass distribution must extend to lower masses. In Fig. 14 we plot the recovered distribution for the minimum BH mass, $m_{\mathrm{min}}$, that characterizes the primary mass scale above which black holes follow the parametrized power-law distribution. The minimum mass is $m_{\rm min} = 2.3^{+0.27}_{-0.23} M_\odot$, with an extremely sharp turn-on of $\delta_m = 0.39^{+1.3}_{-0.36} M_\odot$. By contrast, if we remove the two low-mass events, we find a minimum BH mass of $m_{\rm min} = 5.0^{+0.86}_{-1.7} M_\odot$, which is consistent with a mass gap, and a broader turn-on of $\delta_m = 4.9^{+3.4}_{-3.2} M_\odot$. It is the secondary masses $m_2$ of these events that are in tension with the remainder of the population, as demonstrated in Fig. 14 where the secondary masses are shown by the shaded regions. A single minimum mass is imposed upon all BHs; therefore, the secondary masses of low mass or asymmetric binaries have the strongest impact on our inference of $m_{\rm min}$.

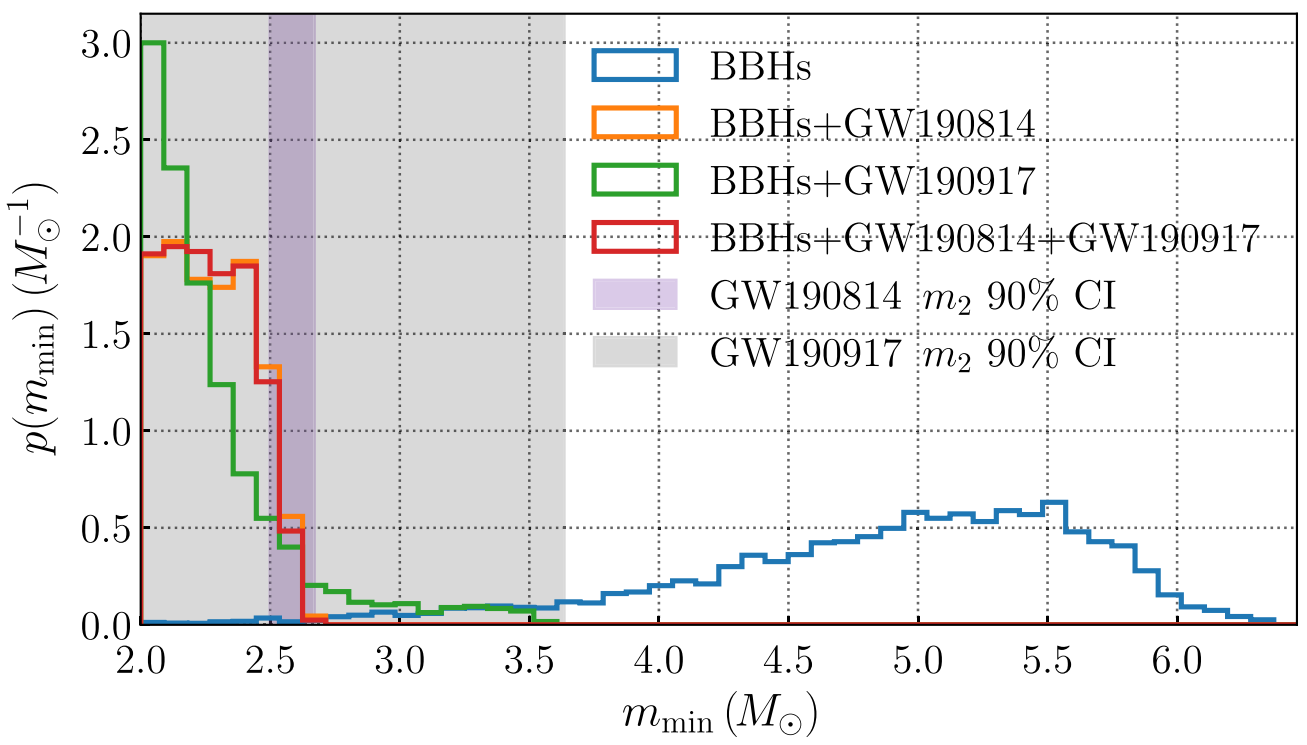

FIG. 14. The posterior distribution on the minimum mass truncation hyperparameter $m_{\rm min}$ inferred with the PP model. The posteriors are shown both including and excluding the two BBH mergers containing low-mass secondaries, GW190814 and GW190917. The cutoff at $m_{\rm min} = 2M_\odot$ corresponds to the lower bound of the prior distribution. The inclusion of either of these two events significantly impacts the distribution. The shaded regions indicate the 90% credible interval on the $m_2$ posterior distribution for the two outlier events GW190814 (purple) and GW190917 (gray).

These analyses imply two key results about the compact binary population. First, the binary black hole population excluding highly asymmetric systems such as GW190814 is well defined, and the analyses carried out in this section are well suited to characterizing the bulk of the BBH population. Second, the detection of GW190814 implies the existence of a subpopulation of highly asymmetric binaries disconnected from the BBH population but potentially connected to the recently identified population of NSBHs. We reach the conclusion that the subpopulation is most likely connected to the recently identified population of NSBHs because GW190814 would be an outlier in the BBH population given the inferred value of $m_2$. Different physical possibilities can potentially account for the existence of such a subpopulation of asymmetric binaries [233–236].

## VII. SPIN DISTRIBUTION OF BLACK HOLES IN BINARIES

Our previous work indicated that black hole spins are small in magnitude and preferentially aligned with their orbital angular momenta [20]. Here, we find two new key conclusions for black hole spins: that the spin distribution broadens above $30M_\odot$, and that the mass ratio and spin are correlated. Adopting previous coarse-grained models, we find consistent conclusions as our analysis of GWTC-2; notably, we still conclude that a fraction of events probably have negative $\chi_{\rm eff}$.

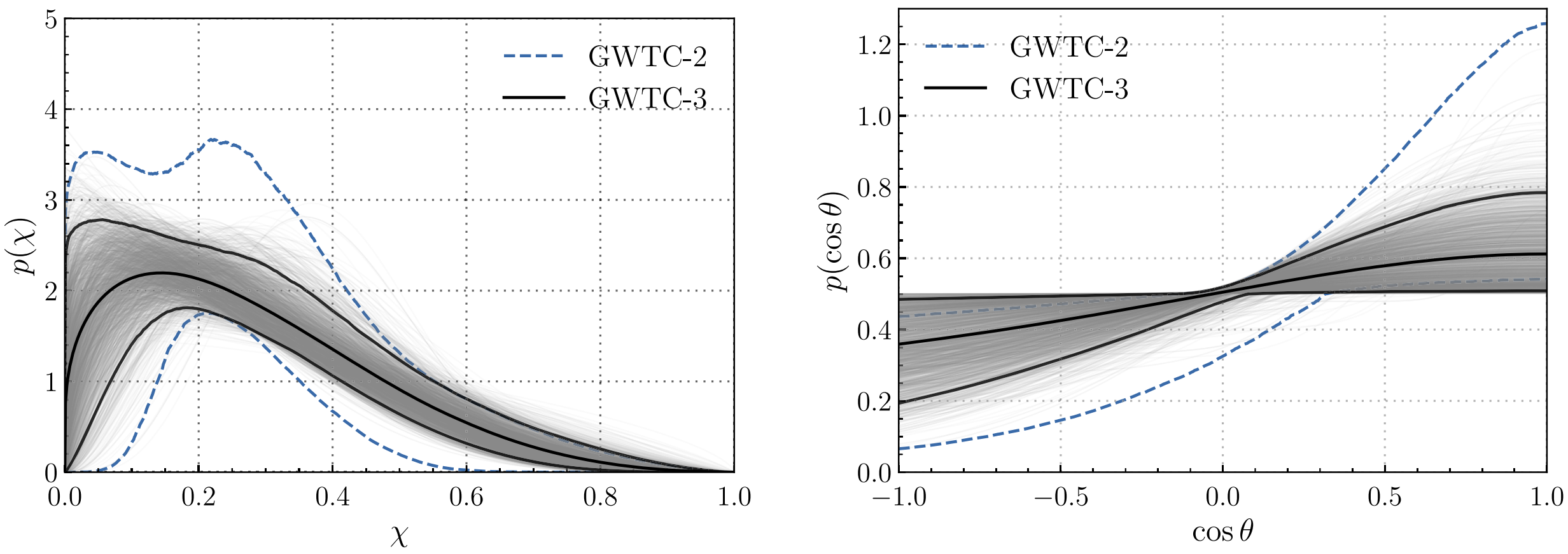

FIG. 15. The distributions of component spin magnitudes $\chi$ (left) and spin-orbit misalignment angles $\theta$ (right) among binary black hole mergers, inferred using the DEFAULT component spin model described further in Sec. 2 a; e.g., both spin magnitudes are drawn from the same distribution. In each figure, solid black lines denote the median and central 90% credible bounds inferred on $p(\chi)$ and $p(\cos\theta)$ using GWTC-3. The light gray traces show individual draws from our posterior distribution on the DEFAULT model parameters, while the blue traces show our previously published results obtained using GWTC-2. As with GWTC-2, in GWTC-3 we conclude that the spin-magnitude distribution peaks at $\chi_i = 0.13^{+0.12}_{-0.11}$, with a tail extending toward larger values. Meanwhile, we now more strongly favor isotropy, obtaining a broad $\cos\theta_i$ distribution that may peak at alignment ($\cos\theta_i = 1$) but that is otherwise largely uniform across all $\cos\theta$.

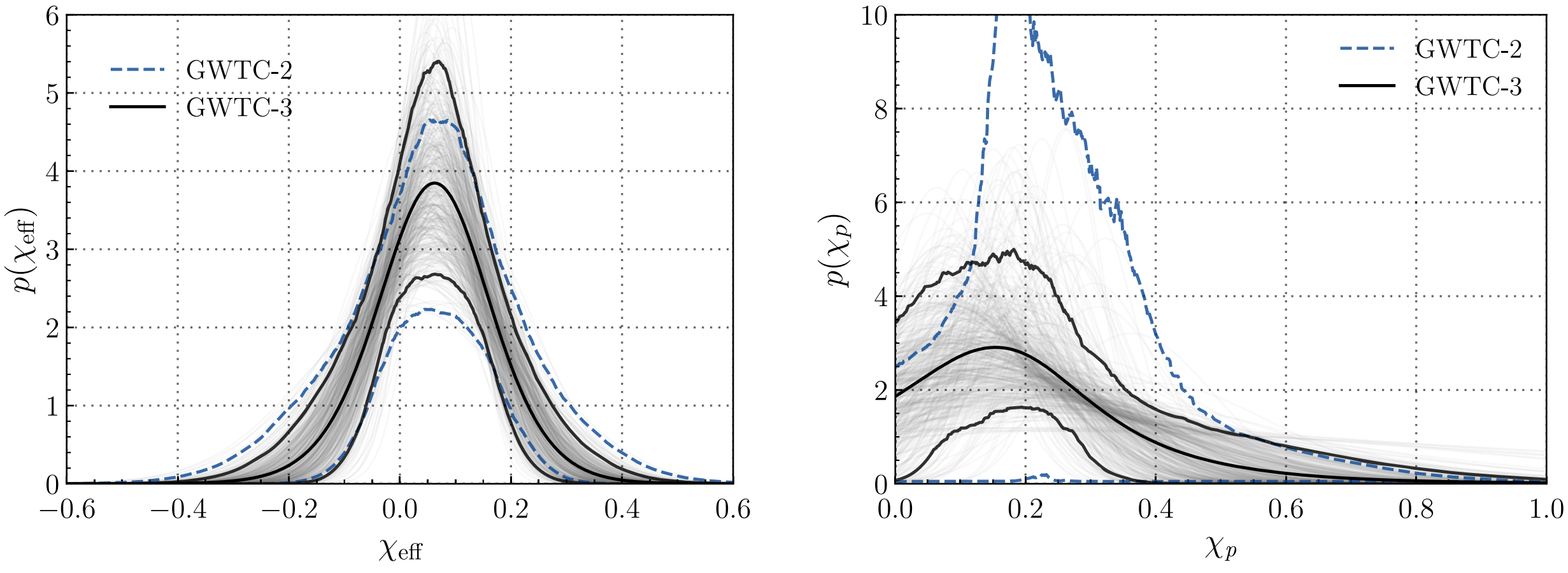

FIG. 16. Left panel: inferred distribution of $\chi_{\mathrm{eff}}$ for our latest full analysis in black. For comparison, the blue distribution and interval shows our inferences derived from GWTC2. Right panel: corresponding result for $\chi_{\mathrm{p}}$. While both panels in this figure are derived using the Gaussian spin model, we find similar conclusions with the other spin models used to analyze GWTC-2.

The component spins of binary black holes may offer vital clues as to the evolutionary pathways that produce merging BBHs [237–244]. The magnitudes of BBH spins are expected to be influenced by the nature of angular momentum transport in stellar progenitors [61–63], natal kicks experienced upon core collapse [239,243–248], processes like tides [65,243,249] and mass transfer that operate in binaries, and the environment in which the binary itself is formed. Their directions, meanwhile, may tell us about the physical processes by which binaries are most often constructed; we expect BBHs born from isolated stellar evolution to possess spins preferentially aligned with their orbital angular momenta, while binaries that are dynamically assembled in dense environments are predicted to exhibit isotropically oriented spins [237,239].

Figure 15 illustrates our constraints on the component spin magnitudes (left) and spin tilts (right) of BBHs under the Default spin model. Using GWTC-3, we make similar conclusions regarding the spin-magnitude distribution as made previously with GWTC-2. In particular, spin magnitudes appear concentrated below $\chi_i \lesssim 0.4$, with a possible tail extending toward large or maximal values. Our understanding of the spin-tilt distribution, in contrast, has evolved with the addition of new BBHs in GWTC-3. As in GWTC-2, we again exclude the case of perfect spin-orbit alignment (corresponding to $\zeta = 1$ and $\sigma_t = 0$). With GWTC-3, however, we more strongly favor a broad or isotropic distribution of spin tilts. This shift is seen in the right-hand side of Fig. 15: Whereas the $\cos\theta$ distribution inferred from GWTC-2 was consistent with tilts concentrated preferentially around $\cos\theta = 1$, evidence for this concentration is now diminished, with O3b results preferring a flatter distribution across $\cos\theta$. Under the Default model, we infer $44^{+6}_{-11}\%$ of black holes in merging binaries to have spins inclined by greater than 90°.

Figure 16 illustrates our updated constraints on the $\chi_{\mathrm{eff}}$ and $\chi_{\mathrm{p}}$ distributions under the Gaussian spin model. As above, our previous results obtained with GWTC-2 are shown in blue, while black curves show our updated measurements with O3b. Measurement of the $\chi_{\mathrm{eff}}$ distribution with GWTC-2 suggest an effective inspiral spin distribution of nonvanishing width centered at $\chi_{\mathrm{eff}} \approx 0.05$, while the $\chi_{\mathrm{p}}$ distribution appears incompatible with a narrow distribution at $\chi_{\mathrm{p}} = 0$, bolstering the conclusion above that the BBH population exhibits a range of nonvanishing spin-tilt misalignment angles. These conclusions are further strengthened when updating our analysis with GWTC-3. We again infer a $\chi_{\mathrm{eff}}$ distribution compatible with small but nonvanishing spins, with a mean centered at $0.06^{+0.04}_{-0.05}$. This inferred distribution extends also to negative values of $\chi_{\mathrm{eff}}$, with $29^{+15}_{-13}\%$ of binaries having $\chi_{\mathrm{eff}} < 0$.

Our updated constraints on the effective precession spin distribution reaffirm the need for nonvanishing $\chi_{\mathrm{p}}$ among the BBH population. The $\chi_{\mathrm{p}}$ measurements made previously with GWTC-2 were consistent with both a broad underlying distribution or a narrow distribution centered at $\chi_{\mathrm{p}} \approx 0.3$; this latter possibility is the source of the apparent jaggedness seen in the GWTC-2 result. We draw similar conclusions with GWTC-3, finding that $\chi_{\mathrm{p}}$ measurements can be explained either by a broad distribution centered at $\chi_{\mathrm{p}} = 0$, or a narrow distribution centered at $\chi_{\mathrm{p}} \approx 0.2$. If we include GW190814 in our sample (which is otherwise excluded by default from our BBH analyses), support for this second mode is diminished, leaving a zero-centered $\chi_{\mathrm{p}}$ distribution with standard deviation $0.16^{+0.15}_{-0.08}$.

In addition to the distributions of effective inspiral spins and component spins of BBHs, respectively, we also explore the distributions of the more and less rapidly spinning components among the BBH population [250]. For a given binary, we define $\boldsymbol{\chi_A} = \max_{|\chi|}(\boldsymbol{\chi_1}, \boldsymbol{\chi_2})$ and $\boldsymbol{\chi_B} = \min_{|\chi|}(\boldsymbol{\chi_1}, \boldsymbol{\chi_2})$ as the component spins with the larger and smaller magnitudes, respectively. As discussed in Sec. IX, some models for stellar evolution and explosion predict that isolated black holes are born effectively nonrotating and that binary black hole systems primarily acquire spin through tidal spin-up of the secondary component by the first-born (nonspinning) black hole. If this is the case, then we expect to observe a nonvanishing distribution of $\boldsymbol{\chi_A}$ but a distribution of $\boldsymbol{\chi_B}$ concentrated at or near zero. Figure 17 shows the resulting distributions of these spin-sorted magnitudes $\chi_A$ (blue) and $\chi_B$ (green), as implied by the DEFAULT model constraints on component spin magnitudes and tilt angles. Light traces show individual posterior draws on these distributions, while dark traces mark our central 90% credible bounds.

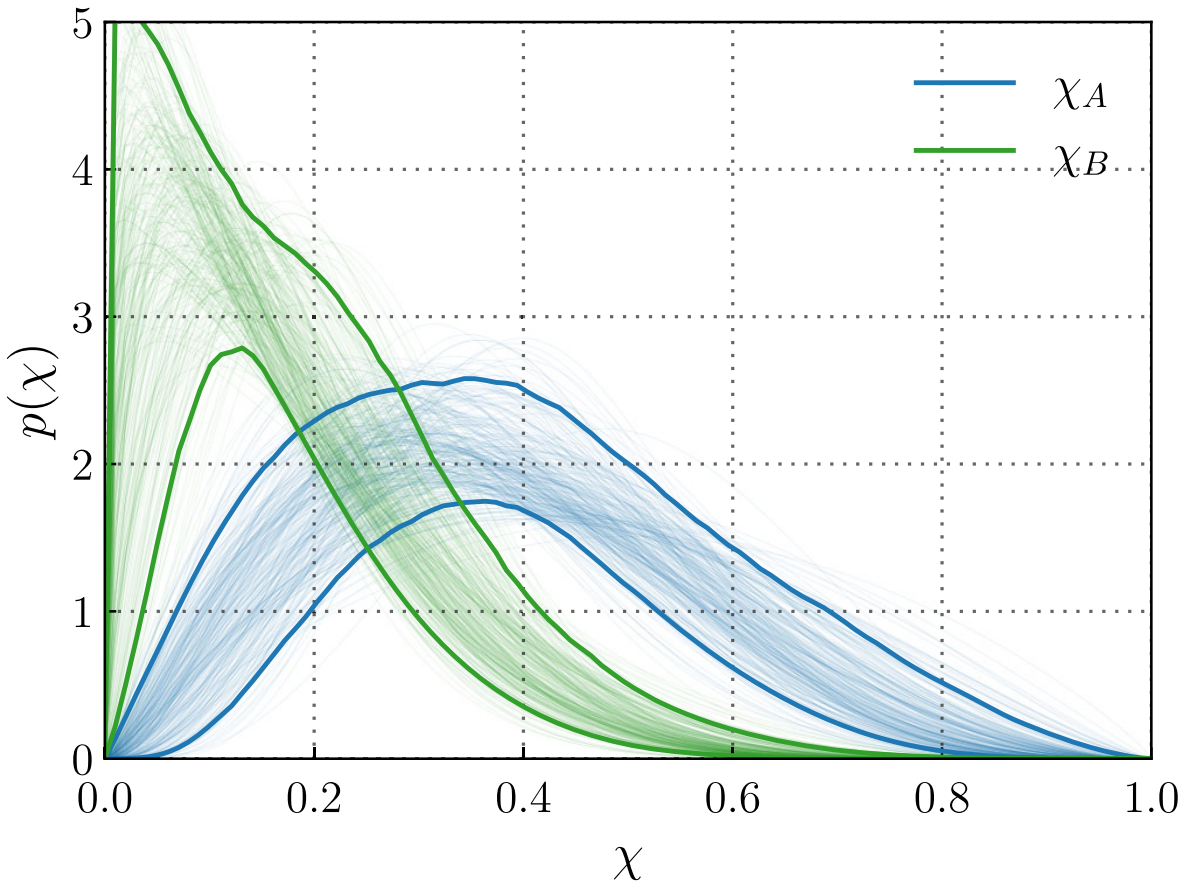

FIG. 17. Distribution of magnitudes of the most ($\chi_A$) and least ($\chi_B$) rapid component spin among BBHs in GWTC-3. Traces show individual draws from our posterior on the spin population under the DEFAULT model, while dark curves indicate 90% credible bounds on $p(\chi_A)$ and $p(\chi_B)$.

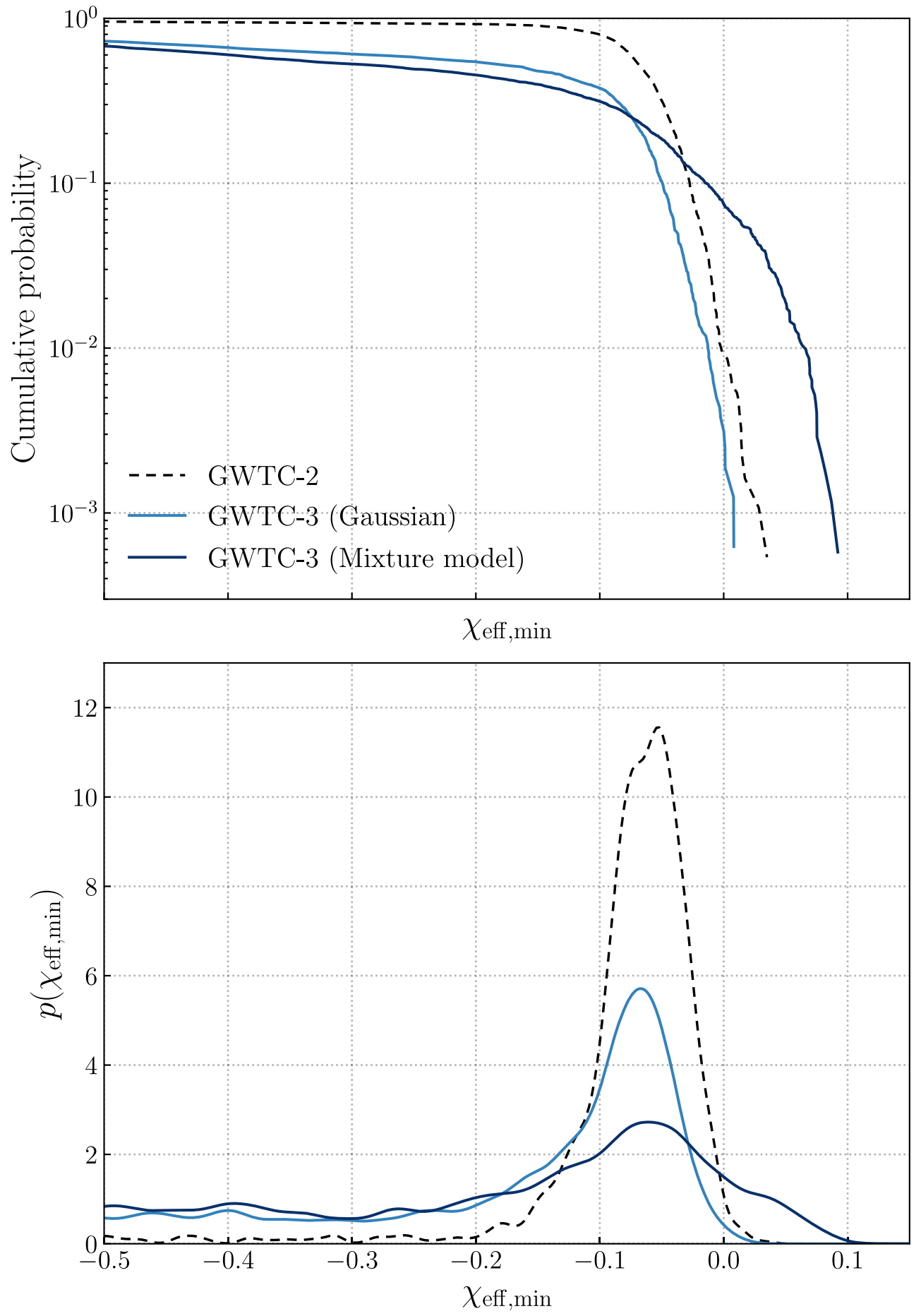

FIG. 18. Cumulative probability (top) and probability density (bottom) of the minimum truncation bound on the $\chi_{\mathrm{eff}}$ distribution as inferred using GWTC-2 and GWTC-3. When modeling the effective inspiral spin distribution as a Gaussian truncated on $\chi_{\mathrm{eff,min}} \le \chi_{\mathrm{eff}} \le 1$, we infer using GWTC-2 that $\chi_{\mathrm{min,eff}} < 0$ at 99.1% credibility, and hence that the data support the existence of BBH mergers with negative effective inspiral spins. Using GWTC-3, this same analysis more strongly infers that $\chi_{\mathrm{min,eff}} < 0$, now at 99.7% credibility. As we discuss further below, evidence for negative effective inspiral spins is diminished under an expanded model that allows for a subset of BBHs to possess vanishing effective inspiral spins. When instead modeling the $\chi_{\mathrm{eff}}$ distribution as a mixture between a broad Gaussian and a narrow Gaussian subpopulation centered at $\chi_{\mathrm{eff}} = 0$ (e.g., the second consistent with zero spin), we infer $\chi_{\mathrm{min,eff}} < 0$ at 92.5% credibility.

The $\chi_A$ distribution, by definition, is concentrated at larger values than the peak seen in Fig. 15 (at $\chi \approx 0.13$). Across the BBH population, these more rapidly spinning components exhibit a distribution that likely peaks near $\chi_A \approx 0.4$, with first and 99th percentiles at $0.07^{+0.05}_{-0.03}$ and $0.8^{+0.08}_{-0.08}$, respectively. Less rapidly spinning components, meanwhile, are centered at or below $\chi_B \lesssim 0.2$, with 99% of values occurring below $0.54^{+0.09}_{-0.08}$.

One significant question noted above and explored in our previous study [20] is the degree to which BBHs exhibit extreme spin-orbit misalignment, with tilt angles exceeding $\theta \ge 90°$ and thus negative effective inspiral spins. Such steeply tilted spins are unlikely for BBH formation from isolated stellar progenitors unless black holes experience stronger-than-expected natal kicks upon formation [239,243–248], and hence might indicate the effects of dynamical interactions during BBH evolution. [As a representative example, one study [66] found a fraction 3%–7% of binaries formed by isolated-binary evolution will have $\chi_{\mathrm{eff}} < 0$, substantially less than the fraction estimated here.] Our GWTC-2 study [20] interpreted the results of the DEFAULT and GAUSSIAN spin analyses as indicating the presence of extremely misaligned spins. As seen in Fig. 15, the component spin-tilt distribution is nonvanishing below $\cos\theta = 0$. Similarly, in Fig. 16 the $\chi_{\mathrm{eff}}$ distribution has significant support at $\chi_{\mathrm{eff}} < 0$. To check whether this requirement for negative $\chi_{\mathrm{eff}}$ was a true feature of the data or an extrapolation of the Gaussian population model (which assumes the existence of extended tails), we extend the Gaussian model to truncate the effective inspiral spin on the range $\chi_{\mathrm{eff,min}} \leq \chi_{\mathrm{eff}} \leq 1$ (rather than $-1 \leq \chi_{\mathrm{eff}} \leq 1$) and hierarchically measure the lower truncation bound $\chi_{\mathrm{eff,min}}$. We find $\chi_{\mathrm{eff,min}} < 0$ at 99.1% credibility, concluding that the data require the presence of negative effective inspiral spins. We obtain consistent results if we perform an identical check with GWTC-3; Fig. 18 illustrates our updated posterior on $\chi_{\mathrm{eff,min}}$, now inferred to be negative at 99.7% credibility.

This interpretation was challenged in other work [251,252], which argued that no evidence for extreme spin misalignment exists if BBH spin models are expanded to allow the existence of a secondary subpopulation with vanishingly small spins. Other avenues of investigation are also in tension with the identification of extreme spin-orbit misalignment. When the $\chi_{\mathrm{eff}}$ distribution is allowed to correlate with other BBH parameters, like the binary mass ratio (see Sec. VII B), evidence for negative $\chi_{\mathrm{eff}}$ values diminishes [253]. Motivated by the concerns raised elsewhere [251,252], we repeat our inference of $\chi_{\mathrm{eff,min}}$ but under an expanded model that allows for a narrow subpopulation of BBH events with extremely small effective inspiral spins:

$$p(\chi_{\mathrm{eff}}|\mu_{\mathrm{eff}},\sigma_{\mathrm{eff}},\chi_{\mathrm{eff,min}}) = \zeta_{\mathrm{bulk}}\mathcal{N}_{[\chi_{\mathrm{eff,min}},1]}(\chi_{\mathrm{eff}}|\mu_{\mathrm{eff}},\sigma_{\mathrm{eff}}) + (1-\zeta_{\mathrm{bulk}})\mathcal{N}_{[-1,1]}(\chi_{\mathrm{eff}}|0,0.01). \tag{18}$$

Here, $\zeta_{\mathrm{bulk}}$ is the fraction of BBHs in the wide bulk population, truncated above $\chi_{\mathrm{eff,min}}$, while $(1-\zeta_{\mathrm{bulk}})$ is the fraction of events residing in the vanishing spin subpopulation, which formally extends from $-1$ to 1. When repeating our inference of $\chi_{\mathrm{eff,min}}$ under this expanded model, our data still prefer a negative $\chi_{\mathrm{eff,min}}$ but with lower significance. As seen in Fig. 18, we now infer that $\chi_{\mathrm{eff,min}} < 0$ at 92.5% credibility. This expanded model allows us to additionally investigate evidence for the existence of a subpopulation of BBHs with vanishingly small spins. GWTC-3 prefers but does not require such a subpopulation to exist. We measure $\zeta_{\mathrm{bulk}} = 0.54^{+0.36}_{-0.26}$, with $\zeta_{\mathrm{bulk}} > 0.2$ at 99% credibility, but also find that our posterior remains consistent with $\zeta_{\mathrm{bulk}} = 1$.

### A. Spin distribution consistent as mass increases

Our previous analysis adopted the same spin distribution at all masses: Given this assumption, the reconstructed spin distribution is dominated by binaries in the bulk of the detected mass distribution. However, both our individual-event parameter estimates and astrophysical formation scenarios suggest possible correlations between black hole masses and spins; such correlations may arise both from isolated binary and dynamical formation [63,254]. In GWTC-3, binaries with the most extreme values of spins have heavier masses: Observations GW170729, GW190517, GW190519, GW190620, GW190706, GW190805, and GW191109 constitute 70% of the binaries with moderate to high spins. This preponderance of massive binaries with large spin suggests that spin distributions are not independent of black hole mass; conversely, such a trend in the detected events may also be due to a higher influence of spin on search sensitivity at the highest masses.

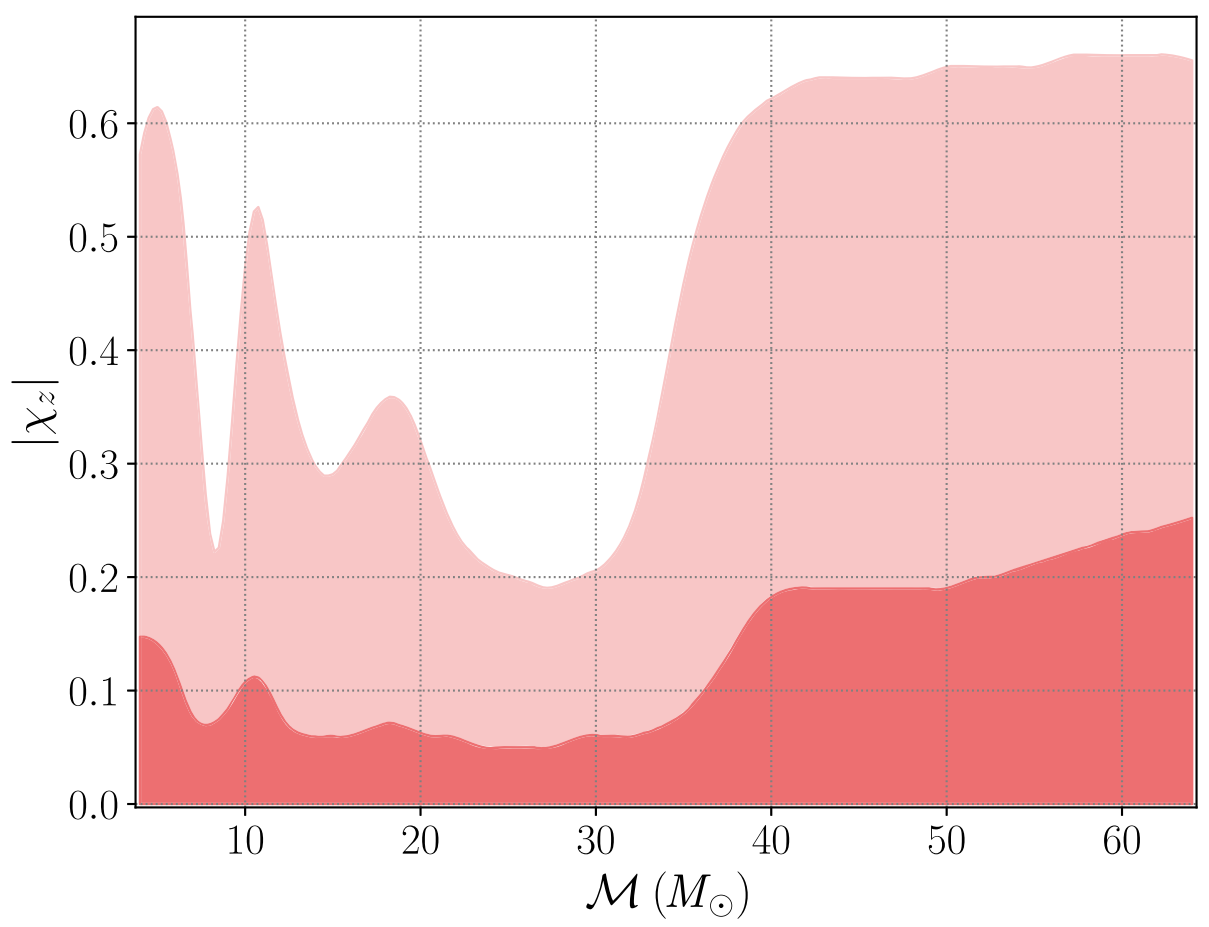

FIG. 19. The dependence of aligned-spin magnitude on the chirp mass. The light (dark) shaded regions are the aligned-spin magnitude at a credibility 90% (50%). The distribution is consistent with small values for lower chirp-mass binaries; however, the spin magnitude is less tightly constrained for chirp masses of $30M_\odot$ and higher.

Using the FM model that identically models the aligned spins $p(\chi_z)$, we reconstruct the trend of $|\chi_z|$ versus mass, accounting for selection effects. Figure 19 shows the aligned-spin-magnitude distribution versus binary chirp mass. At low masses, the aligned spin is consistent with (and well constrained to be close to) zero, (i.e., maximum aligned-spin magnitude averaged over chirp masses $30M_\odot$ or less is 0.38 at 90% credibility). At heavier masses, the aligned spin is still consistent with zero, albeit with larger dispersion (i.e., maximum aligned-spin magnitude averaged over chirp masses $30M_\odot$ or more is 0.5 at 90% credibility). This trend is qualitatively consistent with the relative proportion of events versus chirp mass: Very few observations have high chirp masses, providing relatively little leverage to constrain spins. At high chirp masses, the spin distribution is poorly constrained by only a handful of measurements, closer to our broad prior assumptions, in contrast to the better-constrained distribution at low mass. We have no evidence to support or refute a trend of aligned spin with chirp mass.

Figure 19 suggests the aligned-spin magnitude remains constrained to be close to zero independent of the most well-identified peaks in the mass distribution, contrary to what would be expected from hierarchical formation scenarios for these peaks [34,162,254–257].

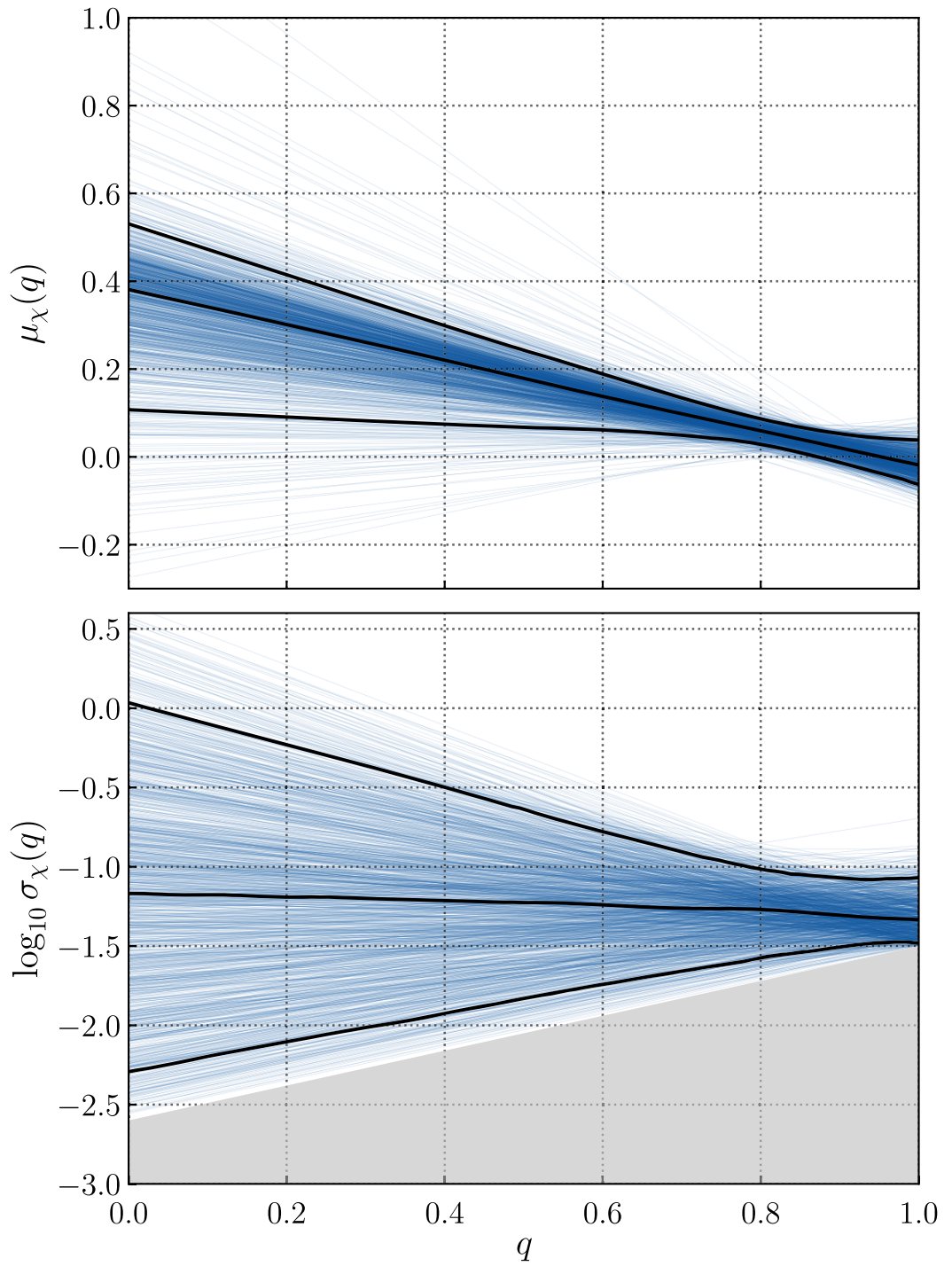

FIG. 20. Posterior constraints on the mean (top) and standard deviation (bottom) of the $\chi_{\rm eff}$ distribution as a function of mass ratio $q$. At 97.5% credibility, we find that the mean of the $\chi_{\rm eff}$ shifts toward larger values for more unequal-mass systems. The gray region in the lower panel shows the area artificially excluded by our prior on the parameters $\sigma_0$ and $\beta$; see Eq. (20).

### B. High spin correlates with asymmetric binaries

BBHs may exhibit an anticorrelation between their mass ratios and spins, such that binaries with $q \sim 1$ favor effective inspiral spin parameters near zero, while binaries with more unequal-mass ratios exhibit preferentially positive $\chi_{\rm eff}$ values [253]. To evaluate the degree to which $q$ and $\chi_{\rm eff}$ are (or are not) correlated, following prior work [253] we adopt a Gaussian model for the $\chi_{\rm eff}$ distribution with a mean and standard deviation that are allowed to evolve with $q$:

$$p(\chi_{\rm eff}|q) \propto \exp\left[-\frac{[\chi_{\rm eff}-\mu(q)]^2}{2\sigma^2(q)}\right], \tag{19}$$

with

$$\mu(q) = \mu_0 + \alpha(q-1), \tag{20a}$$

$$\log_{10}\sigma(q) = \log_{10}\sigma_0 + \beta(q-1). \tag{20b}$$

The new hyperparameters $\alpha$ and $\beta$ measure the extent to which the location or width of the $\chi_{\rm eff}$ distribution changes as a function of mass ratio.

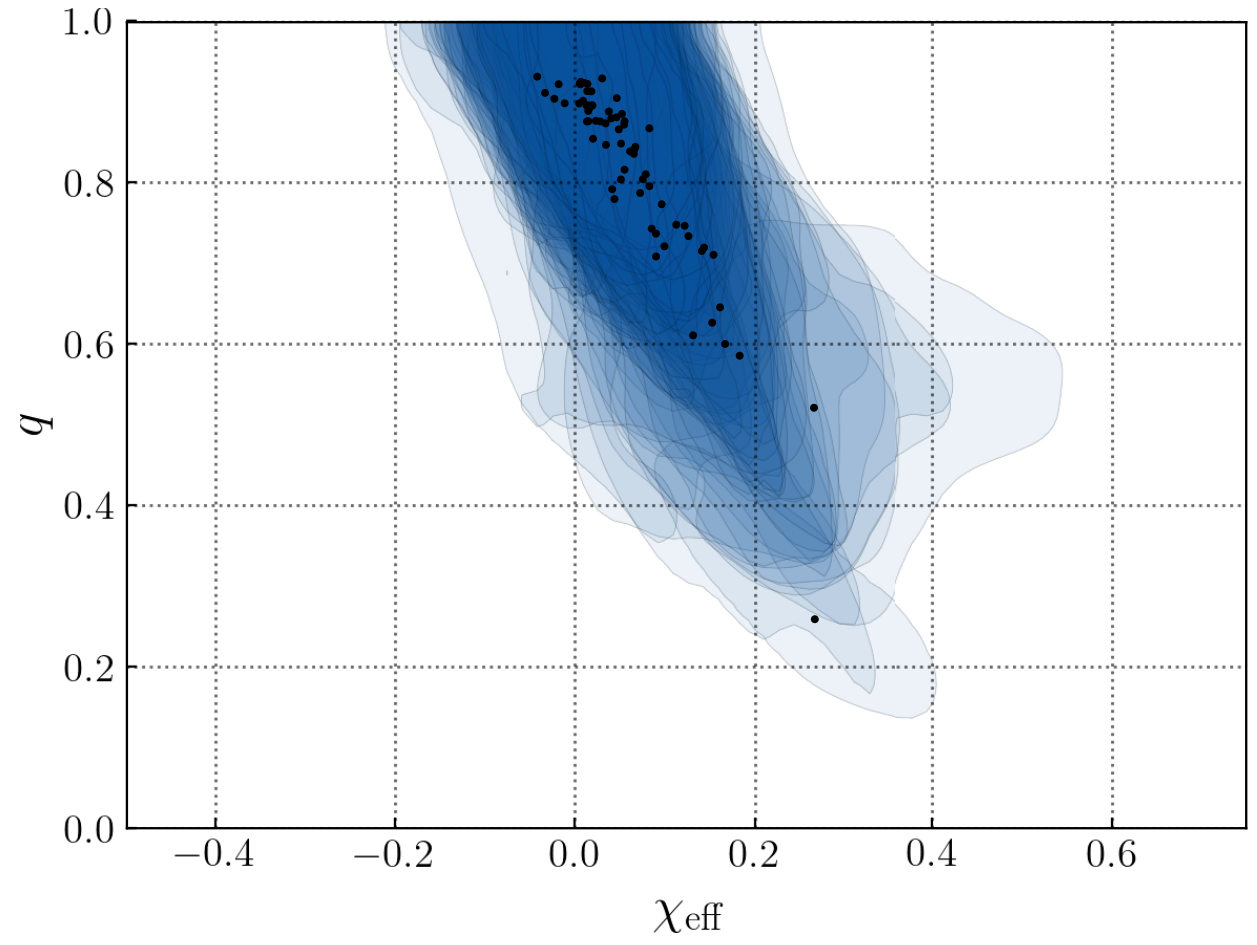

FIG. 21. Posteriors on the mass ratios and effective inspiral spins of BBHs in GWTC-3, reweighted to a population-informed prior allowing for a correlation between $q$ and $\chi_{\rm eff}$. We infer that the mean of the BBH $\chi_{\rm eff}$ distribution shifts toward larger values with decreasing mass ratios. Accordingly, reweighted events shift considerably, such that events with $q \sim 1$ contract about $\chi_{\rm eff} \approx 0$ while events with $q < 1$ shift toward larger effective inspiral spins.

We repeat hierarchical inference of the BBH population, adopting the fiducial model for the primary mass and redshift distribution. At 97.5% credibility, we constrain $\alpha < 0$, indicating that more unequal-mass binaries preferentially possess larger, more positive $\chi_{\rm eff}$. Figure 20 illustrates our constraints on the mean and standard deviation of the $\chi_{\rm eff}$ distribution as a function of mass ratio. Each light trace represents a single sample from our hyperposterior, and the solid black lines denote the median values and central 90% bounds on $\mu(q)$ and $\sigma(q)$ at a given value of $q$. If we adopt these hierarchical results as a new, population-informed prior, Fig. 21 shows the resulting reweighted posteriors for the BBHs among GWTC-3. Each filled contour bounds the central 90% region for a given event in the $q$-$\chi_{\rm eff}$ plane, while black points mark the events' one-dimensional median $q$ and $\chi_{\rm eff}$ measurements.

## VIII. COMPARISON WITH OTHER GW CATALOGS

In this paper, we present population inferences based upon events identified by the LIGO Scientific, Virgo, and KAGRA Collaborations in data taken by the Advanced LIGO and Advanced Virgo instruments during their first three observing runs [3–5]. We impose a FAR threshold of $< 0.25\ {\rm yr}^{-1}$ across all analyses incorporating NS binaries and a lower threshold FAR $< 1\ {\rm yr}^{-1}$ for BBH analyses. This excludes several events which pass the threshold of $p_{\rm astro} > 0.5$ for inclusion in GWTC-3. In addition, a number of analyses of the public GW data from O1, O2, and O3a [71–76] have identified additional candidate binary merger events. In the remainder of the paper, we restrict the primary analysis to events included in GWTC-3. The overriding reason for this is that differences in the analysis methods prevent a detailed evaluation of search sensitivity, as described in Sec. III, which is critical to interpreting the population. In this section, we investigate the consistency of the remaining GWTC-3 events and additional non-GWTC events with the population models inferred in this paper.

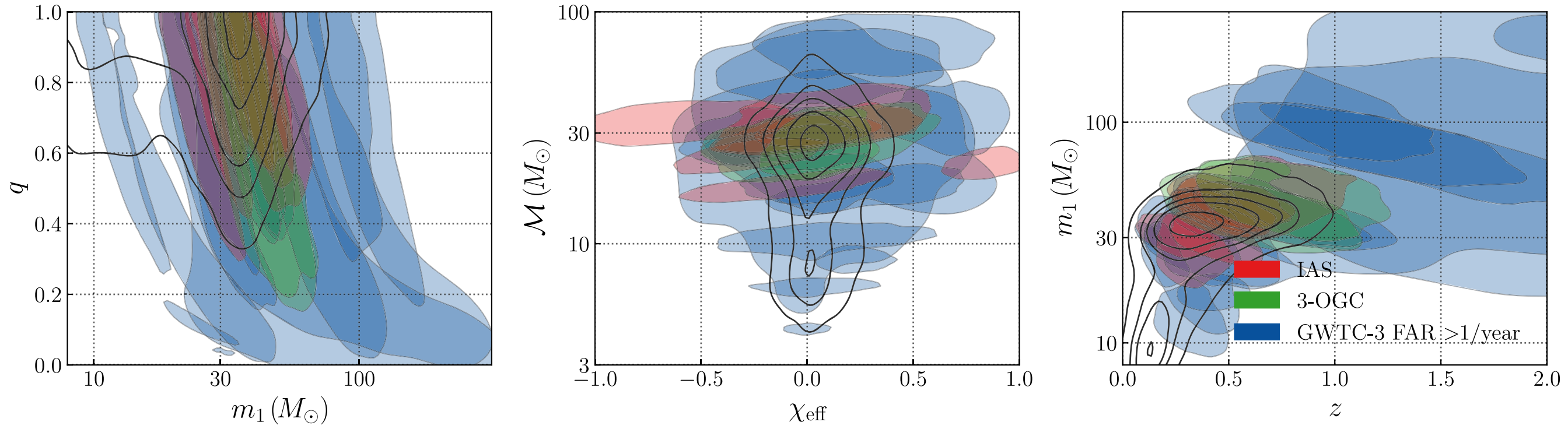

FIG. 22. The measured properties of the BBH candidates not included in the population study presented in this paper (shaded regions), compared to the inferred population from the PP model presented in Sec. VI A (black contours). These include both events which fall below our FAR threshold as well as events identified by other groups. The events are color coded based upon the search which first identified them: catalogs from O1 and O2 [72–74] in red, events from the third Open Gravitational-Wave Catalog (3-OGC) (which incorporates events in O1–O3a) [76] in green, and from GWTC-3 with FAR > 1 $\mathrm{yr}^{-1}$ threshold in blue.

For concreteness, when referring to results reported by external groups, we include all events identified as GWs in their catalogs. In O1, there is one additional event, GW151216, identified in the O1 Institute for Advanced Study (IAS) catalog [72]. The additional events from O2 are GW170304, GW170425, and GW170403 which are identified in the O2 IAS catalog [73,74], and GW170121, GW170202, and GW170727 which were also then independently found in the second Open Gravitational-Wave Catalog (2-OGC) [75]. In O3, we include 16 additional events. These include GW190916_200658 and GW190926_050336 which were originally identified in the third Open Gravitational-Wave Catalog (3-OGC) [76] and independently identified in Gravitational-Wave Transient Catalog 2.1 (GWTC-2.1) [6]; GW190403_051519, GW190426_190642, and GW190514_065416 which are included in GWTC-2.1 but have a FAR below our < 1 $\mathrm{yr}^{-1}$ threshold; GW191113_071753, GW191126_115259, GW191204_110529, GW191219_163120, GW200208_222617, GW200210_092254, GW200220_061928, GW200220_124850, GW200306_093714, GW200308_173609, and GW200322_091133 which are included in GWTC-3 but again have a FAR below our < 1 $\mathrm{yr}^{-1}$ threshold.

In Fig. 22, we show the additional gravitational-wave events which are not included in the sample used in this paper. The additional events are broadly consistent with the population presented here although several events lie at the boundaries of the identified population. Specifically, two of the events have effective inspiral spins that lie outside the inferred population. These are GW151216 with a mean $\chi_{\rm eff} = 0.82$ and GW170403 with a mean $\chi_{\rm eff} = -0.58$. The analysis in Ref. [74] used a prior which is constant in $\chi_{\rm eff}$ which is significantly different from the uniform in the spin-magnitude prior used in the GWTC papers. A reanalysis of GW151216 and GW170403 [258] leads to inferred $\chi_{\rm eff}$ distributions that are more consistent with the population inferred here. Specifically, this gives $\chi_{\rm eff} = 0.5^{+0.2}_{-0.5}$ for GW151216 and $\chi_{\rm eff} = -0.2^{+0.4}_{-0.3}$ for GW170403. In addition, the subthreshold events from GWTC-3 extend the distribution to both higher masses and more asymmetric mass ratios. However, only low-significance events currently populate these regions. Additional observations in future runs will allow us to determine whether these low-significance events were the first hints of a broader population in the mass space.

With regard to events potentially containing NSs, GWTC-3 contains several candidates that do not satisfy our FAR < 1 $\mathrm{yr}^{-1}$ threshold but do have $m_2$ potentially consistent with NS masses, namely, GW191219 and GW200210. Both events are inferred to have highly asymmetric masses and could possibly be an indication of additional NSBH sources, or asymmetric BBHs similar to GW190814. Further observations in future runs will again allow us to investigate these interesting regions of the binary parameter space in greater detail.

## IX. ASTROPHYSICAL INTERPRETATION

### A. Implications for binary black hole formation

#### *1. Mass distribution*

The statistical distribution of BH source properties such as their mass, spin, and redshift can be used to probe the astrophysics of BH binary formation and evolution [28,34,36,63,144,162,164,240,243,244,247,256,259–266]. The analysis performed in Sec. VI identifies structures in the

mass distribution of BBHs that go beyond a standard power-law model and can help to shed light on formation processes. These features were previously identified in an analysis of GWTC-2 [94], but we are now more confident that they are statistically significant (see Sec. VI).

The underlying mass distribution of BBHs inferred in this paper peaks at a primary mass of approximately 10 $M_\odot$, with the majority of BBHs having a primary BH with a mass lower than this value (e.g., Fig. 11). Formation in globular clusters has been long recognized as an important channel for merging BBHs [267–275]. In this scenario, BBHs are assembled during three-body dynamical interactions in a low-metallicity environment. The resulting BH mass distribution is generally predicted to peak at $> 10M_\odot$. Three recent studies of globular cluster formation find that the BBH merger rate is severely suppressed where we observe a peak: One study [35] finds that the BBH merger rate is severely suppressed below about $m \simeq 13M_\odot$ with a corresponding realistic merger rate at this mass value of approximately 0.5 $\mathrm{Gpc}^{-3}\,\mathrm{yr}^{-1}\,M_\odot^{-1}$ (their Fig. 2); another recent study [32] finds similar results, with a peak in their mass distribution at about $m \simeq 15$–$20M_\odot$ (their Fig. 5); a third analysis [275] finds the peak at $m \simeq 20M_\odot$. Taking these results at face value, the inferred high merger rate of sources with $\lesssim 10M_\odot$ may suggest that globular clusters contribute at best subdominantly to the detected population. Current models also disfavor dynamical formation in young clusters to explain the whole BH population at $m \simeq 10M_\odot$. Under the assumption that BHs receive the same kicks as NSs lowered by the fraction of fallback mass, lighter BHs are ejected by supernova kicks and do not participate in the dynamical evolution of the cluster [276–278]. However, the parameter space relevant for merging BHs in dense stellar environments still needs to be fully explored.

Galactic nuclei can produce a BBH population with a much wider mass spectrum than both young and globular clusters [279–285]. Because of their high metallicities and escape velocities, nuclear star clusters can form and retain a significant number of lighter BHs, which can then pair and merge. BBH formation near an active galactic nucleus (AGN) disk can produce a significant population of BBH mergers with a wide mass spectrum [34,221,254,286,287]. In such scenarios, the observed low-mass overdensities without counterparts in spin could be reflections of supernova physics; by contrast, in these hierarchical formation models no evident mechanism can impart them without a corresponding signature in spin. If the BBHs are formed near an AGN disk, this process might select heavier BHs, hardening the BBH mass function and driving the peak of the mass distribution toward values higher than observed [288].

Isolated binary evolution models often predict a peak near $m \simeq 10M_\odot$ [26,63,289–291]. Recent population models find component masses of merging BBHs that peak at $8 - 10M_\odot$ and come from approximately $20 - 30M_\odot$ progenitors [31,63,290]. The overall merger rate normalization of the peak remains, however, poorly constrained. Moreover, the peak of the mass distribution can shift significantly depending on the adopted supernova, natal kick, mass transfer, and wind prescriptions, and star formation history of the Universe [26,31,84,223,276,292,293].

The analysis in Sec. VI suggests the existence of additional peaks in the chirp-mass and primary mass distributions. Assuming these peaks exist, an explanation consistent with our constraints on BH spins is that they originate either from the initial BH mass function, or that they are produced by different populations formed by separate physical processes or formation channels.

The other feature of the inferred BH mass distribution that is shown in our analysis is the apparent lack of truncation at $m \sim 40M_\odot$, which confirms our results based on GWTC-1 and GWTC-2 [4,114]. A mass gap between approximately $50^{+20}_{-10}M_\odot$ and approximately $120M_\odot$ is predicted by stellar evolution models as the result of the pair-instability process in the cores of massive stars [199–205]. However, due to our limited knowledge of the evolution of massive stars, the formation of BHs heavier than approximately $40M_\odot$ from stellar collapse cannot be fully excluded [63,205,211,212,294,295]. The location of the mass gap is sensitive to the uncertain $^{12}\mathrm{C}(\alpha,\gamma)^{16}\mathrm{O}$ reaction rate, which governs the production of oxygen at the expense of carbon [201,211,296]. Moreover, BHs formed from progenitor stars at low metallicities ($Z/Z_\odot \lesssim 0.1$) might avoid altogether the mass limit imposed by pair instability [230,294,295]. The lack of a sharp truncation at high masses might indicate a dynamical process, such as the hierarchical merger of BHs [32,162,206,214,255–257,280,284,297–300] or stars [301–304] in dense clusters or in the gaseous disk surrounding a massive BH [261]. In a hierarchical scenario we would expect the more massive BHs to also have the larger spins [255,256]. While we do observe such a mass-spin correlation above $m \sim 40M_\odot$ (Fig. 19), the binaries with a signature that $\chi_{\mathrm{eff}}$ is not zero all prefer $\chi_{\mathrm{eff}} > 0$ (see Table IV), while hierarchical formation in dynamical environments would lead to isotropically oriented spins. Massive mergers with positive $\chi_{\mathrm{eff}}$ can be obtained through classical isolated-binary evolution [305] if the carbon fusion reaction rate is highly uncertain and allowed to vary to values much lower than typically assumed in binary population codes [211]. BHs can also increase their birth mass beyond the pair-instability mass gap through the efficient accretion of gas from a stellar companion or from a surrounding gaseous disk, although this might be unlikely, as this would require hypercritical accretion (many orders of magnitude grater than the Eddington limit) [67,213,306–308]. Highly coherent accretion on one of the BHs could also explain the negative correlation between $\chi_{\mathrm{eff}}$ and $q$ shown in Fig. 20, although accretion in gas-free scenarios should be highly super-Eddington in order to impart significant spin [67,213]. Alternatively, primordial BHs can have masses above the pair-instability mass threshold, although this most likely requires efficient accretion before the reionization epoch in order not to violate current constraints [309].

### 2. Redshift distribution

In Sec. VI we show that the BBH merger rate increases with redshift, as $(1+z)^{\kappa}$, with $\kappa = 2.9^{+1.7}_{-1.8}$. Although error bars are large, current data prefer a model in which the merger rate evolves steeply with redshift and at a rate that is consistent with the growth in star formation. For binary formation in the field, the predicted value of $\kappa$ is sensitive to the assumed efficiency of common envelope ejection: Values between $\kappa = 0.2$ and 2.5 are all possible, although relatively small values $\kappa \sim 1$ are preferred [31,222,223,276,310]. Delay times in the field are also dependent upon stability of mass transfer [311]. Similarly, $\kappa \lesssim 2$ is often found in models of BBHs formation in open and young clusters [278,312]. Formation through chemically homogeneous evolution predicts a steep redshift dependence $\kappa \sim 3$ [313]. Dynamical formation in globular clusters predicts $\kappa \lesssim 2$ [218,314]; e.g., one study of globular cluster models [35] finds $\kappa = 1.6^{+0.4}_{-0.6}$, and shows that the most important parameter affecting the value of $\kappa$ in the globular cluster scenario is the initial cluster half-mass density, $\rho_h$, while uncertainties in other model parameters (e.g., natal kicks, black hole masses, metallicity) have a small effect. Only models in which globular clusters are formed with a high half-mass density, $\rho_h > 10^5 M_\odot$ pc$^{-3}$, lead to $\kappa \gtrsim 2$ [35]. While uncertainties are large, improved constraints on the merger rate evolution have the potential to unveil important information about the physics of massive binaries [26,63,146,310] and the initial conditions of clusters across cosmic time [218,266,314].

### 3. Spin distribution

We observe evidence that the spin distribution requires both spin-orbit misalignment and also includes events with antialigned spins [$\cos(\theta) < 0$]. BBHs with a large spin-orbit misalignment can be formed in dynamical environments such as globular, nuclear, and young star clusters, or active galactic nuclei [239,257,280,283]. In these systems, two single BHs are paired together during a three-body interaction and/or undergo a number of subsequent dynamical interactions before merging. Their spins have directions that are therefore uncorrelated with each other and with the orbital angular momentum of the binary, leading to an isotropic spin-orbit alignment [237,239]. The evolution of BH spins in AGN disks depends on several uncertain factors, such as the importance of accretion and dynamical encounters, the initial spin orientation, and the efficiency of migration [34,315]. If radial migration of BHs is inefficient, the distribution of $\chi_{\rm eff}$ skews toward higher values because scattering encounters that randomize spin directions become less frequent. On the other hand, efficient migration would imply more frequent dynamical encounters, producing a $\chi_{\rm eff}$ distribution centered on zero [315]. However, the dispersion of the $\chi_{\rm eff}$ distribution also increases characteristically with mass, as with other hierarchical formation scenarios [34,162,254].

Formation from field binaries is thought to produce components with preferentially aligned spins [237,239,243], although $\chi_{\rm eff}$ can extend to negative values with the typical fraction of detectable binaries with negative effective spins ranging from 3% to 10% [243]. Such an alignment, however, is not certain. In fact, all population models of isolated binaries customarily start with the stellar progenitor spins initially perfectly aligned with the orbital angular momentum of the binary. This assumption is made due to simplicity and partly because tidal interactions are thought to quickly remove any spin-orbit misalignment prior to BH formation. On the other hand, there is observational evidence of close and noninteracting or detached massive binaries with highly inclined spin axes, suggesting that massive binaries can form with misaligned spins [316–319]. Other explanations for the misalignment include that tides might not in all cases be able to realign the spins [316–319]. A large spin-orbit misalignment can also be produced if a binary is the inner component of a triple system [314,320–323], where the tertiary component can be either a star, another BH, or even a massive BH [282,324–326]. In this triple scenario, the secular gravitational interaction of the binary with an external companion can randomize the orbital plane of the binary. The complex precessional dynamics of the BH spins in triple systems also changes the spin orientation and leads to a distribution of $\chi_{\rm eff}$ peaked near zero, although with a marginal preference for aligned spins [320,321]. The spin-orbit misalignment found in BH binaries might be explained by a stable episode of mass transfer prior to the formation of the BHs [327]. Finally, the asymmetric mass and neutrino emission during core collapse can also lead to a large spin tilt, possibly probing any value of spin-orbit misalignment (i.e., up to 180°) [246,328]. The magnitude of the tilt and the fraction of misaligned systems depend on the assumed natal kick velocities and directions, which remain poorly constrained observationally [329–331]. Even under optimistic assumptions, however, natal kicks are unlikely to produce a large misalignment for a significant fraction of the population and account for the more isotropic distribution of spin-orbit angles shown in the right panel of our Fig. 15 [242,247,332]. We conclude that the presence of systems with misaligned spins is not in contradiction with a scenario in which the majority of, if not all, BBHs form in the field of galaxies. On the other hand, the fact that the $\chi_{\rm eff}$ distribution is not symmetric around zero, if confirmed, can be used to rule out a model in which all BBHs are formed through dynamical encounters in star clusters [251,262]. Finally, an asymmetric distribution of $\chi_{\rm eff}$ can also be produced through three-body encounters of tidally spun-up binaries in young star clusters, although this process can explain only the spin-orbit misalignment of about 10% of the population [333].

Corroborating our previous conclusion based on GWTC-2 [20], we find that the BH population is typically described by small spins. Predictions for BH spin magnitudes vary depending on the assumptions about stellar winds and their metallicity dependence, tides, and are particularly sensitive to the efficiency of angular momentum transport within the progenitor star [61,62]. If the stellar core remains strongly coupled to the outer envelope during the stellar expansion off the main sequence, then a significant amount of spin can be carried from the core to the envelope. In this case, a BH formed from stellar collapse may be born with nearly zero spins. This implies that the first-born BH will essentially be a Schwarzschild black hole. Nevertheless, isolated-binary evolution can also lead to primary BHs with high spins if both stars have similar masses and both are subject to tidal spin-up or due to mass ratio reversal caused by the Roche-lobe overflow [66]. The second-born BH can more commonly form with significant spin as tidal interactions may realign and increase the spin of its stellar progenitor in between the two supernova explosions [64,334]. If the binary undergoes chemically homogeneous evolution [313,335], its components may both be tidally spun up to near break-up velocity, and keep this rotation rate throughout main sequence evolution, evolving into BHs with large and aligned spins. Black holes that form during the quantum chromodynamic phase transition in the early Universe will all have essentially zero natal spins [336,337]. However, a significant spin can be attained through subsequent gas accretion [309]. Finally, if BBHs are formed or migrate within the accretion disk of a supermassive BH, they can accrete from the surrounding gaseous environment and spin up [315].

We observe neither evidence for nor against an increase in spin magnitude for systems with higher masses [261,338,339] and more unequal-mass ratios [253]. Current stellar evolution models suggest that larger BH masses should correlate with smaller spins because larger BHs originate from more massive stars which undergo more extensive mass loss, carrying away most of the angular momentum and producing BHs with small spins [63,64,243]. A consequence of this should be either a decrease in spin magnitude (and $\chi_{\rm eff}$) with mass above approximately $20 M_\odot$ or no correlation, where the predicted trend depends on the specific stellar evolution models adopted [63,243]. Predictions remain uncertain and are strongly dependent on modeling assumptions about the angular momentum transport within the star, spin dissipation during the supernova, and the treatment of binary interaction prior to BH formation. If future observations identify such a trend, then an increase in spin magnitude with mass might suggest a hierarchical formation scenario. However, as mentioned above, this scenario seems currently at odds with the fact that binaries with more unequal-mass ratios and massive components exhibit preferentially positive $\chi_{\rm eff}$.

### B. Implications for neutron stars

One result from GW observations is in tension with the strong preference for $1.33 M_\odot$ mass objects which has been recovered in Galactic BNSs [92]. Instead, conservatively assuming all objects below the maximum neutron star mass are neutron stars, our unmodeled analysis of the lowest-mass compact objects is consistent with a broad unimodal Gaussian, allowing for highly asymmetric binaries. Our analysis of all individual low-mass (assumed NS) objects suggests a wide NS mass distribution, without the bimodal structure seen in the Galactic NS population. The GW-observed population of low-mass mergers is still small. If this tension persists, however, several avenues exist to explain a discrepancy between Galactic and GW observations, including but not limited to additional formation channels for GW systems, strong observational selection effects, like those used to explain GW190425 [16,340,341] and the smaller body in GW200105, or the prospect that BHs form from nonstellar processes during the quantum chromodynamic phase transition in the early Universe [56,309,342,343].

Our conclusions about the compact object mass spectrum in general and the mass spectrum of NSs in particular will have substantial impact on the understanding of the stellar explosions that generate such compact objects [43,44,84] and the binary interactions that carry these objects toward merger, assuming a stellar origin for low-mass binary mergers.

Our analyses show no evidence for or against the presence of a mass distribution feature closely corresponding to the maximum neutron star mass. Rather, the shape of the neutron star mass distribution, the existence of GW190814, and our results for the mass distribution for compact objects between 3 and $7 M_\odot$ may instead suggest a continuous mass spectrum, albeit strongly suppressed above the masses of known NSs.

Fortunately, the comparatively high prevalence of objects close to the maximum NS mass suggests that we will likely observe several objects near this region in the future. By differentiating objects containing NSs from those without, e.g., through their electromagnetic emission, we will be able to provide several avenues to connect features in the NS mass distribution to fundamental nuclear physics. Our analysis of NSs in merging binaries alone suggests the NS mass distribution likely extends up to the maximum mass $M_{\rm max,TOV}$ expected from studies of the EOS.

Our analyses are also consistent with both symmetric ($q \simeq 0.8$) and significantly asymmetric ($q < 0.8$) binaries containing NSs in BNSs, and modestly ($q \in [0.5, 0.8]$) to highly ($q < 0.5$) asymmetric binaries in NSBHs. Compared to equal-mass mergers [344,345], modestly asymmetric NS mergers (with either NS or BH counterparts) are potentially strong candidates for multimessenger counterparts [346], since an asymmetric merger can eject more mass [347], produce a larger remnant disk [348,349], and potentially produce significant associated gamma ray burst emission [350–352]. For BNSs, our analyses are consistent with a significant fraction of highly asymmetric events. For NSBHs, the discovery of GW200105 and GW200115 demonstrate the existence of asymmetric binaries containing NSs with a range of mass ratios. Based on these events, our inferences about the low-mass compact object distribution suggests that electromagnetically bright NSBH mergers could occur at a small fraction of the overall NSBH rate [353,354].

Generally, a broad mass ratio distribution suggests modestly more favorable prospects for electromagnetic follow-up observations. Conversely, a broad mass ratio distribution complicates simple efforts to interpret existing GW observations which were developed under the assumption that low-mass binary mergers are very frequently of comparable mass [355].

Finally, our analyses here leave GW190814 as an outlier both from BBH systems and from systems that contain a likely NS. Neither component of this binary has exceptional masses; for example, the secondary component could easily be produced from conventional supernova engines [84]. However, based on the merger rates versus mass identified in our study, this system (and the larger sample of asymmetric binaries available at a lower threshold) may require a different formation pathway in order to simultaneously explain the mass ratio, secondary mass, and merger rate [236,283,356].

## X. THE GW BACKGROUND FROM BINARY MERGERS

The observation of binaries with masses in the NSBH range allows us to provisionally complete a census of the different classes of compact binaries that contribute to an astrophysical gravitational-wave background, assuming our existing surveys are sensitive to all relevant sources (i.e., not accounting for frequent subsolar mass mergers). We have previously predicted the contributions of BBH and BNS mergers to the gravitational-wave background, based on the compact binary population observed in GWTC-2 [228]. In Fig. 23, we update this forecast with our latest knowledge of the BBH and BNS population and the newly measured rate of NSBH mergers.

The shaded bands on the left side of Fig. 23 show estimates of and uncertainties on the dimensionless energy-density spectra

$$\Omega(f) = \frac{1}{\rho_c}\frac{d\rho}{d\ln f} \qquad (21)$$

of gravitational waves radiated by each class of compact binary. In Eq. (21), $d\rho$ is the gravitational-wave energy density per logarithmic frequency interval $d\ln f$, while $\rho_c$ is the critical energy density required to close the Universe. We adopt the same model for the merger history of compact binaries used previously [228,358], assuming that compact binary formation rate traces a metallicity-weighted star formation rate model [359–361] with a $p(t_d) \propto t_d^{-1}$ distribution of time delays $t_d$ between binary formation and merger. Time delays are restricted to $20\,\mathrm{Myr} \le t_d \le 13.5\,\mathrm{Gyr}$ for BNS and NSBH mergers and $50\,\mathrm{Myr} \le t_d \le 13.5\,\mathrm{Gyr}$ for BBHs [31,362], with binary formation restricted to redshifts below $z_{\max} = 10$. The birth rate of BBH progenitors is further weighted by the fraction of star formation at metallicities $Z < 0.1Z_{\odot}$ [229,363]. We note that, rather than assuming a form for the binary merger rate $\mathcal{R}(z)$ to predict the gravitational-wave background, one could alternatively leverage constraints on the background to infer $\mathcal{R}(z)$; see Sec. VI D for further details.

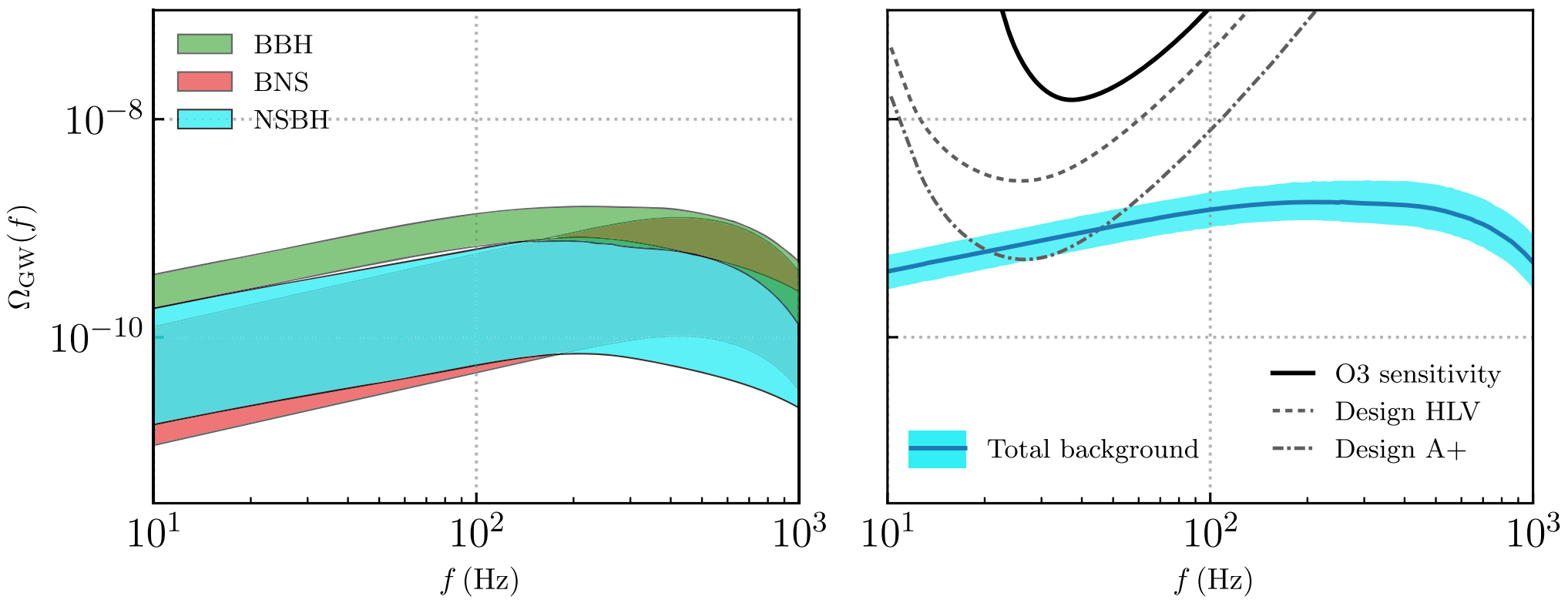

FIG. 23. Forecast of astrophysical gravitational-wave background due to binary mergers following O3. Left: the individual contributions expected from BNS, NSBH, and BBH mergers. While uncertainties on the energy density due to BNS and NSBH are due to Poisson uncertainties in their merger rates, our forecast for the stochastic background due to BBHs additionally includes systematic uncertainties associated with their imperfectly known mass distribution. Right: estimate of the total gravitational-wave background (blue), as well as our experimental current sensitivity (solid black) [228,357]. For comparison, we additionally show the expected sensitivities of the LIGO-Virgo network at design sensitivity, as well as that of LIGO's anticipated A+ configuration.

Within Fig. 23, the stochastic energy density due to BBHs has been marginalized over our uncertainty on both the local merger rate and mass distribution of the BBH population, as measured using the PP mass model. At 25 Hz, we estimate the energy density due to BBHs to be $\Omega_{\mathrm{BBH}}(25\ \mathrm{Hz}) = 5.0^{+1.4}_{-1.8} \times 10^{-10}$. To estimate the contribution due to BNS systems, we adopt the simple rate measurement presented in Sec. IV A under a fixed mass distribution, and correspondingly assume a uniform distribution of neutron star masses between 1 and $2.5M_{\odot}$, giving $\Omega_{\mathrm{BNS}}(25\ \mathrm{Hz}) = 0.6^{+1.7}_{-0.5} \times 10^{-10}$. The contribution due to NSBH systems, meanwhile, is estimated using the BGP rate reported in Table II. For simplicity, we again assume a uniform distribution of neutron star masses between 1 and $2.5M_{\odot}$ and a logarithmically uniform distribution of black hole masses between 5 and $50M_{\odot}$ among NSBH mergers. Under these assumptions, we find $\Omega_{\mathrm{NSBH}}(25\ \mathrm{Hz}) = 0.9^{+2.2}_{-0.7} \times 10^{-10}$.

The blue band on the right side of Fig. 23 denotes the our estimate of the total gravitational-wave background due to the superposition of these three source classes; we expect a total energy density of $\Omega(25\,\mathrm{Hz}) = 6.9^{+3.0}_{-2.1} \times 10^{-10}$. For comparison, the solid black curve marks our present sensitivity to the gravitational-wave background [228,357]. Although our estimate for the background amplitude lies well below current limits, it may be accessible with future detectors, such as the planned A+ LIGO configuration.

## XI. CONCLUSIONS

The third LIGO-Virgo-KAGRA Gravitational-Wave Transient Catalog (GWTC-3) [3] has increased our census of the population of compact mergers by nearly a factor of 2 compared to our analysis of the first half of O3 [20]. We simultaneously employ all observations with FAR $< 0.25\ \mathrm{yr}^{-1}$ to infer the merger rate versus both component masses across the observed mass spectrum. For NSs, we find a broad mass distribution extending up to $2.0^{+0.3}_{-0.3}M_{\odot}$, in contrast to the narrow mass distribution observed for Galactic BNSs. We find the BBH mass distribution is nonuniform, with overdensities at BH masses of 10 and $35M_{\odot}$. These overdensities may reflect the astrophysics associated with generating coalescing binaries, potentially reflecting properties of stellar physics or astrophysical environments. These features may assist future applications of gravitational-wave astronomy. As an example, these sharp features could be redshift independent and, if so, used as standard candles for cosmology [364,365]. We find that the compact object mass distribution exhibits an interval between 2.2 and $6.1M_{\odot}$ where merger rates are suppressed, which could be consistent with past x-ray observations suggesting a mass gap [80–83]. Our analysis lacks sufficient sensitivity to probe the structure of the mass distribution at the highest masses $m_1 > 70M_{\odot}$ in detail; however, so far, we find no evidence for or against an upper mass gap.

We find that observed BH spins are typically small (half less than 0.25). We still conclude that at least some of these spins exhibit substantial spin-orbit misalignment. We corroborate a correlation between BBH effective inspiral spins and mass ratio.

Using parametric models to infer the distribution of BBH merger rate with redshift, we find the BBH merger rate likely increases with redshift; we cannot yet assess more complex models where the shape or extent of the mass distribution changes with redshift.

Analyses presented in our previous work [20] and in a companion paper [366] employ coarse-grained models for the BBH population, smoothing over some of the subtle features identified above. We find that these coarse-grained models draw similar conclusions on current data to our previous studies; see Sec. VI A. Applications that focus on large-scale features of the mass distribution (e.g., the stochastic background, as described in Sec. X) require only these coarse-grained results. Nonetheless, the mass distribution remains a critical source of systematic uncertainty in any merger rate integrated over any mass interval, particularly in mass intervals with few observations. We specifically find the BNS and NSBH merger rates exhibit considerable uncertainty in the mass distribution, with relative merger rate errors within (and between) models far in excess of the expected statistical Poisson error associated with the count of these events. These systematics propagate directly into our most conservative estimates for their merger rates.

The next GW survey could have a BNS detection range increased by approximately 15%–40% [367]. Even without allowing for increased merger rates at higher redshift, the next survey should identify roughly 3 times more events of each class than used in this study, including several new events from the BNS and BHNS category. We continuously revise our assessment of future observing prospects [367].

## ACKNOWLEDGMENTS

This material is based upon work supported by NSF's LIGO Laboratory which is a major facility fully funded by the National Science Foundation. The authors also gratefully acknowledge the support of the Science and Technology Facilities Council (STFC) of the United Kingdom, the Max-Planck-Society, and the State of Niedersachsen/Germany for support of the construction of Advanced LIGO and construction and operation of the GEO600 detector. Additional support for Advanced LIGO was provided by the Australian Research Council. The authors gratefully acknowledge the Italian Istituto Nazionale di Fisica Nucleare (INFN), the French Centre National de la Recherche Scientifique (CNRS), and the Netherlands Organization for Scientific Research for the construction and operation of the Virgo detector and the creation and support of the EGO consortium. The authors also gratefully acknowledge research support from these agencies as well as by the Council of Scientific and Industrial Research of India, the Department of Science and Technology, India, the Science & Engineering Research Board, India, the Ministry of Human Resource Development, India, the Spanish Agencia Estatal de Investigación, the Spanish Ministerio de Ciencia e Innovación and Ministerio de Universidades, the Conselleria de Fons Europeus, Universitat i Cultura and the Direcció General de Política Universitaria i Recerca del Govern de les Illes Balears, the Conselleria d'Innovació, Universitats, Ciència i Societat Digital de la Generalitat Valenciana and the CERCA Programme Generalitat de Catalunya, Spain, the National Science Centre of Poland and the European Union—European Regional Development Fund; Foundation for Polish Science, the Swiss National Science Foundation, the Russian Foundation for Basic Research, the Russian Science Foundation, the European Commission, the European Social Funds, the European Regional Development Funds, the Royal Society, the Scottish Funding Council, the Scottish Universities Physics Alliance, the Hungarian Scientific Research Fund, the French Lyon Institute of Origins, the Belgian Fonds de la Recherche Scientifique, Actions de Recherche Concertées and Fonds Wetenschappelijk Onderzoek—Vlaanderen, Belgium, the Paris Île-de-France Region, the National Research, Development and Innovation Office Hungary, the National Research Foundation of Korea, the Natural Science and Engineering Research Council Canada, Canadian Foundation for Innovation, the Brazilian Ministry of Science, Technology, and Innovations, the International Center for Theoretical Physics South American Institute for Fundamental Research, the Research Grants Council of Hong Kong, the National Natural Science Foundation of China, the Leverhulme Trust, the Research Corporation, the Ministry of Science and Technology, Taiwan, the United States Department of Energy, and the Kavli Foundation. The authors gratefully acknowledge the support of the NSF, STFC, INFN, and CNRS for provision of computational resources. This work was supported by MEXT, JSPS Leading-Edge Research Infrastructure Program, JSPS Grant-in-Aid for Specially Promoted Research, Grant No. 26000005, JSPS Grant-in-Aid for Scientific Research on Innovative Areas 2905, Grants No. JP17H06358, No. JP17H06361, and No. JP17H06364, JSPS Core-to-Core Program A. Advanced Research Networks, JSPS Grant-in-Aid for Scientific Research (S), Grants No. 17H06133 and No. 20H05639, JSPS Grant-in-Aid for Transformative Research Areas (A) 20A203, Grant No. JP20H05854, the joint research program of the Institute for Cosmic Ray Research, University of Tokyo, National Research Foundation and Computing Infrastructure Project of KISTI-GSDC in Korea, Academia Sinica (AS), AS Grid Center and the Ministry of Science and Technology in Taiwan under grants including Grant No. AS-CDA-105-M06, Advanced Technology Center of NAOJ, Mechanical Engineering Center of KEK.

## APPENDIX A: SENSITIVITY ESTIMATION

A key ingredient in Eqs. (3) and (4) is the detection fraction $\xi(\Lambda)$, which estimates the fraction of systems that we expect to successfully detect from some prior volume that extends past our detectors' reach. The detection fraction quantifies selection biases, and so it is critical to accurately characterize. For a population described by parameters $\Lambda$, the detection fraction is

$$\xi(\Lambda) = \int P_{\rm det}(\theta)\pi(\theta|\Lambda)d\theta. \tag{A1}$$

Here, $P_{\rm det}(\theta)$ is the detection probability: The probability that an event with parameters $\theta$ would be detected by a particular search. The detection probability depends on the sky position and orientation of the source binary, and crucially for our purposes, on the masses and redshift of a system, and, to a lesser degree, on the spins.

Given the nonideal nature of the detector data, the variation in network sensitivity over time, and the complexity of both the signal waveforms and the search pipelines, an accurate estimate of $P_{\rm det}(\theta)$ and $\xi(\Lambda)$ requires empirical methods, specifically the use of a large suite of simulated signals added to the data: injections. For analyses that focus on the BBH subpopulation in Sec. VI, we simulate compact binary signals from a reference BBH population and record which ones are successfully detected by the PyCBC, GstLAL, or MBTA search pipelines. We omit the cWB search from our volume estimate, since at present any detection of a binary merger was corroborated by a detection in the remaining pipelines. In addition, we also simulate compact binary signals from reference BNS, NSBH, and IMBH populations. These injections include binaries with component masses in the range $1-600 M_{\odot}$, have spins that are isotropic in orientation, and are uniform-in-comoving volume. Spins are drawn from a distribution that is uniform in the dimensionless spin magnitude up to a maximum of $\chi_{\rm max} = 0.998$ for black holes and $\chi_{\rm max} = 0.4$ for neutron stars. To control computational costs, the expected network signal-to-noise ratio (SNR) of each injection is computed using representative detector power spectral densities for O3. Injections with expected SNR below 6 are assumed not to be detected, and are thus removed from the set analyzed by the search pipelines. A thorough description of the injections and their underlying probability distribution is available as part of our companion data release [368]. These injections are then combined into a single dataset as a mixture model [3,368] in order to assess sensitivity across the entire parameter space and subpopulations. Our analyses in Sec. IV make use of these injections to estimate sensitivity.

Unlike previous synthetic simulation sets used in our population analysis following GWTC-2 [20], the injections used here model spins that are isotropically distributed in orientation and hence allow for orbital precession. Further, the maximum spin magnitude we assume for NS components, 0.4, is significantly larger than for previous injection sets used to estimate BNS merger rates [16]. That said, our injections have an effective $\chi_{\rm eff}$ distribution that is narrow and centered at 0 while analyses using BNS populations with small NS spins inherently have $\chi_{\rm eff} \approx 0$. Because the merger rate depends on spins primarily through the system's $\chi_{\rm eff}$, the specific assumptions made about the spin distribution at low mass have modest impact on the inferred low-mass merger rate.

Following Refs. [144,145,156,369], the point estimate for Eq. (A1) is calculated using a Monte Carlo integral over found injections:

$$\hat{\xi}(\Lambda) = \frac{1}{N_{\rm inj}} \sum_{j=1}^{N_{\rm found}} \frac{\pi(\theta_j|\Lambda)}{p_{\rm draw}(\theta_j)}, \qquad \text{(A2)}$$

where $N_{\rm inj}$ is the total number of injections, $N_{\rm found}$ are the injections that are successfully detected, and $p_{\rm draw}$ is the probability distribution from which the injections are drawn. When using this approach to estimate sensitivity, we marginalize over the uncertainty in $\hat{\xi}(\Lambda)$ and ensure that the effective number of found injections remaining after population reweighting is sufficiently high ($N_{\rm eff} > 4N_{\rm det}$) following Ref. [156]. We also compute (and some analyses like MS employ) semianalytic approximations to the integrated network sensitivity $VT(\theta,\kappa) = \int dtdzdV_c/dz/(1+z)\langle P_{\rm det}(\theta,z)\rangle(1+z)^\kappa$ for fiducial choices of $\kappa$, appropriate to characterize sensitivity to a population with a fixed redshift evolution.

For the O3 observing period, we characterize the found injections as those recovered with a FAR below the corresponding thresholds used in population analyses described in this paper (1 and 0.25 $\rm yr^{-1}$) in either PyCBC, GstLAL, or MBTA. For the O1 and O2 observing periods, we supplement the O3 pipeline injections with mock injections drawn from the same distribution $p_{\rm draw}$ above. For the mock injections, we calculate $P_{\rm det}(m_1, m_2, z, \chi_{1,z}, \chi_{2,z})$ according to the semianalytic approximation used in our analysis of GWTC-2 [114] based on a network signal-to-noise ratio threshold $\rho = 10$ and representative strain noise power spectral densities estimated from data recorded during the O1 and O2 observing runs. We combine O1, O2, and O3 injection sets ensuring a constant rate of injections across the total observing time [370].

## APPENDIX B: POPULATION MODEL DETAILS

In this appendix we provide details about the low-dimensional parametrized population models described above in Sec. III. Each subsection includes a table with a summary of the parameters for that model and the prior distribution used for each parameter. The prior distributions are indicated using abbreviations: For example, $U(0,1)$ translates to uniform on the interval (0,1), $\mathrm{LU}(10^{-6}, 10^5)$ translates to log-uniform on the interval $10^{-6}, 10^5$, and $N(0,1)$ translates to a Gaussian distribution with mean 0 and standard deviation of 1.

Using Monte Carlo summations over samples from each event's posterior distribution to approximate the integral in the likelihood given in Eq. (4) results in statistical error in the likelihood estimates [156,157]. In order to avoid including relics from unconverged Monte Carlo integrals in the posterior distribution, we introduce a data-dependent constraint on the prior determined by the number of effective samples used in the Monte Carlo integral. We define the effective number of samples as

$$N_{\rm eff} = \frac{(\sum_i w_i)^2}{\sum_i w_i^2}, \qquad \text{(B1)}$$

where $w_i$ is the weight for the $i$th event in the Monte Carlo integral.

For the Truncated, Power Law+Peak, Power Law +Spline, Power law+dip+break, Default, Gaussian, and power law population models, we assign only nonzero likelihoods to points in parameter space with an effective sample size of at least the number of observed events in our event list. This is similar to the convergence constraints we enforce when computing sensitivity (see Appendix A) [156].

### 1. Details of mass population models

#### a. Truncated mass model

The Truncated mass model serves as the primary component for some of our mass models. The primary mass distribution for this model follows a power law with spectral index $\alpha$, and with a sharp cutoff at the lower end $m_{\rm min}$ and the upper end of the distribution $m_{\rm max}$:

$$\pi(m_1|\alpha, m_{\rm min}, m_{\rm max}) \propto \begin{cases} m_1^{-\alpha} & m_{\rm min} < m_1 < m_{\rm max}, \\ 0 & \text{otherwise}. \end{cases} \qquad \text{(B2)}$$

Meanwhile, the mass ratio $q \equiv m_2/m_1$ follows a power-law distribution with spectral index $\beta_q$,

TABLE V. Summary of TRUNCATED model parameters.

| Parameter | Description | Prior |
|---|---|---|
| $\alpha$ | Spectral index for the power law of the primary mass distribution | $U(-4, 12)$ |
| $\beta_q$ | Spectral index for the power law of the mass ratio distribution | $U(-2, 7)$ |
| $m_{\min}$ | Minimum mass of the power-law component of the primary mass distribution | $U(2M_\odot, 10M_\odot)$ |
| $m_{\max}$ | Maximum mass of the power-law component of the primary mass distribution | $U(30M_\odot, 100M_\odot)$ |

$$\pi(q|\beta_q, m_{\min}, m_1) \propto \begin{cases} q^{\beta_q} & m_{\min} < m_2 < m_1, \\ 0 & \text{otherwise.} \end{cases} \tag{B3}$$

The parameters for this model are summarized in Table V. For this model, as well as further mass models where a prior on the total merger rate is not specified, the rate prior is proportional to $1/R$, or equivalently to $1/N$ in the notation of Eqs. (3) and (4).

### b. POWER LAW+PEAK mass model

This model is equivalent to Model C from our GWTC-1 analysis [114]. The primary mass distribution is a truncated power law, with the addition of tapering at the lower mass end of the distribution and a Gaussian component:

$$\begin{aligned}\pi(m_1|\lambda_{\rm peak}, \alpha, m_{\min}, \delta_m, m_{\max}, \mu_m, \sigma_m) &= [(1-\lambda_{\rm peak})\mathcal{P}(m_1|-\alpha, m_{\max}) \\ &\quad + \lambda_{\rm peak} G(m_1|\mu_m, \sigma_m)] S(m_1|m_{\min}, \delta_m).\end{aligned} \tag{B4}$$

Here, $\mathcal{P}(m_1|-\alpha, m_{\max})$ is a normalized power-law distribution with spectral index $-\alpha$ and high-mass cutoff $m_{\max}$. Meanwhile, $G(m_1|\mu_m, \sigma_m)$ is a normalized Gaussian distribution with mean $\mu_m$ and width $\sigma_m$. The parameter $\lambda_{\rm peak}$ is a mixing fraction determining the relative prevalence of mergers in $\mathfrak{P}$ and $G$. Finally, $S(m_1, m_{\min}, \delta_m)$ is a smoothing function, which rises from 0 to 1 over the interval $(m_{\min}, m_{\min} + \delta_m)$:

$$S(m|m_{\min}, \delta_m) = \begin{cases} 0 & (m < m_{\min}), \\ [f(m - m_{\min}, \delta_m) + 1]^{-1} & (m_{\min} \le m < m_{\min} + \delta_m), \\ 1 & (m \ge m_{\min} + \delta_m), \end{cases} \tag{B5}$$

with

$$f(m', \delta_m) = \exp\left(\frac{\delta_m}{m'} + \frac{\delta_m}{m' - \delta_m}\right). \tag{B6}$$

The conditional mass ratio distribution in this model also includes the smoothing term:

$$\pi(q|\beta, m_1, m_{\min}, \delta_m) \propto q^{\beta_q} S(qm_1|m_{\min}, \delta_m). \tag{B7}$$

The parameters for this model are summarized in Table VI.

### c. POWER LAW+SPLINE mass model

The POWER LAW+SPLINE mass model explicitly applies a perturbation to a modified version of the fiducial POWER LAW+PEAK model that does not include the Gaussian peak [95]. Let $p(m_1|\alpha, m_{\min}, m_{\max}, \delta_m)$ be the modified POWER LAW+PEAK model without the Gaussian, then the primary mass distribution for the POWER LAW+SPLINE model is given as

$$\begin{aligned}p_{\rm PS}(m_1|\alpha, m_{\min}, m_{\max}, \delta_m, \{f_i\}) \\ = kp(m_1|\alpha, m_{\min}, m_{\max}, \delta_m) \exp[f(m_1|\{f_i\})].\end{aligned} \tag{B8}$$

Above, $k$ is a normalization factor found by numerically integrating $p_{\rm PS}$ over the range of allowed primary masses,

TABLE VI. Summary of POWER LAW+PEAK model parameters.

| Parameter | Description | Prior |
|---|---|---|
| $\alpha$ | Spectral index for the power law of the primary mass distribution | $U(-4, 12)$ |
| $\beta_q$ | Spectral index for the power law of the mass ratio distribution | $U(-2, 7)$ |
| $m_{\min}$ | Minimum mass of the power-law component of the primary mass distribution | $U(2M_\odot, 10M_\odot)$ |
| $m_{\max}$ | Maximum mass of the power-law component of the primary mass distribution | $U(30M_\odot, 100M_\odot)$ |
| $\lambda_{\rm peak}$ | Fraction of BBH systems in the Gaussian component | $U(0, 1)$ |
| $\mu_m$ | Mean of the Gaussian component in the primary mass distribution | $U(20M_\odot, 50M_\odot)$ |
| $\sigma_m$ | Width of the Gaussian component in the primary mass distribution | $U(1M_\odot, 10M_\odot)$ |
| $\delta_m$ | Range of mass tapering at the lower end of the mass distribution | $U(0M_\odot, 10M_\odot)$ |

TABLE VII. Summary of POWER LAW+SPLINE model parameters.

| Parameter | Description | Prior |
|---|---|---|
| $\alpha$ | Spectral index for the power law of the primary mass distribution | $U(-4, 12)$ |
| $\beta_q$ | Spectral index for the power law of the mass ratio distribution | $U(-2, 7)$ |
| $m_{\min}$ | Minimum mass of the power-law component of the primary mass distribution | $U(2M_{\odot}, 10M_{\odot})$ |
| $m_{\max}$ | Maximum mass of the power-law component of the primary mass distribution | $U(30M_{\odot}, 100M_{\odot})$ |
| $\delta_m$ | Range of mass tapering at the lower end of the mass distribution | $U(0M_{\odot}, 10M_{\odot})$ |
| $\{f_i\}$ | Spline perturbation knot heights | $N(0, 1)$ |

$f(m_1|\{f_i\})$ is the perturbation function we model with cubic splines, and $\{f_i\}$ are the heights of the $n$ knots from which $f$ is interpolated. The $n$ knot locations are fixed, spaced linearly in $\log m_1$ space from 2 to $100M_{\odot}$. We additionally restrict the perturbations to converge to the underlying distribution at the boundary nodes by fixing both $f_0$ and $f_{n-1}$ to be 0. We choose $n = 20$ to be the optimal number of knots for this analysis following the same procedure in Ref. [95], which adds a total of 18 additional parameters describing the perturbations to the underlying model. In addition to the primary mass, the conditional mass ratio distribution follows the same form as the POWER LAW+PEAK model defined in Eq. (B7). For each mass distribution inference with the POWER LAW+SPLINE model, we simultaneously fit the spin distribution with the DEFAULT model and the redshift evolution of the merger rate with the POWER LAW evolution model. The parameters and chosen prior distributions for the POWER LAW+SPLINE model are summarized in Table VII.

#### d. FLEXIBLE MIXTURES model

The FLEXIBLE MIXTURES model, Vamana, predicts the population using a sum of weighted components. Each component is composed of a Gaussian, another Gaussian, and a power law to model the chirp mass, the mass ratio, and the components of the spin angular momenta parallel to the orbital angular momentum, respectively. The model is defined as

$$p(\mathcal{M}, q, s_{1z}, s_{2z}|\boldsymbol{\lambda}) = \sum_{i=1}^{N} w_i G(\mathcal{M}|\mu_i^{\mathcal{M}}, \sigma_i^{\mathcal{M}}) G(s_{1z}|\mu_i^{sz}, \sigma_i^{sz}) \times G(s_{2z}|\mu_i^{sz}, \sigma_i^{sz}) \mathcal{P}(q|\alpha_i^q, q_i^{\min}, 1), \tag{B9}$$

where $G$ is the normal distribution and $\mathcal{P}$ is the truncated power law. For the presented analysis we use $N = 11$ components. This choice maximizes the marginal likelihood; however, the predicted population is robust for a wide range of $N$. For a detailed description of this model, see Ref. [98]. The FLEXIBLE MIXTURES model uses a power law to model the redshift evolution of the merger rate, as described in Appendix B 3. The merger rate has a uniform-in-log-distributed prior; the prior distributions for parameters in Eq. (B9) are summarized in Table VIII.

#### e. BINNED GAUSSIAN PROCESS model

The BINNED GAUSSIAN PROCESS models the rate densities $m_1 m_2 dR^i/dm_1 dm_2 = n^i$ as a binned Gaussian process where the index $i$ denotes a particular bin in the two-dimensional $\log m_1 - \log m_2$ parameter space [169,170]. The bin edges in the analysis presented in the paper are located at $[1, 2, 2.5, 3, 4, 5, 6.5, 8, 10, 15, 20, 30, 40, 50, 60, 70, 80, 100]M_{\odot}$ with the assumption that $m_2 \leq m_1$. The probabilistic model for the logarithm of the rate density in each bin is defined as

$$\log n^i \sim N(\mu, \Sigma), \tag{B10}$$

where $\mu$ is the mean of the Gaussian process and $\Sigma$ is the covariance matrix that correlates the bins. Each element of the covariance matrix $\Sigma$ is generated using a squared-exponential kernel $k(x,x')$ which is defined as

$$k(x,x') = \sigma^2 \exp\left(\frac{-(x-x')^2}{2l^2}\right). \tag{B11}$$

TABLE VIII. Summary of FLEXIBLE MIXTURES model parameters. All rates are in $\mathrm{Gpc}^{-3}\,\mathrm{yr}^{-1}$.

| Parameter | Description | Prior |
|---|---|---|
| $w_i$ | Mixing weights | Dirichlet($\boldsymbol{\alpha}$), $\alpha_{1\ldots N} = 1/N$ |
| $\mu_i^{\mathcal{M}}$ | Mean of the normal distribution modeling the chirp mass | LU($5.2M_{\odot}$, $65M_{\odot}$) |
| $\sigma_i^{\mathcal{M}}$ | Scale of the normal distribution modeling the chirp mass | $U(0.02\ \mu_i^{\mathcal{M}},\ 0.18\ \mu_i^{\mathcal{M}})$ |
| $\mu_i^{sz}$ | Mean of the normal distribution modeling the aligned-spin distribution | $U(-0.5,\ 0.5)$ |
| $\sigma_i^{sz}$ | Scale of the normal distribution modeling the aligned-spin distribution | $U(0.05, 0.6)$ |
| $q_i^{\min}$ | Minimum value of the mass ratio | $U(0.1, 0.95)$ |
| $\alpha_i^q$ | Slope of the power law | $U(-7, 2)$ |
| $\mathcal{R}$ | Merger rate | LU(1, 100) |

TABLE IX. Summary of BINNED GAUSSIAN PROCESS model parameters.

| Parameter | Description | Prior |
|---|---|---|
| $\mu$ | Mean log(rate) in each bin | $N(0,10)$ |
| $\sigma$ | Amplitude of the covariance kernel | $N(0,10)$ |
| $\log(l)$ | log(length scale) of the covariance kernel | $N(-0.085,0.93)$ |

For the specific analysis here, we take $x$, $x'$ to be the bin centers in log $m$. The parameter $\sigma$ models the amplitude of the covariances while $l$ is a parameter that defines the length scales over which bins are correlated. The prior distribution chosen here for the length scale is a log-normal distribution with a mean that is the average between the minimum bin spacing

$$\Delta_{\min} \equiv \min_{m_1,m_2} \Delta \log m, \tag{B12}$$

and the maximum bin spacing

$$\Delta_{\max} \equiv \max_{m_1,m_2} \Delta \log m, \tag{B13}$$

with a standard deviation of $(\Delta_{\max} - \Delta_{\min}/4)$. This constrains (at "$2\sigma$" in the prior) the correlation length for the GP to lie between "one bin" and "all the bins." For our analyses presented in the paper, the mean and standard deviation are $-0.085$ and 0.93, respectively. The BINNED GAUSSIAN PROCESS model assumes a redshift distribution such that the overall merger rate of compact binaries is uniform-in-comoving volume. The spin distributions for each component are isotropic in direction and uniform in the spin magnitude with a maximum spin of 0.998 for BHs and 0.4 for NSs; the prior distribution for the relevant parameters in Eqs. (B10) and (B11) is summarized in Table IX.

#### f. POWER LAW+DIP+BREAK model

The POWER LAW+DIP+BREAK model explicitly searches for separation in masses between two subpopulations by employing a broken power law with a dip at the location of the power-law break [159,160]. The dip is modeled by a notch filter with depth $A$, which is fit along with the other model parameters in order to determine the existence and depth of a potential mass gap. No gap corresponds to $A = 0$, whereas $A = 1$ corresponds to precisely zero merger rate over some interval. POWER LAW+DIP +BREAK also employs a low-pass filter at high masses to allow for a tapering of the mass spectrum, which has the effect of a smooth second break to the power law.

The PDB model assumes a merger rate that is uniform-in-comoving volume. It also assumes a spin distribution with isotropically oriented component spins and uniform component spin magnitudes.

The joint mass distribution in this model has the form:

$$p(m_1, m_2) \propto p(m_1)p(m_2)(m_2/m_1)^{\beta}, \tag{B14}$$

$$p(m) \propto p_{\rm pl}(m)n(m)\ell(m), \tag{B15}$$

$$n(m) = 1 - \frac{A}{[1+(M^{\rm gap}_{\rm low}m)^{\eta_{\rm low}}][1+(M^{\rm gap}_{\rm high}m)^{\eta_{\rm high}}]}, \quad \text{and} \tag{B16}$$

$$\ell(m) = \frac{1}{1+(m/m_{\max})^{\eta}}, \tag{B17}$$

where $p_{\rm pl}(m)$ is a broken power law with exponents $\alpha_1$ between $m_{\min}$ and $M^{\rm gap}_{\rm low}$ and $\alpha_2$ between $M^{\rm gap}_{\rm low}$ and $m_{\max}$. The parameters for this model are summarized in Table X.

#### g. Neutron star mass models

The mass models adopted for the BNS and NSBH events in Sec. V assume a basic mass distribution that is common to all NSs, with random pairing into compact binaries. The basic mass distribution is taken to be either a power law or, inspired by the shape of the Galactic BNS mass distribution [91–93], a Gaussian. The BH mass distribution is fixed to be uniform between 3 and $60M_{\odot}$. The NS mass distribution analysis assumes definite source classifications for the events. Thus, the joint mass distribution takes the form

$$p(m_1,m_2) \propto \begin{cases} p(m_1)p(m_2) & \text{if BNS,} \\ U(3M_{\odot},60M_{\odot})p(m_2) & \text{if NSBH,} \end{cases} \tag{B18}$$

with $p(m)$ either a power law with exponent $\alpha$, minimum mass $m_{\min}$ and maximum mass $m_{\max}$, or a Gaussian with a peak of width $\sigma$ at $\mu$, plus sharp minimum and maximum mass cutoffs $m_{\min}$, $m_{\max}$. We call these models POWER and PEAK, respectively. Their hyperparameters and the choices for their prior distributions are listed in Table XI. We additionally impose the constraint $m_{\min} \le \mu \le m_{\max}$ on the PEAK model. Besides the flat $m_{\max}$ prior described in the table, for the analyses excluding GW190814 we use a prior proportional to the cumulative distribution function of $M_{\rm max,TOV}$, i.e., $p(m_{\max}) \propto \int_{m_{\max}}^{\infty} dM_{\rm max,TOV}\, p(M_{\rm max,TOV})$. This enforces our expectation that the NS masses in the gravitational-wave population should not exceed $M_{\rm max,TOV}$.

TABLE X. Summary of Power Law+dip+Break model parameters. The first entries describe the mass distribution parameters, and the last two entries describe the spin distribution parameters.

| Parameter | Description | Prior |
|---|---|---|
| $\alpha_1$ | Spectral index for the power law of the mass distribution at low mass | $U(-8, 2)$ |
| $\alpha_2$ | Spectral index for the power law of the mass distribution at high mass | $U(-3, 2)$ |
| $A$ | Lower mass gap depth | $U(0, 1)$ |
| $M_{\rm low}^{\rm gap}$ | Location of lower end of the mass gap | $U(1.4M_\odot, 3M_\odot)$ |
| $M_{\rm high}^{\rm gap}$ | Location of upper end of the mass gap | $U(3.4M_\odot, 9M\odot)$ |
| $\eta_{\rm low}$ | Parameter controlling how the rate tapers at the low end of the mass gap | 50 |
| $\eta_{\rm high}$ | Parameter controlling how the rate tapers at the low end of the mass gap | 50 |
| $\eta$ | Parameter controlling tapering the truncated power law at high mass | $U(-4, 12)$ |
| $\beta$ | Spectral index for the power-law-in-mass-ratio pairing function | $U(-2, 7)$ |
| $m_{\rm min}$ | Minimum mass of the power-law component of the mass distribution | $U(1M_\odot, 1.4M_\odot)$ |
| $m_{\rm max}$ | Maximum mass of the power-law component of the mass distribution | $U(35M_\odot, 100M_\odot)$ |
| $a_{\rm max,NS}$ | Maximum allowed component spin for objects with mass $< 2.5M_\odot$ | 0.4 |
| $a_{\rm max,BH}$ | Maximum allowed component spin for objects with mass $\geq 2.5M_\odot$ | 1 |

TABLE XI. Summary of Power and Peak NS mass model parameters.

| Parameter | Description | Prior |
|---|---|---|
| $\alpha$ | Spectral index for the power law in the Power NS mass distribution | $U(-4, 12)$ |
| $m_{\rm min}$ | Minimum mass of the NS mass distribution | $U(1.0M_\odot, 1.5M_\odot)$ |
| $m_{\rm max}$ | Maximum mass of the NS mass distribution | $U(1.5M_\odot, 3.0M_\odot)$ |
| $\mu$ | Location of the Gaussian peak in the Peak NS mass distribution | $U(1.0M_\odot, 3.0M_\odot)$ |
| $\sigma$ | Width of the Gaussian peak in the Peak NS mass distribution | $U(0.01M_\odot, 2.00M_\odot)$ |

### 2. Details of spin population models

#### a. Default spin model

This model was introduced in Ref. [114]. Following Ref. [164], the dimensionless spin-magnitude distribution is taken to be a beta distribution,

$$\pi(\chi_{1,2}|\alpha_\chi, \beta_\chi) = \mathrm{Beta}(\alpha_\chi, \beta_\chi), \qquad \text{(B19)}$$

where $\alpha_\chi$ and $\beta_\chi$ are the standard shape parameters that determine the distribution's mean and variance. The beta distribution is convenient because it is bounded on (0,1). The distributions for $\chi_1$ and $\chi_2$ are assumed to be the same. We define $z_i = \cos\theta_i$ as the cosine of the tilt angle between component spin and a binary's orbital angular momentum, and assume that $\mathbf{z}$ is distributed as a mixture of two populations [165]:

$$\pi(\mathbf{z}|\zeta, \sigma_t) = \zeta G_t(\mathbf{z}|\sigma_t) + (1-\zeta)\mathfrak{I}(\mathbf{z}). \qquad \text{(B20)}$$

Here, $\mathfrak{I}(z)$ is an isotropic distribution, while $G_t(z|\sigma_t)$ is a truncated two-dimensional Gaussian peaking at $\mathbf{z} = 0$ (perfect alignment) with width $\sigma_t$. The mixing parameter $\zeta$ controls the relative fraction of mergers drawn from the isotropic distribution and Gaussian subpopulations. The isotropic subpopulation is intended to accommodate dynamically assembled binaries, while $G_t$ is a model for field mergers. The parameters for this model and their priors are summarized in Table XII. Additional constraints to the priors on $\mu_\chi$ and $\sigma_\chi^2$ are applied by setting $\alpha_\chi, \beta_\chi > 1$.

TABLE XII. Summary of Default spin parameters.

| Parameter | Description | Prior |
|---|---|---|
| $\mu_\chi$ | Mean of the beta distribution of spin magnitudes | $U(0, 1)$ |
| $\sigma_\chi^2$ | Variance of the beta distribution of spin magnitudes | $U(0.005, 0.25)$ |
| $\zeta$ | Mixing fraction of mergers from truncated Gaussian distribution | $U(0, 1)$ |
| $\sigma_t$ | Width of truncated Gaussian determining typical spin misalignment | $U(0.1, 4)$ |

TABLE XIII. Summary of GAUSSIAN spin parameters. The $\chi_{\mathrm{eff,min}}$ and $\zeta$ parameters appear only in variants of the GAUSSIAN model, as we discuss below.

| Parameter | Description | Prior |
|---|---|---|
| $\mu_{\mathrm{eff}}$ | Mean of the $\chi_{\mathrm{eff}}$ distribution | $U(-1,1)$ |
| $\sigma_{\mathrm{eff}}$ | Standard deviation of the $\chi_{\mathrm{eff}}$ distribution | $U(0.05,1)$ |
| $\mu_p$ | Mean of the $\chi_{\mathrm{p}}$ distribution | $U(0.05,1)$ |
| $\sigma_p$ | Standard deviation of the $\chi_{\mathrm{p}}$ distribution | $U(0.05,1)$ |
| $\rho$ | Degree of correlation between $\chi_{\mathrm{eff}}$ and $\chi_{\mathrm{p}}$ | $U(-0.75,0.75)$ |
| $\chi_{\mathrm{eff,min}}$ | Lower truncation bound on $\chi_{\mathrm{eff}}$ | $U(-1,\mu_{\mathrm{eff}})$ |
| $\zeta$ | Nonvanishing mixture fraction in Eq. (18) | $U(0,1)$ |

#### b. GAUSSIAN spin model

In addition to the distribution of component spin magnitudes and tilts, we explore the distribution of the effective inspiral spin parameter $\chi_{\mathrm{eff}}$ and the effective precession spin parameter $\chi_{\mathrm{p}}$. In particular, we wish to measure the mean and variance of each parameter, and so model the joint distribution of $\chi_{\mathrm{eff}}$ and $\chi_{\mathrm{p}}$ as a bivariate Gaussian:

$$\pi(\chi_{\mathrm{eff}},\chi_{\mathrm{p}}|\mu_{\mathrm{eff}},\sigma_{\mathrm{eff}},\mu_p,\sigma_p,\rho) \propto G(\chi_{\mathrm{eff}},\chi_{\mathrm{p}}|\boldsymbol{\mu},\boldsymbol{\Sigma}). \tag{B21}$$

The mean of this distribution is $\boldsymbol{\mu} = (\mu_{\mathrm{eff}},\mu_p)$, and its covariance matrix is

$$\boldsymbol{\Sigma} = \begin{pmatrix} \sigma_{\mathrm{eff}}^2 & \rho\sigma_{\mathrm{eff}}\sigma_p \\ \rho\sigma_{\mathrm{eff}}\sigma_p & \sigma_p^2 \end{pmatrix}. \tag{B22}$$

The population parameters governing this model and their corresponding priors are shown in Table XIII. Equation (B21) is truncated to the physically allowed range of each effective spin parameter, with $\chi_{\mathrm{eff}} \in (-1,1)$ and $\chi_{\mathrm{p}} \in (0,1)$. All results in the main text using the GAUSSIAN model are obtained while simultaneously fitting for the BBH mass distribution, assuming the POWER LAW+PEAK model, and the evolving redshift distribution model in Appendix B 3 below.

Two variants of this model are additionally discussed in Sec. VII. In the first, Eq. (B21) is modified such that the effective inspiral spin parameter is truncated not on the interval $(-1,1)$, but on $(\chi_{\mathrm{eff,min}},1)$, where $\chi_{\mathrm{eff,min}}$ is another parameter to be inferred by the data. The second variant inspired by analysis of GWTC-2 [251,252] and defined in Eq. (18), alternatively treats the $\chi_{\mathrm{eff}}$ distribution as a mixture between a bulk component with a variable mean and width and a narrow zero-spin component centered on $\chi_{\mathrm{eff}} = 0$. In this second variant, we measure only the marginal $\chi_{\mathrm{eff}}$ distribution, implicitly assuming that the remaining spin degree of freedom is distributed uniformly and isotropically. As $\chi_{\mathrm{eff}}$ is the primary spin measurable, we do not expect this implicit prior to have a strong effect.

### 3. Redshift evolution model

The POWER LAW redshift evolution model parametrizes the merger rate density per comoving volume and source time as [146]

$$\mathcal{R}(z) = \mathcal{R}_0(1+z)^{\kappa}, \tag{B23}$$

where $\mathcal{R}_0$ denotes the merger rate density at $z = 0$. This implies that the redshift distribution is

$$\frac{dN}{dz} = \mathcal{C}\frac{dV_c}{dz}(1+z)^{\kappa-1}, \tag{B24}$$

where $dV_c/dz$ is the differential comoving volume, and $\mathcal{C}$ is related to $\mathcal{R}_0$ by

$$\mathcal{R}_0 = \mathcal{C}\frac{dV_c}{dz}\left[\int_0^{z_{\mathrm{max}}} dz \frac{dV_c}{dz}(1+z)^{\kappa-1}\right]^{-1}. \tag{B25}$$

We adopt $z_{\mathrm{max}} = 2.3$ as this is a conservative upper bound on the redshift at which we could detect BBH systems during O3, for both detection thresholds used in this work. We employ a uniform prior on $\kappa$ centered at $\kappa = 0$. We take a sufficiently wide prior so that the likelihood is entirely within the prior range, $\kappa \in (-6,6)$.

### 4. Models with multiple independent components

#### a. MULTI SOURCE model

The MULTI SOURCE model [371] extends the MULTISPIN BBH model [20] to include additional subpopulations for BNS and NSBH systems. Each subpopulation (two for BBH, one for BNS, and one for NSBH) is assumed to have an independent rate parameter. Priors for all parameters are given in Table XIV.

The BBH subpopulation is itself a mixture of two subpopulations, (i) a power-law mass distribution $m_1^{-\alpha}q^{\beta}$ truncated to a range $[m_{\mathrm{min,BBH}}, m_{\mathrm{max,BBH}}]$ which is inferred from the data, and (ii) a Gaussian in $(m_1, m_2)$ with independent mean and standard deviation parameters $\mu_{m_1,\mathrm{BBH}}$, $\mu_{m_2,\mathrm{BBH}}$, $\sigma_{m_1,\mathrm{BBH}}$, $\sigma_{m_2,\mathrm{BBH}}$. Both subpopulations and both binary components within them follow independent DEFAULT spin models, with $\zeta \equiv 1$.

TABLE XIV. Summary of MULTI SOURCE model parameters. All rates are in $\mathrm{Gpc}^{-3}\ \mathrm{yr}^{-1}$, and all masses in $M_\odot$. Rate, mass, and spin hyperparameters are separated by horizontal lines.

| Parameter | Description | Prior |
|---|---|---|
| $R_{\mathrm{BBH,pl}}$ | Local merger rate for the BBH power-law subpopulation | $U(0, 1000)$ |
| $R_{\mathrm{BBH},g}$ | Local merger rate for the BBH Gaussian subpopulation | $U(0, 1000)$ |
| $R_{\mathrm{BNS}}$ | Local merger rate for the BNS subpopulation | $U(0, 2000)$ |
| $R_{\mathrm{NSBH}}$ | Local merger rate for the NSBH subpopulation | $U(0, 500)$ |
| $\alpha$ | Primary mass spectral index for the BBH power-law subpopulation | $U(-4, 12)$ |
| $\beta$ | Mass ratio spectral index for the BBH power-law subpopulation | $U(-4, 10)$ |
| $m_{\mathrm{min,BBH,pl}}$ | Minimum mass of the BBH power-law subpopulation | $U(2, 10)$ |
| $m_{\mathrm{max,BBH,pl}}$ | Maximum mass of the BBH power-law subpopulation | $U(30, 100)$ |
| $\mu_{m_1,\mathrm{BBH},g}$ ($\mu_{m_2,\mathrm{BBH},g}$) | Centroid of the primary (secondary) mass distribution for the BBH Gaussian subpopulation | $U(20, 50)$ |
| $\sigma_{m_1,\mathrm{BBH},g}$ ($\sigma_{m_2,\mathrm{BBH},g}$) | Width of the primary (secondary) mass distribution for the BBH Gaussian subpopulation | $U(0.4, 20)$ |
| $\mu_{m,\mathbf{NSBH}}$ | Centroid of the BH mass distribution for the NSBH | $U(3, 50)$ |
| $\sigma_{m,\mathbf{NSBH}}$ | Width of the BH mass distribution for the NSBH | $U(0.4, 20)$ |
| $\mu_{m,\mathrm{NS}}$ | Centroid of the NS mass distribution | $U(1, 3)$ |
| $\sigma_{m,\mathrm{NS}}$ | Width of the NS mass distribution | $U(0.05, 3)$ |
| $m_{\mathrm{max,NS}}$ | Maximum mass of all NSs | $U(2, 3)$ |
| $\mu_{\chi_1,\mathrm{BBH,pl}}$ ($\mu_{\chi_2,\mathrm{BBH,pl}}$) | Mean of the beta distribution of primary (secondary) spin magnitudes for the BBH Gaussian subpopulation | $U(0, 1)$ |
| $\sigma^2_{\chi_1,\mathrm{BBH,pl}}$ ($\sigma^2_{\chi_2,\mathrm{BBH,pl}}$) | Variance of the beta distribution of primary (secondary) spin magnitudes for the BBH Gaussian subpopulation | $U(0, 0.25)$ |
| $\sigma_{t_1,\mathrm{BBH,pl}}$ ($\sigma_{t_2,\mathrm{BBH,pl}}$) | Width of truncated Gaussian, determining typical primary (secondary) spin misalignment for the BBH Gaussian subpopulation | $U(0, 4)$ |
| $\mu_{\chi_1,\mathrm{BBH},g}$ ($\mu_{\chi_2,\mathrm{BBH},g}$) | Mean of the beta distribution of primary (secondary) spin magnitudes for the BBH Gaussian subpopulation | $U(0, 1)$ |
| $\sigma^2_{\chi_1,\mathrm{BBH},g}$ ($\sigma^2_{\chi_2,\mathrm{BBH},g}$) | Variance of the beta distribution of primary (secondary) spin magnitudes for the BBH Gaussian subpopulation | $U(0, 0.25)$ |
| $\sigma_{t_1,\mathrm{BBH},g}$ ($\sigma_{t_2,\mathrm{BBH},g}$) | Width of truncated Gaussian determining typical primary (secondary) spin misalignment for the BBH Gaussian subpopulation | $U(0, 4)$ |
| $\mu_{\chi,\mathrm{NS}\mathbf{BH}}$ | Mean of the beta distribution of spin magnitudes for the BH in the NSBH subpopulation | $U(0, 1)$ |
| $\sigma^2_{\chi,\mathrm{NS}\mathbf{BH}}$ | Variance of the beta distribution of spin magnitudes for the BH in the NSBH subpopulation | $U(0, 0.25)$ |
| $\sigma_{t,\mathrm{NS}\mathbf{BH}}$ | Width of truncated Gaussian determining typical primary (secondary) spin misalignment for the BH in the NSBH subpopulation | $U(0, 4)$ |
| $\mu_{\chi,\mathrm{NS}\mathbf{BH}}$ | Mean of the beta distribution of spin magnitudes for the NS in the NSBH subpopulation | $U(0, 0.05)$ |
| $\sigma^2_{\chi,\mathrm{NS}\mathbf{BH}}$ | Variance of the beta distribution of spin magnitudes for the NS in the NSBH subpopulation | $U(0, 0.0125)$ |
| $\mu_{\chi_1,\mathrm{BNS}}$ ($\mu_{\chi_2,\mathrm{BNS}}$) | Mean of the beta distribution of primary (secondary) spin magnitudes in the BNS subpopulation | $U(0, 0.05)$ |
| $\sigma^2_{\chi_1,\mathrm{BNS}}$ ($\sigma^2_{\chi_2,\mathrm{BNS}}$) | Variance of the beta distribution of primary (secondary) spin magnitudes in the BNS subpopulation | $U(0, 0.0125)$ |

Two more bivariate Gaussians in $m_1$, $m_2$ are used to model BNSs and NSBHs. The BHs in NSBHs follow a Gaussian mass distribution, with free parameters $\mu_{m,\mathrm{NS}\mathbf{BH}}$, $\sigma_{m,\mathrm{NS}\mathbf{BH}}$. As with BBHs, these BHs follow an independent DEFAULT spin model with $\zeta \equiv 1$. All three types of NSs (two in BNS and one in NSBH) follow the same Gaussian mass distribution, with free parameters $\mu_{m,\mathrm{NS}}$, $\sigma_{m,\mathrm{NS}}$, $m_{\mathrm{max,NS}}$ (the minimum mass is assumed to be $1 M_\odot$).

Each type of NS follows an independent Default spin model. To stay within astrophysically plausible spins, the magnitude distributions are scaled down to $\chi_{\max} = 0.05$. Since NS spin tilts are not well measured, we set $\zeta \equiv 0$, assuming they are isotropic, which has the effect of not wasting any samples from parameter estimation. In addition to any mass cutoffs mentioned above, all BHs' component masses are assumed to lie on the range $[2, 100] M_{\odot}$, with those in NSBHs further restricted to $[2, 50] M_{\odot}$ due to our limited injections.

## APPENDIX C: VALIDATION STUDIES

We employ several methods to validate our calculations, notably including comparing results from multiple independent analyses, reproducing previous work through O3a [20], assessing the sensitivity of our results to threshold choices (changing from 1 to 0.2 $\mathrm{yr}^{-1}$ for BBH, or from 0.25 to 1 $\mathrm{yr}^{-1}$ for analyses containing NSs), and performing posterior predictive checks as in our analysis of GWTC-2 [20]. Though these specific technical checks will not be described here, some of these checks can be reproduced with the data release associated with this paper.

Below, we describe additional validation studies we perform to assess whether our results for merger rates are sensitive to the choice of threshold, waveform systematics, or updates to our sensitivity model.

### 1. Effects of the spin distribution on merger rates across all masses

In principle, the mass, spin, and redshift distributions of binaries should be fit simultaneously in order to avoid systematic biases in inferred distribution parameters caused by correlations between measurements of these intrinsic parameters [157,372–378]. However, fixing one or more of these distributions to a realistic form typically introduces biases that have little impact on the parameters of interest. We therefore seek to determine if our choice to fix the spin distribution for the PDB and BGP models has introduced any significant biases in our inference of the mass distribution and overall merger rate.

We compare the PDB analysis presented in Sec. IV with an analysis that utilizes the same mass and redshift distribution but fits for the spin distribution rather than fixing it to one that assumes isotropic and uniformly distributed component spins. For this, we apply the Default [165] spin model described in Sec. III C 2 and Appendix 2 a. The resulting fit is compared to the fiducial analysis in Fig. 24.

We find some differences between the fixed-spin and fit-spin analyses. First, the hyperposterior for the fit-spin analysis is broader than that of the fixed-spin analysis, presumably due to an increase in free parameters. Second, some hyperparameters exhibit a slight shift. The most notable shifts are in the rate and upper gap edge parameters. The shift in the rate is to be expected because the fit to the Default model favors lower spinning objects. Since the detectors are slightly less sensitive to low-spin objects, more support for those objects implies a higher astrophysical rate. Nonetheless, all hyperposterior differences are well within statistical uncertainty, so we conclude that both the fixed-spin and fit-spin cases are acceptable, and use the fixed-spin case for our fiducial results for simplicity.

### 2. NS mass distribution including marginal events

If we loosen the FAR threshold to $< 1\ \mathrm{yr}^{-1}$ so as to include the marginal events GW190917 and GW190426, and repeat the analysis of Sec. V B, the inferred NS mass distribution is virtually unchanged. This can be seen in Fig. 25, which compares the posterior population distributions inferred with and without the marginal events. Traces from the posterior population distribution with respect to the original FAR threshold are also shown. This alternative analysis strongly suggests that substantial uncertainties in the merger rate versus mass dominate our error budget; the handful of observations made to date is not sufficient to overcome the strong impact of our highly uncertain model priors. Moreover, the masses of the NS secondaries in the marginal events are poorly constrained relative to those in GW170817, GW190425, GW200105, and GW200115, such that the FAR $< 0.25\ \mathrm{yr}^{-1}$ events continue to drive the inference.

We also test the robustness of our inferred NS mass distribution against assumptions about the BH mass distribution in NSBHs. In Sec. V B, we fix the mass distribution of BHs in this subpopulation to be uniform. If we repeat the inference using instead the $m^{-3.5}$ power law preferred by the fit to all compact object masses from Sec. IV, we find little change in the inferred NS mass distribution; only the maximum mass increases systematically, but by less than $0.1 M_{\odot}$ in the median. We attribute this to the influence of the BNS observations, which are unaffected by the BH mass model.

### 3. Merger rates including subthreshold triggers

In the main text, our merger rates were calculated after adopting a fixed significance threshold to identify confident events, then fitting population model families to the recovered events' posteriors. By design, such an approach depends on the threshold. Here we employ an alternative threshold-free method of rate estimation which lacks potential biases from an arbitrary choice of significance threshold [70].

We extend methods from GWTC-2.1 [6], also applied to the discovery of GW200105 and GW200115 [248], to estimate the event rate from the full set of triggers (including subthreshold triggers) from a specific binary merger search: here, GstLAL [99–101]. In doing so, we allow for population distributions that fit our observations and account for still-considerable uncertainty in the mass distribution, rather than adopting a fixed population model with fixed model hyperparameters. Compared to previous publications, the results presented in this section update the BBH merger rates presented in GWTC-2.1 by including O3b events [3]. We also update the NSBH rate quoted in Ref. [248] by incorporating all O3 triggers, rather than as previously truncating to the first nine months of O3.

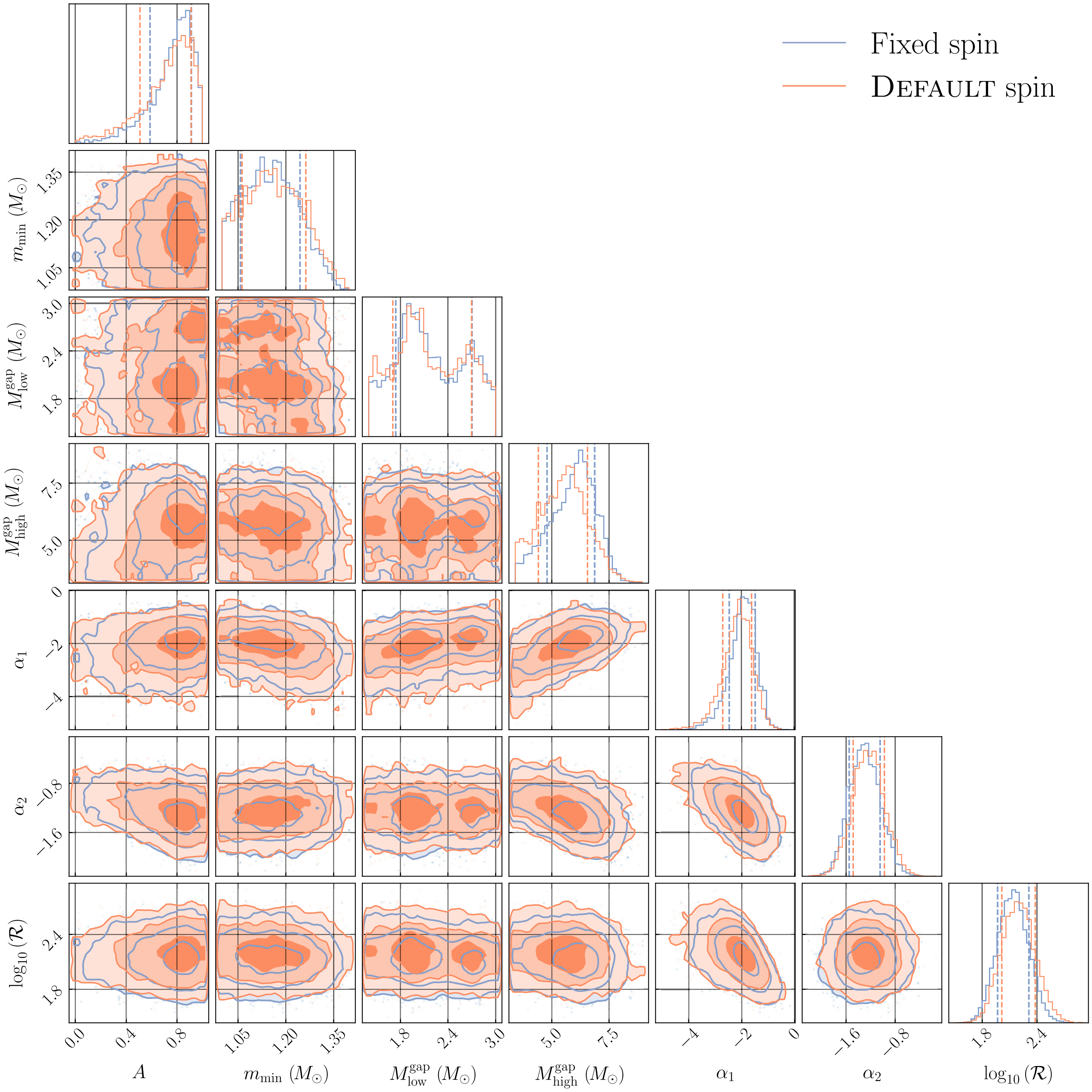

FIG. 24. Corner plot of inferred PDB mass and rate hyperparameters under an analysis that fixes the spin distribution (blue) and simultaneously fits the spin distribution using the DEFAULT model (orange). The fit-spin hyperposterior is slightly shifted and widened when compared to the fixed-spin case, but all changes are within statistical uncertainties.

We use a multicomponent mixture model [172] to construct the posterior of astrophysical counts of CBC events by assuming that foreground and background events are independent Poisson processes. We then estimate the spacetime volume sensitivity of the pipeline using simulated events which are re-scaled to an astrophysical population model [369]. We then compute the rates as the ratio of the counts to $VT$. In order to marginalize over population hyperparameters we compute several $VT$'s, each corresponding to a population hyperparameter sample drawn from the inferred hyperposterior for the astrophysical population model. Finally, we integrate over the count posterior obtained for each of these samples with an appropriate weight, effectively marginalizing over the population hyperparameters:

$$\begin{aligned} p(R|\vec{x}) &= \int p(R|\vec{\Lambda},\vec{x})p(\Lambda|\{d\})d\vec{\Lambda} \\ &= \int VT\, p(N|\vec{x})p(VT|\vec{\Lambda})p(\vec{\Lambda}|\{d\})d(VT)d\vec{\Lambda} \\ &= \sum_{i,j} VT_{ij}\times p(N_{ij}|\vec{x}), \end{aligned} \tag{C1}$$

where $\vec{x}$ is the complete set of triggers (including subthreshold triggers) and $\{d\}$ is the set of data from gravitational-wave detections used in population model inference, as in Sec. III B. The astrophysical count posterior is given by $p(N|\vec{x})$, where $N = R\times VT$; we evaluate by sampling via $N_{ij} = R\times VT_{ij}$ where $VT_{ij}$ is the $i$th $VT$ sample drawn from $p(VT|\vec{\Lambda}_j)$ for the $j$th hyperparameter sample $\vec{\Lambda}_j$ drawn from the inferred hyperposterior $p(\vec{\Lambda}|\{d\})$ [the extra $VT$ factor in Eq. (C1) arises from the Jacobian $dN/dR$]. Following Ref. [172], we take the distribution $p(VT|\vec{\Lambda}_j)$ to be

$$p(VT|\vec{\Lambda}_j) = \frac{1}{VT\sqrt{2\pi\sigma^2}}\exp\left[-\frac{[\ln VT - \ln\langle VT\rangle(\vec{\Lambda}_j)]^2}{2\sigma^2}\right], \tag{C2}$$

where $\langle VT(\vec{\Lambda}_j)\rangle$ is calculated by reweighting simulated sources to an astrophysical population with hyperparameter $\vec{\Lambda}_j$, and $\sigma$ is the quadrature sum of a calibration error of 10% [141] and Monte Carlo uncertainty.

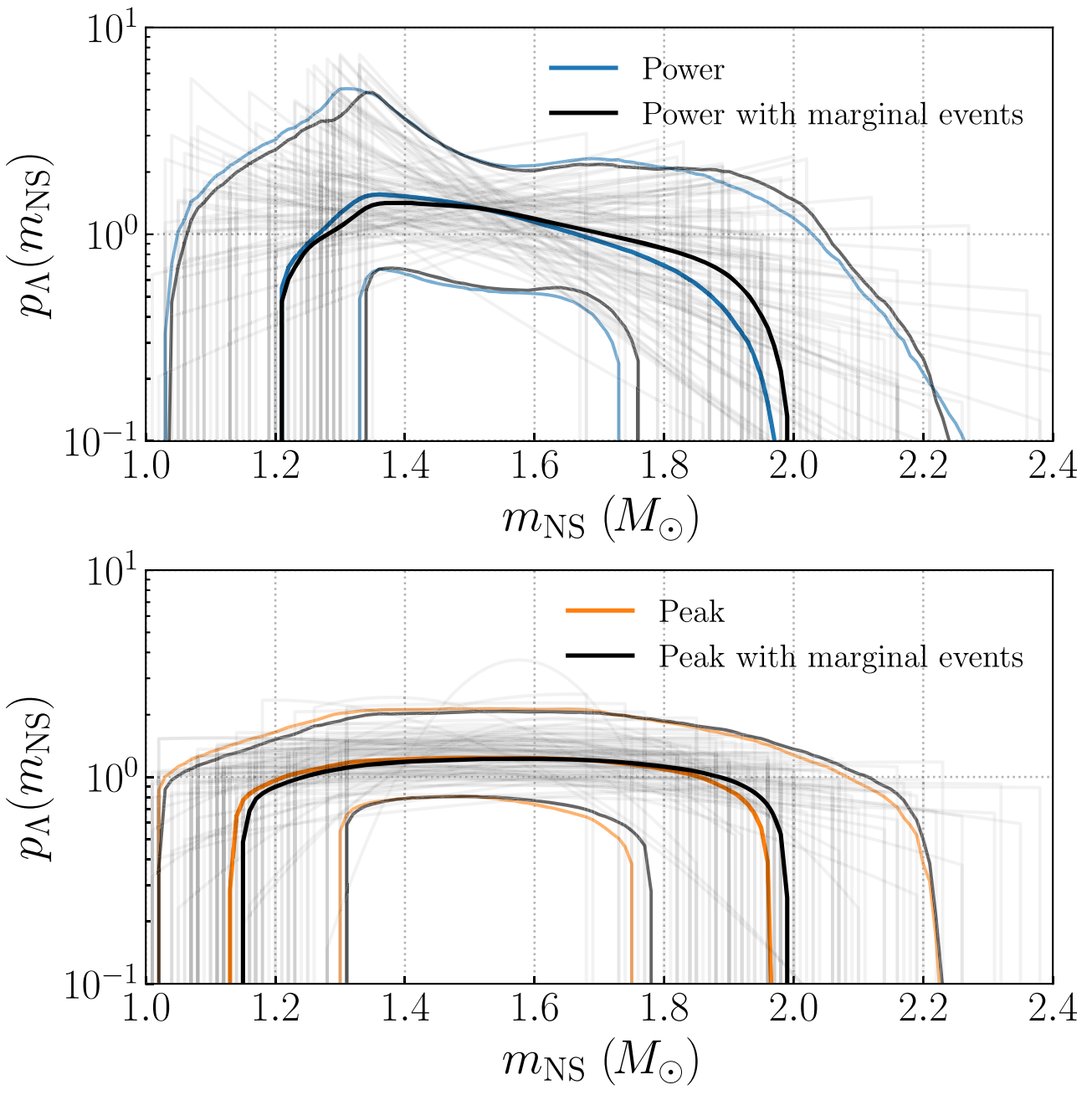

FIG. 25. Inferred neutron star mass distribution with and without the marginal events GW190426 and GW190917. Top: median and 90% confidence region of the inferred NS mass distribution for the POWER model, using the event list at a FAR threshold of 0.25 $\mathrm{yr}^{-1}$ (blue) and 1 $\mathrm{yr}^{-1}$ (orange). Traces from the posterior population distribution with respect to the stricter FAR threshold are plotted in gray. Bottom: same as the top panel but for the PEAK model. The inclusion of the marginal events has a negligible impact on the inferred mass distribution.

Using hyperparameter samples from the posterior inferred using the PP model with data through the end of O3, as in Sec. VI and imposing a Jeffreys prior $\propto N^{-1/2}$ on the astrophysical counts, we compute a BBH merger rate of 24.81–63.58 $\mathrm{Gpc}^{-3}\,\mathrm{yr}^{-1}$. A similar calculation for the BGP model, again with a Jeffreys prior $\propto N^{-1/2}$ imposed on the astrophysical counts, yields a NSBH merger rate of 14.57–187.96 $\mathrm{Gpc}^{-3}\,\mathrm{yr}^{-1}$, which is consistent with 11–140 $\mathrm{Gpc}^{-3}\,\mathrm{yr}^{-1}$, the joint inference for the NSBH merger rate presented in the main text. We also compute a BNS merger rate using a fixed population of BNSs, distributed uniformly in component masses that lie within 1 to $2.5M_\odot$. This uses the same multicomponent mixture model [172] as described above, with the only difference being, that instead of marginalizing over population hyperparameters like with the BBH and NSBH merger rates, for BNSs, we use a fixed population. Hence, it updates the BNS merger rate reported in GWTC-2.1 [6] by including all of O1 through O3 instead of truncating at O3a. We report a BNS merger rate of 28.76–462.23 $\mathrm{Gpc}^{-3}\,\mathrm{yr}^{-1}$ which is consistent with the GWTC 2.1 rate as well as the other BNS rates quoted in this paper that were computed from only high-significance triggers.

### 4. Effect of waveform systematics on population

All O3b BBH events analyzed in this paper have source properties inferred using two different waveform models: SEOBNRv4PHM [137] and IMRPhenomXPHM [138], both of which include effects of higher-order multipole moments and spin precession. The posterior distribution for each event is then checked for consistency between waveform models before use in our analyses [3].

The event GW200129 is the highest SNR event exhibiting notable inconsistencies between the source properties inferred with the two waveform models. The event analysis using IMRPhenomXPHM infers much more support for unequal masses and precessing spins relative to the analysis using SEOBNRv4PHM.

To test if the inferred BBH spin population depends on the waveform model chosen for this event, we repeat our O3 population inference using the PP model for three different choices of waveform model for GW200129:

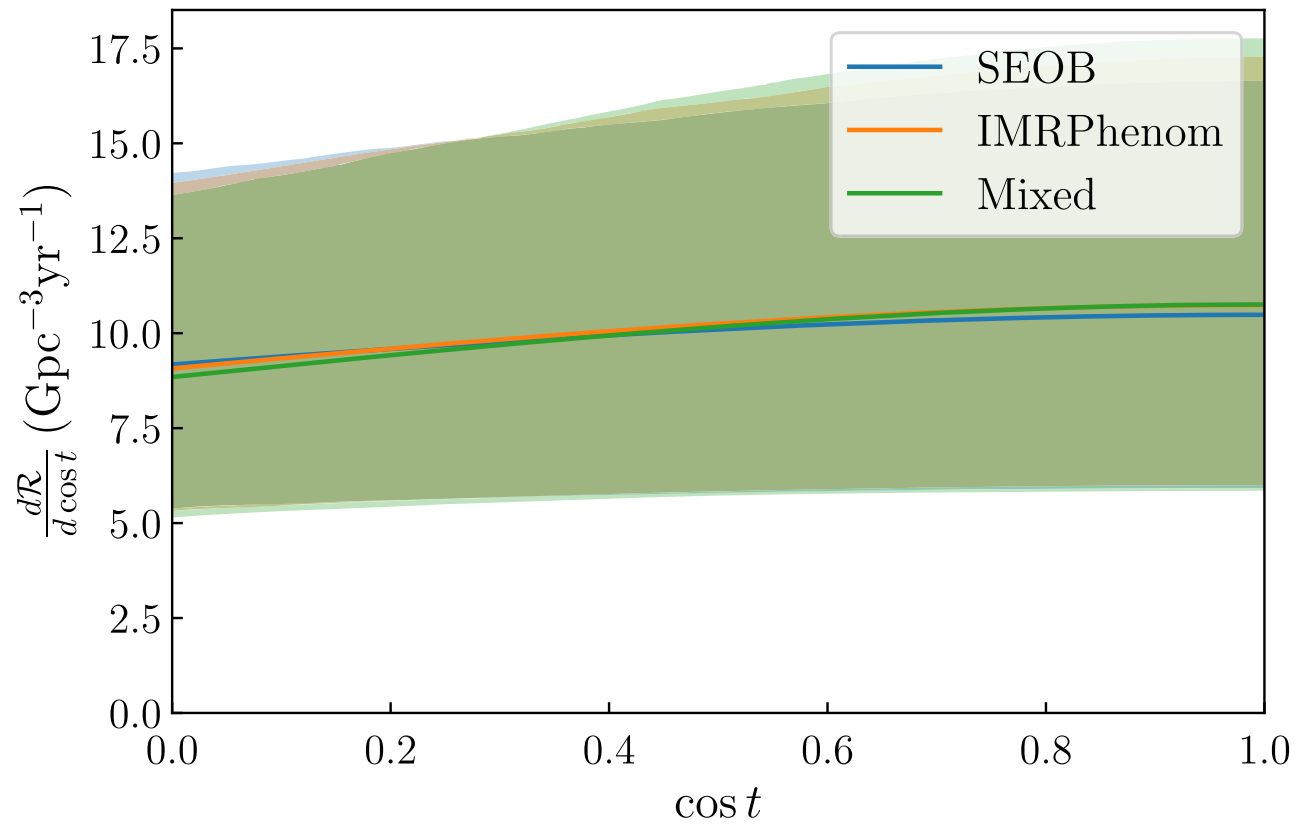

FIG. 26. Inferred differential merger rate as a function of the cosine of the tilt angle ($t_i$), where $i$ indexes the body of the binary. We demonstrate that the differences in the posterior distribution for GW200129's spin parameters have a minimal effect on the inferred spin-tilt population. The population is inferred using posterior distributions for GW200129 using the IMRPhenomXPHM waveform model (orange), SEOBNRv4PHM model (blue), and a mixture of both (green). Dashed lines are 90% credible intervals.

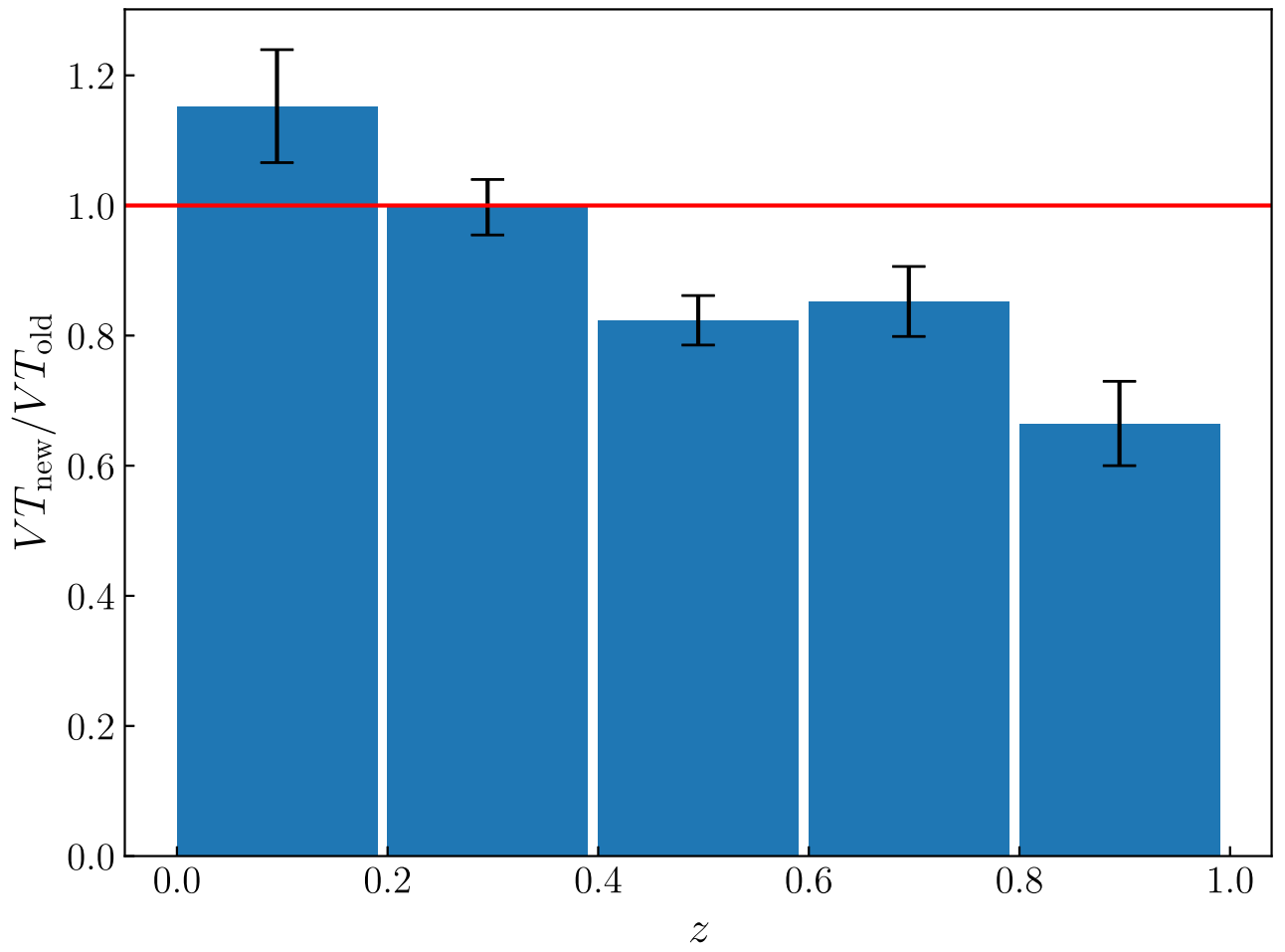

FIG. 27. Comparison of our current BBH merger sensitivity estimate in the O3a observing run ($VT_{\rm new}$) to that used in Ref. [20] ($VT_{\rm old}$) as a function of redshift, for events with chirp masses between 20 and $50 M_\odot$. Our current sensitivity model differs from what was used in Ref. [20] in two important ways: We use updated detection pipelines relative to those used in our GWTC-2 [20] and we use injections which include spin precession. There is a relative increase (decrease) in sensitivity at low (high) redshift. Computed by reweighting injections to a fiducial population for each of the two injection sets.

IMRPhenomXPHM, SEOBNRv4PHM, and a mix of the two. As shown in Fig. 26, the inferred spin population is not significantly affected by changes in the waveform model for this event.

### 5. Impact of sensitivity on redshift evolution inference

As noted in Sec. II, one change in the sensitivity estimation procedure between this work and our previous study of GWTC-2 [20] is the use of injections that account for the effect of precession and as well as updates to our detection pipelines as detailed in Ref. [3]. Since precession was not included in the injections used in Ref. [20], the full spin distribution could not be reweighted to calculate the sensitivity via Eq (A2), and thus, for the purposes of sensitivity estimation, an approximation was made that $S_{x,y} \in (-0.5, 0.5)$. Since we now use precessing injections, we do the reweighting procedure including the full spin distribution as a function of $\Lambda$. To test if this difference in our sensitivity estimation procedure is responsible for the change in the inferred redshift evolution, we repeat the population analysis reported in Sec. VI, using our updated sensitivity model, but including only events analyzed in the GWTC-2 populations study [20]. From this analysis, we infer $\kappa > 0$ at 97.6% credibility, as opposed to the 85% credibility reported previously [20], indicating a much stronger preference for a merger rate increasing with redshift. We conclude that the differences between our current results for the evolution of the BBH merger rate and those reported previously [20] are due to improvements to our sensitivity model rather than the presence of the additional events in GWTC-3.

In Fig. 27 we compare the redshift dependence of our current sensitivity model to that of the sensitivity model used previously [20]. To make this comparison, we reweight the injections used previously [20] to the same spin distribution assumed in that study, and assuming a fiducial PP and POWER LAW model for the mass and redshift distributions, respectively. We reweight the current injections to this same mass and redshift distribution, but reweight them to the median inferred spin distribution obtained previously [20] to mimic a astrophysically realistic population. Both injection sets cover only the observing times of the O3a observing run. Taking the ratio of the corresponding sensitivities, we find our sensitivity increases for low-redshift events and decreases for high-redshift events, relative to the sensitivity model used in our prior work [20]. We expect to see an increase in sensitivity between our previous analysis [20] and our current calculation due to updates to the detection pipelines. The relative decrease in sensitivity at higher redshifts indicates a bias in the previous sensitivity estimate, implying that the BBH merger rate at high redshift was underestimated in our earlier study [20]. Accounting for the shift in sensitivity as a function of redshift causes a relative decrease in local BBH merger rate and a relative increase in high-redshift BBH merger rate, leading to a higher inferred value for $\kappa$.

One possible explanation for the shift in sensitivity is that the use of precession in the injections for sensitivity estimation caused a nontrivial change in the inferred sensitive hypervolume, given that we do observe precession in the BBH population. Our current detection pipelines use template banks that include only aligned-spin components;

TABLE XV. Bayes factors for each of the previously used phenomenological mass models relative to the model with highest marginal likelihood, POWER LAW+PEAK. The previous results from GWTC-2 are shown in the second column with the updated catalog results in the third column.

| Model | GWTC-2 $\log_{10}\mathcal{B}$ | GWTC-3 $\log_{10}\mathcal{B}$ |
|---|---|---|
| POWER LAW + PEAK | 0.0 | 0.0 |
| BROKEN POWER LAW + PEAK | −0.11 | −0.46 |
| MULTI PEAK | −0.3 | −0.22 |
| BROKEN POWER LAW | −0.92 | −2.0 |

this can result in up to tens of percent reduced sensitivity to a population of BBHs with spin precession, depending on the degree of precession possible [379–381]. The farthest precessing sources, which, due to their distances, correspond to FARs closest to the detection threshold, are therefore the most susceptible to dropping below the detection threshold with our current pipelines, causing us to see a decrease in sensitivity to a population of BBHs with precession relative to a strictly nonprecessing population. Additionally, both the use of population-informed reweighting of the spin distribution to calculate sensitivity to a population and the incorporation of additional detection pipelines may have contributed to a more accurate estimate of our sensitivity across parameter space.

## APPENDIX D: ADDITIONAL STUDIES OF THE BINARY BLACK HOLE DISTRIBUTION

### 1. Analyses from GWTC-2

We report updated Bayes factor comparisons for these various models in Table XV, showing that the POWER LAW +PEAK model is favored over the other BBH mass distribution models tested. We highlight the key differences between the model priors for GWTC-2 compared to GWTC-3: The prior on $\beta_q$ is changed from $U(-4, 12)$ to $U(-2, 7)$, and the main population results now include an evolving redshift model where the prior on $\kappa$ is changed from 0 to $U(-10, 10)$.

In addition to these analyses, we use a variation of the MULTIPEAK model to study the feature in the mass distribution at approximately $10M_\odot$. In GWTC-2 the prior on the mean of the peaks are $U(20, 50)$ and $U(50, 100)$ for the lower and upper mass peaks, respectively. We modify these priors to be $U(5, 20)$ and $U(20, 100)$. This updated MULTIPEAK model is the most preferred model with a $\log_{10}\mathcal{B}$ of 1.0 compared to the POWER LAW+PEAK model. This further supports our findings of the peaklike feature at approximately $10M_\odot$ in the mass distribution.

### 2. Comprehensive BBH merger rates

In Table IV, we evaluate BBH merger rates over targeted mass subsets of the whole BBH space, using models specifically targeted to reproduce new features of the binary black hole mass distribution. For broader context, Table XVI also provides the corresponding merger rates in these intervals from all the models presented in this work.

## APPENDIX E: POPULATION-WEIGHTED POSTERIORS

With an increasing number of events, we can use the distribution of the population of compact binaries to inform our priors for parameter estimation. By reweighting the initial analysis of compact binaries with the population distribution we can obtain posterior distribution for the events in GWTC-3 with population-informed priors. Using our population analysis with models POWER LAW+PEAK and FLEXIBLE MIXTURES we provide population-weighted posteriors (Fig. 28) for $m_1$, $q$, and $\chi_{\rm eff}$ for the BBHs population (69 events).

TABLE XVI. Merger rates in $\mathrm{Gpc^{-3}\,yr^{-1}}$ for black hole binaries quoted at the 90% credible interval. Rates are given for three ranges of primary mass $m_1$, as well as for the entire population. The PDB, MS, and BGP merger rates are derived assuming the merger rate does not increase with redshift, using a threshold FAR $< 0.25\ \mathrm{yr^{-1}}$ (Sec. IV). For FM, PS, and PP, merger rates are reported at $z = 0.2$ estimated using a threshold FAR $< 1\ \mathrm{yr^{-1}}$ (Sec. VI). The merged rates reported in the MERGED row are the union of the preceding three rows, which all account for distance-dependent merger rate and adopt a consistent threshold. The final row shows merger rates deduced from our analysis of GWTC-2 [20], which assumed a redshift-independent merger rate. Compare to Table IV.

| | $m_1 \in [5, 20]M_\odot$ $m_2 \in [5, 20]M_\odot$ | $m_1 \in [20, 50]M_\odot$ $m_2 \in [5, 50]M_\odot$ | $m_1 \in [50, 100]M_\odot$ $m_2 \in [5, 100]M_\odot$ | All BBHs |
|---|---|---|---|---|
| PDB (pair) | $17^{+10}_{-6.0}$ | $6.8^{+2.2}_{-1.7}$ | $0.68^{+0.42}_{-0.29}$ | $25^{+10}_{-7.0}$ |
| PDB (ind) | $9.4^{+5.6}_{-3.7}$ | $11^{+3.0}_{-2.0}$ | $1.6^{+0.9}_{-0.7}$ | $22^{+8.0}_{-6.0}$ |
| MS | $30^{+23}_{-13}$ | $6.6^{+2.9}_{-2.3}$ | $0.73^{+0.87}_{-0.52}$ | $37^{+24}_{-13}$ |
| BGP | $20.0^{+11.0}_{-8.0}$ | $6.3^{+3.0}_{-2.2}$ | $0.75^{+1.1}_{-0.46}$ | $33.0^{+16.0}_{-10.0}$ |
| PS | $27^{+12}_{-8.8}$ | $3.5^{+1.5}_{-1.1}$ | $0.19^{+0.16}_{-0.09}$ | $31^{+13}_{-9.2}$ |
| FM | $21.1^{+11.6}_{-7.8}$ | $4.3^{+2.0}_{-1.4}$ | $0.2^{+0.2}_{-0.1}$ | $26.5^{+11.7}_{-8.6}$ |
| PP | $23.6^{+13.7}_{-9.0}$ | $4.5^{+1.7}_{-1.3}$ | $0.2^{+0.1}_{-0.1}$ | $28.3^{+13.9}_{-9.1}$ |
| MERGED | 13.3–39 | 2.5–6.3 | 0.099–0.4 | 17.9–44 |
| PP (O3a) | $16.0^{+13.0}_{-7.7}$ | $6.8^{+2.7}_{-1.9}$ | $0.5^{+0.4}_{-0.3}$ | $25.3^{+16.1}_{-9.9}$ |

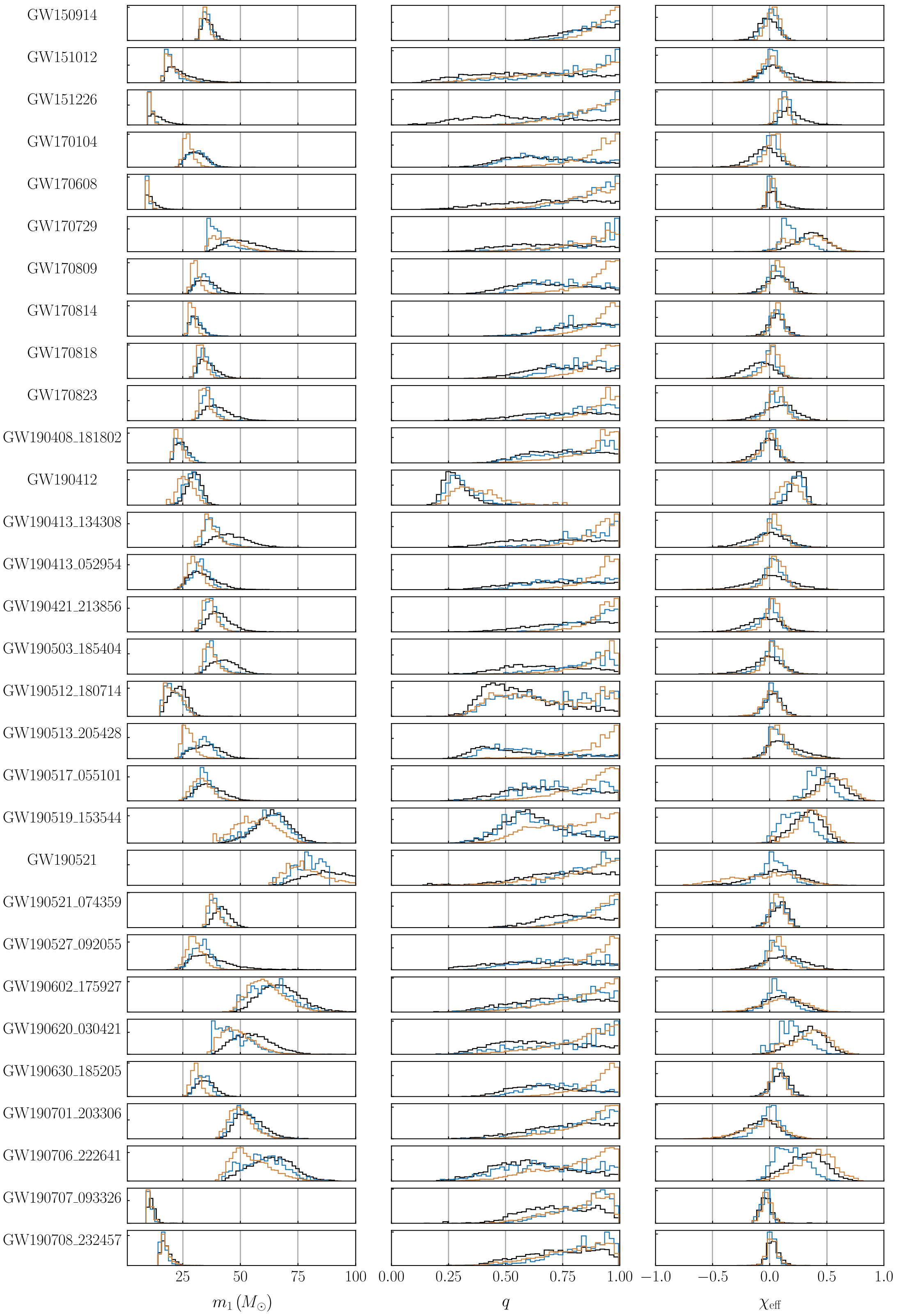

FIG. 28. Posterior distributions (black) for binary black hole events weighted by the population results from POWER LAW+PEAK (blue) and FLEXIBLE MIXTURES (orange).

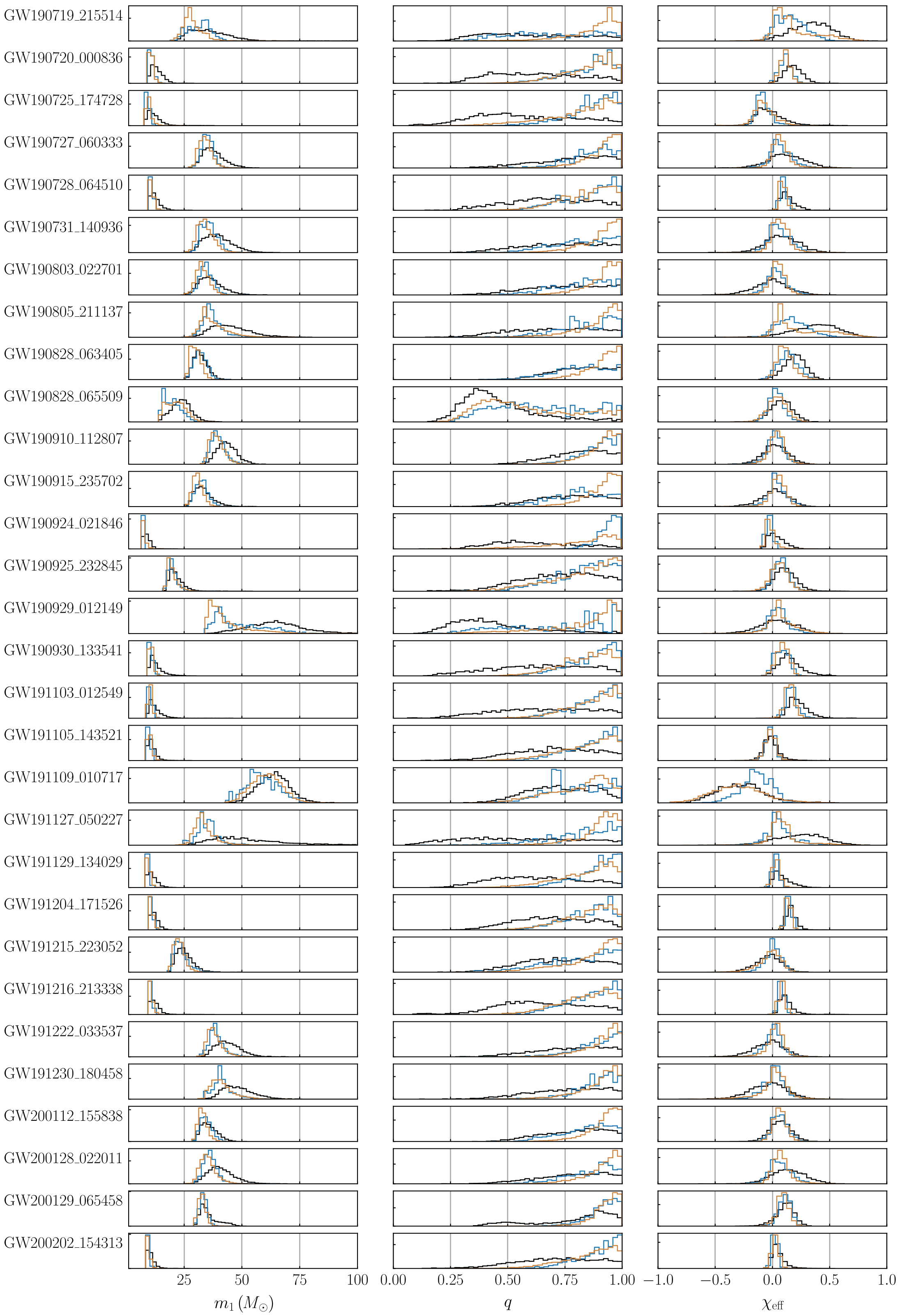

FIG. 28. (Continued).

Some of our analyses will show apparent changes in the inferences about the mass ratio. These seemingly substantive changes reflect the relatively weak constraints provided by the fiducial parameter inferences used as input and shown in black. Specifically, several low-amplitude or low-mass events have extremely weak constraints on mass ratios, with posterior support extending to $q < 0.4$. This extended feature reflects the prior distribution on component masses, conditioned on modest constraints on chirp mass. To be concrete, using the corresponding prior distributions for these events, conditioned on a suitable chirp-mass interval, we often find a posterior distribution with comparable support for $q = 1$ (i.e., the Savage-Dickey estimate of the Bayes factor for unequal mass would be nearly unity).

Examining Fig. 28 in light of this caveat, we find that our population models and the fiducial model agree. For these events, the population reweightings, as expected, strongly favor symmetric component masses (e.g., GW190503_185404, GW190720_000836, GW191127_050227). For a few binaries, however, the two population reweightings disagree. The most notable example is GW190513_205428, where the FLEXIBLE MIXTURES model pulls the posterior distribution to more symmetric component masses and a lower primary mass. Both population models also pull the majority of the posteriors closer to $\chi_{\rm eff} \sim 0$. However, given the FLEXIBLE MIXTURES analysis models spins as dependent on chirp mass the events with higher mass and higher spin do not drawn to $\chi_{\rm eff} \sim 0$ as strongly as the POWER LAW+PEAK model (e.g., GW191109_010717) and in some cases the FLEXIBLE MIXTURES reweighted posterior moves to higher $\chi_{\rm eff}$ values (e.g., GW190706_222641).

R. Abbott,[1] T. D. Abbott,[2] F. Acernese,[3,4] K. Ackley,[5] C. Adams,[6] N. Adhikari,[7] R. X. Adhikari,[1] V. B. Adya,[8] C. Affeldt,[9,10] D. Agarwal,[11] M. Agathos,[12,13] K. Agatsuma,[14] N. Aggarwal,[15] O. D. Aguiar,[16] L. Aiello,[17] A. Ain,[18] P. Ajith,[19] T. Akutsu,[20,21] P. F. de Alarcón,[22] S. Akcay,[13,23] S. Albanesi,[24] A. Allocca,[25,4] P. A. Altin,[8] A. Amato,[26] C. Anand,[5] S. Anand,[1] A. Ananyeva,[1] S. B. Anderson,[1] W. G. Anderson,[7] M. Ando,[27,28] T. Andrade,[29] N. Andres,[30] T. Andrić,[31] S. V. Angelova,[32] S. Ansoldi,[33,34] J. M. Antelis,[35] S. Antier,[36] F. Antonini,[17] S. Appert,[1] Koji Arai,[1] Koya Arai,[37] Y. Arai,[37] S. Araki,[38] A. Araya,[39] M. C. Araya,[1] J. S. Areeda,[40] M. Arène,[36] N. Aritomi,[27] N. Arnaud,[41,42] M. Arogeti,[43] S. M. Aronson,[2] K. G. Arun,[44] H. Asada,[45] Y. Asali,[46] G. Ashton,[5] Y. Aso,[47,48] M. Assiduo,[49,50] S. M. Aston,[6] P. Astone,[51] F. Aubin,[30] C. Austin,[2] S. Babak,[36] F. Badaracco,[52] M. K. M. Bader,[53] C. Badger,[54] S. Bae,[55] Y. Bae,[56] A. M. Baer,[57] S. Bagnasco,[24] Y. Bai,[1] L. Baiotti,[58] J. Baird,[36] R. Bajpai,[59] M. Ball,[60] G. Ballardin,[42] S. W. Ballmer,[61] A. Balsamo,[57] G. Baltus,[62] S. Banagiri,[63] D. Bankar,[11] J. C. Barayoga,[1] C. Barbieri,[64,65,66] B. C. Barish,[1] D. Barker,[67] P. Barneo,[29] F. Barone,[68,4] B. Barr,[69] L. Barsotti,[70] M. Barsuglia,[36] D. Barta,[71] J. Bartlett,[67] M. A. Barton,[69,20] I. Bartos,[72] R. Bassiri,[73] A. Basti,[74,18] M. Bawaj,[75,76] J. C. Bayley,[69] A. C. Baylor,[7] M. Bazzan,[77,78] B. Bécsy,[79] V. M. Bedakihale,[80] M. Bejger,[81] I. Belahcene,[41] V. Benedetto,[82] D. Beniwal,[83] T. F. Bennett,[84] J. D. Bentley,[14] M. BenYaala,[32] F. Bergamin,[9,10] B. K. Berger,[73] S. Bernuzzi,[13] C. P. L. Berry,[15,69] D. Bersanetti,[85] A. Bertolini,[53] J. Betzwieser,[6] D. Beveridge,[86] R. Bhandare,[87] U. Bhardwaj,[88,53] D. Bhattacharjee,[89] S. Bhaumik,[72] I. A. Bilenko,[90] G. Billingsley,[1] S. Bini,[91,92] R. Birney,[93] O. Birnholtz,[94] S. Biscans,[1,70] M. Bischi,[49,50] S. Biscoveanu,[70] A. Bisht,[9,10] B. Biswas,[11] M. Bitossi,[42,18] M.-A. Bizouard,[95] J. K. Blackburn,[1] C. D. Blair,[86,6] D. G. Blair,[86] R. M. Blair,[67] F. Bobba,[96,97] N. Bode,[9,10] M. Boer,[95] G. Bogaert,[95] M. Boldrini,[98,51] L. D. Bonavena,[77] F. Bondu,[99] E. Bonilla,[73] R. Bonnand,[30] P. Booker,[9,10] B. A. Boom,[53] R. Bork,[1] V. Boschi,[18] N. Bose,[100] S. Bose,[11] V. Bossilkov,[86] V. Boudart,[62] Y. Bouffanais,[77,78] A. Bozzi,[42] C. Bradaschia,[18] P. R. Brady,[7] A. Bramley,[6] A. Branch,[6] M. Branchesi,[31,101] J. Brandt,[43] J. E. Brau,[60] M. Breschi,[13] T. Briant,[102] J. H. Briggs,[69] A. Brillet,[95] M. Brinkmann,[9,10] P. Brockill,[7] A. F. Brooks,[1] J. Brooks,[42] D. D. Brown,[83] S. Brunett,[1] G. Bruno,[52] R. Bruntz,[57] J. Bryant,[14] T. Bulik,[103] H. J. Bulten,[53] A. Buonanno,[104,105] R. Buscicchio,[14] D. Buskulic,[30] C. Buy,[106] R. L. Byer,[73] L. Cadonati,[43] G. Cagnoli,[26] C. Cahillane,[67] J. Calderón Bustillo,[107,108] J. D. Callaghan,[69] T. A. Callister,[109,110] E. Calloni,[25,4] J. Cameron,[86] J. B. Camp,[111] M. Canepa,[112,85] S. Canevarolo,[113] M. Cannavacciuolo,[96] K. C. Cannon,[114] H. Cao,[83] Z. Cao,[115] E. Capocasa,[20] E. Capote,[61] G. Carapella,[96,97] F. Carbognani,[42] J. B. Carlin,[116] M. F. Carney,[15] M. Carpinelli,[117,118,42] G. Carrillo,[60] G. Carullo,[74,18] T. L. Carver,[17] J. Casanueva Diaz,[42] C. Casentini,[119,120] G. Castaldi,[121] S. Caudill,[53,113] M. Cavaglià,[89] F. Cavalier,[41] R. Cavalieri,[42] M. Ceasar,[122] G. Cella,[18] P. Cerdá-Durán,[123] E. Cesarini,[120] W. Chaibi,[95] K. Chakravarti,[11] S. Chalathadka Subrahmanya,[124] E. Champion,[125] C.-H. Chan,[126] C. Chan,[114] C. L. Chan,[108] K. Chan,[108] M. Chan,[127] K. Chandra,[100] P. Chanial,[42] S. Chao,[126] C. E. A. Chapman-Bird,[69] P. Charlton,[128] E. A. Chase,[15]

E. Chassande-Mottin,[36] C. Chatterjee,[86] Debarati Chatterjee,[11] Deep Chatterjee,[7] M. Chaturvedi,[87] S. Chaty,[36] K. Chatziioannou,[1] C. Chen,[129,130] H. Y. Chen,[70] J. Chen,[126] K. Chen,[131] X. Chen,[86] Y.-B. Chen,[132] Y.-R. Chen,[133] Z. Chen,[17] H. Cheng,[72] C. K. Cheong,[108] H. Y. Cheung,[108] H. Y. Chia,[72] F. Chiadini,[134,97] C-Y. Chiang,[135] G. Chiarini,[78] R. Chierici,[136] A. Chincarini,[85] M. L. Chiofalo,[74,18] A. Chiummo,[42] G. Cho,[137] H. S. Cho,[138] R. K. Choudhary,[86] S. Choudhary,[11] N. Christensen,[95] H. Chu,[131] Q. Chu,[86] Y-K. Chu,[135] S. Chua,[8] K. W. Chung,[54] G. Ciani,[77,78] P. Ciecielag,[81] M. Cieślar,[81] M. Cifaldi,[119,120] A. A. Ciobanu,[83] R. Ciolfi,[139,78] F. Cipriano,[95] A. Cirone,[112,85] F. Clara,[67] E. N. Clark,[140] J. A. Clark,[1,43] L. Clarke,[141] P. Clearwater,[142] S. Clesse,[143] F. Cleva,[95] E. Coccia,[31,101] E. Codazzo,[31] P.-F. Cohadon,[102] D. E. Cohen,[41] L. Cohen,[2] M. Colleoni,[144] C. G. Collette,[145] A. Colombo,[64] M. Colpi,[64,65] C. M. Compton,[67] M. Constancio Jr.,[16] L. Conti,[78] S. J. Cooper,[14] P. Corban,[6] T. R. Corbitt,[2] I. Cordero-Carrión,[146] S. Corezzi,[76,75] K. R. Corley,[46] N. Cornish,[79] D. Corre,[41] A. Corsi,[147] S. Cortese,[42] C. A. Costa,[16] R. Cotesta,[105] M. W. Coughlin,[63] J.-P. Coulon,[95] S. T. Countryman,[46] B. Cousins,[148] P. Couvares,[1] D. M. Coward,[86] M. J. Cowart,[6] D. C. Coyne,[1] R. Coyne,[149] J. D. E. Creighton,[7] T. D. Creighton,[150] A. W. Criswell,[63] M. Croquette,[102] S. G. Crowder,[151] J. R. Cudell,[62] T. J. Cullen,[2] A. Cumming,[69] R. Cummings,[69] L. Cunningham,[69] E. Cuoco,[42,152,18] M. Curyło,[103] P. Dabadie,[26] T. Dal Canton,[41] S. Dall'Osso,[31] G. Dálya,[153] A. Dana,[73] L. M. DaneshgaranBajastani,[84] B. D'Angelo,[112,85] B. Danila,[154] S. Danilishin,[155,53] S. D'Antonio,[120] K. Danzmann,[9,10] C. Darsow-Fromm,[124] A. Dasgupta,[80] L. E. H. Datrier,[69] S. Datta,[11] V. Dattilo,[42] I. Dave,[87] M. Davier,[41] G. S. Davies,[156] D. Davis,[1] M. C. Davis,[122] E. J. Daw,[157] R. Dean,[122] D. DeBra,[73] M. Deenadayalan,[11] J. Degallaix,[158] M. De Laurentis,[25,4] S. Deléglise,[102] V. Del Favero,[125] F. De Lillo,[52] N. De Lillo,[69] W. Del Pozzo,[74,18] L. M. DeMarchi,[15] F. De Matteis,[119,120] V. D'Emilio,[17] N. Demos,[70] T. Dent,[107] A. Depasse,[52] R. De Pietri,[159,160] R. De Rosa,[25,4] C. De Rossi,[42] R. DeSalvo,[121] R. De Simone,[134] S. Dhurandhar,[11] M. C. Díaz,[150] M. Diaz-Ortiz Jr.,[72] N. A. Didio,[61] T. Dietrich,[105,53] L. Di Fiore,[4] C. Di Fronzo,[14] C. Di Giorgio,[96,97] F. Di Giovanni,[123] M. Di Giovanni,[31] T. Di Girolamo,[25,4] A. Di Lieto,[74,18] B. Ding,[145] S. Di Pace,[98,51] I. Di Palma,[98,51] F. Di Renzo,[74,18] A. K. Divakarla,[72] A. Dmitriev,[14] Z. Doctor,[60] L. D'Onofrio,[25,4] F. Donovan,[70] K. L. Dooley,[17] S. Doravari,[11] I. Dorrington,[17] M. Drago,[98,51] J. C. Driggers,[67] Y. Drori,[1] J.-G. Ducoin,[41] P. Dupej,[69] O. Durante,[96,97] D. D'Urso,[117,118] P.-A. Duverne,[41] S. E. Dwyer,[67] C. Eassa,[67] P. J. Easter,[5] M. Ebersold,[161] T. Eckhardt,[124] G. Eddolls,[69] B. Edelman,[60] T. B. Edo,[1] O. Edy,[156] A. Effler,[6] S. Eguchi,[127] J. Eichholz,[8] S. S. Eikenberry,[72] M. Eisenmann,[30] R. A. Eisenstein,[70] A. Ejlli,[17] E. Engelby,[40] Y. Enomoto,[27] L. Errico,[25,4] R. C. Essick,[162] H. Estellés,[144] D. Estevez,[163] Z. Etienne,[164] T. Etzel,[1] M. Evans,[70] T. M. Evans,[6] B. E. Ewing,[148] V. Fafone,[119,120,31] H. Fair,[61] S. Fairhurst,[17] A. M. Farah,[162] S. Farinon,[85] B. Farr,[60] W. M. Farr,[109,110] N. W. Farrow,[5] E. J. Fauchon-Jones,[17] G. Favaro,[77] M. Favata,[165] M. Fays,[62] M. Fazio,[166] J. Feicht,[1] M. M. Fejer,[73] E. Fenyvesi,[71,167] D. L. Ferguson,[168] A. Fernandez-Galiana,[70] I. Ferrante,[74,18] T. A. Ferreira,[16] F. Fidecaro,[74,18] P. Figura,[103] I. Fiori,[42] M. Fishbach,[15] R. P. Fisher,[57] R. Fittipaldi,[169,97] V. Fiumara,[170,97] R. Flaminio,[30,20] E. Floden,[63] H. Fong,[114] J. A. Font,[123,171] B. Fornal,[172] P. W. F. Forsyth,[8] A. Franke,[124] S. Frasca,[98,51] F. Frasconi,[18] C. Frederick,[173] J. P. Freed,[35] Z. Frei,[153] A. Freise,[174] R. Frey,[60] P. Fritschel,[70] V. V. Frolov,[6] G. G. Fronzé,[24] Y. Fujii,[175] Y. Fujikawa,[176] M. Fukunaga,[37] M. Fukushima,[21] P. Fulda,[72] M. Fyffe,[6] H. A. Gabbard,[69] B. U. Gadre,[105] J. R. Gair,[105] J. Gais,[108] S. Galaudage,[5] R. Gamba,[13] D. Ganapathy,[70] A. Ganguly,[19] D. Gao,[177] S. G. Gaonkar,[11] B. Garaventa,[85,112] F. García,[36] C. García-Núñez,[93] C. García-Quirós,[144] F. Garufi,[25,4] B. Gateley,[67] S. Gaudio,[35] V. Gayathri,[72] G.-G. Ge,[177] G. Gemme,[85] A. Gennai,[18] J. George,[87] R. N. George,[168] O. Gerberding,[124] L. Gergely,[154] P. Gewecke,[124] S. Ghonge,[43] Abhirup Ghosh,[105] Archisman Ghosh,[178] Shaon Ghosh,[7,165] Shrobana Ghosh,[17] B. Giacomazzo,[64,65,66] L. Giacoppo,[98,51] J. A. Giaime,[2,6] K. D. Giardina,[6] D. R. Gibson,[93] C. Gier,[32] M. Giesler,[179] P. Giri,[18,74] F. Gissi,[82] J. Glanzer,[2] A. E. Gleckl,[40] P. Godwin,[148] J. Golomb,[1] E. Goetz,[180] R. Goetz,[72] N. Gohlke,[9,10] B. Goncharov,[5,31] G. González,[2] A. Gopakumar,[181] M. Gosselin,[42] R. Gouaty,[30] D. W. Gould,[8] B. Grace,[8] A. Grado,[182,4] M. Granata,[158] V. Granata,[96] A. Grant,[69] S. Gras,[70] P. Grassia,[1] C. Gray,[67] R. Gray,[69] G. Greco,[75] A. C. Green,[72] R. Green,[17] A. M. Gretarsson,[35] E. M. Gretarsson,[35] D. Griffith,[1] W. Griffiths,[17] H. L. Griggs,[43] G. Grignani,[76,75] A. Grimaldi,[91,92] S. J. Grimm,[31,101] H. Grote,[17] S. Grunewald,[105] P. Gruning,[41] D. Guerra,[123] G. M. Guidi,[49,50] A. R. Guimaraes,[2] G. Guixé,[29] H. K. Gulati,[80] H.-K. Guo,[172] Y. Guo,[53] Anchal Gupta,[1] Anuradha Gupta,[183] P. Gupta,[53,113] E. K. Gustafson,[1] R. Gustafson,[184] F. Guzman,[185] S. Ha,[186] L. Haegel,[36] A. Hagiwara,[37,187] S. Haino,[135] O. Halim,[34,188] E. D. Hall,[70] E. Z. Hamilton,[161] G. Hammond,[69] W.-B. Han,[189] M. Haney,[161] J. Hanks,[67] C. Hanna,[148] M. D. Hannam,[17] O. Hannuksela,[113,53] H. Hansen,[67] T. J. Hansen,[35] J. Hanson,[6] T. Harder,[95] T. Hardwick,[2] K. Haris,[53,113] J. Harms,[31,101] G. M. Harry,[190] I. W. Harry,[156] D. Hartwig,[124] K. Hasegawa,[37] B. Haskell,[81] R. K. Hasskew,[6] C.-J. Haster,[70] K. Hattori,[191] K. Haughian,[69] H. Hayakawa,[192] K. Hayama,[127] F. J. Hayes,[69] J. Healy,[125] A. Heidmann,[102] A. Heidt,[9,10] M. C. Heintze,[6] J. Heinze,[9,10] J. Heinzel,[193] H. Heitmann,[95] F. Hellman,[194] P. Hello,[41] A. F. Helmling-Cornell,[60] G. Hemming,[42]

M. Hendry,[69] I. S. Heng,[69] E. Hennes,[53] J. Hennig,[195] M. H. Hennig,[195] A. G. Hernandez,[84] F. Hernandez Vivanco,[5] M. Heurs,[9,10] S. Hild,[155,53] P. Hill,[32] Y. Himemoto,[196] A. S. Hines,[185] Y. Hiranuma,[197] N. Hirata,[20] E. Hirose,[37] S. Hochheim,[9,10] D. Hofman,[158] J. N. Hohmann,[124] D. G. Holcomb,[122] N. A. Holland,[8] I. J. Hollows,[157] Z. J. Holmes,[83] K. Holt,[6] D. E. Holz,[162] Z. Hong,[198] P. Hopkins,[17] J. Hough,[69] S. Hourihane,[132] E. J. Howell,[86] C. G. Hoy,[17] D. Hoyland,[14] A. Hreibi,[9,10] B-H. Hsieh,[37] Y. Hsu,[126] G-Z. Huang,[198] H-Y. Huang,[135] P. Huang,[177] Y-C. Huang,[133] Y.-J. Huang,[135] Y. Huang,[70] M. T. Hübner,[5] A. D. Huddart,[141] B. Hughey,[35] D. C. Y. Hui,[199] V. Hui,[30] S. Husa,[144] S. H. Huttner,[69] R. Huxford,[148] T. Huynh-Dinh,[6] S. Ide,[200] B. Idzkowski,[103] A. Iess,[119,120] B. Ikenoue,[21] S. Imam,[198] K. Inayoshi,[201] C. Ingram,[83] Y. Inoue,[131] K. Ioka,[202] M. Isi,[70] K. Isleif,[124] K. Ito,[203] Y. Itoh,[204,205] B. R. Iyer,[19] K. Izumi,[206] V. JaberianHamedan,[86] T. Jacqmin,[102] S. J. Jadhav,[207] S. P. Jadhav,[11] A. L. James,[17] A. Z. Jan,[125] K. Jani,[208] J. Janquart,[113,53] K. Janssens,[209,95] N. N. Janthalur,[207] P. Jaranowski,[210] D. Jariwala,[72] R. Jaume,[144] A. C. Jenkins,[54] K. Jenner,[83] C. Jeon,[211] M. Jeunon,[63] W. Jia,[70] H.-B. Jin,[212,213] G. R. Johns,[57] A. W. Jones,[86] D. I. Jones,[214] J. D. Jones,[67] P. Jones,[14] R. Jones,[69] R. J. G. Jonker,[53] L. Ju,[86] P. Jung,[56] k. Jung,[186] J. Junker,[9,10] V. Juste,[163] K. Kaihotsu,[203] T. Kajita,[215] M. Kakizaki,[191] C. V. Kalaghatgi,[17,113] V. Kalogera,[15] B. Kamai,[1] M. Kamiizumi,[192] N. Kanda,[204,205] S. Kandhasamy,[11] G. Kang,[216] J. B. Kanner,[1] Y. Kao,[126] S. J. Kapadia,[19] D. P. Kapasi,[8] S. Karat,[1] C. Karathanasis,[217] S. Karki,[89] R. Kashyap,[148] M. Kasprzack,[1] W. Kastaun,[9,10] S. Katsanevas,[42] E. Katsavounidis,[70] W. Katzman,[6] T. Kaur,[86] K. Kawabe,[67] K. Kawaguchi,[37] N. Kawai,[218] T. Kawasaki,[27] F. Kéfélian,[95] D. Keitel,[144] J. S. Key,[219] S. Khadka,[73] F. Y. Khalili,[90] S. Khan,[17] E. A. Khazanov,[220] N. Khetan,[31,101] M. Khursheed,[87] N. Kijbunchoo,[8] C. Kim,[221] J. C. Kim,[222] J. Kim,[223] K. Kim,[224] W. S. Kim,[225] Y.-M. Kim,[226] C. Kimball,[15] N. Kimura,[187] M. Kinley-Hanlon,[69] R. Kirchhoff,[9,10] J. S. Kissel,[67] N. Kita,[27] H. Kitazawa,[203] L. Kleybolte,[124] S. Klimenko,[72] A. M. Knee,[180] T. D. Knowles,[164] E. Knyazev,[70] P. Koch,[9,10] G. Koekoek,[53,155] Y. Kojima,[227] K. Kokeyama,[228] S. Koley,[31] P. Kolitsidou,[17] M. Kolstein,[217] K. Komori,[70,27] V. Kondrashov,[1] A. K. H. Kong,[229] A. Kontos,[230] N. Koper,[9,10] M. Korobko,[124] K. Kotake,[127] M. Kovalam,[86] D. B. Kozak,[1] C. Kozakai,[47] R. Kozu,[192] V. Kringel,[9,10] N. V. Krishnendu,[9,10] A. Królak,[231,232] G. Kuehn,[9,10] F. Kuei,[126] P. Kuijer,[53] S. Kulkarni,[183] A. Kumar,[207] P. Kumar,[179] Rahul Kumar,[67] Rakesh Kumar,[80] J. Kume,[28] K. Kuns,[70] C. Kuo,[131] H-S. Kuo,[198] Y. Kuromiya,[203] S. Kuroyanagi,[233,234] K. Kusayanagi,[218] S. Kuwahara,[114] K. Kwak,[186] P. Lagabbe,[30] D. Laghi,[74,18] E. Lalande,[235] T. L. Lam,[108] A. Lamberts,[95,236] M. Landry,[67] P. Landry,[237] B. B. Lane,[70] R. N. Lang,[70] J. Lange,[168] B. Lantz,[73] I. La Rosa,[30] A. Lartaux-Vollard,[41] P. D. Lasky,[5] M. Laxen,[6] A. Lazzarini,[1] C. Lazzaro,[77,78] P. Leaci,[98,51] S. Leavey,[9,10] Y. K. Lecoeuche,[180] H. K. Lee,[238] H. M. Lee,[137] H. W. Lee,[222] J. Lee,[137] K. Lee,[239] R. Lee,[133] J. Lehmann,[9,10] A. Lemaître,[240] M. Leonardi,[20] N. Leroy,[41] N. Letendre,[30] C. Levesque,[235] Y. Levin,[5] J. N. Leviton,[184] K. Leyde,[36] A. K. Y. Li,[1] B. Li,[126] J. Li,[15] K. L. Li,[241] T. G. F. Li,[108] X. Li,[132] C-Y. Lin,[242] F-K. Lin,[135] F-L. Lin,[198] H. L. Lin,[131] L. C.-C. Lin,[186] F. Linde,[243,53] S. D. Linker,[84] J. N. Linley,[69] T. B. Littenberg,[244] G. C. Liu,[129] J. Liu,[9,10] K. Liu,[126] X. Liu,[7] F. Llamas,[150] M. Llorens-Monteagudo,[123] R. K. L. Lo,[1] A. Lockwood,[245] M. Loh,[40] L. T. London,[70] A. Longo,[246,247] D. Lopez,[161] M. Lopez Portilla,[113] M. Lorenzini,[119,120] V. Loriette,[248] M. Lormand,[6] G. Losurdo,[18] T. P. Lott,[43] J. D. Lough,[9,10] C. O. Lousto,[125] G. Lovelace,[40] J. F. Lucaccioni,[173] H. Lück,[9,10] D. Lumaca,[119,120] A. P. Lundgren,[156] L.-W. Luo,[135] J. E. Lynam,[57] R. Macas,[156] M. MacInnis,[70] D. M. Macleod,[17] I. A. O. MacMillan,[1] A. Macquet,[95] I. Magaña Hernandez,[7] C. Magazzù,[18] R. M. Magee,[1] R. Maggiore,[14] M. Magnozzi,[85,112] S. Mahesh,[164] E. Majorana,[98,51] C. Makarem,[1] I. Maksimovic,[248] S. Maliakal,[1] A. Malik,[87] N. Man,[95] V. Mandic,[63] V. Mangano,[98,51] J. L. Mango,[249] G. L. Mansell,[67,70] M. Manske,[7] M. Mantovani,[42] M. Mapelli,[77,78] F. Marchesoni,[250,75,251] M. Marchio,[20] F. Marion,[30] Z. Mark,[132] S. Márka,[46] Z. Márka,[46] C. Markakis,[12] A. S. Markosyan,[73] A. Markowitz,[1] E. Maros,[1] A. Marquina,[146] S. Marsat,[36] F. Martelli,[49,50] I. W. Martin,[69] R. M. Martin,[165] M. Martinez,[217] V. A. Martinez,[72] V. Martinez,[26] K. Martinovic,[54] D. V. Martynov,[14] E. J. Marx,[70] H. Masalehdan,[124] K. Mason,[70] E. Massera,[157] A. Masserot,[30] T. J. Massinger,[70] M. Masso-Reid,[69] S. Mastrogiovanni,[36] A. Matas,[105] M. Mateu-Lucena,[144] F. Matichard,[1,70] M. Matiushechkina,[9,10] N. Mavalvala,[70] J. J. McCann,[86] R. McCarthy,[67] D. E. McClelland,[8] P. K. McClincy,[148] S. McCormick,[6] L. McCuller,[70] G. I. McGhee,[69] S. C. McGuire,[252] C. McIsaac,[156] J. McIver,[180] T. McRae,[8] S. T. McWilliams,[164] D. Meacher,[7] M. Mehmet,[9,10] A. K. Mehta,[105] Q. Meijer,[113] A. Melatos,[116] D. A. Melchor,[40] G. Mendell,[67] A. Menendez-Vazquez,[217] C. S. Menoni,[166] R. A. Mercer,[7] L. Mereni,[158] K. Merfeld,[60] E. L. Merilh,[6] J. D. Merritt,[60] M. Merzougui,[95] S. Meshkov,[1] C. Messenger,[69] C. Messick,[168] P. M. Meyers,[116] F. Meylahn,[9,10] A. Mhaske,[11] A. Miani,[91,92] H. Miao,[14] I. Michaloliakos,[72] C. Michel,[158] Y. Michimura,[27] H. Middleton,[116] L. Milano,[25] A. L. Miller,[52] A. Miller,[84] B. Miller,[88,53] S. Miller,[1] M. Millhouse,[116] J. C. Mills,[17] E. Milotti,[188,34] O. Minazzoli,[95,253] Y. Minenkov,[120] N. Mio,[254] Ll. M. Mir,[217] M. Miravet-Tenés,[123] C. Mishra,[255] T. Mishra,[72] T. Mistry,[157] S. Mitra,[11]

V. P. Mitrofanov,[90] G. Mitselmakher,[72] R. Mittleman,[70] O. Miyakawa,[192] A. Miyamoto,[204] Y. Miyazaki,[27] K. Miyo,[192] S. Miyoki,[192] Geoffrey Mo,[70] L. M. Modafferi,[22] E. Moguel,[173] K. Mogushi,[89] S. R. P. Mohapatra,[70] S. R. Mohite,[7] I. Molina,[40] M. Molina-Ruiz,[194] M. Mondin,[84] M. Montani,[49,50] C. J. Moore,[14] D. Moraru,[67] F. Morawski,[81] A. More,[11] C. Moreno,[35] G. Moreno,[67] Y. Mori,[203] S. Morisaki,[7] Y. Moriwaki,[191] G. Morrás,[256] B. Mours,[163] C. M. Mow-Lowry,[14,174] S. Mozzon,[156] F. Muciaccia,[98,51] Arunava Mukherjee,[257] D. Mukherjee,[148] Soma Mukherjee,[150] Subroto Mukherjee,[80] Suvodip Mukherjee,[88] N. Mukund,[9,10] A. Mullavey,[6] J. Munch,[83] E. A. Muñiz,[61] P. G. Murray,[69] R. Musenich,[85,112] S. Muusse,[83] S. L. Nadji,[9,10] K. Nagano,[206] S. Nagano,[258] A. Nagar,[24,259] K. Nakamura,[20] H. Nakano,[260] M. Nakano,[37] R. Nakashima,[218] Y. Nakayama,[203] V. Napolano,[42] I. Nardecchia,[119,120] T. Narikawa,[37] L. Naticchioni,[51] B. Nayak,[84] R. K. Nayak,[261] R. Negishi,[197] B. F. Neil,[86] J. Neilson,[82,97] G. Nelemans,[262] T. J. N. Nelson,[6] M. Nery,[9,10] P. Neubauer,[173] A. Neunzert,[219] K. Y. Ng,[70] S. W. S. Ng,[83] C. Nguyen,[36] P. Nguyen,[60] T. Nguyen,[70] L. Nguyen Quynh,[263] W.-T. Ni,[212,177,133] S. A. Nichols,[2] A. Nishizawa,[28] S. Nissanke,[88,53] E. Nitoglia,[136] F. Nocera,[42] M. Norman,[17] C. North,[17] S. Nozaki,[191] J. F. Nuño Siles,[256] L. K. Nuttall,[156] J. Oberling,[67] B. D. O'Brien,[72] Y. Obuchi,[21] J. O'Dell,[141] E. Oelker,[69] W. Ogaki,[37] G. Oganesyan,[31,101] J. J. Oh,[225] K. Oh,[199] S. H. Oh,[225] M. Ohashi,[192] N. Ohishi,[47] M. Ohkawa,[176] F. Ohme,[9,10] H. Ohta,[114] M. A. Okada,[16] Y. Okutani,[200] K. Okutomi,[192] C. Olivetto,[42] K. Oohara,[197] C. Ooi,[27] R. Oram,[6] B. O'Reilly,[6] R. G. Ormiston,[63] N. D. Ormsby,[57] L. F. Ortega,[72] R. O'Shaughnessy,[125] E. O'Shea,[179] S. Oshino,[192] S. Ossokine,[105] C. Osthelder,[1] S. Otabe,[218] D. J. Ottaway,[83] H. Overmier,[6] A. E. Pace,[148] G. Pagano,[74,18] M. A. Page,[86] G. Pagliaroli,[31,101] A. Pai,[100] S. A. Pai,[87] J. R. Palamos,[60] O. Palashov,[220] C. Palomba,[51] H. Pan,[126] K. Pan,[133,229] P. K. Panda,[207] H. Pang,[131] P. T. H. Pang,[53,113] C. Pankow,[15] F. Pannarale,[98,51] B. C. Pant,[87] F. H. Panther,[86] F. Paoletti,[18] A. Paoli,[42] A. Paolone,[51,264] A. Parisi,[129] H. Park,[7] J. Park,[265] W. Parker,[6,252] D. Pascucci,[53] A. Pasqualetti,[42] R. Passaquieti,[74,18] D. Passuello,[18] M. Patel,[57] M. Pathak,[83] B. Patricelli,[42,18] A. S. Patron,[2] S. Paul,[60] E. Payne,[5] M. Pedraza,[1] M. Pegoraro,[78] A. Pele,[6] F. E. Peña Arellano,[192] S. Penn,[266] A. Perego,[91,92] A. Pereira,[26] T. Pereira,[267] C. J. Perez,[67] C. Périgois,[30] C. C. Perkins,[72] A. Perreca,[91,92] S. Perriès,[136] J. Petermann,[124] D. Petterson,[1] H. P. Pfeiffer,[105] K. A. Pham,[63] K. S. Phukon,[53,243] O. J. Piccinni,[51] M. Pichot,[95] M. Piendibene,[74,18] F. Piergiovanni,[49,50] L. Pierini,[98,51] V. Pierro,[82,97] G. Pillant,[42] M. Pillas,[41] F. Pilo,[18] L. Pinard,[158] I. M. Pinto,[82,97,268] M. Pinto,[42] B. Piotrzkowski,[7] K. Piotrzkowski,[52] M. Pirello,[67] M. D. Pitkin,[269] E. Placidi,[98,51] L. Planas,[144] W. Plastino,[246,247] C. Pluchar,[140] R. Poggiani,[74,18] E. Polini,[30] D. Y. T. Pong,[108] S. Ponrathnam,[11] P. Popolizio,[42] E. K. Porter,[36] R. Poulton,[42] J. Powell,[142] M. Pracchia,[30] T. Pradier,[163] A. K. Prajapati,[80] K. Prasai,[73] R. Prasanna,[207] G. Pratten,[14] M. Principe,[82,268,97] G. A. Prodi,[270,92] L. Prokhorov,[14] P. Prosposito,[119,120] L. Prudenzi,[105] A. Puecher,[53,113] M. Punturo,[75] F. Puosi,[18,74] P. Puppo,[51] M. Pürrer,[105] H. Qi,[17] V. Quetschke,[150] R. Quitzow-James,[89] F. J. Raab,[67] G. Raaijmakers,[88,53] H. Radkins,[67] N. Radulesco,[95] P. Raffai,[153] S. X. Rail,[235] S. Raja,[87] C. Rajan,[87] K. E. Ramirez,[6] T. D. Ramirez,[40] A. Ramos-Buades,[105] J. Rana,[148] P. Rapagnani,[98,51] U. D. Rapol,[271] A. Ray,[7] V. Raymond,[17] N. Raza,[180] M. Razzano,[74,18] J. Read,[40] L. A. Rees,[190] T. Regimbau,[30] L. Rei,[85] S. Reid,[32] S. W. Reid,[57] D. H. Reitze,[1,72] P. Relton,[17] A. Renzini,[1] P. Rettegno,[272,24] A. Reza,[53] M. Rezac,[40] F. Ricci,[98,51] D. Richards,[141] J. W. Richardson,[1] L. Richardson,[185] G. Riemenschneider,[272,24] K. Riles,[184] S. Rinaldi,[18,74] K. Rink,[180] M. Rizzo,[15] N. A. Robertson,[1,69] R. Robie,[1] F. Robinet,[41] A. Rocchi,[120] S. Rodriguez,[40] L. Rolland,[30] J. G. Rollins,[1] M. Romanelli,[99] R. Romano,[3,4] C. L. Romel,[67] A. Romero-Rodríguez,[217] I. M. Romero-Shaw,[5] J. H. Romie,[6] S. Ronchini,[31,101] L. Rosa,[4,25] C. A. Rose,[7] D. Rosińska,[103] M. P. Ross,[245] S. Rowan,[69] S. J. Rowlinson,[14] S. Roy,[113] Santosh Roy,[11] Soumen Roy,[273] D. Rozza,[117,118] P. Ruggi,[42] K. Ryan,[67] S. Sachdev,[148] T. Sadecki,[67] J. Sadiq,[107] N. Sago,[274] S. Saito,[21] Y. Saito,[192] K. Sakai,[275] Y. Sakai,[197] M. Sakellariadou,[54] Y. Sakuno,[127] O. S. Salafia,[66,65,64] L. Salconi,[42] M. Saleem,[63] F. Salemi,[91,92] A. Samajdar,[53,113] E. J. Sanchez,[1] J. H. Sanchez,[40] L. E. Sanchez,[1] N. Sanchis-Gual,[276] J. R. Sanders,[277] A. Sanuy,[29] T. R. Saravanan,[11] N. Sarin,[5] B. Sassolas,[158] H. Satari,[86] B. S. Sathyaprakash,[148,17] S. Sato,[278] T. Sato,[176] O. Sauter,[72] R. L. Savage,[67] T. Sawada,[204] D. Sawant,[100] H. L. Sawant,[11] S. Sayah,[158] D. Schaetzl,[1] M. Scheel,[132] J. Scheuer,[15] M. Schiworski,[83] P. Schmidt,[14] S. Schmidt,[113] R. Schnabel,[124] M. Schneewind,[9,10] R. M. S. Schofield,[60] A. Schönbeck,[124] B. W. Schulte,[9,10] B. F. Schutz,[17,9,10] E. Schwartz,[17] J. Scott,[69] S. M. Scott,[8] M. Seglar-Arroyo,[30] T. Sekiguchi,[28] Y. Sekiguchi,[279] D. Sellers,[6] A. S. Sengupta,[273] D. Sentenac,[42] E. G. Seo,[108] V. Sequino,[25,4] A. Sergeev,[220] Y. Setyawati,[113] T. Shaffer,[67] M. S. Shahriar,[15] B. Shams,[172] L. Shao,[201] A. Sharma,[31,101] P. Sharma,[87] P. Shawhan,[104] N. S. Shcheblanov,[240] S. Shibagaki,[127] M. Shikauchi,[114] R. Shimizu,[21] T. Shimoda,[27] K. Shimode,[192] H. Shinkai,[280] T. Shishido,[48] A. Shoda,[20] D. H. Shoemaker,[70] D. M. Shoemaker,[168] S. ShyamSundar,[87] M. Sieniawska,[103] D. Sigg,[67] L. P. Singer,[111] D. Singh,[148] N. Singh,[103] A. Singha,[155,53] A. M. Sintes,[144] V. Sipala,[117,118] V. Skliris,[17] B. J. J. Slagmolen,[8] T. J. Slaven-Blair,[86] J. Smetana,[14] J. R. Smith,[40] R. J. E. Smith,[5] J. Soldateschi,[281,282,50] S. N. Somala,[283] K. Somiya,[218] E. J. Son,[225] K. Soni,[11]

S. Soni,[2] V. Sordini,[136] F. Sorrentino,[85] N. Sorrentino,[74,18] H. Sotani,[284] R. Soulard,[95] T. Souradeep,[271,11] E. Sowell,[147] V. Spagnuolo,[155,53] A. P. Spencer,[69] M. Spera,[77,78] R. Srinivasan,[95] A. K. Srivastava,[80] V. Srivastava,[61] K. Staats,[15] C. Stachie,[95] D. A. Steer,[36] J. Steinhoff,[105] J. Steinlechner,[155,53] S. Steinlechner,[155,53] S. P. Stevenson,[142] D. J. Stops,[14] M. Stover,[173] K. A. Strain,[69] L. C. Strang,[116] G. Stratta,[285,50] A. Strunk,[67] R. Sturani,[267] A. L. Stuver,[122] S. Sudhagar,[11] V. Sudhir,[70] R. Sugimoto,[286,206] H. G. Suh,[7] A. G. Sullivan,[46] T. Z. Summerscales,[287] H. Sun,[86] L. Sun,[8] S. Sunil,[80] A. Sur,[81] J. Suresh,[114,37] P. J. Sutton,[17] Takamasa Suzuki,[176] Toshikazu Suzuki,[37] B. L. Swinkels,[53] M. J. Szczepańczyk,[72] P. Szewczyk,[103] M. Tacca,[53] H. Tagoshi,[37] S. C. Tait,[69] H. Takahashi,[288] R. Takahashi,[20] A. Takamori,[39] S. Takano,[27] H. Takeda,[27] M. Takeda,[204] C. J. Talbot,[32] C. Talbot,[1] H. Tanaka,[289] Kazuyuki Tanaka,[204] Kenta Tanaka,[289] Taiki Tanaka,[37] Takahiro Tanaka,[274] A. J. Tanasijczuk,[52] S. Tanioka,[20,48] D. B. Tanner,[72] D. Tao,[1] L. Tao,[72] E. N. Tapia San Martín,[53,20] C. Taranto,[119] J. D. Tasson,[193] S. Telada,[290] R. Tenorio,[144] J. E. Terhune,[122] L. Terkowski,[124] M. P. Thirugnanasambandam,[11] L. Thomas,[14] M. Thomas,[6] P. Thomas,[67] J. E. Thompson,[17] S. R. Thondapu,[87] K. A. Thorne,[6] E. Thrane,[5] Shubhanshu Tiwari,[161] Srishti Tiwari,[11] V. Tiwari,[17] A. M. Toivonen,[63] K. Toland,[69] A. E. Tolley,[156] T. Tomaru,[20] Y. Tomigami,[204] T. Tomura,[192] M. Tonelli,[74,18] A. Torres-Forné,[123] C. I. Torrie,[1] I. Tosta e Melo,[117,118] D. Töyrä,[8] A. Trapananti,[250,75] F. Travasso,[75,250] G. Traylor,[6] M. Trevor,[104] M. C. Tringali,[42] A. Tripathee,[184] L. Troiano,[291,97] A. Trovato,[36] L. Trozzo,[4,192] R. J. Trudeau,[1] D. S. Tsai,[126] D. Tsai,[126] K. W. Tsang,[53,292,113] T. Tsang,[293] J-S. Tsao,[198] M. Tse,[70] R. Tso,[132] K. Tsubono,[27] S. Tsuchida,[204] L. Tsukada,[114] D. Tsuna,[114] T. Tsutsui,[114] T. Tsuzuki,[21] K. Turbang,[294,209] M. Turconi,[95] D. Tuyenbayev,[204] A. S. Ubhi,[14] N. Uchikata,[37] T. Uchiyama,[192] R. P. Udall,[1] A. Ueda,[187] T. Uehara,[295,296] K. Ueno,[114] G. Ueshima,[297] C. S. Unnikrishnan,[181] F. Uraguchi,[21] A. L. Urban,[2] T. Ushiba,[192] A. Utina,[155,53] H. Vahlbruch,[9,10] G. Vajente,[1] A. Vajpeyi,[5] G. Valdes,[185] M. Valentini,[91,92] V. Valsan,[7] N. van Bakel,[53] M. van Beuzekom,[53] J. F. J. van den Brand,[155,298,53] C. Van Den Broeck,[113,53] D. C. Vander-Hyde,[61] L. van der Schaaf,[53] J. V. van Heijningen,[52] J. Vanosky,[1] M. H. P. M. van Putten,[299] N. van Remortel,[209] M. Vardaro,[243,53] A. F. Vargas,[116] V. Varma,[179] M. Vasúth,[71] A. Vecchio,[14] G. Vedovato,[78] J. Veitch,[69] P. J. Veitch,[83] J. Venneberg,[9,10] G. Venugopalan,[1] D. Verkindt,[30] P. Verma,[232] Y. Verma,[87] D. Veske,[46] F. Vetrano,[49] A. Viceré,[49,50] S. Vidyant,[61] A. D. Viets,[249] A. Vijaykumar,[19] V. Villa-Ortega,[107] J.-Y. Vinet,[95] A. Virtuoso,[188,34] S. Vitale,[70] T. Vo,[61] H. Vocca,[76,75] E. R. G. von Reis,[67] J. S. A. von Wrangel,[9,10] C. Vorvick,[67] S. P. Vyatchanin,[90] L. E. Wade,[173] M. Wade,[173] K. J. Wagner,[125] R. C. Walet,[53] M. Walker,[57] G. S. Wallace,[32] L. Wallace,[1] S. Walsh,[7] J. Wang,[177] J. Z. Wang,[184] W. H. Wang,[150] R. L. Ward,[8] J. Warner,[67] M. Was,[30] T. Washimi,[20] N. Y. Washington,[1] J. Watchi,[145] B. Weaver,[67] S. A. Webster,[69] M. Weinert,[9,10] A. J. Weinstein,[1] R. Weiss,[70] C. M. Weller,[245] F. Wellmann,[9,10] L. Wen,[86] P. Weßels,[9,10] K. Wette,[8] J. T. Whelan,[125] D. D. White,[40] B. F. Whiting,[72] C. Whittle,[70] D. Wilken,[9,10] D. Williams,[69] M. J. Williams,[69] A. R. Williamson,[156] J. L. Willis,[1] B. Willke,[9,10] D. J. Wilson,[140] W. Winkler,[9,10] C. C. Wipf,[1] T. Wlodarczyk,[105] G. Woan,[69] J. Woehler,[9,10] J. K. Wofford,[125] I. C. F. Wong,[108] C. Wu,[133] D. S. Wu,[9,10] H. Wu,[133] S. Wu,[133] D. M. Wysocki,[7] L. Xiao,[1] W-R. Xu,[198] T. Yamada,[289] H. Yamamoto,[1] Kazuhiro Yamamoto,[191] Kohei Yamamoto,[289] T. Yamamoto,[192] K. Yamashita,[203] R. Yamazaki,[200] F. W. Yang,[172] L. Yang,[166] Y. Yang,[300] Yang Yang,[72] Z. Yang,[63] M. J. Yap,[8] D. W. Yeeles,[17] A. B. Yelikar,[125] M. Ying,[126] K. Yokogawa,[203] J. Yokoyama,[28,27] T. Yokozawa,[192] J. Yoo,[179] T. Yoshioka,[203] Hang Yu,[132] Haocun Yu,[70] H. Yuzurihara,[37] A. Zadrożny,[232] M. Zanolin,[35] S. Zeidler,[301] T. Zelenova,[42] J.-P. Zendri,[78] M. Zevin,[162] M. Zhan,[177] H. Zhang,[198] J. Zhang,[86] L. Zhang,[1] T. Zhang,[14] Y. Zhang,[185] C. Zhao,[86] G. Zhao,[145] Y. Zhao,[20] Yue Zhao,[172] Y. Zheng,[89] R. Zhou,[194] Z. Zhou,[15] X. J. Zhu,[5] Z.-H. Zhu,[115] A. B. Zimmerman,[168] Y. Zlochower,[125] M. E. Zucker,[1,70] and J. Zweizig[1]

(LIGO Scientific Collaboration, Virgo Collaboration, and KAGRA Collaboration)

[1]*LIGO Laboratory, California Institute of Technology, Pasadena, California 91125, USA*
[2]*Louisiana State University, Baton Rouge, Louisiana 70803, USA*
[3]*Dipartimento di Farmacia, Università di Salerno, I-84084 Fisciano, Salerno, Italy*
[4]*INFN, Sezione di Napoli, Complesso Universitario di Monte S. Angelo, I-80126 Napoli, Italy*
[5]*OzGrav, School of Physics & Astronomy, Monash University, Clayton 3800, Victoria, Australia*
[6]*LIGO Livingston Observatory, Livingston, Louisiana 70754, USA*
[7]*University of Wisconsin-Milwaukee, Milwaukee, Wisconsin 53201, USA*
[8]*OzGrav, Australian National University, Canberra, Australian Capital Territory 0200, Australia*
[9]*Max Planck Institute for Gravitational Physics (Albert Einstein Institute), D-30167 Hannover, Germany*
[10]*Leibniz Universität Hannover, D-30167 Hannover, Germany*
[11]*Inter-University Centre for Astronomy and Astrophysics, Pune 411007, India*

[12]University of Cambridge, Cambridge CB2 1TN, United Kingdom
[13]Theoretisch-Physikalisches Institut, Friedrich-Schiller-Universität Jena, D-07743 Jena, Germany
[14]University of Birmingham, Birmingham B15 2TT, United Kingdom
[15]Center for Interdisciplinary Exploration & Research in Astrophysics (CIERA), Northwestern University, Evanston, Illinois 60208, USA
[16]Instituto Nacional de Pesquisas Espaciais, 12227-010 São José dos Campos, São Paulo, Brazil
[17]Gravity Exploration Institute, Cardiff University, Cardiff CF24 3AA, United Kingdom
[18]INFN, Sezione di Pisa, I-56127 Pisa, Italy
[19]International Centre for Theoretical Sciences, Tata Institute of Fundamental Research, Bengaluru 560089, India
[20]Gravitational Wave Science Project, National Astronomical Observatory of Japan (NAOJ), Mitaka City, Tokyo 181-8588, Japan
[21]Advanced Technology Center, National Astronomical Observatory of Japan (NAOJ), Mitaka City, Tokyo 181-8588, Japan
[22]Universitat de les Illes Balears, IAC3–IEEC, E-07122 Palma de Mallorca, Spain
[23]University College Dublin, Dublin 4, Ireland
[24]INFN Sezione di Torino, I-10125 Torino, Italy
[25]Università di Napoli “Federico II”, Complesso Universitario di Monte S. Angelo, I-80126 Napoli, Italy
[26]Université de Lyon, Université Claude Bernard Lyon 1, CNRS, Institut Lumière Matière, F-69622 Villeurbanne, France
[27]Department of Physics, The University of Tokyo, Bunkyo-ku, Tokyo 113-0033, Japan
[28]Research Center for the Early Universe (RESCEU), The University of Tokyo, Bunkyo-ku, Tokyo 113-0033, Japan
[29]Institut de Ciències del Cosmos (ICCUB), Universitat de Barcelona, C/ Martí i Franquès 1, Barcelona, 08028, Spain
[30]Laboratoire d’Annecy de Physique des Particules (LAPP), Univ. Grenoble Alpes, Université Savoie Mont Blanc, CNRS/IN2P3, F-74941 Annecy, France
[31]Gran Sasso Science Institute (GSSI), I-67100 L’Aquila, Italy
[32]SUPA, University of Strathclyde, Glasgow G1 1XQ, United Kingdom
[33]Dipartimento di Scienze Matematiche, Informatiche e Fisiche, Università di Udine, I-33100 Udine, Italy
[34]INFN, Sezione di Trieste, I-34127 Trieste, Italy
[35]Embry-Riddle Aeronautical University, Prescott, Arizona 86301, USA
[36]Université de Paris, CNRS, Astroparticule et Cosmologie, F-75006 Paris, France
[37]Institute for Cosmic Ray Research (ICRR), KAGRA Observatory, The University of Tokyo, Kashiwa City, Chiba 277-8582, Japan
[38]Accelerator Laboratory, High Energy Accelerator Research Organization (KEK), Tsukuba City, Ibaraki 305-0801, Japan
[39]Earthquake Research Institute, The University of Tokyo, Bunkyo-ku, Tokyo 113-0032, Japan
[40]California State University Fullerton, Fullerton, California 92831, USA
[41]Université Paris-Saclay, CNRS/IN2P3, IJCLab, 91405 Orsay, France
[42]European Gravitational Observatory (EGO), I-56021 Cascina, Pisa, Italy
[43]School of Physics, Georgia Institute of Technology, Atlanta, Georgia 30332, USA
[44]Chennai Mathematical Institute, Chennai 603103, India
[45]Department of Mathematics and Physics, Gravitational Wave Science Project, Hirosaki University, Hirosaki City, Aomori 036-8561, Japan
[46]Columbia University, New York, New York 10027, USA
[47]Kamioka Branch, National Astronomical Observatory of Japan (NAOJ), Kamioka-cho, Hida City, Gifu 506-1205, Japan
[48]The Graduate University for Advanced Studies (SOKENDAI), Mitaka City, Tokyo 181-8588, Japan
[49]Università degli Studi di Urbino “Carlo Bo”, I-61029 Urbino, Italy
[50]INFN, Sezione di Firenze, I-50019 Sesto Fiorentino, Firenze, Italy
[51]INFN, Sezione di Roma, I-00185 Roma, Italy
[52]Université catholique de Louvain, B-1348 Louvain-la-Neuve, Belgium
[53]Nikhef, Science Park 105, 1098 XG Amsterdam, Netherlands
[54]King’s College London, University of London, London WC2R 2LS, United Kingdom
[55]Korea Institute of Science and Technology Information (KISTI), Yuseong-gu, Daejeon 34141, Republic of Korea
[56]National Institute for Mathematical Sciences, Yuseong-gu, Daejeon 34047, Republic of Korea
[57]Christopher Newport University, Newport News, Virginia 23606, USA
[58]International College, Osaka University, Toyonaka City, Osaka 560-0043, Japan

[59]School of High Energy Accelerator Science, The Graduate University for Advanced Studies (SOKENDAI), Tsukuba City, Ibaraki 305-0801, Japan
[60]University of Oregon, Eugene, Oregon 97403, USA
[61]Syracuse University, Syracuse, New York 13244, USA
[62]Université de Liège, B-4000 Liège, Belgium
[63]University of Minnesota, Minneapolis, Minnesota 55455, USA
[64]Università degli Studi di Milano-Bicocca, I-20126 Milano, Italy
[65]INFN, Sezione di Milano-Bicocca, I-20126 Milano, Italy
[66]INAF, Osservatorio Astronomico di Brera sede di Merate, I-23807 Merate, Lecco, Italy
[67]LIGO Hanford Observatory, Richland, Washington 99352, USA
[68]Dipartimento di Medicina, Chirurgia e Odontoiatria "Scuola Medica Salernitana," Università di Salerno, I-84081 Baronissi, Salerno, Italy
[69]SUPA, University of Glasgow, Glasgow G12 8QQ, United Kingdom
[70]LIGO Laboratory, Massachusetts Institute of Technology, Cambridge, Massachusetts 02139, USA
[71]Wigner RCP, RMKI, H-1121 Budapest, Konkoly Thege Miklós út 29-33, Hungary
[72]University of Florida, Gainesville, Florida 32611, USA
[73]Stanford University, Stanford, California 94305, USA
[74]Università di Pisa, I-56127 Pisa, Italy
[75]INFN, Sezione di Perugia, I-06123 Perugia, Italy
[76]Università di Perugia, I-06123 Perugia, Italy
[77]Università di Padova, Dipartimento di Fisica e Astronomia, I-35131 Padova, Italy
[78]INFN, Sezione di Padova, I-35131 Padova, Italy
[79]Montana State University, Bozeman, Montana 59717, USA
[80]Institute for Plasma Research, Bhat, Gandhinagar 382428, India
[81]Nicolaus Copernicus Astronomical Center, Polish Academy of Sciences, 00-716, Warsaw, Poland
[82]Dipartimento di Ingegneria, Università del Sannio, I-82100 Benevento, Italy
[83]OzGrav, University of Adelaide, Adelaide, South Australia 5005, Australia
[84]California State University, Los Angeles, 5151 State University Drive, Los Angeles, California 90032, USA
[85]INFN, Sezione di Genova, I-16146 Genova, Italy
[86]OzGrav, University of Western Australia, Crawley, Western Australia 6009, Australia
[87]RRCAT, Indore, Madhya Pradesh 452013, India
[88]GRAPPA, Anton Pannekoek Institute for Astronomy and Institute for High-Energy Physics, University of Amsterdam, Science Park 904, 1098 XH Amsterdam, Netherlands
[89]Missouri University of Science and Technology, Rolla, Missouri 65409, USA
[90]Faculty of Physics, Lomonosov Moscow State University, Moscow 119991, Russia
[91]Università di Trento, Dipartimento di Fisica, I-38123 Povo, Trento, Italy
[92]INFN, Trento Institute for Fundamental Physics and Applications, I-38123 Povo, Trento, Italy
[93]SUPA, University of the West of Scotland, Paisley PA1 2BE, United Kingdom
[94]Bar-Ilan University, Ramat Gan, 5290002, Israel
[95]Artemis, Université Côte d'Azur, Observatoire de la Côte d'Azur, CNRS, F-06304 Nice, France
[96]Dipartimento di Fisica "E.R. Caianiello", Università di Salerno, I-84084 Fisciano, Salerno, Italy
[97]INFN, Sezione di Napoli, Gruppo Collegato di Salerno, Complesso Universitario di Monte S. Angelo, I-80126 Napoli, Italy
[98]Università di Roma "La Sapienza", I-00185 Roma, Italy
[99]Univ Rennes, CNRS, Institut FOTON—UMR6082, F-3500 Rennes, France
[100]Indian Institute of Technology Bombay, Powai, Mumbai 400 076, India
[101]INFN, Laboratori Nazionali del Gran Sasso, I-67100 Assergi, Italy
[102]Laboratoire Kastler Brossel, Sorbonne Université, CNRS, ENS-Université PSL, Collège de France, F-75005 Paris, France
[103]Astronomical Observatory Warsaw University, 00-478 Warsaw, Poland
[104]University of Maryland, College Park, Maryland 20742, USA
[105]Max Planck Institute for Gravitational Physics (Albert Einstein Institute), D-14476 Potsdam, Germany
[106]L2IT, Laboratoire des 2 Infinis—Toulouse, Université de Toulouse, CNRS/IN2P3, UPS, F-31062 Toulouse Cedex 9, France
[107]IGFAE, Campus Sur, Universidade de Santiago de Compostela, 15782 Spain
[108]The Chinese University of Hong Kong, Shatin, NT, Hong Kong
[109]Stony Brook University, Stony Brook, New York 11794, USA
[110]Center for Computational Astrophysics, Flatiron Institute, New York, New York 10010, USA
[111]NASA Goddard Space Flight Center, Greenbelt, Maryland 20771, USA

[112]Dipartimento di Fisica, Università degli Studi di Genova, I-16146 Genova, Italy
[113]Institute for Gravitational and Subatomic Physics (GRASP), Utrecht University, Princetonplein 1, 3584 CC Utrecht, Netherlands
[114]RESCEU, University of Tokyo, Tokyo, 113-0033, Japan
[115]Department of Astronomy, Beijing Normal University, Beijing 100875, China
[116]OzGrav, University of Melbourne, Parkville, Victoria 3010, Australia
[117]Università degli Studi di Sassari, I-07100 Sassari, Italy
[118]INFN, Laboratori Nazionali del Sud, I-95125 Catania, Italy
[119]Università di Roma Tor Vergata, I-00133 Roma, Italy
[120]INFN, Sezione di Roma Tor Vergata, I-00133 Roma, Italy
[121]University of Sannio at Benevento, I-82100 Benevento, Italy and INFN, Sezione di Napoli, I-80100 Napoli, Italy
[122]Villanova University, 800 Lancaster Ave, Villanova, Pennsylvania 19085, USA
[123]Departamento de Astronomía y Astrofísica, Universitat de València, E-46100 Burjassot, València, Spain
[124]Universität Hamburg, D-22761 Hamburg, Germany
[125]Rochester Institute of Technology, Rochester, New York 14623, USA
[126]National Tsing Hua University, Hsinchu City, 30013 Taiwan, Republic of China
[127]Department of Applied Physics, Fukuoka University, Jonan, Fukuoka City, Fukuoka 814-0180, Japan
[128]OzGrav, Charles Sturt University, Wagga Wagga, New South Wales 2678, Australia
[129]Department of Physics, Tamkang University, Danshui Dist., New Taipei City 25137, Taiwan
[130]Department of Physics and Institute of Astronomy, National Tsing Hua University, Hsinchu 30013, Taiwan
[131]Department of Physics, Center for High Energy and High Field Physics, National Central University, Zhongli District, Taoyuan City 32001, Taiwan
[132]CaRT, California Institute of Technology, Pasadena, California 91125, USA
[133]Department of Physics, National Tsing Hua University, Hsinchu 30013, Taiwan
[134]Dipartimento di Ingegneria Industriale (DIIN), Università di Salerno, I-84084 Fisciano, Salerno, Italy
[135]Institute of Physics, Academia Sinica, Nankang, Taipei 11529, Taiwan
[136]Université Lyon, Université Claude Bernard Lyon 1, CNRS, IP2I Lyon/IN2P3, UMR 5822, F-69622 Villeurbanne, France
[137]Seoul National University, Seoul 08826, Republic of Korea
[138]Pusan National University, Busan 46241, Republic of Korea
[139]INAF, Osservatorio Astronomico di Padova, I-35122 Padova, Italy
[140]University of Arizona, Tucson, Arizona 85721, USA
[141]Rutherford Appleton Laboratory, Didcot OX11 0DE, United Kingdom
[142]OzGrav, Swinburne University of Technology, Hawthorn VIC 3122, Australia
[143]Université libre de Bruxelles, Avenue Franklin Roosevelt 50–1050 Bruxelles, Belgium
[144]Universitat de les Illes Balears, IAC3—IEEC, E-07122 Palma de Mallorca, Spain
[145]Université Libre de Bruxelles, Brussels 1050, Belgium
[146]Departamento de Matemáticas, Universitat de València, E-46100 Burjassot, València, Spain
[147]Texas Tech University, Lubbock, Texas 79409, USA
[148]The Pennsylvania State University, University Park, Pennsylvania 16802, USA
[149]University of Rhode Island, Kingston, Rhode Island 02881, USA
[150]The University of Texas Rio Grande Valley, Brownsville, Texas 78520, USA
[151]Bellevue College, Bellevue, Washington 98007, USA
[152]Scuola Normale Superiore, Piazza dei Cavalieri, 7–56126 Pisa, Italy
[153]MTA-ELTE Astrophysics Research Group, Institute of Physics, Eötvös University, Budapest 1117, Hungary
[154]University of Szeged, Dóm tér 9, Szeged 6720, Hungary
[155]Maastricht University, P.O. Box 616, 6200 MD Maastricht, Netherlands
[156]University of Portsmouth, Portsmouth, PO1 3FX, United Kingdom
[157]The University of Sheffield, Sheffield S10 2TN, United Kingdom
[158]Université Lyon, Université Claude Bernard Lyon 1, CNRS, Laboratoire des Matériaux Avancés (LMA), IP2I Lyon/IN2P3, UMR 5822, F-69622 Villeurbanne, France
[159]Dipartimento di Scienze Matematiche, Fisiche e Informatiche, Università di Parma, I-43124 Parma, Italy
[160]INFN, Sezione di Milano Bicocca, Gruppo Collegato di Parma, I-43124 Parma, Italy
[161]Physik-Institut, University of Zurich, Winterthurerstrasse 190, 8057 Zurich, Switzerland
[162]University of Chicago, Chicago, Illinois 60637, USA

[163]Université de Strasbourg, CNRS, IPHC UMR 7178, F-67000 Strasbourg, France
[164]West Virginia University, Morgantown, West Virginia 26506, USA
[165]Montclair State University, Montclair, New Jersey 07043, USA
[166]Colorado State University, Fort Collins, Colorado 80523, USA
[167]Institute for Nuclear Research, Hungarian Academy of Sciences, Bem t'er 18/c, H-4026 Debrecen, Hungary
[168]Department of Physics, University of Texas, Austin, Texas 78712, USA
[169]CNR-SPIN, c/o Università di Salerno, I-84084 Fisciano, Salerno, Italy
[170]Scuola di Ingegneria, Università della Basilicata, I-85100 Potenza, Italy
[171]Observatori Astronòmic, Universitat de València, E-46980 Paterna, València, Spain
[172]The University of Utah, Salt Lake City, Utah 84112, USA
[173]Kenyon College, Gambier, Ohio 43022, USA
[174]Vrije Universiteit Amsterdam, 1081 HV, Amsterdam, Netherlands
[175]Department of Astronomy, The University of Tokyo, Mitaka City, Tokyo 181-8588, Japan
[176]Faculty of Engineering, Niigata University, Nishi-ku, Niigata City, Niigata 950-2181, Japan
[177]State Key Laboratory of Magnetic Resonance and Atomic and Molecular Physics, Innovation Academy for Precision Measurement Science and Technology (APM), Chinese Academy of Sciences, Xiao Hong Shan, Wuhan 430071, China
[178]Universiteit Gent, B-9000 Gent, Belgium
[179]Cornell University, Ithaca, New York 14850, USA
[180]University of British Columbia, Vancouver, BC V6T 1Z4, Canada
[181]Tata Institute of Fundamental Research, Mumbai 400005, India
[182]INAF, Osservatorio Astronomico di Capodimonte, I-80131 Napoli, Italy
[183]The University of Mississippi, University, Mississippi 38677, USA
[184]University of Michigan, Ann Arbor, Michigan 48109, USA
[185]Texas A&M University, College Station, Texas 77843, USA
[186]Department of Physics, Ulsan National Institute of Science and Technology (UNIST), Ulju-gun, Ulsan 44919, Republic of Korea
[187]Applied Research Laboratory, High Energy Accelerator Research Organization (KEK), Tsukuba City, Ibaraki 305-0801, Japan
[188]Dipartimento di Fisica, Università di Trieste, I-34127 Trieste, Italy
[189]Shanghai Astronomical Observatory, Chinese Academy of Sciences, Shanghai 200030, China
[190]American University, Washington, D.C. 20016, USA
[191]Faculty of Science, University of Toyama, Toyama City, Toyama 930-8555, Japan
[192]Institute for Cosmic Ray Research (ICRR), KAGRA Observatory, The University of Tokyo, Kamioka-cho, Hida City, Gifu 506-1205, Japan
[193]Carleton College, Northfield, Minnesota 55057, USA
[194]University of California, Berkeley, California 94720, USA
[195]Maastricht University, 6200 MD, Maastricht, Netherlands
[196]College of Industrial Technology, Nihon University, Narashino City, Chiba 275-8575, Japan
[197]Graduate School of Science and Technology, Niigata University, Nishi-ku, Niigata City, Niigata 950-2181, Japan
[198]Department of Physics, National Taiwan Normal University, sec. IV, Taipei 116, Taiwan
[199]Astronomy & Space Science, Chungnam National University, Yuseong-gu, Daejeon 34134, Republic of Korea, Republic of Korea
[200]Department of Physics and Mathematics, Aoyama Gakuin University, Sagamihara City, Kanagawa 252-5258, Japan
[201]Kavli Institute for Astronomy and Astrophysics, Peking University, Haidian District, Beijing 100871, China
[202]Yukawa Institute for Theoretical Physics (YITP), Kyoto University, Sakyou-ku, Kyoto City, Kyoto 606-8502, Japan
[203]Graduate School of Science and Engineering, University of Toyama, Toyama City, Toyama 930-8555, Japan
[204]Department of Physics, Graduate School of Science, Osaka City University, Sumiyoshi-ku, Osaka City, Osaka 558-8585, Japan
[205]Nambu Yoichiro Institute of Theoretical and Experimental Physics (NITEP), Osaka City University, Sumiyoshi-ku, Osaka City, Osaka 558-8585, Japan
[206]Institute of Space and Astronautical Science (JAXA), Chuo-ku, Sagamihara City, Kanagawa 252-0222, Japan
[207]Directorate of Construction, Services & Estate Management, Mumbai 400094, India

[208]*Vanderbilt University, Nashville, Tennessee 37235, USA*
[209]*Universiteit Antwerpen, Prinsstraat 13, 2000 Antwerpen, Belgium*
[210]*University of Białystok, 15-424 Białystok, Poland*
[211]*Department of Physics, Ewha Womans University, Seodaemun-gu, Seoul 03760, Republic of Korea*
[212]*National Astronomical Observatories, Chinese Academic of Sciences, Chaoyang District, Beijing, China*
[213]*School of Astronomy and Space Science, University of Chinese Academy of Sciences, Chaoyang District, Beijing, China*
[214]*University of Southampton, Southampton SO17 1BJ, United Kingdom*
[215]*Institute for Cosmic Ray Research (ICRR), The University of Tokyo, Kashiwa City, Chiba 277-8582, Japan*
[216]*Chung-Ang University, Seoul 06974, Republic of Korea*
[217]*Institut de Física d'Altes Energies (IFAE), Barcelona Institute of Science and Technology, and ICREA, E-08193 Barcelona, Spain*
[218]*Graduate School of Science, Tokyo Institute of Technology, Meguro-ku, Tokyo 152-8551, Japan*
[219]*University of Washington Bothell, Bothell, Washington 98011, USA*
[220]*Institute of Applied Physics, Nizhny Novgorod, 603950, Russia*
[221]*Ewha Womans University, Seoul 03760, Republic of Korea*
[222]*Inje University Gimhae, South Gyeongsang 50834, Republic of Korea*
[223]*Department of Physics, Myongji University, Yongin 17058, Republic of Korea*
[224]*Korea Astronomy and Space Science Institute, Daejeon 34055, Republic of Korea*
[225]*National Institute for Mathematical Sciences, Daejeon 34047, Republic of Korea*
[226]*Ulsan National Institute of Science and Technology, Ulsan 44919, Republic of Korea*
[227]*Department of Physical Science, Hiroshima University, Higashihiroshima City, Hiroshima 903-0213, Japan*
[228]*School of Physics and Astronomy, Cardiff University, Cardiff, CF24 3AA, United Kingdom*
[229]*Institute of Astronomy, National Tsing Hua University, Hsinchu 30013, Taiwan*
[230]*Bard College, 30 Campus Rd, Annandale-On-Hudson, New York 12504, USA*
[231]*Institute of Mathematics, Polish Academy of Sciences, 00656 Warsaw, Poland*
[232]*National Center for Nuclear Research, 05-400 Świerk-Otwock, Poland*
[233]*Instituto de Fisica Teorica, 28049 Madrid, Spain*
[234]*Department of Physics, Nagoya University, Chikusa-ku, Nagoya, Aichi 464-8602, Japan*
[235]*Université de Montréal/Polytechnique, Montreal, Quebec H3T 1J4, Canada*
[236]*Laboratoire Lagrange, Université Côte d'Azur, Observatoire Côte d'Azur, CNRS, F-06304 Nice, France*
[237]*Canadian Institute for Theoretical Astrophysics, University of Toronto, Toronto, Ontario M5S 3H8, Canada*
[238]*Department of Physics, Hanyang University, Seoul 04763, Republic of Korea*
[239]*Sungkyunkwan University, Seoul 03063, Republic of Korea*
[240]*NAVIER, École des Ponts, Univ Gustave Eiffel, CNRS, Marne-la-Vallée, France*
[241]*Department of Physics, National Cheng Kung University, Tainan City 701, Taiwan*
[242]*National Center for High-performance computing, National Applied Research Laboratories, Hsinchu Science Park, Hsinchu City 30076, Taiwan*
[243]*Institute for High-Energy Physics, University of Amsterdam, Science Park 904, 1098 XH Amsterdam, Netherlands*
[244]*NASA Marshall Space Flight Center, Huntsville, Alabama 35811, USA*
[245]*University of Washington, Seattle, Washington 98195, USA*
[246]*Dipartimento di Matematica e Fisica, Università degli Studi Roma Tre, I-00146 Roma, Italy*
[247]*INFN, Sezione di Roma Tre, I-00146 Roma, Italy*
[248]*ESPCI, CNRS, F-75005 Paris, France*
[249]*Concordia University Wisconsin, Mequon, Wisconsin 53097, USA*
[250]*Università di Camerino, Dipartimento di Fisica, I-62032 Camerino, Italy*
[251]*School of Physics Science and Engineering, Tongji University, Shanghai 200092, China*
[252]*Southern University and A&M College, Baton Rouge, Los Angeles 70813, USA*
[253]*Centre Scientifique de Monaco, 8 quai Antoine Ier, MC-98000, Monaco*
[254]*Institute for Photon Science and Technology, The University of Tokyo, Bunkyo-ku, Tokyo 113-8656, Japan*
[255]*Indian Institute of Technology Madras, Chennai 600036, India*
[256]*Instituto de Fisica Teorica, Universidad Autónoma de Madrid, 28049 Madrid, Spain*
[257]*Saha Institute of Nuclear Physics, Bidhannagar, West Bengal 700064, India*

[258]*The Applied Electromagnetic Research Institute, National Institute of Information and Communications Technology (NICT), Koganei City, Tokyo 184-8795, Japan*
[259]*Institut des Hautes Etudes Scientifiques, F-91440 Bures-sur-Yvette, France*
[260]*Faculty of Law, Ryukoku University, Fushimi-ku, Kyoto City, Kyoto 612-8577, Japan*
[261]*Indian Institute of Science Education and Research, Kolkata, Mohanpur, West Bengal 741252, India*
[262]*Department of Astrophysics/IMAPP, Radboud University Nijmegen, P.O. Box 9010, 6500 GL Nijmegen, Netherlands*
[263]*Department of Physics, University of Notre Dame, Notre Dame, Indiana 46556, USA*
[264]*Consiglio Nazionale delle Ricerche—Istituto dei Sistemi Complessi, Piazzale Aldo Moro 5, I-00185 Roma, Italy*
[265]*Korea Astronomy and Space Science Institute (KASI), Yuseong-gu, Daejeon 34055, Republic of Korea*
[266]*Hobart and William Smith Colleges, Geneva, New York 14456, USA*
[267]*International Institute of Physics, Universidade Federal do Rio Grande do Norte, Natal RN 59078-970, Brazil*
[268]*Museo Storico della Fisica e Centro Studi e Ricerche "Enrico Fermi", I-00184 Roma, Italy*
[269]*Lancaster University, Lancaster LA1 4YW, United Kingdom*
[270]*Università di Trento, Dipartimento di Matematica, I-38123 Povo, Trento, Italy*
[271]*Indian Institute of Science Education and Research, Pune, Maharashtra 411008, India*
[272]*Dipartimento di Fisica, Università degli Studi di Torino, I-10125 Torino, Italy*
[273]*Indian Institute of Technology, Palaj, Gandhinagar, Gujarat 382355, India*
[274]*Department of Physics, Kyoto University, Sakyou-ku, Kyoto City, Kyoto 606-8502, Japan*
[275]*Department of Electronic Control Engineering, National Institute of Technology, Nagaoka College, Nagaoka City, Niigata 940-8532, Japan*
[276]*Departamento de Matemática da Universidade de Aveiro and Centre for Research and Development in Mathematics and Applications, Campus de Santiago, 3810-183 Aveiro, Portugal*
[277]*Marquette University, 11420 W. Clybourn St., Milwaukee, Wisconsin 53233, USA*
[278]*Graduate School of Science and Engineering, Hosei University, Koganei City, Tokyo 184-8584, Japan*
[279]*Faculty of Science, Toho University, Funabashi City, Chiba 274-8510, Japan*
[280]*Faculty of Information Science and Technology, Osaka Institute of Technology, Hirakata City, Osaka 573-0196, Japan*
[281]*Università di Firenze, Sesto Fiorentino I-50019, Italy*
[282]*INAF, Osservatorio Astrofisico di Arcetri, Largo E. Fermi 5, I-50125 Firenze, Italy*
[283]*Indian Institute of Technology Hyderabad, Sangareddy, Khandi, Telangana 502285, India*
[284]*iTHEMS (Interdisciplinary Theoretical and Mathematical Sciences Program), The Institute of Physical and Chemical Research (RIKEN), Wako, Saitama 351-0198, Japan*
[285]*INAF, Osservatorio di Astrofisica e Scienza dello Spazio, I-40129 Bologna, Italy*
[286]*Department of Space and Astronautical Science, The Graduate University for Advanced Studies (SOKENDAI), Sagamihara City, Kanagawa 252-5210, Japan*
[287]*Andrews University, Berrien Springs, Michigan 49104, USA*
[288]*Research Center for Space Science, Advanced Research Laboratories, Tokyo City University, Setagaya, Tokyo 158-0082, Japan*
[289]*Institute for Cosmic Ray Research (ICRR), Research Center for Cosmic Neutrinos (RCCN), The University of Tokyo, Kashiwa City, Chiba 277-8582, Japan*
[290]*National Metrology Institute of Japan, National Institute of Advanced Industrial Science and Technology, Tsukuba City, Ibaraki 305-8568, Japan*
[291]*Dipartimento di Scienze Aziendali—Management and Innovation Systems (DISA-MIS), Università di Salerno, I-84084 Fisciano, Salerno, Italy*
[292]*Van Swinderen Institute for Particle Physics and Gravity, University of Groningen, Nijenborgh 4, 9747 AG Groningen, Netherlands*
[293]*Faculty of Science, Department of Physics, The Chinese University of Hong Kong, Shatin, N.T., Hong Kong*
[294]*Vrije Universiteit Brussel, Boulevard de la Plaine 2, 1050 Ixelles, Belgium*
[295]*Department of Communications Engineering, National Defense Academy of Japan, Yokosuka City, Kanagawa 239-8686, Japan*
[296]*Department of Physics, University of Florida, Gainesville, Florida 32611, USA*
[297]*Department of Information and Management Systems Engineering, Nagaoka University of Technology, Nagaoka City, Niigata 940-2188, Japan*

[298]*Vrije Universiteit Amsterdam, 1081 HV Amsterdam, Netherlands*
[299]*Department of Physics and Astronomy, Sejong University, Gwangjin-gu, Seoul 143-747, Republic of Korea*
[300]*Department of Electrophysics, National Chiao Tung University, Hsinchu, Taiwan*
[301]*Department of Physics, Rikkyo University, Toshima-ku, Tokyo 171-8501, Japan*